\documentclass[a4paper,fleqn]{cas-sc}

\usepackage[numbers,sort&compress]{natbib}

\usepackage{epstopdf}
\newcommand{\be}{\begin{equation}}
\newcommand{\ee}{\end{equation}}
\newcommand{\bea}{\begin{eqnarray}}
\newcommand{\eea}{\end{eqnarray}}

\newcommand{\eqn}[1]{(\ref{#1})}
\newcommand{\bel}[1]{\be\label{#1}}
\newcommand{\ba}{\begin{array}{c}}
\newcommand{\bat}{\begin{array}{cc}}
\newcommand{\ea}{\end{array}}
\newcommand{\dis}{\displaystyle}
\newcommand{\bi}{\begin{itemize}}
\newcommand{\ei}{\end{itemize}}

\newcommand{\cK}{{\cal K}}
\newcommand{\cL}{{\cal L}}
\newcommand{\cJ}{{\cal J}}

\newcommand{\cO}{{\cal O}}

\newcommand{\cC}{{\cal C}}
\newcommand{\cG}{{\cal G}}

\newcommand{\ie}{{\it i.e.},\ }
\def\no{\nonumber}

\def\slashchar#1{\setbox0=\hbox{$#1$}\dimen0=\wd0%
\setbox1=\hbox{/}\dimen1=\wd1%
\ifdim\dimen0>\dimen1%
\rlap{\hbox to \dimen0{\hfil/\hfil}}#1\else \rlap{\hbox to
\dimen1{\hfil$#1$\hfil}}/\fi}
\begin{document}
\let\WriteBookmarks\relax
\def\floatpagepagefraction{1}
\def\textpagefraction{.001}
\shorttitle{Precision QCD physics}
\shortauthors{A Pich}

\title [mode = title]{Precision physics with inclusive QCD processes}                      

\author{Antonio Pich}[orcid=0000-0002-8019-5463]
\ead{antonio.pich@ific.uv.es}

\address{Departament de F\'\i sica Te\`orica, IFIC, Universitat de Val\`encia -- CSIC, Parc Científic,  Catedrático José Beltrán 2, E-46980 Paterna, Spain}

\begin{abstract}
The inclusive production of hadrons through electroweak currents can be rigorously analysed with short-distance theoretical tools. The associated observables are insensitive to the involved infrared behaviour of the strong interaction, allowing for very precise tests of Quantum Chromodynamics. The theoretical predictions for $\sigma(e^+e^-\to\mathrm{hadrons})$ and the hadronic decay widths of the $\tau$ lepton and the $Z$, $W$ and Higgs bosons have reached an impressive accuracy of $\cO(\alpha_s^4)$. Precise experimental measurements of the $Z$ and $\tau$ hadronic widths have made possible the accurate determination of the strong coupling at two very different energy scales, providing a highly significant experimental verification of asymptotic freedom. A detailed discussion of the theoretical description of these processes and their current phenomenological status is presented. The most precise determinations of $\alpha_s$ from other sources are also briefly reviewed and compared with the fully-inclusive results.  
\end{abstract}


\maketitle

\tableofcontents

\section{Introduction}

Quantum Chromodynamics (QCD) \cite{Fritzsch:1972jv,Fritzsch:1973pi}
provides a successful description of the strong interaction in terms of a single parameter: the strong coupling constant 
$\alpha_s$. This beautiful gauge theory has been precisely tested in may
different processes and over a very broad range of mass scales. 
Although many aspects of the hadronic world need still to be better understood, the overwhelming consistency of all experimental results has established beyond any doubt that QCD is the right dynamical theory of the strong force. 

At low energies, the growing of the effective running coupling generates a complicated non-perturbative regime, responsible for the hadronization of quarks and gluons into a rich variety of colour-singlet composite particles. A precise quantitative description of the hadron formation and dynamics remains unfortunately as an important open problem, which so far has been only partially approached through effective field theory descriptions and numerical tools. In fact, a complete analytical proof of confinement has not yet been accomplished, in spite of the many efforts performed along the years. 
Nevertheless, all theoretical studies and the large amount of data accumulated
indicate that confinement is a truly fundamental property of QCD. Colourful objects have never been observed as asymptotic states.

Assuming that confinement is exact, one can perform very precise predictions
for the inclusive production of hadrons in processes that do not contain strongly-interacting particles in the initial state, such as $e^+e^-\to\mathrm{hadrons}$, $Z\to\mathrm{hadrons}$, $W^\pm\to\mathrm{hadrons}$,
$\tau^-\to\nu_\tau + \mathrm{hadrons}$ or $H\to\mathrm{hadrons}$. Since
the separate identity of the produced hadrons is not specified, one just needs to compute the total production of quarks and gluons, summing over all possible configurations. Confinement guarantees that the computed QCD cross section or decay width will be identical to the corresponding inclusive hadronic production because the total probability that quarks and gluons hadronize is just one.

Pure perturbative calculations are usually enough to achieve accurate descriptions of high-energy inclusive processes. At low energies they need to be complemented with non-perturbative corrections that scale as powers of $\Lambda_{\mathrm{QCD}}^{2n}/s^n$, starting with $n=2$. Using short-distance operator-product-expansion (OPE) \cite{Wilson:1969zs} techniques, one can control rigorously these power
corrections and determine above which scales their numerical impact becomes negligible.

The following sections present a detailed discussion of the theoretical tools involved in the analysis of inclusive processes and the current status of the resulting predictions.
The running QCD coupling and quark masses are introduced in Section~\ref{sec:running}, which describes their associated $\beta$ and $\gamma$ functions that are currently known to $\cO(\alpha_s^5)$. Section~\ref{sec:correlators} discusses the two-point correlation functions of the vector, axial-vector, scalar and pseudoscalar QCD currents, summarizing our present $\cO(\alpha_s^4)$ knowledge of these important dynamical objects. The inclusive high-energy observables are analysed in Section~\ref{sec:InclusiveObservables}, which contains the QCD predictions for the $e^+e^-$ annihilation cross section into hadrons, and the hadronic widths of the electroweak $Z$, $W$ and Higgs bosons. Section~\ref{sec:Rtau} reviews the theoretical analysis of the $\tau$ hadronic width, including perturbative and non-perturbative contributions, and updates its current phenomenological status. The highly-precise four-loop determinations of the strong coupling from the $Z$ and $\tau$ hadronic widths are compared in Section~\ref{sec:alphas} with the most accurate values of $\alpha_s$ extracted from other sources, exhibiting the great success of QCD in correctly describing strong-interacting phenomena over a very broad range of energy scales. A few summarizing comments are finally given in Section~\ref{sec:summary}. Some complementary technical details are compiled in appendices.

\section{The QCD running coupling}
\label{sec:running}

The unique coupling constant of QCD, $\alpha_s\equiv g_s^2/(4\pi)$, is obviously the critical parameter governing all phenomena associated with the strong interaction. The renormalized coupling $\alpha_s(\mu^2)$ depends on the chosen renormalization scheme and scale. The $\overline{\mathrm{MS}}$ scheme \cite{Bardeen:1978yd} is the conventionally adopted choice, while the dependence on the scale $\mu$ is determined by the renormalization-group equation
\begin{equation}\label{eq:beta-rge}
\mu\,\frac{d\alpha_s}{d\mu}\; =\; \alpha_s\;\beta(\alpha_s)\, ,
\qquad\qquad\qquad\qquad
\beta(\alpha_s)\; =\;\sum_{n=1}\, \beta_n\, 
\left(\frac{\alpha_s}{\pi}\right)^n\, ,
\end{equation}
which defines the so-called $\beta$ function.\footnote{
Notice that several different conventions are used in the literature for the normalization of the expansion coefficients $\beta_n$ (global sign, factors of $\pi$, $n\ge 0$, etc.). The same comment applies to the $\gamma_n$ coefficients in Eq.~\eqn{eq:runningMass}.
}
The perturbative expansion of $\beta(\alpha_s)$ in powers of $a_s\equiv \alpha_s/\pi$ is already known to an impressive accuracy of five loops. The first two coefficients are independent of the chosen (mass-independent) renormalization scheme \cite{Espriu:1981eh}:
\bel{eq:beta1-2}
\beta_1\, =\,\frac{1}{3}\, n_f - \frac{11}{2}\, ,
\qquad\qquad\qquad
\beta_2\, =\, -\frac{51}{4} + \frac{19}{12} \, n_f \, ,
\ee
where $n_f$ is the number of quark flavours.
The negative value of $\beta_1$ (for $n_f\le 16$) demonstrates that QCD is an asymptotically-free quantum field theory \cite{Gross:1973id,Politzer:1973fx}, \ie that the coupling decreases for increasing values of the renormalization scale $\mu$. This behaviour is reinforced by the two-loop contribution to the $\beta$ function \cite{Caswell:1974gg,Jones:1974mm}, which satisfies $\beta_2<0$, provided $n_f\le 8$. In the $\overline{\mathrm{MS}}$ scheme, the three-loop coefficient \cite{Tarasov:1980au},
remains also negative for $n_f\le 5$, while $\beta_4$ \cite{vanRitbergen:1997va,Czakon:2004bu} and $\beta_5$~\cite{Baikov:2016tgj,Luthe:2016ima,Herzog:2017ohr,Luthe:2017ttc,Luthe:2017ttg,Chetyrkin:2017bjc} are always negative numbers (except at very large values of $n_f$ for $\beta_5$), independently of the number of flavours considered:
\begin{eqnarray}
\label{eq:beta3}
\beta_3 &\!\!\! =&\!\!\! {1\over 64}\left[ -2857 + {5033\over 9} \, n_f
- {325\over 27} \, n_f^2 \right]\,  ,
\\[5pt]
\label{eq:beta4}
\beta_4 &\!\!\! =&\!\!\!
\frac{-1}{128}\, \left[
\frac{149753}{6}+3564 \,\zeta_3 
-\left( \frac{1078361}{162}+\frac{6508}{27}\,\zeta_3 \right)\, n_f
+ \left( \frac{50065}{162}+\frac{6472}{81}\,\zeta_3 \right)\, n_f^2
+\frac{1093}{729}\, n_f^3 \right]\, ,
\\[5pt]
\beta_5 &\!\!\! = &\!\!\!
-\frac{1}{512}\,\Biggl\{
\frac{8157455}{16} +\frac{621885}{2} \,\zeta_{3} -\frac{88209}{2} \,\zeta_{4}
-288090 \,\zeta_{5}
\nonumber\\
&& \hskip .8cm \mbox{} +\, n_f\:\left[
-\frac{336460813}{1944} 
-\frac{4811164}{81}  \, \zeta_{3}
+\frac{33935}{6}  \, \zeta_{4}
+\frac{1358995}{27}  \, \zeta_{5}
\right]
\nonumber\\
&& \hskip .8cm \mbox{} +\, n_f^2\:\left[
\frac{25960913}{1944} 
+\frac{698531}{81}  \, \zeta_{3}
-\frac{10526}{9}  \, \zeta_{4}
-\frac{381760}{81}  \, \zeta_{5}
\right]
\nonumber\\
&& \hskip .8cm \mbox{} +\, n_f^3\:\left[
-\frac{630559}{5832} 
-\frac{48722}{243}  \, \zeta_{3}
+\frac{1618}{27}  \, \zeta_{4}
+\frac{460}{9}  \, \zeta_{5} \right]
\, +\, n_f^4\:\left[ \frac{1205}{2916} -\frac{152}{81}  \, \zeta_{3} \right]\,
\Biggr\}\, .
\end{eqnarray}
The numerical constants $\zeta_3 = 1.202056903\ldots$, $\zeta_4 = \pi^4/90$ 
and $\zeta_5 = 1.036927755\ldots $ are the values of
the Riemann zeta function \
$\zeta_p\equiv\zeta(p) = \sum_{n=1}^\infty \frac{1}{n^p}$ \ at $p=3$, 4 and 5, respectively.
The very modest growth of the computed coefficients $\beta_n$ with the perturbative order gives rise to a surprisingly smooth power expansion. 
Taking $n_f=5$ as a representative value, one gets
\bel{eq:betaGrowing}
\beta(\alpha_s)\, =\,  \beta_1 a_s\,\left( 1 + 1.2609\, a_s + 1.4748\, a_s^2 + 9.8359\, a_s^3 + 7.8825\, a_s^4\right)\, .
\ee

The current five-loop knowledge of the $\beta$ function provides a very precise perturbative control of the scale dependence of $\alpha_s$.
Integrating the renormalization-group equation \eqn{eq:beta-rge}, 
\bel{eq:IntRGE}
\log{\left(\frac{\mu}{\mu_0}\right)}\, =\, \int_{\mu_0}^\mu \frac{d\mu}{\mu}\, =\, \int_{\alpha_s(\mu_0^2)}^{\alpha_s(\mu^2)} \frac{d\alpha_s}{\alpha_s\, \beta(\alpha_s)}\, ,
\ee
one obtains an implicit relation between the QCD coupling at an arbitrary renormalization scale $\mu$ and its value at some other reference scale $\mu_0$.
Expanding the integrand perturbatively,
$a_s(\mu^2)\equiv \alpha_s(\mu^2)/\pi$ can be easily expressed as an expansion in powers of $a_0\equiv a_s(\mu_0^2)$, which is governed by powers of the coefficients $\beta_n$ and $L\equiv\log{(\mu/\mu_0)}$:
\bea\label{eq:alphaRunPert}
a_s(\mu^2) &\!\!\! = &\!\!\! 
a_0 \left\{ 1 + \beta_1 L a_0
+ \left( \beta_1^2 L^2 +\beta_2 L\right) a_0^2
+ \left( \beta_1^3 L^3 + \frac{5}{2}\, \beta_1 \beta_2 L^2 + \beta_3 L\right) a_0^3\right.
\no\\ &\!\!\! + &\!\!\!
\left[ \beta_1^4 L^4  + \frac{13}{3}\, \beta_1^2 \beta_2 L^3
+ \left(\frac{3}{2}\, \beta_2^2 + 3\, \beta_1 \beta_3\right) L^2 + \beta_4 L \right] a_0^4
\no\\ &\!\!\! + &\!\!\!\left.
\left[ \beta_1^5 L^5  + \frac{77}{12}\, \beta_1^3 \beta_2 L^4
+ \frac{1}{6} \left( 35\, \beta_1 \beta_2^2 + 36\, \beta_1^2 \beta_3\right) L^3 + \frac{7}{2} \left( \beta_1 \beta_4 + \beta_2 \beta_3\right) L^2 + \beta_5 L \right] a_0^5 + \cdots\right\}\, .
\eea
When the scales $\mu$ and $\mu_0$ are widely separated, the integral \eqn{eq:IntRGE} must be solved exactly, at a given order in the perturbative expansion of $\beta(\alpha_s)$, because the logarithm $L$ is large. At two loops, one gets the compact analytical expression
\bel{eq:alphaRun-2loops}
\alpha_s(\mu^2)\, =\, \frac{\alpha_s(\mu_0^2)}{1- \log{\left(\frac{\mu}{\mu_0}\right)}\,\beta_1 \frac{\alpha_s(\mu_0^2)}{\pi} \left[ 1 + \frac{\beta_2}{\beta_1} \,\frac{\alpha_s(\mu_0^2)}{\pi}\right]}\, ,
\ee
which resums all leading (LO), $a_0^n L^n$, and next-to-leading order (NLO), $a_0^{n+1} L^n$, logarithmic corrections in the perturbative series~\eqn{eq:alphaRunPert}. 
The present 5-loop accuracy in the $\beta$ function entails a resummation of N${}^4$LO logarithmic contributions to the running of $\alpha_s(\mu^2)$, {\it i.e.}, corrections of the form $a_0^{n+4} L^n$. This can be easily achieved, solving numerically the integral~\eqn{eq:IntRGE}.\footnote{There exist, however, excellent analytical approximations at four loops \cite{Rodrigo:1997zd} that can be generalized to higher loop orders.}

\begin{figure}[t]
\centering
\includegraphics[width=0.4\textwidth,clip]{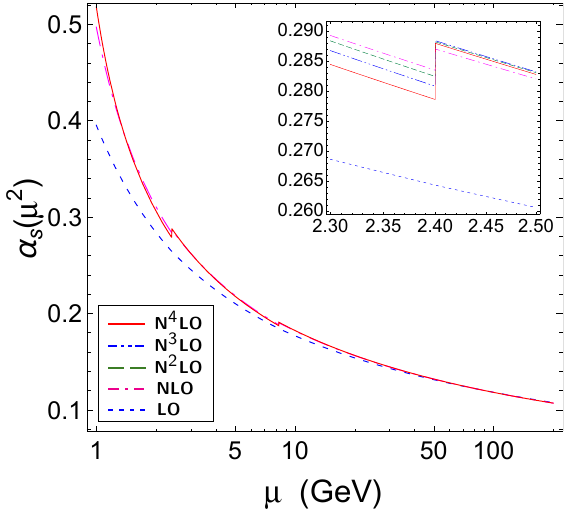}
\caption{Scale dependence of $\alpha_s$ at different perturbative orders, fixing its value at $M_Z$.}
\label{fig:running}
\end{figure}

The running of the strong coupling over a wide range of renormalization scales, fixing its value at $M_Z$, is shown in figure~\ref{fig:running} with different levels of approximation. Owing to the fast convergence of the $\beta$ function, the NLO resummation gives already an excellent approximation to 
$\alpha_s(\mu^2)$ over the entire plotted region. Higher-order corrections are obviously more visible at low values of $\mu$, where the coupling is larger. However, the four and five loop contributions are so small that their effects can only be seen with a big magnification, as shown in the figure inset.

\subsection{Quark masses}
\label{subsec:QuarkMasses}

The quark masses constitute additional QCD parameters that need to be properly defined. Confinement implies that quarks are not asymptotic states and, therefore, an on-shell mass emerging as a physical pole singularity in the quark propagator, the so-called pole quark mass, only exists in (truncated) perturbation theory. Moreover, the perturbative
loop expansion of such a pole quark mass is badly behaved, being very sensitive to renormalon singularities. 

A much better treatment of quark masses is to consider them as additional couplings of the QCD Lagrangian. As any other quantum-field-theory couplings, the renormalized quark masses need to be specified in a given renormalization scheme and at some renormalization scale and, therefore, they are not physical quantities by themselves. The so-called running quark mass $m_q(\mu^2)$ satisfies the renormalization-group equation
\bel{eq:runningMass}
\mu\,\frac{d m_q}{d\mu}\; =\; -m_q \;\gamma(\alpha_s)\, ,
\qquad\qquad\qquad\qquad
\gamma(\alpha_s)\; =\;\sum_{n=1}\, \gamma_n \,
\left(\frac{\alpha_s}{\pi}\right)^n\, ,
\ee
which is governed by the $\gamma$ function, also known as the quark-mass anomalous dimension. The dependence of $\gamma(\alpha_s)$ with the strong coupling has been also computed to five loops.
Similarly to the $\beta$ function, the first two terms in the perturbative expansion of $\gamma(\alpha_s)$ are independent of the chosen (mass-independent) renormalization scheme \cite{Espriu:1981eh,Tarrach:1980up}:
\bel{eq:gamma-1-2}
\gamma_1\, =\, 2\, ,
\qquad\qquad\qquad
\gamma_2\, =\, \frac{101}{12} - \frac{5}{18} \, n_f \, ,
\ee
In the $\overline{\mathrm{MS}}$ scheme, the values of the three \cite{Tarasov:1982gk,Larin:1993tq,Tarasov:2019rwk}, four \cite{Chetyrkin:1997dh,Vermaseren:1997fq} and five-loop \cite{Baikov:2014qja,Luthe:2016xec,Baikov:2017ujl} coefficients are:
\bea\label{eq:gamma3}
\gamma_3 &\!\!\! = &\!\!\!
\frac{1}{24}\left[ \frac{3747}{4} - \left(\frac{554}{9} + 40\, \zeta_3\right)\, n_f - \frac{35}{27}\, n_f^2\right]\, ,
\\[5pt]
\gamma_4 &\!\!\! = &\!\!\!
\frac{1}{128}\left\{ \frac{4603055}{162} + \frac{135680}{27}\,\zeta_3 -8800\,\zeta_5
\, +\, n_f\:
\left[-\frac{91723}{27}-\frac{34192}{9}\,\zeta_3 + 880\,\zeta_4 + \frac{18400}{9}\,\zeta_5\right]
\right.\nonumber\\
&& \hskip .4cm \mbox{} \left. +\, n_f^2\:
\left[\frac{5242}{243} + \frac{800}{9}\,\zeta_3 - \frac{160}{3}\,\zeta_4\right]
+ n_f^3\left[-\frac{332}{243}+ \frac{64}{27}\,\zeta_3\right]\right\}\, ,
\\[5pt]
\gamma_5 &\!\!\! = &\!\!\!
\frac{1}{512}\Biggl\{ 
\frac{99512327}{162} + \frac{46402466}{243}\,\zeta_3 + 96800\,\zeta_3^2 -\frac{698126}{9}\,\zeta_4
-\frac{231757160}{243}\,\zeta_5 + 242000\,\zeta_6 + 412720\,\zeta_7
\nonumber\\
&& \hskip .45cm \mbox{}
+\, n_f\:\left[ -\frac{150736283}{1458} -\frac{12538016}{81}\,\zeta_3 -\frac{75680}{9}\,\zeta_3^2 + \frac{2038742}{27}\,\zeta_4 + \frac{49876180}{243}\,\zeta_5 -\frac{638000}{9}\,\zeta_6
\right.\nonumber\\
&& \hskip 1.3cm\left. \mbox{}
-\frac{1820000}{27}\,\zeta_7\right]
\nonumber\\
&& \hskip .45cm \mbox{}
+\, n_f^2\:\left[ \frac{1320742}{729} + \frac{2010824}{243}\,\zeta_3 +\frac{46400}{27}\,\zeta_3^2 -\frac{166300}{27}\,\zeta_4 -\frac{264040}{81}\,\zeta_5 + \frac{92000}{27}\,\zeta_6\right]
\nonumber\\
&& \hskip .45cm \mbox{}
+\, n_f^3\:\left[ \frac{91865}{1458} +\frac{12848}{81}\,\zeta_3 + \frac{448}{9}\,\zeta_4 -\frac{5120}{27}\,\zeta_5\right]
\, +\, n_f^4\:\left[-\frac{260}{243} -\frac{320}{243}\,\zeta_3 + \frac{64}{27}\,\zeta_4\right]
\Biggr\} .
\eea
In addition to $\zeta_{3,4,5}$, already present in the known coefficients of the $\beta$ function, $\gamma_5$ involves also the numerical factors $\zeta_6 = \pi^6/945$ and $\zeta_7 = 1.008349277\ldots$

Similarly to what happens with the $\beta$ function, the five computed terms of the $\gamma$ function exhibit a very modest growth with the perturbative order. The resulting expansion in powers of $\alpha_s$ is very smooth, indicating a surprisingly good perturbative convergence. With $n_f=5$ flavours, 
\bel{eq:gamma_exp}
\gamma(\alpha_s) \, =\,  \gamma_1 a_s\,\left( 1 + 3.5139\, a_s + 7.4199\, a_s^2 + 11.0343\, a_s^3 + 41.8205\, a_s^4\right)\, .
\ee
Thus, perturbation theory appears to give an excellent description even at large values of $\alpha_s\sim 0.5$, corresponding to very low renormalization scales $\mu\sim 1~$~GeV. The solution of the renormalization-group equation \eqn{eq:runningMass} is given in Appendix~\ref{app:masses}, which also contains the relation between the running and pole quark masses.

\subsection{Quark mass thresholds and effective QCD theories}
\label{eq:matching}

The explicit dependence on $n_f$ exhibited by the perturbative coefficients of the $\beta$ and $\gamma$ functions implies that the values of the running strong coupling and quark masses depend on the considered number of `active' quark flavours. Thus, one needs to properly define the matter content of the quantum field theory that is being used to describe physics. This is particularly important in mass-independent renormalization schemes such as the $\overline{\mathrm{MS}}$ one, because heavy particles with masses $M\gg \mu$ do not decouple \cite{Appelquist:1974tg}. They contribute to the $\beta$ and $\gamma$ functions and, moreover, induce dangerous quantum corrections involving large logarithms that grow as $\log{(M^2/\mu^2)}$. It is then convenient to remove (`integrate out') the heavy states from the Lagrangian and work with an effective field theory that has a reduced matter content with only light particles \cite{Weinberg:1980wa,Hall:1980kf,Ovrut:1980uv,Ovrut:1981ue}.

At very high energies above the top mass scale, QCD contains the six known quark flavours. At lower energies, one usually removes the heavy top quark and defines and effective five-flavour theory, which has slightly different values for $\alpha_s$ and the light quark masses. As one goes further down in energy, the bottom and charm quarks can also be removed, giving rise to effective QCD theories with $n_f=4$ and 3, respectively. 
The effective theories with $n_f$ and $n_f-1$ flavours are related by the condition that they should generate the same physical predictions in their common range of validity:
\bel{eq:QCDnf}
\cL_{\mathrm{QCD}}^{(n_f)}\quad\Longleftrightarrow\quad
\cL_{\mathrm{QCD}}^{(n_f-1)}\, +\, \sum_{d_i>4}\frac{\tilde c_i}{M_q^{d_i-4}}\; O_i\, .
\ee
Here $M_q$ denotes the mass of the heavy quark that has been removed and $O_i$ stands for local gauge-invariant operators of dimension $d_i>4$, constructed with the light-quark and gluon fields. The two Lagrangians are formally identical (they only differ in the number of quark flavours), up to corrections suppressed by inverse powers of $M_q$. However, since quantum corrections are different, the numerical values of their couplings need to be different also. They are related by matching conditions that can be written as perturbative expansions in powers of the original strong coupling:
\bea
\alpha_s^{(n_f-1)}(\mu^2) & = & \alpha_s^{(n_f)}(\mu^2)\,\left\{
1 + \sum_{k=1} \sum_{n=0}^k d_{kn}\; \left[a_s^{(n_f)}(\mu^2)\right]^k\, 
\log^n{(\mu^2/M_q^2)}\right\} ,
\\
m_q^{(n_f-1)}(\mu^2) & = & m_q^{(n_f)}(\mu^2)\,\left\{
1 + \sum_{k=2} \sum_{n=0}^k h_{kn}\; \left[a_s^{(n_f)}(\mu^2)\right]^k\, 
\log^n{(\mu^2/M_q^2)}\right\} ,
\eea
where $M_q\equiv M_q(\mu^2)$ is the running  mass of the heavy quark that has been integrated out and $a_s^{(n_f)}\equiv \alpha_s^{(n_f)}/\pi$.

Taking the logarithmic derivative of these equations with respect to the renormalization scale, using the renormalization group equations \eqn{eq:beta-rge} and \eqn{eq:runningMass}, and identifying both sides order by order in $a_s^{(n_f)}$, one can determine the logarithmic coefficients $d_{kn}$ and $h_{kn}$ with $n\not=0$ in terms of $d_{k'0}$, $h_{k'0}$, $\beta_{k'}$ and $\gamma_{k'}$, with $k'\le k$ \cite{Rodrigo:1997zd}. The non-logarithmic coefficients $d_{k0}$ and $h_{k0}$ need to be evaluated explicitly; they are currently known to four loops \cite{Schroder:2005hy,Chetyrkin:2005ia,Kniehl:2006bg,Liu:2015fxa}. The explicit values of all these coefficients are compiled in Appendix~\ref{app:matching}.
The small discontinuities on the curves plotted in figure~\ref{fig:running} reflect the crossing of the charm and bottom thresholds where the different QCD${}_{n_f}$ effective theories have been matched.

\section{Current correlators}
\label{sec:correlators}

Inclusive observables, such as $\sigma(e^+e^-\to\mathrm{hadrons})$, $\Gamma(Z\to\mathrm{hadrons})$ or $\Gamma(W\to\mathrm{hadrons})$ proceed through the colour-singlet vector $\, V^{\mu}_{ij} = \bar{q}_j \gamma^{\mu} q_i \, $
and axial-vector
$\, A^{\mu}_{ij} = \bar{q}_j \gamma^{\mu} \gamma_5 q_i \,$
quark currents ($i,j=u,d,s\ldots$).
The QCD dynamics is then encoded in the two-point correlation functions
\begin{equation}\label{eq:correlators}
\Pi^{\mu \nu}_{ij,\cJ}(q)\; \equiv\;
 i \int d^4x \;\, \mathrm{e}^{iqx}\,
\langle 0|T(\cJ^{\mu}_{ij}(x)\, \cJ^{\nu}_{ij}(0)^\dagger)|0\rangle
\; =\;
\left( -g^{\mu\nu} q^2 + q^{\mu} q^{\nu}\right) \, \Pi_{ij,\,\cJ}^{T}(q^2)
 +   q^{\mu} q^{\nu}\, \Pi_{ij,\,\cJ}^{L}(q^2) \, ,
\end{equation}
where $\cJ=V,A$ and the superscript in the transverse and longitudinal components
denotes the corresponding angular momentum $J=1$ (T) and $J=0$ (L)
in the hadronic rest frame ($\vec q = \vec 0$). 

For physical values of the momentum transfer $q^\mu$ ($q^2>0$), these correlators acquire absorptive parts that correspond to the measurable hadronic spectral distributions with the given quantum numbers:
\bel{eq:ImPi}
\mathrm{Im}\, \Pi^{\mu \nu}_{ij,\,\cJ}(q)\, =\, \frac{1}{2i}\left[ \Pi^{\mu \nu}_{ij,\,\cJ}(q)\ - \Pi^{\mu \nu}_{ij,\,\cJ}(q)^\dagger\right]\, =\,\frac{1}{2}\, (2\pi)^4\,\sum_n \delta^{(4)}(q-p_n)\; \langle 0|\cJ^\mu_{ij}(0)|n\rangle\,\langle n|\cJ^\nu_{ij}(0)^\dagger|0\rangle\, .
\ee
This relation is easily obtained inserting between the two currents the completeness relation $1 = \sum_n |n\rangle \langle n|$, and using translation invariance to integrate the space--time coordinate. The right-hand side involves a sum over all physical states (hadrons) that can be produced through the considered quark current, including their corresponding phase-space integration. Thus, it is directly related to the observable width or cross section.

The definition of the two-point functions $\Pi^{\mu \nu}_{ij,\cJ}(q)$ does not involve any hadrons. It only contains a vacuum matrix element of the T-product of two quark currents, which can be calculated in terms of the quark and gluon fields of QCD. At high-enough values of $q^2$, it is then possible to analyse these correlators with perturbative tools.
As shown in Eq.~\eqn{eq:ImPi}, computing a given correlator and taking its absorptive part, which is generated by the sum of all possible cuts in the corresponding Feynman diagrams, is equivalent to the evaluation of the sum of squared current matrix elements for all possible quark and gluon final states, including the corresponding phase-space integrations. In spite of having one additional loop, the calculation of the vacuum-polarization topologies contributing to these two-point functions is much easier than the direct computation of the rates \cite{Chetyrkin:1994js}. Moreover, infrared divergences are naturally avoided since they automatically cancel within each diagram (virtual and bremsstrahlung contributions just correspond to different cuts of the same diagram).

In the limit of zero quark masses, the vector and axial-vector currents are conserved ($\partial_\mu \cJ^\mu =0$), which implies that $q^2\, \Pi_{ij,\,\cJ}^{L}(q^2)=0$, and therefore
\begin{equation}\label{eq:correlators2}
\left.\Pi^{\mu \nu}_{ij,\cJ}(q)\right|_{m_q=0}\; =\;
\left( -g^{\mu\nu} q^2 + q^{\mu} q^{\nu}\right) \left[ \Pi_{ij,\,\cJ}^{L}(q^2)
+ \Pi_{ij,\,\cJ}^{T}(q^2)\right]
\,\equiv\,
\left( -g^{\mu\nu} q^2 + q^{\mu} q^{\nu}\right) \, \Pi_{ij,\,\cJ}^{L+T}(q^2)
\, .
\end{equation}
%
\begin{figure}[t]
\centering
\begin{minipage}[c]{0.4\textwidth}\centering
\includegraphics[width=0.7\textwidth,clip]{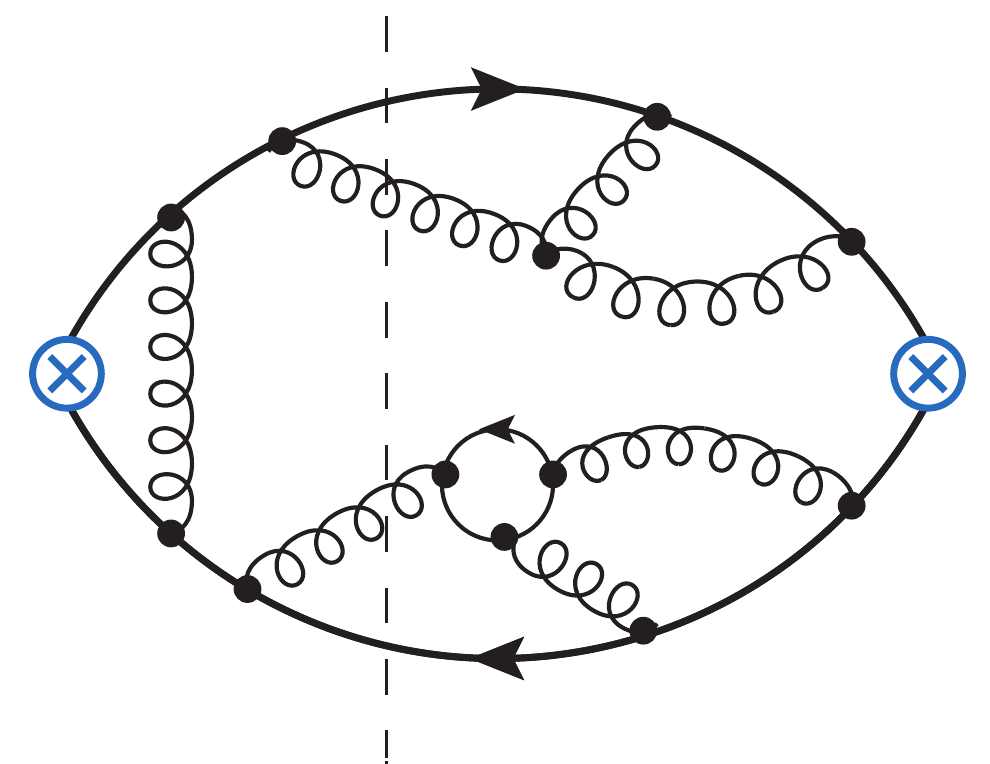}
\end{minipage}
\hskip 1.4cm
\begin{minipage}[c]{0.4\textwidth}\centering
\includegraphics[width=0.9\textwidth,clip]{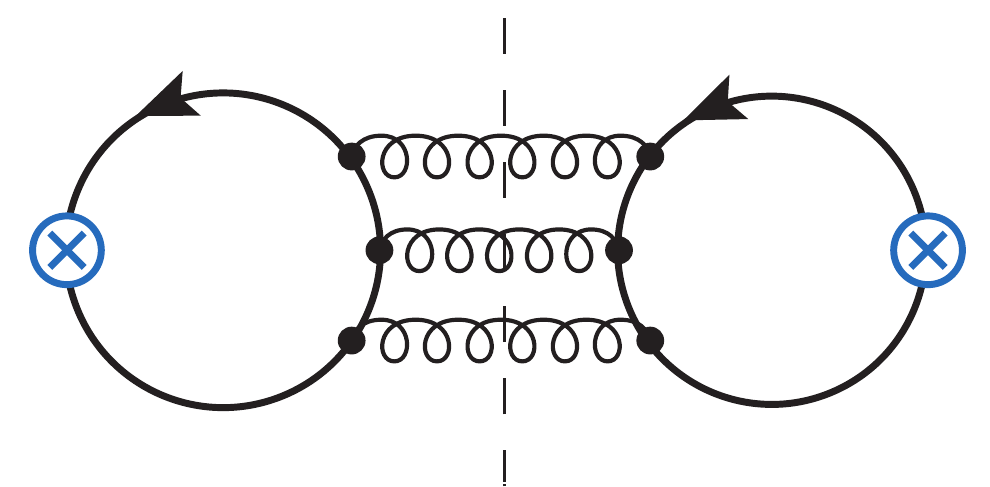}
\end{minipage}
\caption{Examples of non-singlet (left) and singlet (right) Feynman diagrams 
contributing to the current correlation functions. The crossed vertices indicate the current insertions. The vertical dashed lines show two possible absorptive cuts, corresponding to $q\bar q gg$ (left) and $ggg$ (right) on-shell intermediate states.} 
\label{fig:topologies}
\end{figure}
%
When $i\not= j$, the two quark currents must necessarily be connected through a quark loop, with the non-singlet topology shown in the left diagram of figure~\ref{fig:topologies}. This results in identical perturbative contributions to the vector and axial massless correlators:\footnote{This equality is obvious at lowest order (naive quark loop) because the $\gamma_5$ factor of the axial vertex can be anticommuted with the $\slashchar{k}$ of the internal quark propagator: $\gamma^\mu\gamma_5\slashchar{k}\gamma^\nu\gamma_5 = \gamma^\mu\slashchar{k}\gamma^\nu$. Each insertion of a gluon vertex on the quark line introduces a $\gamma^\mu$ and one additional quark propagator, {\it i.e.}, two $\gamma^{\mu_i}$ matrices. Thus, at any order in $\alpha_s$, the $\gamma_5$ from one vertex can always be moved to the other vertex with an even number of anticonmutations, and $\gamma_5^2 =1$.\label{foot:Non-Singlet_Corr}
}
\be
\Pi(s)\,\equiv\, \left.\Pi_{i\not=j,V}^{L+T}(s)\right|_{m_q=0}\, =\, \left.\Pi_{i\not=j,A}^{L+T}(s)\right|_{m_q=0}\, .
\ee
The (massless) perturbative function $\Pi(s)$ does not depend on the quark flavour indices $i,j$ because gluonic interactions are flavour blind. It is convenient to take its logarithmic derivative and define the Euclidean  
($Q^2=-q^2$) Adler function~\cite{Adler:1974gd}  
\bea\label{eq:Adler}
D(Q^2)\; \equiv\; -Q^2\frac{d}{dQ^2}\Pi(Q^2) &\!\! = &\!\! 
\frac{N_C}{12\pi^2}\;\left\{
1 + \sum_{n=1}\sum_{p=0}^{n-1}\; K_{n,p}\,
\left( {\alpha_s(\mu^2)\over \pi}\right)^n\,\log^p{(Q^2/\mu^2)}\right\}
\no\\ &\!\! = &\!\!
\frac{1}{4\pi^2}\;\left\{
1 + \sum_{n=1}\; K_{n,0}\,
\left( {\alpha_s(Q^2)\over \pi}\right)^n\right\}\, ,
\eea
with $N_C=3$ the number of quark colours. This eliminates unwanted (renormalization-scheme and scale dependent) subtraction constants, which do not contribute to any physical observable, so that $D(Q^2)$ satisfies an homogenous renormalization-group equation:
\be
\mu\,\frac{d D(Q^2)}{d\mu}\, =\, 0
\qquad\longrightarrow\qquad
\left\{ \mu\,\frac{\partial}{\partial\mu} + \beta\!\left(\alpha_s(\mu^2)\right)\,\alpha_s(\mu^2)\;\frac{\partial}{\partial\alpha_s(\mu^2)}\right\} D(Q^2)\, =\, 0\, .
\ee
In the second line of Eq.~\eqn{eq:Adler}, all logarithmic corrections have been summed up into the running coupling $\alpha_s(Q^2)$, with the choice of renormalization scale $\mu^2=Q^2$. Using Eq.~\eqn{eq:alphaRunPert} to expand $\alpha_s(Q^2)$ in powers of $\alpha_s(\mu^2)$, one recovers the naive perturbative expansion in the first line. Thus, all coefficients $K_{n,p}$ with $p\not=0$ are functions of $K_{m<n,0}$ and $\beta_{m<n}$.

The Adler function is currently known to $\mathcal{O}(\alpha_s^4)$. In the $\overline{\mathrm{MS}}$ scheme, the coefficients $K_{m<5,0}$ have the values:
\begin{eqnarray}
\lefteqn{\hskip -.5cm K_{1,0} \, =\, 1 \qquad\mbox{\cite{Appelquist:1973uz,Zee:1973sr}}\, ,
\qquad\qquad
K_{2,0}\, =\, \frac{365}{24} - 11\,\zeta_3 +  \left( \frac{2}{3}\,\zeta_3 - \frac{11}{12}\right) n_f\qquad
\mbox{\cite{Chetyrkin:1979bj,Dine:1979qh}}
\, , }&&
\nonumber\\
\lefteqn{\hskip -.5cm 
K_{3,0}\, =\, 
\frac{87029}{288} - \frac{1103}{4}\, \zeta_3 + \frac{275}{6}\, \zeta_5 + 
      \left(-\frac{7847}{216} + \frac{262}{9}\, \zeta_3 - \frac{25}{9}\, \zeta_5\right) n_f + 
     \left(\frac{151}{162} - \frac{19}{27}\, \zeta_3\right) n_f^2 
\qquad
\mbox{\cite{Gorishnii:1990vf,Surguladze:1990tg,Chetyrkin:1996ez}}
\, ,}&& 
\nonumber\\
\lefteqn{\hskip -.5cm 
K_{4,0} \, =\, 
\frac{144939499}{20736} - \frac{5693495}{864}\,\zeta_3 + \frac{5445}{8}\,
\zeta^2_3 + \frac{65945}{288}\,\zeta_5 - \frac{7315}{48}\,\zeta_7
}&&
\nonumber\\
\lefteqn{\hskip .3cm\mbox{}
+ \left(  -\frac{13044007}{10368} + \frac{12205}{12}\,\zeta_3 - 55\,\zeta^2_3 + \frac{29675}{432}\,\zeta_5 + \frac{665}{72}\,\zeta_7\right) n_f
}&&
\nonumber\\
\lefteqn{\hskip .3cm\mbox{}
+ \left( \frac{1045381}{15552} - \frac{40655}{864}\,\zeta_3 + \frac{5}{6}\,\zeta^2_3 - \frac{260}{27}\,\zeta_5\right) n_f^2
+ \left( -\frac{6131}{5832} + \frac{203}{324}\,\zeta_3 + \frac{5}{18}\,\zeta_5
\right) n_f^3
%
\qquad\mbox{\cite{Baikov:2008jh,Baikov:2010je,Herzog:2017dtz}}\, .
}&&
\end{eqnarray}
Although $D(Q^2)$ does not depend on renormalization conventions, its truncated perturbative expansion at $\mathcal{O}(\alpha_s^n)$ contains a residual dependence on the adopted renormalization scale and scheme of $\mathcal{O}(\alpha_s^{n+1})$. Theoretical predictions are usually done at $\mu^2=Q^2$ in order to avoid large $\log^p{(Q^2/\mu^2)}$ corrections which could deteriorate the convergence of the perturbative series. The numerical sensitivity to the choice of $\mu$, within a physically reasonable range around the physical scale $Q^2$, gives a useful assessment of the perturbative uncertainty. A conventional range of variation is $Q^2/2 \le \mu^2 \le 2 Q^2$.

Integrating Eq.~\eqn{eq:Adler} one immediately obtains $\Pi(Q^2)$ up to an irrelevant integration constant. The perturbative logarithmic corrections induce imaginary parts in the physical Minkowskian region $q^2=-Q^2>0$, where the logarithmic cut generates a discontinuity between the values of $\Pi(q^2)$ above and below the real axis:
\bel{eq:SpectralFunction}
\mathrm{Im}\, \Pi(q^2+i\epsilon)\; =\; \theta(q^2)\;\frac{1}{4\pi}\;\left\{ 1 +
\sum_{n=1} F_n^{\mathrm{NS}}\, \left(\frac{\alpha_s(q^2)}{\pi}\right)^n\right\} .
\ee
This perturbative quark-level expression corresponds to the physical hadronic spectral function that can be experimentally accessed. The expansion coefficients are easily found to be:
\bea
\lefteqn{F_1^{\mathrm{NS}}\, =\, K_{1,0} \, =\, 1\, ,
\qquad\qquad\qquad\qquad
F_2^{\mathrm{NS}}\, =\, K_{2,0}\, =\, 1.98571 -0.115295\; n_f\, ,} &&
\nonumber\\
\lefteqn{F_3^{\mathrm{NS}}\, =\, K_{3,0} - \frac{\pi^2}{12}\,\beta_1^2 K_{1,0}\, =\, -6.63694-1.20013\; n_f - 0.00517836\; n_f^2\, ,} &&
\nonumber\\
\lefteqn{F_4^{\mathrm{NS}} \, =\, K_{4,0} - \frac{\pi^2}{12}\left[3 \beta_1^2 K_{2,0} + \frac{5}{2}\,\beta_1\beta_2 K_{1,0}\right]\, =\, -156.608 + 18.7748\; n_f - 0.797434\; n_f^2 + 0.0215161\; n_f^3\, .} &&
\eea
The analytical continuation from the Euclidean region ($Q^2=-q^2>0$), where $\alpha_s(Q^2)$ is defined, to Minkowskian  values of $q^2>0$ generates the additional corrections proportional to $\pi^2$ \cite{Pennington:1981cw} in $F_3^{\mathrm{NS}}$ and $F_4^{\mathrm{NS}}$.

The neutral-current correlators ($i=j$) receive additional singlet contributions where each current couples to a different quark loop (right diagram in figure~\ref{fig:topologies}). Since gluons have $J^{PC}=1^{--}$ and colour, these topologies start to contribute at $\mathcal{O}(\alpha_s^3)$ and $\mathcal{O}(\alpha_s^2)$, respectively, for the vector and axial-vector currents: 
\begin{eqnarray}
\Delta^\mathrm{S} D_V(Q^2)\; =\; \frac{N_C}{12\pi^2}\;\sum_{n=3}\; d^V_n
\left( {\alpha_s(Q^2)\over \pi}\right)^n ,
\qquad\quad
\Delta^\mathrm{S} D_A(Q^2)\; =\; \frac{N_C}{12\pi^2}\;\sum_{n=2}\; d^{A}_n
\left( {\alpha_s(Q^2)\over \pi}\right)^n .
\end{eqnarray}
In the $\overline{\mathrm{MS}}$ scheme, the $\cO(\alpha_s^3)$
\cite{Gorishnii:1990vf,Surguladze:1990tg,Chetyrkin:1996ez} and $\cO(\alpha_s^4)$ \cite{Herzog:2017dtz,Baikov:2012zn} vector-current singlet coefficients are:
\be
d_3^V\, =\, \frac{5}{9} \left(\frac{11}{24} - \zeta_3\right)\, , 
\qquad\quad
d_4^V = \frac{5795}{576} -\frac{8245}{432}\,\zeta_3 - \frac{55}{12}\,\zeta_3^2 + \frac{2825}{216}\,\zeta_5
+ \left(-\frac{745}{1296} + \frac{65}{72}\,\zeta_3 + \frac{5}{18}\,\zeta_3^2 -\frac{25}{36}\,\zeta_5\right) n_f\, .
\ee
%
%

The singlet axial topologies are much more subtle because they contain a single $\gamma_5$ matrix within each of the two separate fermion loops attached to the currents. The LO contribution involves a two-gluon exchange between two (anomalous) triangular fermion graphs. 
In the standard theory of electroweak interactions, the axial couplings to the $Z$ boson of up-type and down-type quarks have opposite signs and equal strength. Therefore, the singlet axial contribution of a given electroweak doublet to the hadronic $Z$ width vanishes for massless (or equal-mass) quarks.
The only relevant contribution originates from the third fermion family and will be discussed in Section~\ref{subsec:Zwidth}.

\subsection{Quark-mass corrections}

When quark masses are taken into account, we must distinguish the separate Adler functions for the vector and axial-vector correlators, and their non-zero longitudinal components \cite{Pich:1998yn,Pich:1999hc}:
\bel{eq:Adler_def_L_L+T}
D^{L+T}_{ij,\,\cJ}(Q^2)\equiv -Q^2 \frac{{\rm d}}{{\rm d}Q^2} \left[\Pi^{L+T}_{ij,\,\cJ}(Q^2)\right]
\, , \qquad\qquad\qquad
D^{L}_{ij,\,\cJ}(Q^2)\equiv 
\frac{\dis {\rm d}}{\dis {\rm d}Q^2}  \left[Q^2\, \Pi^{L}_{ij,\,\cJ}(Q^2)\right] \, .
\ee
At large values of $Q^2\gg m_q^2$, the quark-mass corrections can be computed as an expansion in powers of $m_q^2/Q^2$. In terms of  the running quark masses 
$\overline{m}_i\equiv m_i(Q^2)$ and the running coupling $a(Q^2)\equiv\alpha_s(Q^2)/\pi$, the $\cO(m_q^2)$ and $\cO(m_q^4)$ contributions to the non-singlet correlators have the following flavour structure~\cite{Pich:1999hc}:
\begin{eqnarray}
\label{DLTij}
\Delta_{m_q} D^{L+T}_{ij,\,\cJ}(Q^2)
&\!\!\! = &\!\!\! 
-\frac{3}{4 \pi^2 Q^2} \biggl\{ 
\left(\overline{m}_j^{\, 2} + \overline{m}_i^{\, 2}\right)  \sum_{n=0}
c^{L+T}_{n}  a^n(Q^2) 
\,\pm\, \overline{m}_j \overline{m}_i \, \sum_{n=1}
e^{L+T}_{n}  a^n(Q^2) 
 \, +\, \Bigl(\sum_k \overline{m}_k^{\, 2}\Bigr)  \sum_{n=2}  
f^{L+T}_{n}  a^n(Q^2) \biggr\}  
\nonumber\\ &\!\!\! - &\!\!\! 
\frac{3}{\pi^2 (Q^2)^2} \,\sum_{n=0} a^n(Q^2)\; \biggl\{
\left( \overline{m}_j^{\, 4} + \overline{m}_i^{\, 4}\right)\,  h_n^{L+T}
\,\pm\, \frac{5}{3}\,\overline{m}_j \overline{m}_i
\left( \overline{m}_j^{\, 2} + \overline{m}_i^{\, 2}\right)\,  k_n^{L+T}
\, -\,\overline{m}_j^{\, 2} \overline{m}_i^{\, 2}\,  g_n^{L+T}
\biggr.\nonumber \\ &&\hskip 2.6cm\biggl.
+ \, \Bigl(\sum_k \overline{m}^{\, 4}_k \Bigr) \,   j_n^{L+T}
\, +\, 2 \, \Bigl(\sum_{k\neq l}  \overline{m}^{\, 2}_k  \overline{m}^{\, 2}_l \Bigr) \,  
 u_n^{L+T} 
\biggr\} 
\nonumber\\ &\!\!\! + &\!\!\!
 \cO[m_q^6/(Q^2)^3]\, ,
\end{eqnarray}
\begin{eqnarray}
\label{DLij}
D^{L}_{ij,\,\cJ}(Q^2)
&\!\!\! = &\!\!\! 
\frac{3}{8 \pi^2 Q^2} 
\,\left(\overline{m}_j \mp \overline{m}_i \right)^2
\, \sum_{n=0}  \, d^{L}_{n}   \, a^n(Q^2) 
\nonumber \\ &\!\!\! + &\!\!\!
\frac{3}{2\pi^2 (Q^2)^2} 
\,\left(\overline{m}_j \mp \overline{m}_i \right)^2
\, \sum_{n=0}  \, a^n(Q^2) \; \biggl\{
\left( \overline{m}_j^{\, 2} + \overline{m}_i^{\, 2}\right)\, h_n^{L}
\,\pm\, {3\over 2}\, \overline{m}_j\,\overline{m}_i\,\, k_n^{L}\,
+  \,\Bigl(\sum_{k}  \overline{m}^{\, 2}_k \Bigr)  \,\, j_n^{L} 
\biggr\} 
\nonumber \\ &\!\!\! + &\!\!\!
\cO[m_q^6/(Q^2)^3]\, ,
\end{eqnarray}
where the upper signs correspond to $\cJ=V$ and the lower ones to $\cJ=A$.
Notice that the non-singlet axial-vector results are easily obtained from the vector ones by reversing the sign of either $m_i$ or $m_j$.\footnote{This can be easily understood applying an argument analogous to the one in footnote~\ref{foot:Non-Singlet_Corr}.}
The dependences on the masses of active quarks other than $i$ and $j$
are generated by internal quark loops coupled to gluons, which give rise to sums over all quark masses (QCD is flavour blind). This type of structures
start to contribute at $\cO(\alpha_s^2)$.
The longitudinal Adler function is proportional to the global factor
$(\overline{m}_i \mp \overline{m}_j)^2$ and vanishes for massless quarks. Moreover, the conservation of the vector current for equal quark masses implies $D^{L}_{ij,\,V}(Q^2)=0$ when $m_i = m_j$.

The known values of the perturbative coefficients $c^{L+T}_n$,  $e^{L+T}_n$, $f^{L+T}_n$, $d^{L}_n$, $h^{L+T,L}_n$, $k^{L+T,L}_n$, $g^{L+T}_n$, $j^{L+T,L}_n$ and $u^{L+T}_n$ are collected in Appendix \ref{app:QuarkMasses}.
In Eqs.~\eqn{DLTij} and \eqn{DLij} all perturbative logarithms have been summed up into the running masses and strong coupling by choosing $\mu^2=Q^2$.
The dependence on the renormalization scale is governed by the homogeneous renormalization-group equations satisfied by the Adler functions. Explicit expressions for generic values of $\mu$ can be found in Ref.~\cite{Pich:1999hc}.

\subsection{Operator product expansion}

At short distances ($x^\mu\to 0$), the T-product of two currents in Eq.~\eqn{eq:correlators} can be expanded in a series of gauge-invariant local operators, defined at $x^\mu = 0$, with c-number coefficients which are functions of $x^2$ scaling as $\cC(x)\sim (x^2)^{-3+D/2}$ where $D$ is the dimension of the corresponding operator \cite{Wilson:1969zs}.  Although the series involves an infinite number of terms, only a finite number of them contribute at any finite order in $x^2$. Once the space--time integration is performed, the resulting expression becomes a series in inverse powers of $Q^2$:
\be\label{eq:OPE}
\left.\Pi_{ij,\,\cJ}^{L/T}(Q^2)\right|^{\mathrm{OPE}}\, =\, \sum_{D=2n}\,
\frac{1}{(Q^2)^{D/2}}\,\sum_{\mathrm{dim}\, O =D}
\cC_{ij,\,\cJ}^{L/T}(Q^2,\mu^2) \;\langle 0| O(\mu^2) |0\rangle\, .
\ee
The arbitrary factorization scale $\mu$ separates the short-distance contributions from scales higher than $\mu$, which are absorbed into the dimensionless Wilson coefficients $\cC_{ij,\,\cJ}^{L/T}(Q^2,\mu^2)$, and the long-distance effects from lower scales that remain in the matrix elements of the local operators.

A standard perturbative calculation gives rise to normal-ordered operators that have a null expectation value in the perturbative vacuum. Thus, the perturbative results discussed before correspond to the Wilson coefficient of the dimension-zero identity operator in Eq.~\eqn{eq:OPE}. The QCD vacuum is however non-perturbative and generates non-zero vacuum expectation values for many composite operators such as the quark condensate $\langle 0|\bar q q|0\rangle$, responsible for the breaking of chiral symmetry, or the gluon condensate $\langle 0|\frac{\alpha_s}{\pi}\, G_{\mu\nu}G^{\mu\nu}|0\rangle$ that breaks the scale invariance of massless QCD.
The OPE allows us to incorporate these non-perturbative dynamical contributions through a series of power corrections in $1/Q^2$, governed by the vacuum expectation values of all possible gauge- and Lorentz-invariant operators, the so-called condensates \cite{Shifman:1978bx,Shifman:1978by,Shifman:1978bw,Novikov:1980uj}. 
These condensates are universal quantities, independent of the particular process or correlator being investigated, that parametrize the dynamical properties of the QCD vacuum.

With quark and gluon fields, it is not possible to build gauge-invariant scalar operators with dimension $D=2$. The first contributions to the OPE of current correlators can only originate in $D=4$ operators: the gluon condensate and 
$m_q\,\langle 0|\bar q q|0\rangle$.\footnote{The quark mass is needed to cancel the renormalization-scale dependence of the $\bar q q$ operator.} Thus, the leading non-perturbative corrections are suppressed by a factor $(1/Q^2)^2$ and fade away very fast when the momentum transfer increases. The numerical size of the condensates is determined by the appropriate dimensional powers of the QCD scale $\Lambda_{\mathrm{QCD}}\sim 300$~MeV. For instance, the most recent lattice compilation quotes $(-\langle 0|\bar u u|0\rangle)^{1/3} = (272\pm 5)$~MeV, at $\mu=2$~GeV, in the $\overline{\mathrm{MS}}$ scheme  with $m_{u,d}=0$ \cite{Aoki:2019cca}. Therefore, for $Q^2$ values well above the 1~GeV region, the correlation functions~\eqn{eq:correlators} can be theoretically predicted with high accuracy.
The most relevant power corrections to the vector and axial-vector correlators are compiled in Appendix~\ref{app:D=4}.

The OPE is rigorously defined for Euclidean values of $q^2=-Q^2<0$, where the passage from the limit $x^\mu\to 0$ to the limit $q^\mu\to\infty$ in the space--time integral \eqn{eq:correlators} is strictly correct and implies $Q^2\to\infty$. The result can be analytically continued to the complex $q^2$ plane, except for the singularities of the correlation function. This excludes the physical Minkowskian region in the positive real axis ($q^2>0$), where $\Pi_{ij,\,\cJ}^{L/T}(Q^2)$ has a logarithmic cut. In fact, the perturbative spectral function \eqn{eq:SpectralFunction} contains quark-antiquark and multi-gluon thresholds which, owing to confinement, are not present in the measurable spectral function that exhibits instead multi-hadron thresholds and resonance structures. Nevertheless, the short-distance approach to inclusive quantities is expected to be valid, provided both the data and the theory are smeared over a suitable energy range that minimizes the sensitivity to resonances and threshold effects~\cite{Sakurai:1973rh,Poggio:1975af}.

\subsection{Scalar and pseudoscalar two-point functions}

The massless QCD Lagrangian is invariant under independent $\mathrm{SU}(n_f)$ flavour transformations of the left- and right-handed quark chiralities, which entails the conservation of the corresponding vector and axial-vector Noether currents. The explicit breaking of chiral symmetry induced by the quark masses generates non-zero divergences for these currents, involving the scalar $J_{ij}^{S} = \bar q_j q_i$ and pseudoscalar $J_{ij}^{P} = \bar q_j\gamma_5 q_i$ quark currents:
\bel{eq:CurrentDivergences} 
\partial_\mu V^\mu_{ij}(x)\, = \, i\,\left( m_j - m_i\right)\, J_{ij}^{S}(x)\, ,
\qquad\qquad\qquad
\partial_\mu A^\mu_{ij}(x) \, = \, i\,\left( m_j + m_i\right)\, J_{ij}^{P}(x)\, .
\ee
Current conservation implies that the vector and axial-vector currents do not get renormalized, {\it i.e.}, their associated anomalous dimensions are identically zero.
Their divergences are also renormalization-group invariant quantities. Therefore, the scalar and pseudoscalar currents must depend on the renormalization scale in such a way that the product $(m_j\mp m_i)\, J_{ij}^{S/P}$ remains invariant. Thus,
\be
\mu\,\frac{d}{d\mu}\, J_{ij}^{S/P}\; =\; \gamma(\alpha_s)\; J_{ij}^{S/P}\, ,
\ee
with $\gamma(\alpha_s)$ the quark-mass anomalous dimension. Since they depend on renormalization conventions, the scalar and pseudoscalar currents cannot be physical observables by themselves. Only renormalization-group invariant products such as $(m_j\mp m_i)\, J_{ij}^{S/P}$ can appear in measurable quantities.

Analysing the correlation functions of two current divergences, it is possible to derive the Ward identity \cite{Becchi:1980vz,Broadhurst:1981jk}
\bel{eq:WardIdentity}
q_\mu q_\mu \,\Pi^{\mu\nu}_{ij,\,V/A}(q)\, =\, (q^2)^2\, \Pi^L_{ij,\,V/A}(q^2)
\, =\, (m_j\mp m_i)^2\, \Pi_{ij,\, S/P}(q^2) +
(m_j\mp m_i)\, \langle 0|\bar q_j q_j \mp \bar q_i q_i |0\rangle\, ,
\ee
which relates the longitudinal correlators with the scalar and pseudoscalar two-point functions, 
\be 
\Pi_{ij,\, S/P}(q^2)\; \equiv\;  i \int d^4x \;\, \mathrm{e}^{iqx}\, 
\langle 0|T(\cJ^{S/P}_{ij}(x)\, \cJ^{S/P}_{ij}(0)^\dagger)|0\rangle\, .
\ee
An explicit proof is given in Appendix~\ref{app:WardIdentity}.
The upper and lower signs refer to the vector--scalar  and axial--pseudoscalar relations, respectively.
This identity shows explicitly that the quark condensate is an order parameter of the QCD chiral symmetry breaking. The last term is in fact related with the corresponding Goldstone-boson masses \cite{GellMann:1968rz},
\be 
(m_u + m_d)\,  \langle 0|\bar uu + \bar dd |0\rangle\, =\, -2\,m_\pi^2 F_\pi^2\, .
\ee

For massless quarks, the dynamical breaking of chiral symmetry by the QCD vacuum implies the presence of a Goldstone pole at $q^2=0$ in the axial two-point function \cite{Pich:2018ltt}: 
\be 
\Pi_{ij,A}^{\mu\nu}(q)\, =\, -q^\mu q^\nu\; \frac{2\, F^2}{q^2+i\epsilon}\, +\,\cdots\, ,
\ee
where $F$ is the pion decay constant $F_\pi$ in the chiral limit:  $\langle 0 | A^\mu_{ij}|\pi_{ij}(q)\rangle = i\sqrt{2} F_\pi\, q^\mu$. Obviously, this Goldstone contribution to $\Pi(q^2)$ cannot be unambiguously separated in transverse and longitudinal components. Nevertheless, Eq.~\eqn{eq:WardIdentity} guarantees that $(q^2)^2\, \Pi^L_{ij,\,\cJ}(q^2) = 0$ in the zero-mass limit.

The quark-mass contributions to $\Pi^L_{ij,\,\cJ}(q^2)$ are determined by the scalar and pseudoscalar correlators. Of special phenomenological interest are the perturbative contributions to $\Pi_{ij,\, S/P}(q^2)$, which generate the absorptive spectral function
\bel{eq:ScalarSpectralFunction}
\mathrm{Im}\, \Pi_{ij,\, S/P}(q^2)\; =\; \theta(q^2)\; q^2\;\frac{3}{8\pi}\;\left\{ 1 +
\sum_{n=1} G_n\, \left(\frac{\alpha_s(q^2)}{\pi}\right)^n +\, \cO(m_q^2/s)\right\} .
\ee
This is the relevant dynamical information that governs the hadronic width of the Higgs boson. The expansion coefficients are given by \cite{Herzog:2017dtz,Chetyrkin:1996sr,Baikov:2005rw}:
\bea\label{eq:ScalarSpectralFunction2}
\lefteqn{G_1\, =\, d_1^L \, =\, \frac{17}{3}\, ,}&&
\nonumber\\
\lefteqn{G_2\, =\, d_2^L - \frac{\pi^2}{12}\, \gamma_1\, (2\gamma_1-\beta_1)\, d_0^L\; =\;
35.93996 - 1.35865\; n_f\, ,} &&
\nonumber\\
\lefteqn{G_3\, =\, d_3^L - \frac{\pi^2}{12}\,\left\{ \left[2\gamma_2\, (2\gamma_1-\beta_1)-\gamma_1\beta_2\right]\, d_0^L + (\gamma_1 - \beta_1)\, (2\gamma_1 - \beta_1)\, d_1^L\right\}
}&&\nonumber\\ \lefteqn{\mbox{}\hskip .45cm =\,
164.13921 - 25.77119\; n_f + 0.258974\; n_f^2\, ,
}&&
\nonumber\\
\lefteqn{G_4\, =\, d_4^L 
-\frac{\pi^2}{12}\,\left\{ \left[ \gamma_3\, (4\gamma_1-3\beta_1) + 2 \gamma_2\, (\gamma_2-\beta_2)-\gamma_1\beta_3  
-\frac{\pi^2}{40}\, \gamma_1\, (\
\gamma_1-\beta_1)\, (2\gamma_1-\beta_1)\, (2\gamma_1-3\beta_1)\right] d_0^L \right. }&&
\nonumber\\
\lefteqn{\mbox{}\hskip 2.25cm\left.
+\, \left[ 4\gamma_2\, (\gamma_1-\beta_1) -\frac{1}{2}\,\beta_2\, (6\gamma_1-5\beta_1)\right] d_1^L
+ (\gamma_1-\beta_1)\, (2\gamma_1-3\beta_1)\, d_2^L\right\}
}&&
\nonumber\\
\lefteqn{\mbox{}\hskip .45cm =\, 39.33687 - 220.92924\, n_f + 9.68481\, n_f^2 - 0.02046\, n_f^3
\, ,}&&
\eea
where $d_n^L$ are the $\cO(m_q^2 \alpha_s^n)$ corrections to $D^L_{ij,\,\cJ}(Q^2)$ in Eq.~\eqn{DLij}. Notice the appearance of additional contributions proportional to $\pi^2$ (and $\pi^4$ in $G_4$), generated by the analytical continuation of the perturbative logarithms to the Minkowskian region. These $\pi^{2n}$ terms turn out to be quite important in this case, generating a sizeable numerical cancellation with the $d_n^L$ expansion coefficients. Taking $n_f=5$, $d_2^L = 42.03$ gets reduced to $G_2 = 29.15$, $d_3^L = 353.23$ converts into $G_3 = 41.76$, and $d_4^L = 3512.2$ changes to $G_4 = -825.75$.

\section{Inclusive observables} 
\label{sec:InclusiveObservables}

At high energies, naive perturbation theory is usually adopted to predict quantities such as $R_{e^+e^-}(s)$ and the hadronic decay widths of the $Z$, $W$ and Higgs bosons. These observables correspond to hadronic spectral functions (correlator discontinuities) on the physical cut, where the OPE is not justified. However, non-perturbative effects are still assumed to be suppressed by the factor $(\Lambda_{\mathrm{QCD}}/\sqrt{s})^4$, provided the physical scale $s$ is large enough and far away from thresholds and hadronic resonance structures.
Thus, one identifies the physical correlator $\Pi_{ij,\,\cJ}(s)$, corresponding to the analysed observable, with its pure perturbative approximation in terms of quarks and gluons:
\bel{eq:LocalDuality}
\mathrm{Im} \,\Pi_{ij,\,\cJ}(s)\, =\, 
\mathrm{Im} \,\Pi_{ij,\,\cJ}^{\mathrm{OPE}}(s)\, .
\ee
This is a a strong assumption, known as local quark--hadron duality, which is expected to be well satisfied at $\sqrt{s}\gtrsim M_W$ where non-perturbative corrections to the inclusive observables can indeed be neglected. Thus, one can perform precise predictions for the hadronic widths of the Standard Model electroweak bosons that have masses at the electroweak scale. However, the validity of this assumption for $R_{e^+e^-}(s)$ at much lower energies, and how fast the theoretical accuracy deteriorates, are open questions that can only be currently answered through explicit data analyses.

\subsection[$\sigma(e^+e^-\to\mathrm{hadrons})$]{$\boldsymbol{\sigma(e^+e^-\to\mathbf{hadrons})}$}
\label{subsec:Ree}

At lowest order, the inclusive production of hadrons in $e^+e^-$ annihilation proceeds through the exchange of virtual photons and $Z$ bosons. Well below the $Z$ peak, we can focus on the $\gamma$-exchange amplitude that involves the electromagnetic vector current $V^\mu_{\mathrm{e.m.}} = \sum_i Q_i\,\bar q_i\gamma^\mu q_i$ (electroweak corrections can be easily added whenever needed).
In order to better analyse the QCD dynamics, it is convenient to consider the
ratio of the electromagnetic $e^+e^-\to\mathrm{hadrons}$ and $e^+e^-\to\mu^+\mu^-$ cross sections, where many common factors cancel out. Using Eq.~\eqn{eq:ImPi}, this ratio can be easily written in terms of the non-singlet and singlet spectral functions of the electromagnetic current correlator:
\begin{eqnarray}\label{eq:R_ee}
R_{e^+e^-}(s)& \equiv&
\frac{\sigma(e^+e^-\to \mathrm{hadrons})}{\sigma(e^+e^-\to\mu^+\mu^-)}\;
=\; 12 \pi \;\left\{ \sum_i Q_i^2\;\, \mathrm{Im}\, \Pi(s)
+ \biggl(\sum_i Q_i\biggr)^{\! 2}\; \mathrm{Im}\, \Delta^\mathrm{S}\Pi_V(s)
\; + \; \mathcal{O}\left(\frac{m_q^2}{s}\right) \right\}
\nonumber\\
& =& \sum_{i} Q_i^2\; N_C \; \left\{ 1\, +\,
\sum_{n\geq 1} F_n \,\left({\alpha_s(s)\over\pi}\right)^{\! n} \right\}
\, + \, \mathcal{O}\left(\frac{m_q^2}{s}\right)\, .
\end{eqnarray}
The singlet contribution is strongly suppressed by the sum over quark electric charges of different signs, and it has been included as a small correction to the non-singlet coefficients,
\be
F_{n<3}\, =\, F_n^{\mathrm{NS}}\, ,
\qquad\qquad\qquad
F_{n\ge 3}\, =\,  F_n^{\mathrm{NS}} + F_n^{\mathrm{S}}\; \frac{\left(\sum_i Q_i\right)^{\! 2}}{\sum_i Q_i^2}\, ,
\ee
where $F_3^{\mathrm{S}} = d_3^V$ and $F_4^{\mathrm{S}} = d_4^V$.
For $n_f=5$ flavours, one gets\ $F_1 = 1$, $F_2 = 1.40923$,
$F_3 = -12.8046$ and $F_4 = -80.4337$ \cite{Baikov:2010je,Baikov:2012zn,Baikov:2012er}.

\begin{figure}[t]
\centering
\includegraphics[width=12cm,clip]{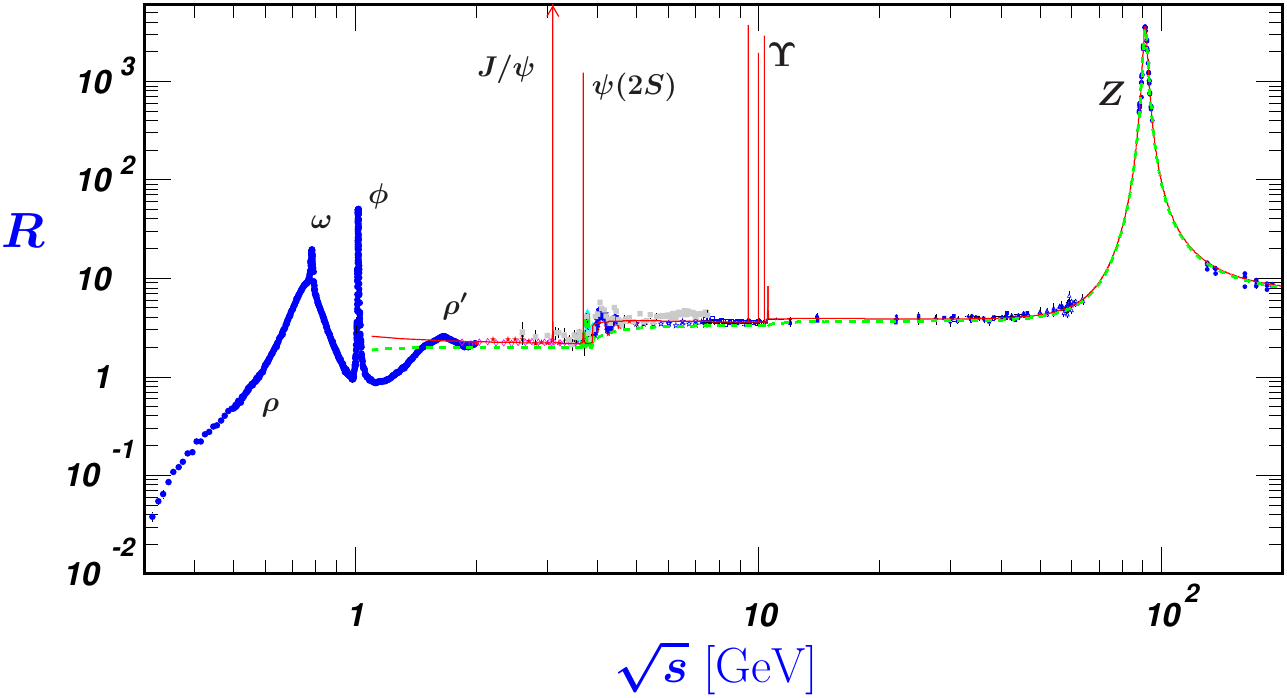}
\caption{World data on $R_{e^+e^-}(s)$, compared with the naive quark-model prediction  (broken green line) and the perturbative QCD result (solid red line). Data errors above 2~GeV are only statistical. Figure taken from Ref.~\cite{Zyla:2020zbs}.} 
\label{fig:Ree}
\end{figure}

Figure~\ref{fig:Ree} displays the available experimental measurements of the total hadronic production cross section, including the $Z$-exchange contribution, normalized to the electromagnetic muon-production cross section $\sigma(e^+e^-\to\mu^+\mu^-) = 4\pi \alpha^2(s)/(3s)$~\cite{Zyla:2020zbs}. As expected, the perturbative QCD prediction (continuous red line) agrees very well with the data in those regions where the cross section is smooth, {\it i.e.}, where multi-hadron thresholds and resonances are smeared out. Moreover, in the resonance region the correct average value is also reproduced. 
Around the $c\bar c$ and $b\bar b$ thresholds, one can also appreciate the jumps from $n_f$ equal three to four and from four to five. For comparison, the dashed green lines show the naive quark-model expectation ($\alpha_s=0$) that only counts the number of active quarks in a given region, weighted by their charges squared.


At high energies, the very precise theoretical prediction could be used to perform a N${}^3$LO determination of $\alpha_s(s)$. Unfortunately, since the QCD contribution only amounts to a small few per-cent correction, the experimental uncertainties are too large to achieve a competitive result.

\subsection[$\Gamma(Z\to\mathrm{hadrons})$]{$\boldsymbol{\Gamma(Z\to\mathrm{\bf hadrons})}$}
\label{subsec:Zwidth}

The electroweak neutral current involves a combination of the vector and axial-vector currents, weighted with the corresponding quark couplings to the $Z$ boson:
\be
J^\mu_Z\, =\, \sum_i (g_V^i V_{ii}^\mu - g_A^i A_{ii}^\mu)\, ,
\qquad\qquad
g_V^i\, =\, 2\, T^3_i \left( 1 - 4\, |Q_i|\,\sin^2{\theta_W}\right)\, ,
\qquad\qquad
g_A^i\, =\, 2\, T^3_i \, ,
\ee
with $T^3_i$ the third component of weak isospin. The hadronic decay width of the $Z$ is then governed by the absorptive part of the correlation function of two $J^\mu_Z$ currents (only the transverse piece contributes). It is given by
\be
\Gamma(Z\to\mathrm{hadrons}) \; =\; \frac{G_F M_Z^3}{24\pi\sqrt{2}}\;
N_C\;\left\{
\sum_i \left( |g_V^i|^2 + |g_A^i|^2\right) r_{\mathrm{NS}}
\, +\, \sum_i  |g_V^i|^2  r_{\mathrm{S}}^V\, +\, r_{\mathrm{S;t,b}}^A\, +\, \Delta_Z
\right\}
\, ,
\ee
where the sum over $i = u, d, s, c, b$ includes the five kinematically allowed  flavour decay channels, and
\be 
r_{\mathrm{NS}}\, =\, 1 + \sum_{n=1}\;  F_n^{\mathrm{NS}}\, a_Z^n\, ,
\qquad\qquad\qquad
r_{\mathrm{S}}^V\, =\, \sum_{n=3}\;  F_n^{\mathrm{S}}\, a_Z^n\, ,
\qquad\qquad\qquad
a_Z \,\equiv\, \frac{\alpha_s(M_Z^2)}{\pi}\, ,
\ee
with $n_f=5$.
The non-singlet QCD contributions are included in the perturbative series $r_{\mathrm{NS}}$, while $r_{\mathrm{S}}^V$ incorporates the corrections from singlet vector topologies.
The singlet contributions to the axial correlator, generated by the bottom and top quarks, are given by $r_{\mathrm{S;t,b}}^A$ and have been calculated in the limit of a heavy top quark mass \cite{Baikov:2012er,Chetyrkin:1993jm,Larin:1993ju,Chetyrkin:1993ug}: 
\bea 
r_{\mathrm{S;t,b}}^A & =& \left( \ell_t -3.0833 \right)  a_Z^2
\, +\, \left( 1.9167\,\ell_t^2 + 3.7222\,\ell_t -15.9877 \right) a_Z^3
\no\\ & + &
\left( 3.6736\,\ell_t^3+  14.6597\,\ell_t^2 -17.6637\,\ell_t +49.0309 \right)  a_Z^4
\, +\, \cO(\alpha_s^5)\, ,
\eea
where $\ell_t = \log{(M_Z^2/M_t^2)}$ with $M_t$ the pole top mass.
Owing to their non-decoupling behaviour \cite{Veltman:1977kh,Bernabeu:1987me,Bernabeu:1990ws}, top-quark loops induce corrections that are not suppressed by inverse powers of the top mass and need to be taken explicitly into account in the $n_f=5$ QCD theory \cite{Kniehl:1989bb,Kniehl:1989qu}.
The additional term $\Delta_Z$ includes $m_b^2/M_Z^2$ and $m_b^4/M_Z^4$
corrections, which are known to $\cO(\alpha_s^4)$ \cite{Baikov:2004ku} and $\cO(\alpha_s^3)$ \cite{Chetyrkin:2000zk}, respectively, 
QCD contributions proportional to inverse powers of the top quark mass up to
$\cO(\alpha_s^3 M_Z^6/m_t^6)$ 
\cite{Kniehl:1989bb,Kniehl:1989qu,Chetyrkin:1993tt,Larin:1994va}, 
one-loop \cite{Bernabeu:1987me,Akhundov:1985fc,Beenakker:1988pv}, two-loop \cite{Dubovyk:2018rlg,Dubovyk:2019szj} and leading fermionic three-loop \cite{Chen:2020xzx} electroweak corrections, as well as mixed QCD-electroweak contributions \cite{Czarnecki:1996ei,Fleischer:1999iq,Freitas:2014hra}.

\begin{figure}[t]
\centering
\includegraphics[width=0.5\textwidth,clip]{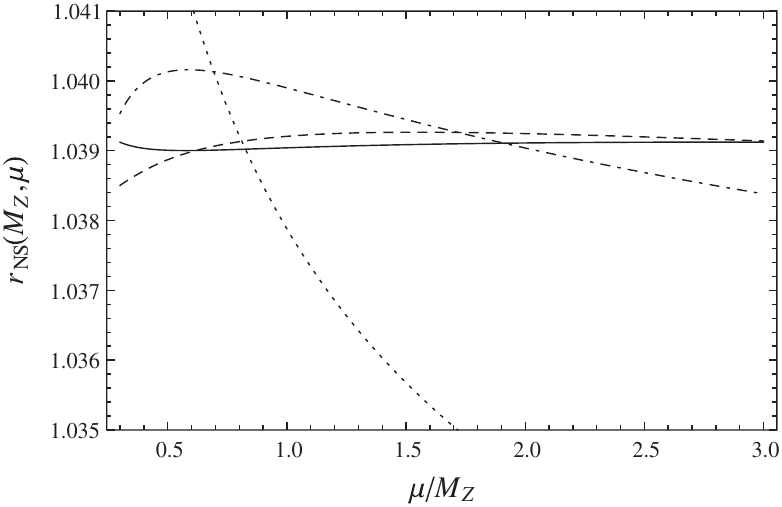}
\caption{Renormalization-scale dependence of the non-singlet perturbative correction $r_{\mathrm{NS}}$ at $\cO(\alpha_s)$ (dotted), $\cO(\alpha_s^2)$ (dash-dotted), $\cO(\alpha_s^3)$ (dashed) and $\cO(\alpha_s^4)$ (solid curve), with $\alpha_s(M_Z^2) = 0.1190$. Figure taken from Ref.~\cite{Baikov:2012er}.} 
\label{fig:rNS_mu}
\end{figure}

The achieved $\cO(\alpha_s^4)$ accuracy implies a very good theoretical control of the perturbative QCD series. Figure~\ref{fig:rNS_mu} displays the renormalization-scale dependence of $r_{\mathrm{NS}}$ at different loop approximations~\cite{Baikov:2012er}, exhibiting a clear stabilization of the result as the perturbative order increases. A similar reduction of the sensitivity to $\mu$ is observed in the much smaller corrections $r^V_{\mathrm{S}}$ and $r_{\mathrm{S;t,b}}^A$.  Although the $\cO(\alpha_s)$ result has a strong logarithmic dependence with $\mu$, it already gives a very good approximation for $\mu = M_Z$, as expected. The four-loop result has a very low sensitivity to the renormalization scale, which puts the corresponding scale uncertainty well below the per-mille level. Taking into account all other computed corrections, the total theoretical uncertainty remains around a factor of four smaller than the current experimental error \cite{dEnterria:2020cpv}.
Non-perturbative uncertainties can be safely neglected at the $Z$ mass scale. 

The hadronic width of the $Z$ boson was precisely measured by the LEP and SLC experiments \cite{ALEPH:2005ab}. The current world average value, $\Gamma(Z\to\mathrm{ hadrons})= (1744.4\pm 2.0)$~MeV \cite{Zyla:2020zbs}, can then be used to determine the strong coupling.\footnote{The actual phenomenological analyses are based on the precisely measured $Z$ pseudo-observables $\Gamma_Z^{\mathrm{tot}}$, $R_Z\equiv \Gamma(Z\to\mathrm{hadrons})/\Gamma(Z\to e^+e^-)$ and $\sigma_Z^{\mathrm{had}}$.}
Assuming the validity of the electroweak Standard Model, this observable is included in the global fit to electroweak precision data, which provides a very  
accurate value of $\alpha_s(M_Z^2)$ \cite{Zyla:2020zbs,dEnterria:2020cpv,Baak:2014ora,Haller:2018nnx}:
\begin{equation}\label{eq:alpha_Z}
\alpha_s^{(n_f=5)}(M_Z^2)\; \equiv\; \alpha_s(M_Z^2)\; =\; 0.1199\pm 0.0029\, .
\end{equation}

\subsection[$\Gamma(W\to\mathrm{hadrons})$]{$\boldsymbol{\Gamma(W\to\mathrm{\bf hadrons})}$}
\label{subsec:Wwidth}

The hadronic width of the $W$ boson involves the transverse component of the two-point correlation function of two charged left currents, {\it i.e.} the sum of the non-singlet vector and axial-vector correlators, weighted with the corresponding Cabibbo-Kobayashi-Maskawa (CKM) \cite{Cabibbo:1963yz,Kobayashi:1973fv} quark-mixing factors:
\be 
\Gamma(W\to\mathrm{hadrons})\, =\, \frac{G_F M_W^3}{6\pi\sqrt{2}}\;
N_C\; \sum_{i=u,c}\sum_{j=d,s,b} |V_{ij}|^2\;\left\{ r_{\mathrm{NS}}^W \, +\, \Delta_W
\right\} ,
\ee
with ($n_f=5$)
\be 
r_{\mathrm{NS}}^W\, =\, 1 \, +\, \sum_{n=1}\;  F_n^{\mathrm{NS}}\, 
\left(\frac{\alpha_s(M_W^2)}{\pi}\right)^n\, .
\ee
The factor $\Delta_W$ incorporates the small $m_q^2/M_W^2$ and $m_q^4/M_W^4$
QCD contributions, together with the electroweak \cite{Chang:1981qq,Bardin:1986fi,Denner:1990tx,Denner:1991kt,Kniehl:2000rb} and mixed QCD-electroweak \cite{Kara:2013dua} corrections.

In spite of its very accurate theoretical prediction, the $W$ hadronic width does not provide at present a competitive determination of the strong coupling.
Since $\alpha_s(M_W^2)$ is a small parameter, the size of the QCD correction $r_{\mathrm{NS}}^W-1$ amounts only to a $\sim 3\%$ effect. Therefore, below per-mille experimental accuracies would be required, which is far away from the precision of the currently measured $W$ observables \cite{Zyla:2020zbs}:
\be 
R_W\, \equiv\,\frac{\Gamma(W\to\mathrm{hadrons})}{\Gamma(W\to e^+e^-)}\, =\,
2.069\pm 0.019\, ,
\qquad\qquad\qquad
\Gamma_W^{\mathrm{had}}\, =\, (1405\pm 29)~\mathrm{MeV}\, .
\ee 
Assuming the unitarity of the quark mixing matrix, a recent combined analysis of the available $W$ data \cite{dEnterria:2020cpv,dEnterria:2016rbf} obtains the value
\be 
\label{eq:alpha_W}
\alpha_s(M_Z^2)\; =\; 0.101\pm 0.027\, ,
\ee
with an uncertainty one order of magnitude larger than \eqn{eq:alpha_Z}.
Alternatively, one can use the world average value of the strong coupling to perform a quantitative unitarity test of the CKM matrix \cite{Zyla:2020zbs}:
\be 
\sum_{j=d,s,b} \left( |V_{uj}|^2 + |V_{cj}|^2\right) \, =\, 2.002\pm 0.027\, .
\ee

\subsection[$\Gamma(H\to\mathrm{hadrons})$]{$\boldsymbol{\Gamma(H\to\mathrm{\bf hadrons})}$}
\label{subsec:Hwidth}

\begin{figure}[t]
\centering
\begin{minipage}[c]{0.4\textwidth}\centering
\includegraphics[width=0.7\textwidth,clip]{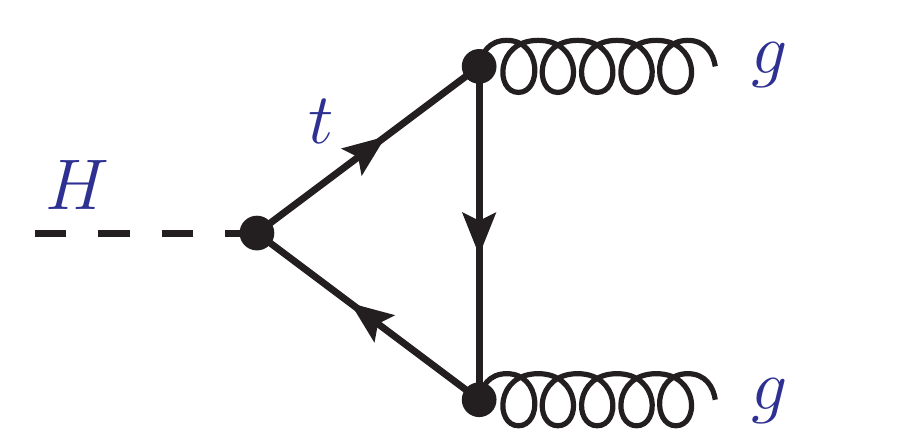}
\end{minipage}
\hskip 1.4cm
\begin{minipage}[c]{0.4\textwidth}\centering
\includegraphics[width=0.6\textwidth,clip]{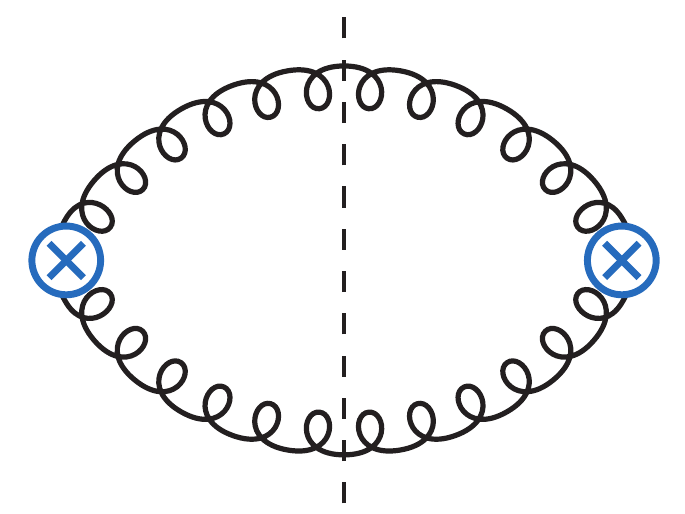}
\end{minipage}
\caption{Lowest-order top contribution to the effective $Hgg$ vertex (left) and correlation function of two $G^2$ operators at LO  (right). The crossed vertices denote the  insertions of the $G^a_{\mu\nu}G^{a\mu\nu}$ operator. The vertical dashed line indicates the two-gluon absorptive cut.} 
\label{fig:Hgg}
\end{figure}

Owing to the non-decoupling behaviour of the Yukawa couplings, the top quark generates through quantum corrections an effective Higgs coupling to gluons, 
$\cL_{Hgg} = -\frac{1}{v}\,  C_{Hgg}\, H
\,G^a_{\mu\nu}G^{a\mu\nu}$, 
with a coupling strength $C_{Hgg}$ which is a function of $M_H^2/m_t^2$ that becomes independent of the top quark mass in the limit $M_H\ll 2m_t$.  Figure~\ref{fig:Hgg} (left) displays the LO Feynman graph contributing to $C_{Hgg}$. This effective interaction completely dominates the decay width of the Higgs into two gluons because the other quark contributions are suppressed by their much lighter masses. In the heavy $m_t$ limit, the coefficient function $C_{Hgg}$ is known at N${}^4$LO \cite{Schroder:2005hy,Chetyrkin:2005ia,Herzog:2017dtz,Chetyrkin:1997un}.

Therefore, in the $n_f=5$ theory, the Higgs couples to strongly-interacting particles through the effective  Lagrangian
\bel{eq:Hcouplings}
\cL_H\, =\, -\frac{1}{v}\, H\,\left\{ \sum_i m_i \cJ_{ii}^S\, +\, 
C_{Hgg}\, G^a_{\mu\nu}G^{a\mu\nu}\right\}\, ,
\ee
where $i=u,d,s,c,b$, and $v = (\sqrt{2} G_F)^{-1/2} = 246$~GeV is the Higgs vacuum expectation value.
The total hadronic decay width of the Higgs can then be written in the form
\bea\label{eq:Hhadronicwidth}
\Gamma (H\to\mathrm{hadrons})&\!\! =&\!\! \frac{\sqrt{2}G_F}{M_H}\;\left\{\,  
\sum_i\, m_i^2\;  \mathrm{Im}\,\Pi_{ii,\, S}(M_H^2)
\, +\, C_{Hgg}^2\; \mathrm{Im}\,\Pi_{G^2}(M_H^2)
\, +\, 2\, C_{Hgg}\; \mathrm{Im}\,\Pi_{G^2S}(M_H^2)
\right\}
\nonumber\\ &\!\! = &\!\!
\frac{3\sqrt{2}G_F}{8\pi}\, M_H\,  \left\{
\sum_i\, \overline{m}_i^{\, 2}  
\left[\, 1 + \sum_{n=1} G_n\, a_H^n
\right]
\, + \,\frac{M_H^2}{27}\; a_H^2
\; K_{gg}
\, +\, \cO\left(\frac{m_i^2}{M_H^2}\right)
\right\}\, ,
\eea
with $\overline{m}_i \equiv m_i(M_H^2)$ and $a_H \equiv \alpha_s(M_H^2)/\pi$ the running quark masses and strong coupling at $\mu^2=M_H^2$, defined in $n_f=5$ QCD. The first term contains the absorptive contribution of the scalar quark correlator in Eq.~\eqn{eq:ScalarSpectralFunction}, which is generated by the five open $H\to q_i\bar q_i (g)$
decay channels. Since the Higgs Yukawa couplings are proportional to the quark masses, the hadronic width is obviously dominated by the $H\to b\bar b$ contribution. The dependence on the quark masses is fully known up to and including the $\cO(\alpha_s^2)$
contributions \cite{Harlander:1997xa,Chetyrkin:1997mb,Chetyrkin:1998ix}.

The second term in \eqn{eq:Hhadronicwidth} contains the gluonic component $\Gamma(H\to gg)$, {\it i.e.}, the top-induced contribution to the Higgs width, which can be extracted from the absorptive part of the two-point correlation function of two $G^a_{\mu\nu}G^{a\mu\nu}$ operators, $\Pi_{G^2}(q^2)$, at $q^2=M_H^2$ (figure~\ref{fig:Hgg}, right). It starts to contribute at $\cO(\alpha_s^2)$ and has been computed up to $\cO(\alpha_s^6)$ \cite{Herzog:2017dtz,Chetyrkin:1997un,Inami:1982xt,Djouadi:1991tka,Chetyrkin:1997iv,Chetyrkin:1997vj,Baikov:2006ch}:
\bea
K_{gg}&\!\! = &\!\! 1\, +\, 17.9167\; a_H \, +\, \left(156.81 - 5.7083\;\ell_{tH}\right)\; a_H^2\, 
+ \,\left( 452.46 -122.44\;\ell_{tH} + 10.94\;\ell_{tH}^2\right)\; a_H^3
\nonumber\\
&\!\! + &\!\! \left( -6502.1 - 1106.1\;\ell_{tH} + 284.09\;\ell_{tH}^2 -20.97\;\ell_{tH}^3\right)\; a_H^4 \, +\, \cO(a_H^5)
\, ,
\eea
where $\ell_{tH}\equiv \log{(\mu_t^2/M_H^2)}$ with $\mu_t = m_t(m_t^2)$ the running top quark mass evaluated at its own mass scale.
Taking $\mu_t = 164$~GeV and $\alpha_s(M_Z^2) = 0.118\pm 0.012$, one finds $K_{gg} = 1.844\pm 0.046$ \cite{Herzog:2017dtz}, where the error is dominated by the assumed uncertainty in $\alpha_s$. The dependence on the numerical value of $\mu_t$ is very weak; a change of $\mu_t$ by 4 GeV only modifies the result by 0.04\%.
Taking into account the QCD correction to the $b\bar b$ width,
$K_{b\bar b} = 1 + \sum_n G_n a_H^n = 1.241\pm 0.031$, the relative weight of the 2-gluon and $b\bar b$ components is given  by
\be 
\frac{\Gamma(H\to gg)}{\Gamma(H\to b\bar b)}\, \approx\, \frac{a_H^2 M_H^2}{27\, \overline{m}_b^{\, 2}}\;\frac{K_{gg}}{K_{b\bar b}}\, =\, 0.14 \, ,
\ee
for $\overline{m}_b = 2.773$~GeV [$m_b(m_b^2) = 4.163$~GeV], up to $\cO(m_b^2/M_H^2)$ corrections.

There is in addition a mixed correlation function $\Pi_{G^2S}$ between the two terms in Eq.~\eqn{eq:Hcouplings}, which also starts at $\cO(\alpha_s^2)$. However, since the bottom quark contribution to the $Hgg$ vertex is suppressed by a factor $m_b^2/M_H^2$, it has a smaller impact on the total Higgs hadronic width. This contribution and other small corrections suppressed by powers of $m_t$ have been analysed in Refs.~\cite{Chetyrkin:1995pd,Larin:1995sq,Schreck:2007um,Davies:2017xsp}.

\begin{figure}[t]
\centering
\includegraphics[width=0.45\textwidth,clip]{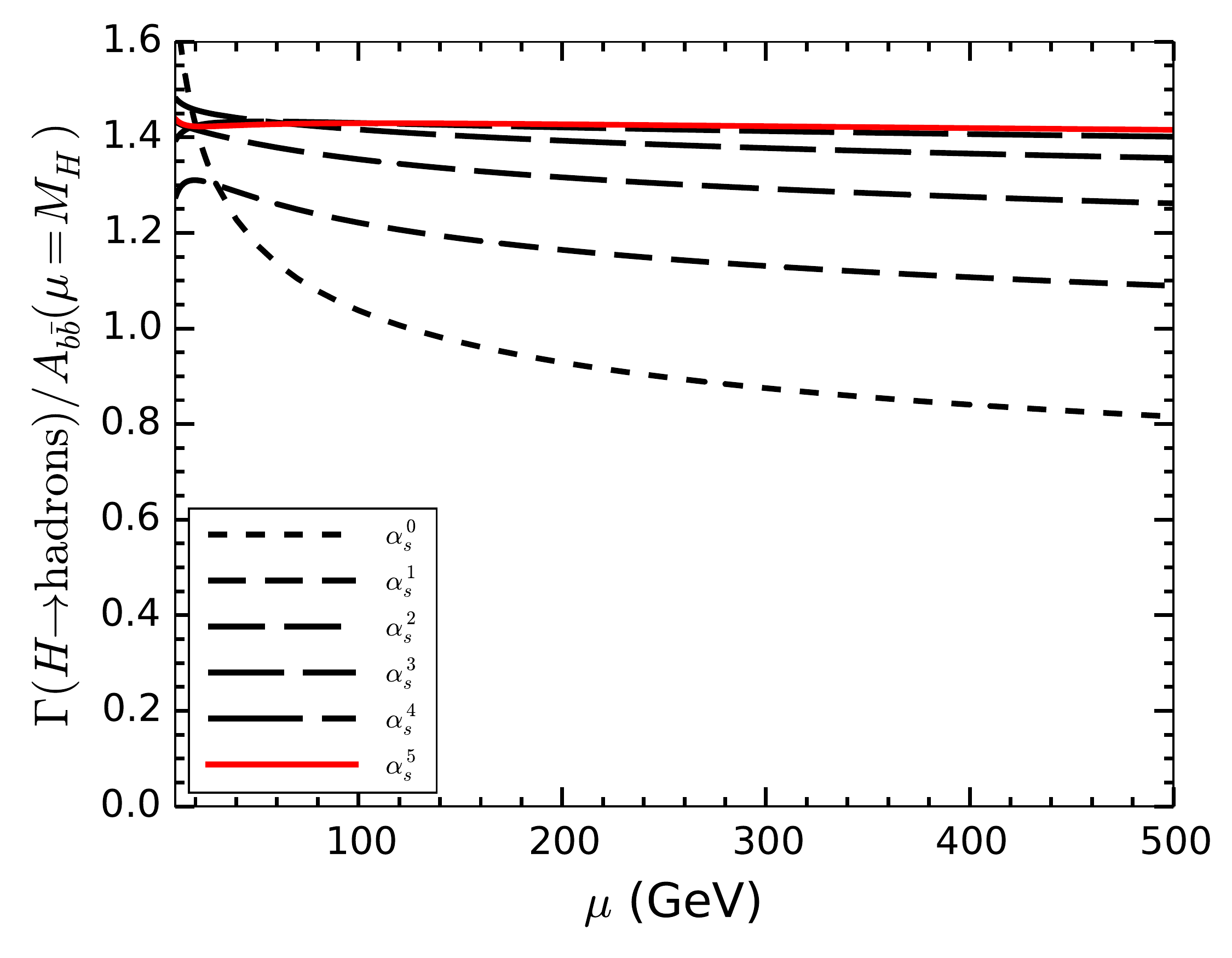}
\caption{Renormalization scale dependence of the QCD corrections to the total Higgs hadronic width at different perturbative orders. Figure taken from Ref.~\cite{Davies:2017xsp}.} 
\label{fig:Hwidth_mu}
\end{figure}

The different QCD contributions can be all combined in a global correction factor $\cK(\mu)$ to the LO $H\to b\bar b$ decay width:
\be 
\Gamma (H\to\mathrm{hadrons}) \, =\, A_{b\bar b}(\mu^2) \;\cK(\mu^2) \,\equiv\,
\frac{3\sqrt{2}G_F}{8\pi}\, M_H\, m_b^2(\mu^2)\; \cK(\mu^2)\, .
\ee
The lowest-order expression  $A_{b\bar b}(\mu^2)$ reabsorbs the leading QCD logarithmic corrections into the running bottom mass squared \cite{Braaten:1980yq}. Therefore, it has a strong dependence on the renormalization scale $\mu$ that is cancelled by the perturbative QCD correction $\cK(\mu^2)$, which is fully known at N${}^4$LO, {\it i.e.}, $\cO(\alpha_s^4)$. Figure~\ref{fig:Hwidth_mu}~\cite{Davies:2017xsp} displays the variation of the product $\cK(\mu^2)\,  m_b^2(\mu^2)/ m_b^2(M_H^2)$ over a broad range of $\mu$ values reaching up to 500~GeV, at the different perturbative orders. One observes a steady flattening of the curves as the precision increases, reaching an almost $\mu$-independent  N${}^4$LO result. The continuous line includes also the $\cO(\alpha_s^5)$ corrections to $K_{gg}$.
%
%

The final theoretical prediction for the Higgs hadronic width \cite{deFlorian:2016spz,Spira:2016ztx},
\be
\Gamma (H\to b\bar b)\, =\, (2.38\pm 0.06)\; \mathrm{MeV}\, ,
\qquad\qquad\qquad
\Gamma (H\to gg)\, =\, (0.34\pm 0.12)\; \mathrm{MeV}\, ,
\ee
includes electroweak corrections \cite{Fleischer:1980ub,Bardin:1990zj,Dabelstein:1991ky,Kniehl:1991ze} and mixed electroweak-QCD contributions \cite{Kwiatkowski:1994cu,Kniehl:1994ju,Chetyrkin:1996wr,Mihaila:2015lwa}. 
The uncertainties are obviously dominated by the current experimental error on the Higgs mass. The main QCD uncertainties originate in the input values of $\alpha_s$ and the bottom quark mass.

\section{The hadronic width of the $\boldsymbol{\tau}$ lepton}
\label{sec:Rtau}

\begin{figure}[t]
\centering
\includegraphics[width=0.35\textwidth,clip]{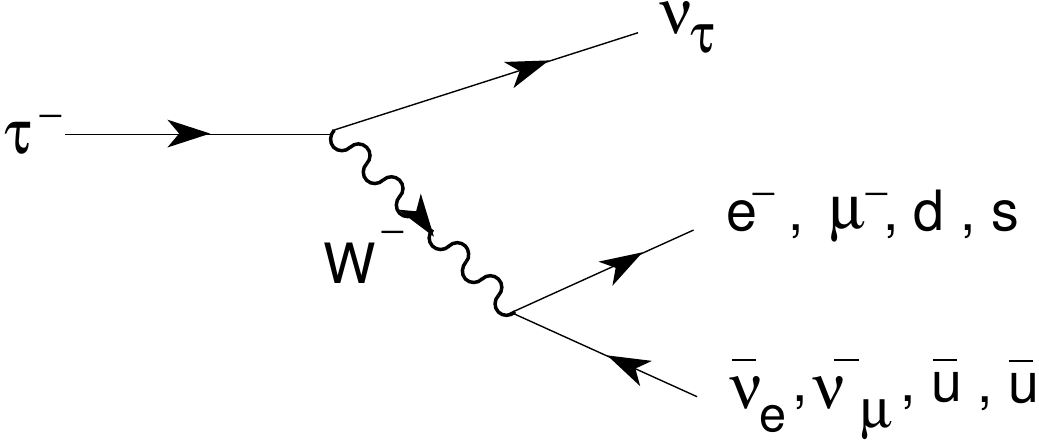}
\caption{Tree-level Feynman diagram generating the decays $\tau^-\to\nu_\tau X^-$ ($X^- =  e^-\bar\nu_e,\, \mu^-\bar\nu_\mu,\, d\bar u,\, s\bar u$).} 
\label{fig:TauDecay}
\end{figure}

The $\tau$ lepton decays through the emission of a virtual $W$ boson that
generates four possible $\nu_\tau X^-$ final states, with $X^- =  e^-\bar\nu_e,\, \mu^-\bar\nu_\mu,\, d\bar u,\, s\bar u$ (figure~\ref{fig:TauDecay}).
If final fermion masses and QCD effects are ignored, the universality of the $W$ couplings implies that the four decay modes have equal probabilities, except for an additional global factor $N_C |V_{ui}|^2$ ($i=d,s$) in the two semileptonic channels. Since $|V_{ud}|^2 + |V_{us}|^2 = 1 - |V_{ub}|^2 \approx 1$, the total $\tau$ hadronic width is then predicted to be
a factor of $N_C=3$ larger than $\Gamma(\tau^-\to\nu_\tau e^-\bar\nu_e)$. Experimentally the ratio of the hadronic and electronic decay widths is around $3.6$ \cite{Zyla:2020zbs,Pich:2013lsa,Amhis:2019ckw}.
The missing QCD corrections enhance the hadronic $\tau$ decay width by about 20\%, which is a sizeable effect but much smaller than the naive expectation for a low-energy observable at the scale $m_\tau = 1.777$~GeV.

The hadronic decay rate of the $\tau$ can be written as an integral of the  left-handed $ud$ and $us$ spectral functions over the total invariant mass $s$ of the final-state hadrons \cite{Braaten:1991qm}:
\be
\label{eq:R_tau}
R_{\tau} \, \equiv\,
\frac{ \Gamma [\tau^- \to \nu_\tau +\mathrm{hadrons}]}{ \Gamma [\tau^- \to \nu_\tau e^- {\bar \nu}_e]}
\; =\; 12 \pi\, S_{\mathrm{EW}} \int^{m_\tau^2}_0 {ds \over m_\tau^2 } \,
 \left(1-{s \over m_\tau^2}\right)^2
\biggl[ \left(1 + 2 {s \over m_\tau^2}\right)\,
 \mathrm{Im}\, \Pi^{T}_\tau(s)
 \, +\, \mathrm{Im}\, \Pi^{L}_\tau(s) \biggr]\,  ,
\ee
with ($J=T,L$)
\be\label{eq:pi}
\Pi^{J}_\tau(s)  \; \equiv  \;
  |V_{ud}|^2 \, \left( \Pi^{J}_{ud,V}(s) + \Pi^{J}_{ud,A}(s) \right)
\, + \,
|V_{us}|^2 \, \left( \Pi^{J}_{us,V}(s) + \Pi^{J}_{us,A}(s) \right)\, .
\ee
The global factor $S_{\mathrm{EW}}=1.0201\pm 0.0003$ accounts for the (renormalization-group improved) electroweak radiative corrections \cite{Marciano:1988vm,Braaten:1990ef,Erler:2002mv}.

Experimentally, it is possible to separate the total hadronic width into inclusive contributions associated with the different quark currents:
\be\label{eq:r_tau_v,a,s}
 R_\tau \; = \; R_{\tau,V} + R_{\tau,A} + R_{\tau,S}\, .
\ee
The first two terms in \eqn{eq:pi} correspond to $R_{\tau,V}$ and $R_{\tau,A}$, while the remaining Cabibbo-suppressed contributions are included in $R_{\tau,S}$. Non-strange decays into an even or odd number of pions belong to the vector ($R_{\tau,V}$) or axial-vector ($R_{\tau,A}$) widths, respectively. Strange decays ($R_{\tau,S}$) are characterized by an odd number of kaons in the final state.

In the observables discussed in the previous section, the invariant mass of the hadronic final states had a fixed value determined by the mass of the decaying boson, or by the centre-of-mass energy of the electron--positron beams in $R_{e^+e^-}(s)$. However, owing to the emitted $\tau$ neutrino, the hadronic $\tau$ decay width in Eq.~\eqn{eq:R_tau} involves an integral over the whole kinematical range allowed for the momentum flowing along the virtual $W$ propagator. Thus, $R_\tau$ is a much more inclusive observable.
Although the hadronic spectral functions cannot be predicted at present from first principles in the low-$s$ region
entering the integration~\eqn{eq:R_tau}, the integral itself can be rigorously calculated, thanks to the analyticity properties of the correlators $\Pi^{J}_{ij,\,\cJ}(s)$. They are analytic functions of $s$ in the entire complex plane, except along the positive real $s$ axis where their imaginary parts have discontinuities. Using the closed contour in figure~\ref{fig:contour},
$R_\tau$ can then be expressed as a contour integral in the complex $s$ plane running counter-clockwise around the circle $|s|=m_\tau^2$ \cite{Braaten:1991qm,Braaten:1988hc,Narison:1988ni,Braaten:1988ea}:
%
\begin{figure}[tb]\centering
\includegraphics[width=6.4cm,clip]{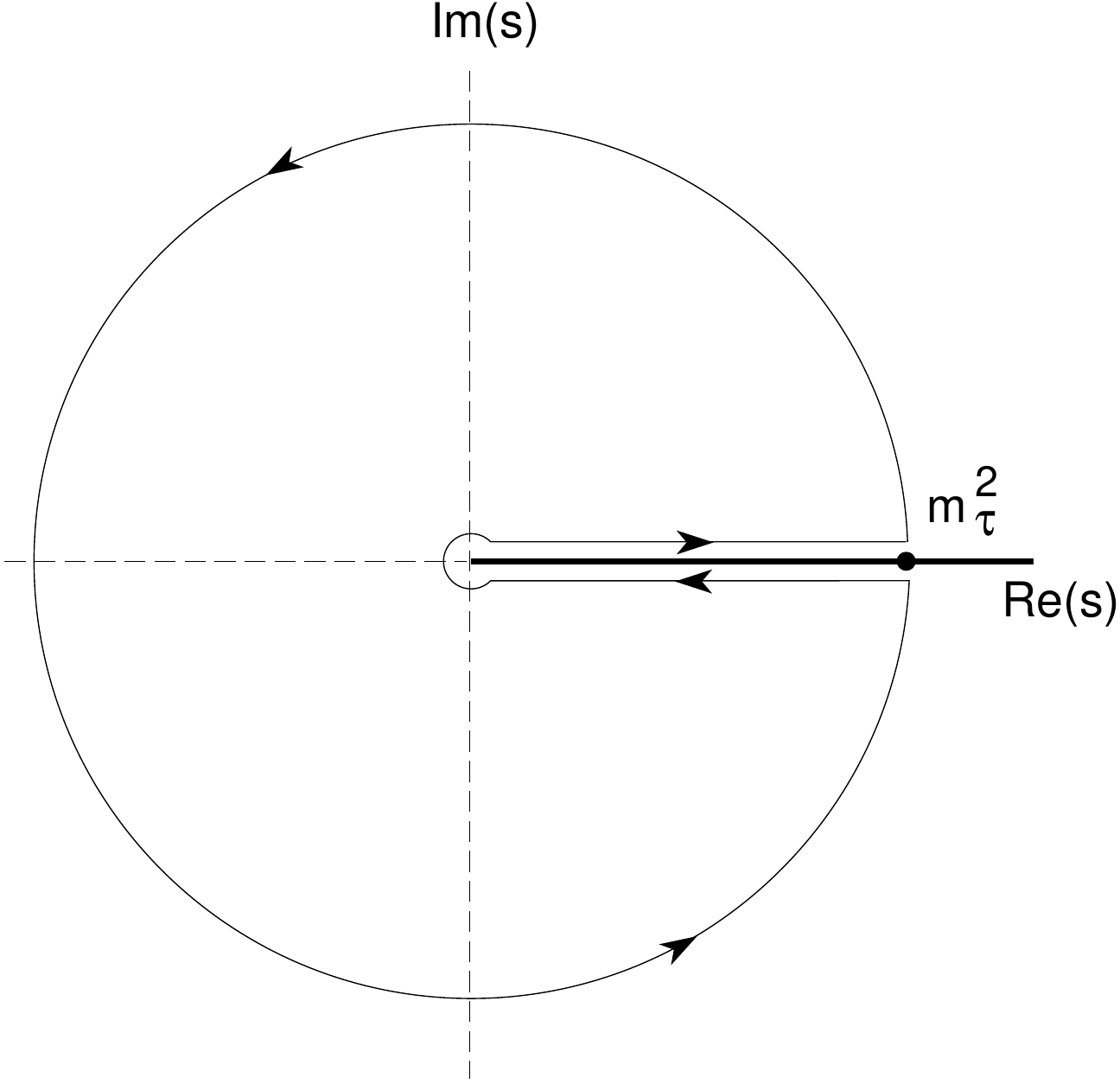}
\caption{Integration contour in the complex $s$ plane, used to obtain
Eq.~\protect\eqn{eq:circle}.}
\label{fig:contour}
\end{figure}
%
\bel{eq:circle}
 R_\tau\;  =\; 6 \pi i\, \, S_{\mathrm{EW}}\, \oint_{|s|=m_\tau^2} {ds \over m_\tau^2} \;
 \left(1 - {s \over m_\tau^2}\right)^2\,
 \left[ \left(1 + 2\, {s \over m_\tau^2}\right)\, \Pi^{L+T}_\tau(s)
         - 2\, {s \over m_\tau^2}\, \Pi^{L}_\tau(s) \right] \, .
\ee
The integral along the whole complex contour in figure~\ref{fig:contour} vanishes, because there are no singularities enclosed within the contour. Moreover, Schwarz's reflection principle implies that $\Pi^{J}_{\tau}(s+i\epsilon) - \Pi^{J}_{\tau}(s-i\epsilon) = 2 i\, \mathrm{Im}\, \Pi^{J}_{\tau}(s+i\epsilon)$, so that the sum of the integrals below and above the real axis is equal to Eq.~\eqn{eq:R_tau} up to a global minus sign. The mathematical identity \eqn{eq:circle} then follows.\footnote{Instead of the circle $|s|= m_\tau^2$, the integration \eqn{eq:circle} could obviously be made along any complex contour starting and finishing at $s=m_\tau^2$.}  

Eq.~\eqn{eq:circle} only requires the correlators for complex values of $s$, with $|s|= m_\tau^2$ that is significantly larger than the scale associated with non-perturbative effects in QCD. The OPE is valid in the whole integration range except, perhaps, the point $s=m_\tau^2$ where the circuit touches the real axis. However, the integrand contains a kinematical double zero, precisely at this point, heavily suppressing the contribution from
the region near the branch cut.
Moreover, the combinations $\Pi^{L+T}_\tau(s)$ and $s\,\Pi^{L}_\tau(s)$, can be unambiguously calculated using the OPE, which allows us to perform a rigorous theoretical prediction of $R_\tau$, organized as a systematic expansion in inverse powers of $m_\tau^2$ with coefficients that depend only logarithmically on $m_\tau$ \cite{Braaten:1991qm}:
\be\label{eq:r_v}
R_{\tau,V/A}\;  = \; {3 \over 2}\, |V_{ud}|^2\,
   S_{\mathrm{EW}}\; \left( 1 + \delta_{\mathrm{P}} +
      \sum_{D=2,4\ldots} \delta^{(D)}_{ud,V/A} \right)\,  ,
\ee
\be \label{eq:r_s}
R_{\tau,S}\;  =\;
 3\, |V_{us}|^2\, S_{\mathrm{EW}}\; \left( 1 + \delta_{\mathrm{P}} +
  \sum_{D=2,4\ldots} \delta^{(D)}_{us} \right)\, .
\ee
The corrections to $R_\tau$ from dimension-$D$ operators have been expressed in terms of the fractional corrections $\delta^{(D)}_{ij,\,\cJ}$ to the LO contribution from the current with quantum numbers $\{ ij,\cJ\}$, and 
\be
\delta^{(D)}_{ij}\, =\, (\delta^{(D)}_{ij,V} + \delta^{(D)}_{ij,A})/2
\ee
is the average of the vector and axial-vector corrections. 
The purely perturbative QCD correction, neglecting quark masses, $\delta_{\mathrm{P}} = \delta^{(0)}_{ij,\,\cJ}$ is the same for all the components of $R_\tau$.

If the light quark masses are neglected, $s \,\Pi_\tau^{L}(s) = 0$ and Eq.~(\ref{eq:circle}) only receives contributions from the correlator
$\Pi^{L+T}_\tau(s)$, multiplied by a global coefficient factor
$(1 - x)^2 (1 + 2 x) = 1 - 3 x^2 + 2x^3$  with $x\equiv s/m_\tau^2$.
According to Cauchy's theorem, the only non-perturbative contributions to the circle integration in (\ref{eq:circle}) originate then from operators of dimensions $D=6$ and 8, up to tiny logarithmic running corrections. The leading non-perturbative operators of dimension four can only contribute to $R_\tau$ with an additional suppression factor of $\cO(\alpha_s^2)$, which makes their effect negligible \cite{Braaten:1991qm}.

The Cabibbo-allowed component of the $\tau$ hadronic width,
$ R_{\tau,V+A} = R_{\tau,V} + R_{\tau,A}$,
is then a gold-plated observable to test perturbative QCD. Quark mass effects are tiny (smaller than $10^{-4}$) and the non-perturbative correction is heavily suppressed by six powers of the $\tau$ mass. Since the strong coupling is large at the $\tau$ mass scale, the perturbative contribution is very sizeable, $\delta_{\mathrm{P}}\sim 20\%$, and dominates the theoretical prediction, making possible to perform an accurate determination of the fundamental QCD coupling \cite{Braaten:1991qm,Narison:1988ni}.

\subsection[Perturbative contribution to $R_\tau$]{Perturbative contribution to $\boldsymbol{R_\tau}$}

The QCD correlation function of two left-handed charged currents only receives contributions from non-singlet topologies. Using integration by parts, $R_\tau$ can be more conveniently expressed in terms of the Adler functions $D^{L+T}_{ij,\,\cJ}(Q^2)$ and $D^{L}_{ij,\,\cJ}(Q^2)$ \cite{Pich:1998yn,Pich:1999hc}. In the massless quark limit, the perturbative contribution is given by
\be
\label{deltaP}
1+\delta_{\mathrm{P}}\, =\, -2\pi i \oint_{|s|=m_\tau^2} \, \frac{{\rm d}s}{s}\;
\left(1-{s\over m_\tau^2}\right)^3  
\left( 1 + {s\over m_\tau^2}\right)\,  D(s) \, .  
\ee
Inserting the expansion of $D(s)$ in powers of $\alpha_s(-s)$ in Eq.~\eqn{eq:Adler}, $\delta_{\mathrm{P}}$ can be written in the form
\be\label{eq:r_k_exp}
\delta_{\mathrm{P}} \; =\;
\sum_{n=1}  K_n \, A^{(n)}(\alpha_s)
\; =\; \sum_{n=1}\,  (K_n + g_n) \, a_\tau^n \;\equiv\;
\sum_{n=1}\,  r_n \, a_\tau^n \, ,
\ee
where $K_n$ are the Adler-function coefficients $K_{n,0}$ for $n_f=3$ flavours,\footnote{The $\tau$ hadronic width does not involve any charmed particles and, therefore, it is better described in the $n_f=3$ QCD effective theory with only the three light quark flavours. Since a virtual charm quark could only appear in internal fermion loops, the charm corrections to $R_\tau$ are suppressed by a factor $\alpha_s^2 m_\tau^2/(4 m_c^2)$ \cite{Chetyrkin:1993tt}. They are currently known up to and including the
$\cO[\alpha_s^3 m_\tau^6/(2 m_c)^6]$ \cite{Larin:1994va} and induce a very tiny numerical effect, $\Delta_c\delta_{\mathrm{P}}\sim 0.0004$, which is much smaller than the expected size of the unknown $\cO(\alpha_s^5)$ contribution.}
and the functions \cite{LeDiberder:1992jjr}
\be\label{eq:a_xi}
A^{(n)}(\alpha_s) \; = \; {1\over 2 \pi i}\,
\oint_{|s| = m_\tau^2} {ds \over s} \;
  \left({\alpha_s(-s)\over\pi}\right)^n\;
 \left( 1 - 2 {s \over m_\tau^2} + 2 {s^3 \over m_\tau^6}
         - {s^4 \over  m_\tau^8} \right)
\; = \;      a_\tau^n\, +\, \cO(a_\tau^{n+1})    
\ee
are contour integrals in the complex plane, which only depend on
$a_\tau\equiv\alpha_s(m_\tau^2)/\pi$. 
Expanding the integrals in powers of $a_\tau$ \cite{Braaten:1991qm}, 
one obtains a perturbative series for $\delta_{\mathrm{P}}$ with coefficients $r_n\equiv K_n + g_n$. The additional contributions $g_n$ are generated
by the running of the strong coupling along the integration contour and turn out to be rather large:
\bea
\lefteqn{
g_1 \, =\, 0\, ,
\qquad\qquad\qquad
g_2 \, =\, -\frac{19}{24}\,\beta_1\, K_1\, ,
\qquad\qquad\qquad
g_3 \, =\, \left[\left(\frac{265}{288}-\frac{\pi^2}{12}\right)\,\beta_1^2  -\frac{19}{24}\,\beta_2\right] K_1
-\frac{19}{12}\,\beta_1\, K_2
\, ,
}&&
\nonumber\\
\lefteqn{
g_4 \, =\, \left[\left(-\frac{3355}{2304}+\frac{19}{96}\,\pi^2\right)\beta_1^3 +\left(\frac{1325}{576}-\frac{5}{24}\,\pi^2\right)\beta_1\beta_2-\frac{19}{24}\,\beta_3\right] K_1
+\left[\left(\frac{265}{96}-\frac{\pi^2}{4}\right)\beta_1^2-\frac{19}{12}\,\beta_2\right] K_2 -\frac{19}{8}\,\beta_1\, K_3
\, ,}&&\hskip 15.7cm\mbox{}
\nonumber\\
\lefteqn{
g_5 \, =\,
\left[\left(\frac{41041}{13824}-\frac{265}{576}\,\pi^2 +\frac{\pi^4}{80}\right)\beta_1^4 
-\left(\frac{43615}{6912}-\frac{247}{288}\,\pi^2\right)\beta_1^2\beta_2
+\left(\frac{265}{96}-\frac{\pi^2}{4}\right)\left(\beta_1\beta_3+\frac{1}{2}\,\beta_2^2\right)
-\frac{19}{24}\,\beta_4
\right] K_1}&&
\nonumber\\
\lefteqn{\phantom{g_5}\, +\,
\left[\left(-\frac{3355}{576}+\frac{19}{24}\,\pi^2\right)\beta_1^3
+\left(\frac{1855}{288} -\frac{7}{12}\,\pi^2\right)\beta_1\beta_2-\frac{19}{12}\,\beta_3\right] K_2
+ \left[\left(\frac{265}{48} -\frac{\pi^2}{2}\right)\beta_1^2-\frac{19}{8}\,\beta_2
\right] K_3 -\frac{19}{6}\,\beta_1\, K_4\, .
}&& 
\nonumber\\
\eea
Their numerical values are compared in table~\ref{tab:Kcoeff} with the $K_n$ contributions from the original Adler function. Clearly, the running corrections dominate the final result. Although the five-loop coefficient $K_5$ has not been yet computed, the running factor $g_5$ is fully determined by the known values of $K_{m<5}$ and $\beta_{m<5}$, and turns out to be large and positive.

%
\begin{table}[width=.68\linewidth,cols=6,htb]   
\caption{Perturbative coefficients of the Adler function and the FOPT approximation to $\delta_{\mathrm{P}}$.}
\label{tab:Kcoeff}
\begin{tabular*}{\tblwidth}{@{} LLLLLL@{} } 
\toprule 
$n$ & 1 & 2 & 3 & 4 & 5
\\ \midrule 
$K_n$ & 1 & $1.63982$ & $6.37101$ & $49.0757$ &
\\
$g_n$ & 0 & $3.56250$ & $19.9949$ & $78.0029$ & $307.783$
\\
$r_n$ & 1 & 5.20232 & $26.3659$ & $127.079$ &
\\ \bottomrule 
\end{tabular*}
\end{table}
%

The integrals $A^{(n)}(\alpha_s)$ can be computed numerically with very high accuracy,
using the exact solution (up to unknown $\beta_{n>5}$ contributions) for $\alpha_s(-s)$
given by the renormalization-group $\beta$-function equation \cite{LeDiberder:1992jjr}.
Table~\ref{tab:Afun} shows the numerical values
for $A^{(n)}(\alpha_s)$ with $n\le 5$, obtained at the one-, two-, three-, four- and five-loop approximations, together with the corresponding {\it contour-improved perturbation theory} (CIPT) \cite{LeDiberder:1992jjr,Pivovarov:1991rh} result  $\delta_{\mathrm{P}} = \sum_{n=1}^4\, K_n\, A^{(n)}(\alpha_s)$, taking $\alpha_s(m_\tau^2) = 0.33$.
The numbers in the table exhibit a very good perturbative convergence. The value
of $\delta_{\mathrm{P}}$ predicted at four loops only differs by 0.5\% from the one-loop result. Adding the five-loop $\beta$-function coefficient, only modifies the fourth significant digit of $\delta_{\mathrm{P}}$.

%
\begin{table}[width=\linewidth,cols=7,htb] 
\caption{ Exact results
for $A^{(n)}(\alpha_s)$ ($n\le 5$) at different $\beta$-function approximations,
and corresponding values of \ $\delta_{\mathrm{P}} = \sum_{n=1}^4\, K_n\, A^{(n)}(\alpha_s)$,
for $\alpha_s(m_\tau^2) =0.33$. The last row shows the truncated FOPT estimates at $\cO(a_\tau^4)$.}
\label{tab:Afun}
\begin{tabular*}{\tblwidth}{@{} LLLLLLL@{} }   
\toprule
& $A^{(1)}(\alpha_s)$ & $A^{(2)}(\alpha_s)$ & $A^{(3)}(\alpha_s)$
 & $A^{(4)}(\alpha_s)$ & $A^{(5)}(\alpha_s)$ &
 $\delta_{\mathrm{P}}$
\\ \midrule
$\beta_{n>1}=0$ & $0.14041$ & $0.01745$ & $0.00196$ & $0.00020$ & $0.000019$ & 
$0.19143$ \\
$\beta_{n>2}=0$ & $0.14321$ & $0.01740$ & $0.00186$ & $0.00018$ & $0.000014$ &
$0.19215$ \\
$\beta_{n>3}=0$ & $0.14326$ & $0.01724$ & $0.00181$ & $0.00017$ & $0.000013$ &
$0.19123$ \\
$\beta_{n>4}=0$ & $0.14305$ & $0.01712$ & $0.00178$ & $0.00016$ & $0.000013$ &
$0.19046$
\\
$\beta_{n>5}=0$ & $0.14294$ & $0.01708$ & $0.00178$ & $0.00016$ & $0.000012$ &
$0.19017$
\\ \midrule
$\cO(a_\tau^4)$ FOPT & $0.15148$ & $0.02156$ & $0.00246$ & $0.00012$ & \hfill ---\hfill &
$0.20847$
\\ \bottomrule
\end{tabular*}
\end{table}
%

The last row in table~\ref{tab:Afun} displays the results obtained with 
the {\it fixed-order perturbation theory} (FOPT) approximation
$\delta_{\mathrm{P}} = \sum_{n=1}^4\, r_n\, a_\tau^n$, {\it i.e.}, with
the truncated expansion in powers of $a_\tau$ at $\cO(a_\tau^4)$, showing that they approximate the integrals $A^{(n)}(\alpha_s)$ rather badly and overestimate $\delta_{\mathrm{P}}$ by 10\% at $\alpha_s(m_\tau^2) = 0.33$. Therefore, for a given measurement of $\delta_P$, FOPT leads to a smaller fitted value of the strong coupling than CIPT.

FOPT generates a slowly-converging series because of the large coefficients $r_n$.
The bad perturbative behaviour originates in the long logarithmic running of $\alpha_s(-s)$ along the circle $|s|=m_\tau^2$, with $\log{(-s/m_\tau^2)} = i\phi$ ($\phi\in [-\pi, \pi]$). This can be easily understood analytically at one-loop \cite{LeDiberder:1992jjr}, where one makes within the contour integral the series expansion
\bel{eq:fopt_ap}
\frac{\alpha_s(-s)}{\pi} \;\approx\; \frac{a_\tau}{1-i\beta_1 a_\tau\phi/2}\;\approx\; a_\tau \sum_n \left(\frac{i}{2}\beta_1 a_\tau\phi\right)^n \, .
\ee
When $|\phi|\sim\cO(\pi)$ this expansion only converges for $a_\tau < 0.14$. At the four-loop level the radius of convergence becomes slightly smaller than the physical value of $a_\tau$. The non-convergent behaviour of the truncated series induces a large renormalization-scale dependence in the FOPT result. In contrast, the resummation of all these large logarithms with the CIPT prescription gives rise to a well-behaved perturbative series with a very mild dependence on the renormalization scale.

The high four-loop accuracy achieved for $R_\tau$ and its strong sensitivity to the QCD coupling have triggered much effort to estimate the size of the unknown higher-order corrections  \cite{LovettTurner:1994hx,Ball:1995ni,Kataev:1995vh,LovettTurner:1995ti,Neubert:1995gd,Maxwell:1996he,Maxwell:1997yw,Maxwell:2001he,Davier:2005xq,Jamin:2005ip,Davier:2008sk,Pich:2010xb,Cvetic:2010ut,Pich:2011cj,Groote:2012jq,Boito:2016pwf,Boito:2018rwt,Wu:2019mky}, which constitute the dominant uncertainty in the theoretical prediction. In fact, since $\alpha_s$ is large at the scale $m_\tau$, the $\tau$ hadronic width could be sensitive to the expected asymptotic behaviour of perturbative series in quantum field theories. Different Borel summations, based on renormalon models \cite{Beneke:1998ui}, have been advocated \cite{Altarelli:1994vz,Cvetic:2001ws,Beneke:2008ad,Menke:2009vg,Caprini:2009vf,DescotesGenon:2010cr,Caprini:2011ya,Abbas:2012py,Beneke:2012vb,Abbas:2012fi,Abbas:2013usa,Caprini:2019kwp,Caprini:2020lff}, with and without contour improvement. FOPT has been often claimed to approach faster the Borel sum, but it has been shown recently that CIPT and FOPT lead in fact to two different Borel sums \cite{Hoang:2020mkw}.
In any case, the currently known $K_n$ coefficients do not show yet any renormalonic behaviour, and the QCD $\beta$ and $\gamma$ functions exhibit a surprisingly good perturbative convergence. Thus, at the achieved loop precision, the perturbative expansion of the dynamical Adler correlator does not seem to be close to an asymptotic-series regime.

The current experimental value of the Cabibbo-allowed $\tau$ hadronic width,
\cite{Zyla:2020zbs,Pich:2013lsa,Amhis:2019ckw},
\bel{eq:RtauV+Aexp}
R_{\tau,V+A}\, =\, 3.4709\pm 0.0079\, ,
\ee
implies that
$\delta_{\mathrm{P}} + \delta_{\mathrm{NP}} = 0.1963\pm 0.0025$, where $\delta_{\mathrm{NP}} = \sum_{D\ge 2} \delta^{(D)}_{ud}$ contains the small non-perturbative corrections and tiny light-quark mass contributions. The expected size of the power corrections is $|\delta_{\mathrm{NP}}|\sim (\Lambda_{\mathrm{QCD}}/m_\tau)^6 < 1\%$. Taking this conservative upper bound as an additional uncertainty, one gets $\alpha_s(m_\tau^2) = 0.334\pm 0.015$ with CIPT and
$\alpha_s(m_\tau^2) = 0.315\pm 0.012$ with FOPT. 
The perturbative error has been estimated including the fifth-order term $K_5 A^{(5)}(\alpha_s)$ with $K_5 = 275\pm 400$ and varying the renormalization scale in the range $\mu^2/m_\tau^2\in [0.5,2]$. The chosen central value of $K_5$ is in the range advocated by renormalon models, but the generous uncertainty allows for a correction of opposite sign. The smaller $\mu$ dependence of CIPT gets compensated by an increased sensitivity to $K_5$, which results in a slightly larger final uncertainty than FOPT.

\subsection{Hadronic invariant-mass distribution}
\label{subsec:InvariantMass}

\begin{figure}[t]
\centerline{
\includegraphics[width=0.34\textwidth]{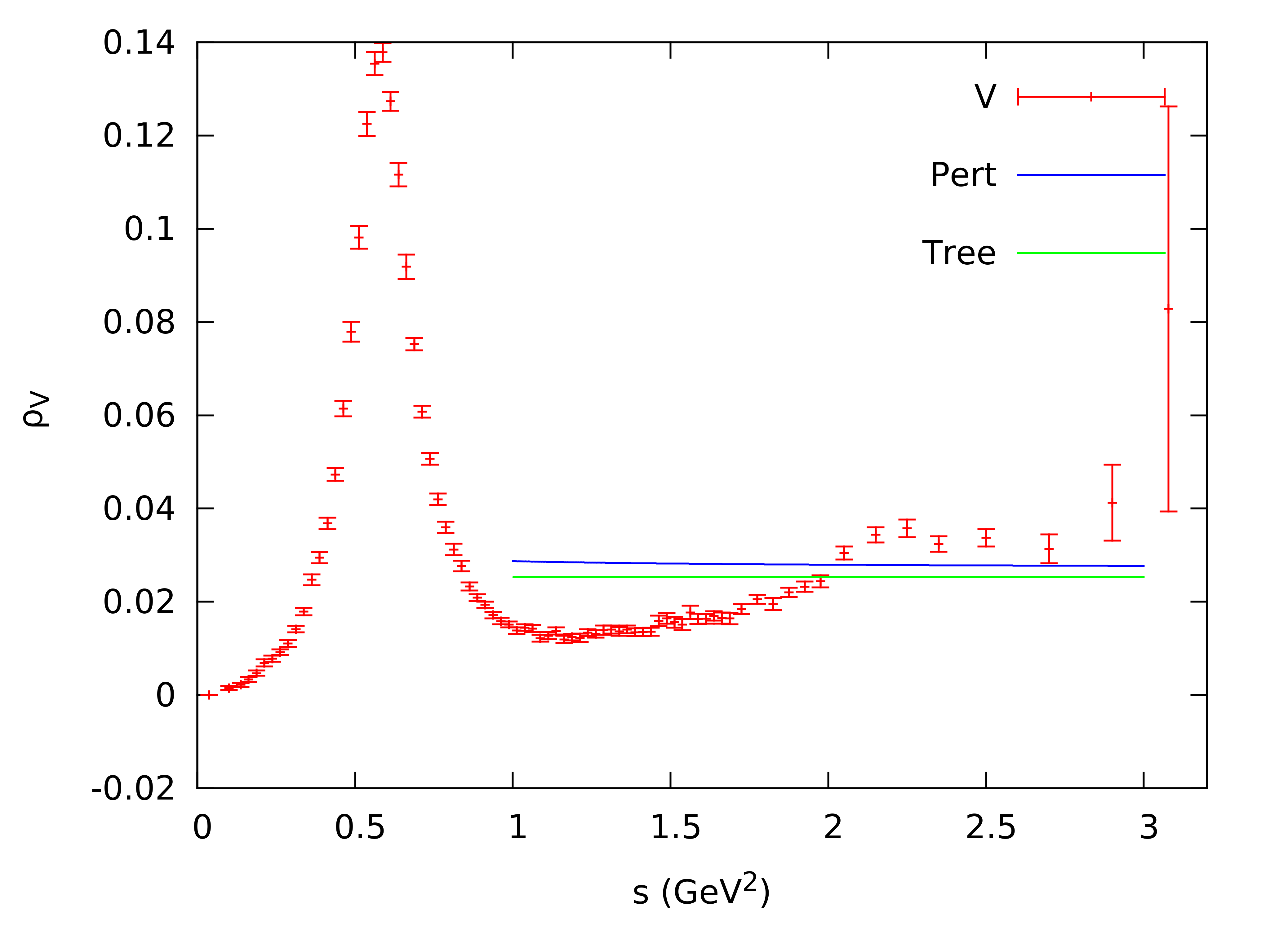}\hskip -.15cm
\includegraphics[width=0.34\textwidth]{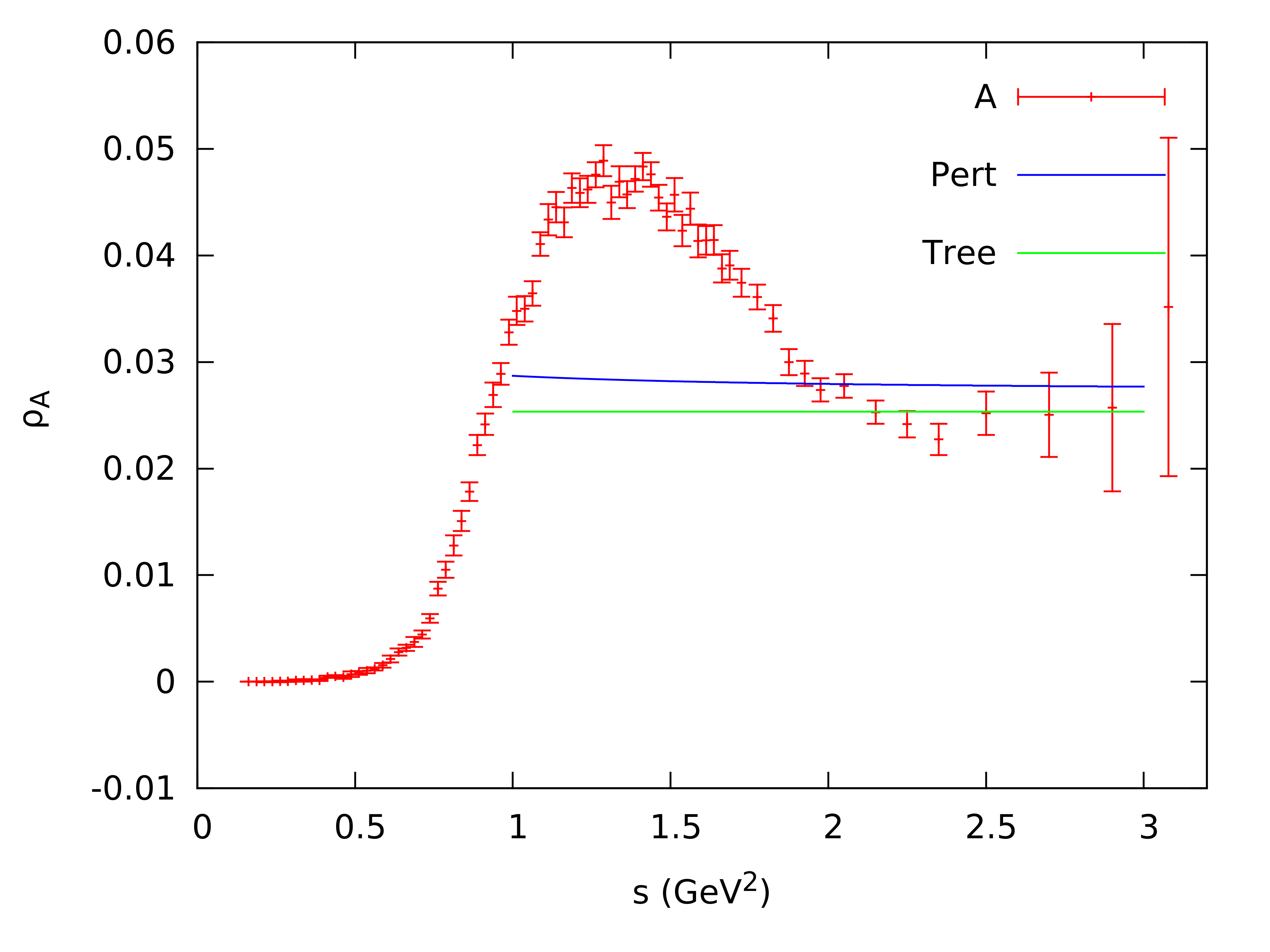}\hskip -.15cm
\includegraphics[width=0.34\textwidth]{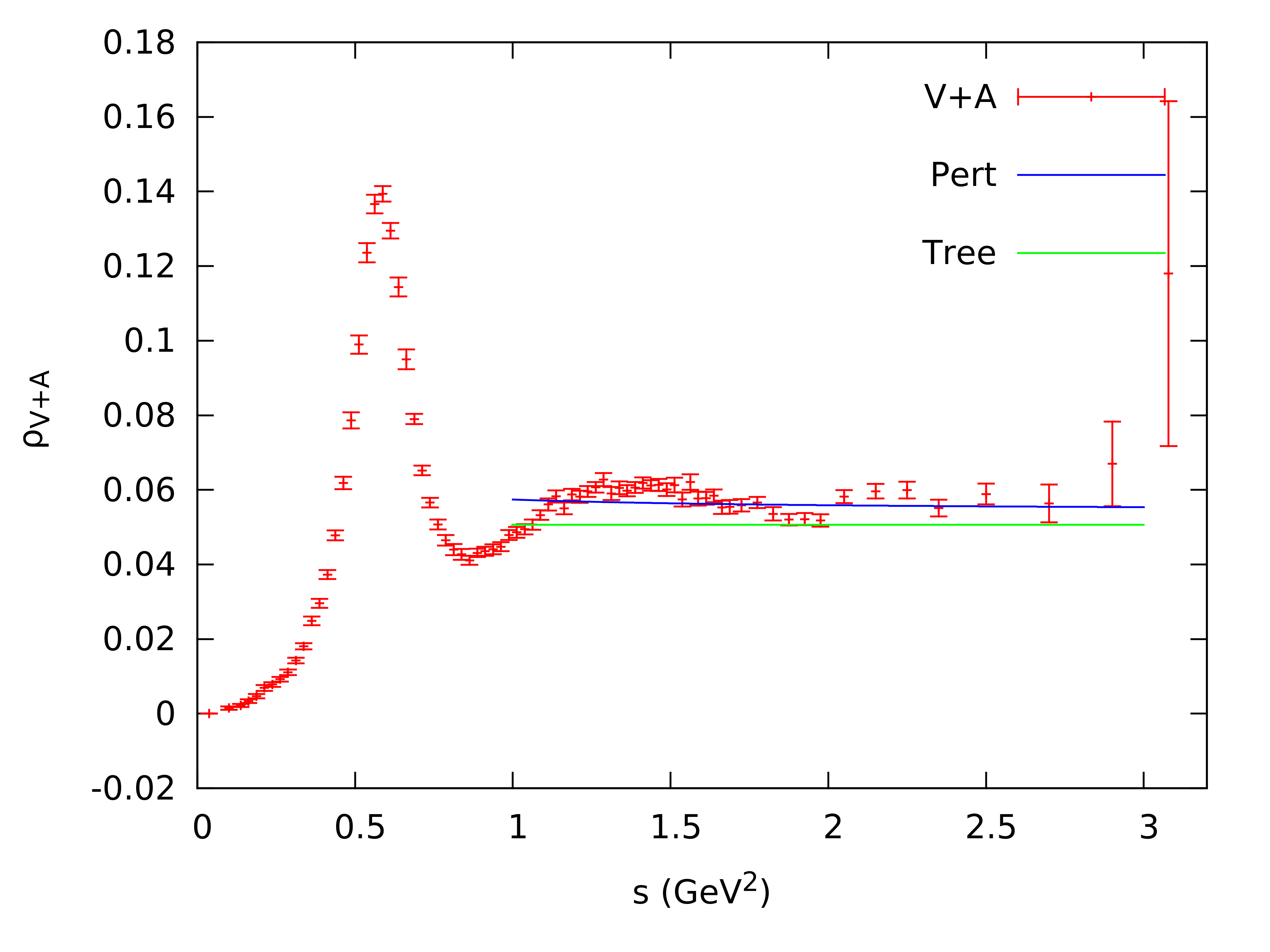}
}
\caption{Spectral functions for the Cabibbo-allowed vector ($V$), axial-vector ($A$) and $V+A$ channels \cite{Pich:2016bdg}, determined from ALEPH $\tau$ data \cite{Davier:2013sfa}.}
\label{fig:SpectralFunction}
\end{figure}

The experimental invariant-mass distribution of the final hadrons in the Cabibbo-allowed $\tau$ decays determines the spectral functions $ \frac{1}{\pi}\,\mathrm{Im}\, \Pi_{ud,\,\cJ}^{L+T}(s)$, shown in figure~\ref{fig:SpectralFunction}, while the only relevant contribution to $\mathrm{Im}\, \Pi^{L}_{ud,V+A}(s)$ is the $\pi^-$ final state at $s=m_\pi^2$:
\be 
\frac{1}{\pi}\,\mathrm{Im} \,\Pi_{ud,\, V+A}^{L}(s)\, =\, 2\, F_\pi^2\;\delta(s-m_\pi^2)\, +\, \cO(m_\pi^2/m_\tau^2)\, .
\ee
The measured distributions contain precious dynamical information that makes possible to 
extract the small non-perturbative corrections to $R_\tau$ from the data themselves  \cite{LeDiberder:1992zhd}.

The same analyticity argument leading to Eq.~\eqn{eq:circle} can be applied to any weighted integral of the hadronic spectral functions \cite{LeDiberder:1992zhd,Pich:1989pq},
\bel{eq:weighted_integrals}
\int_0^{s_0} ds\; \omega (s)\; \frac{1}{\pi}\,\mathrm{Im}\,\Pi^{L+T}_{ij,\,\cJ}(s)\; =\;
\frac{i}{2\pi}\; \oint_{|s|=s_0} ds\; \omega (s) \;\Pi^{L+T}_{ij,\,\cJ}(s)\, ,
\ee
where $\omega (s)$ is an arbitrary weight function without singularities in the
region $|s|\leq s_0$. The left-hand-side integral is directly determined by the experimental data, while the OPE can be used to express the right-hand-side as an expansion in inverse powers of $s_0$.
Weighting the spectral distribution with different functional dependences on $s$, one becomes sensitive to the power corrections in the OPE. For instance, weights of the form $(s/m_\tau^2)^n$ project the OPE contribution of dimension $D=2n+2$.
The theoretical precision is not as good as the one in $R_\tau$ because non-perturbative effects are less suppressed, but the added information substantially increases the final accuracy of the $\alpha_s$ determination.

The detailed experimental studies performed by the ALEPH \cite{Buskulic:1993sv,Barate:1998uf,Schael:2005am}, CLEO \cite{Coan:1995nk} and OPAL \cite{Ackerstaff:1998yj} collaborations confirmed long time ago that the non-perturbative corrections to $R_\tau$ are below 1\%, {\it i.e.}, that they are smaller than the perturbative uncertainties. The most recent and precise experimental analysis, carried out with the ALEPH data, obtains
$\delta_{\mathrm{NP}}  = -0.0064\pm 0.0013$ \cite{Davier:2005xq,Davier:2008sk,Davier:2013sfa},
in good agreement with the theoretical expectations \cite{Braaten:1991qm} and previous experimental determinations. This analysis concludes that \cite{Davier:2013sfa}
\bel{eq:AlphasALEPH}
\alpha_s^{(n_f=3)}(m_\tau^2)\, =\, 0.332\pm 0.005_{\mathrm{exp}}\pm 0.011_{\mathrm{th}}
\qquad\qquad \text{(Davier et al.)}\, ,
\ee
where the second uncertainty takes into account the different central values obtained with the CIPT ($0.341$) and FOPT ($0.324$) prescriptions, adding quadratically half their difference as an additional systematic error. 

The ALEPH analysis is based on the weights $\omega_{kl}(x) = (1-x)^{2+k} x^l (1+2x)$
\cite{LeDiberder:1992zhd}, which incorporate the phase-space and spin-1 factors in \eqn{eq:R_tau} so that one can directly use the measured distribution. A more complete phenomenological analysis of the same experimental data \cite{Pich:2016bdg,Pich:2016mgv} has recently investigated the stability of the results under changes of the chosen weights and has explored a large variety of alternative methodologies, including the dependence on the upper integration limit $s_0$ \cite{Davier:2005xq,Narison:1993sx,Girone:1995xb} 
that was fixed at $m_\tau^2$ in  \cite{Davier:2013sfa}.
The most reliable determinations, summarized in table~\ref{tab:AlphaMoments}, are extracted with the weights $\omega_{kl}(x)$, $\hat\omega_{kl}(x)=(1-x)^{2+k} x^l$, $\omega^{(2,m)}(x) = 1 - (m+2)\, x^{m+1} + (m+1)\, x^{m+2}$
and $\omega_a^{(1,m)}(x) = (1-x^{m+1})\, \mathrm{e}^{-ax}$. In addition to the perturbative errors, all quoted results include as an additional theoretical uncertainty the variations under various modifications of the fit procedures. The table displays a very consistent set of results, obtained with different numerical approaches that have different sensitivities to potential non-perturbative corrections. The excellent overall agreement, and the many complementary tests successfully performed, demonstrate their robustness and reliability.
From the results quoted in the table, one gets the final combined value \cite{Pich:2016bdg}:
\bel{eq:alphaTauMtau} 
\alpha_s^{(n_f=3)}(m_\tau^2)\, =\, 0.328 \pm 0.013\, ,
\ee
in excellent agreement with the previous result in Eq.~\eqn{eq:AlphasALEPH}.

\begin{table}[width=.8\linewidth,cols=4,pos=t]
\caption{Determinations of $\alpha_{s}^{(n_f=3)}(m_{\tau}^{2})$ from the $V+A$ spectral distribution of Cabibbo-allowed $\tau$ decays, with different methods \protect\cite{Pich:2016bdg}.}
\label{tab:AlphaMoments}
\begin{tabular*}{\tblwidth}{@{} LLLL@{} }
\toprule 
 & CIPT & FOPT & Average
\\ \midrule 
$\omega_{kl}(x)$ weights & $0.339 \,{}^{+\, 0.019}_{-\, 0.017}$ &
$0.319 \,{}^{+\, 0.017}_{-\, 0.015}$ & $0.329 \,{}^{+\, 0.020}_{-\, 0.018}$
\\[3pt]
$\hat\omega_{kl}(x)$ weights  & $0.338 \,{}^{+\, 0.014}_{-\, 0.012}$ &
$0.319 \,{}^{+\, 0.013}_{-\, 0.010}$ & $0.329 \,{}^{+\, 0.016}_{-\, 0.014}$
\\[3pt]
$\omega^{(2,m)}(x)$ weights  & $0.336 \,{}^{+\, 0.018}_{-\, 0.016}$ &
$0.317 \,{}^{+\, 0.015}_{-\, 0.013}$ & $0.326 \,{}^{+\, 0.018}_{-\, 0.016}$
\\[2pt]
$s_0$ dependence  & $0.335 \pm 0.014$ &
$0.323 \pm 0.012$ & $0.329 \pm 0.013$
\\[2pt]
$\omega_a^{(1,m)}(x)$ weights  & $0.328 \, {}^{+\, 0.014}_{-\, 0.013}$ &
$0.318 \, {}^{+\, 0.015}_{-\, 0.012}$ & $0.323 \, {}^{+\, 0.015}_{-\, 0.013}$
\\  \midrule
Average & $0.335\pm 0.013$ & $0.320\pm 0.012$ & $0.328\pm 0.013$
\\ \bottomrule
\end{tabular*}
\end{table}

\begin{figure}[hbt]
\centering
\includegraphics[width=0.41\textwidth,clip]{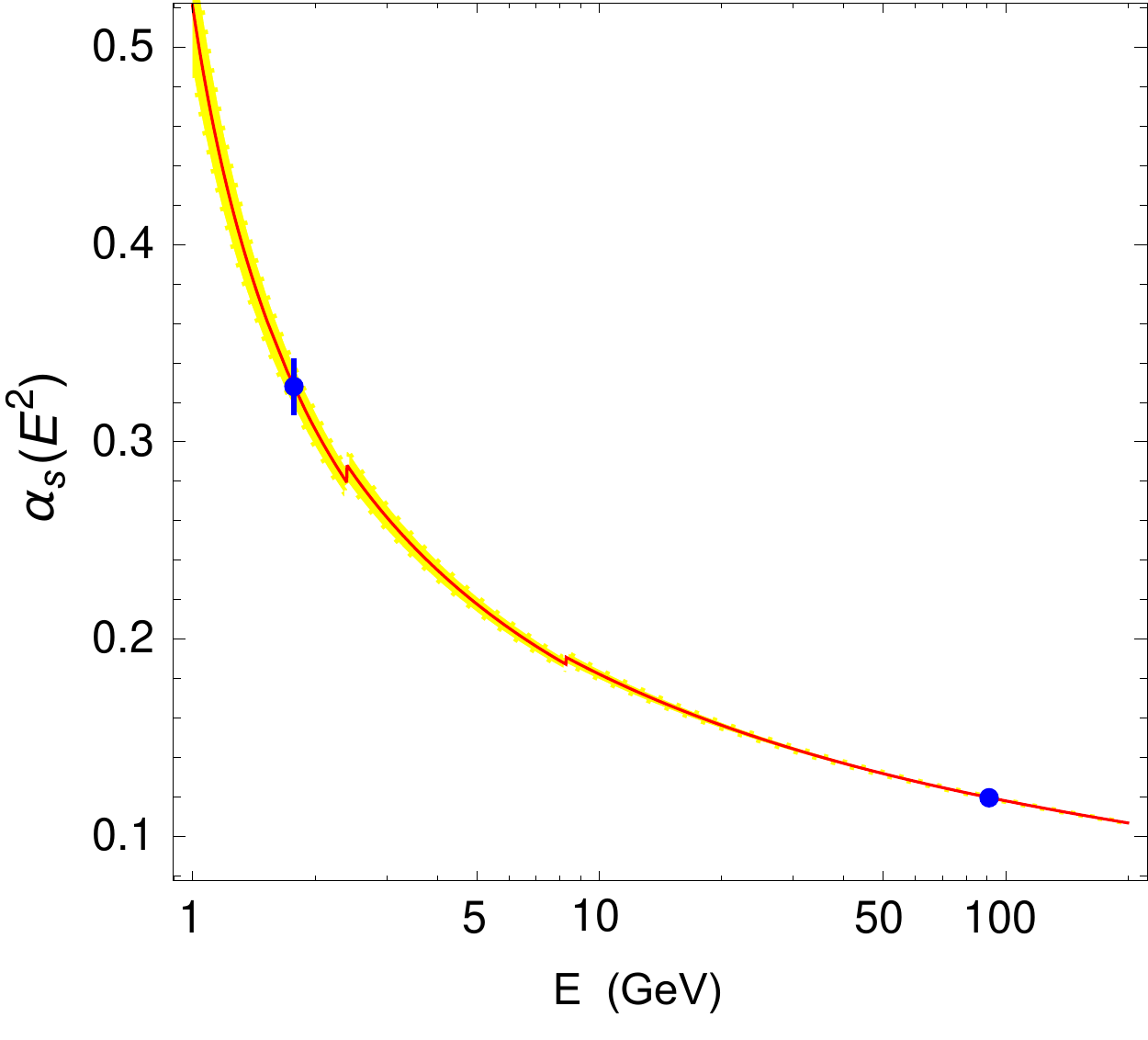}
\caption{The five-loop evolution of $\alpha_s(m_\tau^2)=0.328\pm 0.013$, determined from hadronic $\tau$ decays, is compared with the measurement of $\alpha_s(M_Z^2)$ from $\Gamma_Z$. The yellow band indicates how the error shrinks at higher scales.} 
\label{fig:runningTau}
\end{figure}

The value of the strong coupling at the $\tau$ mass scale is significantly larger ($\sim 16\sigma$) than the result extracted from the $Z$ hadronic width in Eq.~\eqn{eq:alpha_Z}. Evolving the $\tau$ decay determination to the scale $M_Z$,  the strong coupling decreases to 
\bel{eq:AlphaTauMZ}
\alpha_s^{(n_f=5)}(M_Z^2)\, =\, 0.1197 \pm 0.0015\, ,
\ee
which nicely agrees with the direct measurement at the $Z$ peak and has an even smaller uncertainty. As shown by the yellow band in figure~\ref{fig:runningTau}, the running from $m_\tau$ to $M_Z$ decreases the error of $\alpha_s$ by a factor $\alpha_s^2(M_Z^2)/\alpha_s^2(m_\tau^2)\sim 0.1$, reflecting the larger sensitivity to the strong coupling at low energies. The comparison of the $\tau$ and $Z$ determinations of $\alpha_s$ at a common scale,
\bel{eq:AFtest}
\left.\alpha_s^{(n_f=5)}(M_Z^2)\right|_Z -
\left.\alpha_s^{(n_f=5)}(M_Z^2)\right|_\tau\, =\, 0.0002\pm 0.0029_Z\pm 0.0015_\tau\, ,
\ee
confirms the predicted QCD running at the five-loop level, providing a precise verification of asymptotic freedom.

\subsection{Sensitivity to the vacuum structure}

The precise determination of $\alpha_s(m_\tau^2)$ is done with carefully-chosen weights that minimize the non-perturbative contaminations. These weights vanish at the point $s=s_0$ where the complex contour touches the real axis, in order to suppress contributions from the region near the branch cut, and the non-perturbative corrections to their weighted integrals remain much smaller than the perturbative results. Moreover, the strong coupling is extracted from the total $V+A$ distribution which, as shown in figure~\ref{fig:SpectralFunction}, is remarkably flat and approaches very fast the QCD predictions. The opening of high-multiplicity hadronic thresholds dilutes very soon the prominent $\rho(2\pi)$ and $a_1(3\pi)$ resonance peaks. The more inclusive nature of the $V+A$ spectral function, compared to the separate $V$ and $A$ distributions, is also reflected in smaller non-perturbative corrections to its weighted integrals \cite{Pich:2016bdg,Pich:2016mgv}, as expected from the predicted opposite signs of the $D=6$ power corrections in the vector and axial-vector channels \cite{Braaten:1991qm}. Nevertheless, the independent analyses of the $V$ and $A$ distributions also provide consistent values of $\alpha_s(m_\tau^2)$, although they have larger systematic uncertainties that need to be carefully assessed. Taking into account as an additional error the spread of central values obtained with the different weights analysed  \cite{Pich:2016bdg}, one gets
$\alpha_s(m_\tau^2) = 0.328\pm 0.018$ from the vector spectral function, while the axial one leads to  $\alpha_s(m_\tau^2) = 0.322\pm 0.018$.

The availability of good experimental data offers us the possibility to explore the strong-coupling regime of QCD, analysing observables that maximize the non-perturbative effects. For instance, taking weights that do not vanish at $s=s_0$ and, therefore, are more exposed to potential failures of the OPE near the real axis (duality violations), or using much lower values of $s_0$ with enhanced power corrections.
This is obviously not a good strategy to perform clean and accurate determinations of $\alpha_s$, but it provides an interesting way to investigate the QCD vacuum structure and try to better understand the complicated dynamics involved. 

Violations of quark--hadron duality have been analysed through a direct fit of the vector spectral function from $\hat s_0=1.55\;\mathrm{GeV}^2$ to $m_\tau^2$ with an ad hoc four-parameter functional ansatz \cite{Cata:2008ru}, together with the integrated distribution (without any weight) below $\hat s_0$ \cite{Boito:2014sta}. Although the OPE is not valid in the real axis and the absence of weighting makes the low-energy integral up to $\hat s_0$ very exposed to uncontrolled effects, specially at such low $\hat s_0$, a quite reasonable value of $\alpha_s$ is extracted: $\alpha_s(m_\tau^2) = 0.301\pm 0.012$.\footnote{
Slightly higher but less precise values were obtained before from a similar analysis with OPAL data \cite{Boito:2012cr}. Averaging the two results, one finds 
$\alpha_s(m_\tau^2) = 0.309\pm 0.012$ 
\cite{Boito:2014sta}.
}
The uncertainties are however, largely underestimated, since the fitted coupling strongly depends on the chosen value of $\hat s_0$ and the assumed spectral function ansatz; small variations of the adopted choices lead to fluctuations larger than $3\sigma$ \cite{Pich:2016bdg,Pich:2016mgv,Pich:2018jiy}. The large correlations among all fitted quantities convert $\alpha_s$ into an additional effective-model parameter, incorporating unaccounted systematics.
The $\tau$ data has been also employed to constrain other low-energy modellings of $\alpha_s$, which extrapolate the perturbative behaviour into the confinement regime by using effective descriptions of the strong coupling that are well-defined in the infrared domain (freezing, analytic coupling, light-front holography, etc.) \cite{Nesterenko:2013vja,Deur:2016tte,Ayala:2017tco,Cvetic:2020naz}.

Of special interest are the weighted integrals of the non-strange $V-A$ distribution because, owing to the chiral invariance of massless QCD, the associated correlation function vanishes identically to all orders in perturbation theory, in the limit of zero quark masses. The difference between the measured vector and axial-vector distributions can then be used to test the non-perturbative QCD dynamics without any contamination from perturbation theory. The non-zero value of $\Pi_{ud,V-A}^{L+T}(Q^2)$ originates in the spontaneous breaking of chiral symmetry by the QCD vacuum. The lowest-dimensional operators contributing to this correlator have $D\ge 6$, which implies that at large momenta it scales as $1/(Q^2)^3$  and, therefore, must satisfy the two super-convergent Weinberg sum rules \cite{Weinberg:1967kj}:
\bea\label{eq:WSR}
\int_{4m_\pi^2}^\infty ds\;\frac{1}{\pi}\,\mathrm{Im}\,\Pi^{L+T}_{ud,V-A}(s) &\!\! =&\!\! 2\, F_\pi^2\, ,
\no\\
\int_{4m_\pi^2}^\infty ds\; s\; \frac{1}{\pi}\,\mathrm{Im}\,\Pi^{L+T}_{ud,V-A}(s) &\!\! =&\!\! 2\, F_\pi^2 m_\pi^2\, .
\eea
With non-zero quark masses taken into account, the first relation is still exact, while the second gets a negligible correction of $\cO(m_q^2)$ \cite{Floratos:1978jb}. This entails a very strong theoretical restriction on the $V-A$ spectral function that complements very efficiently the available experimental information. 

Chiral perturbation theory ($\chi$PT) \cite{Weinberg:1978kz,Gasser:1983yg,Gasser:1984gg}, the low-energy effective field theory of the QCD Goldstone bosons ($\pi$, $K$, $\eta$) \cite{Ecker:1994gg,Pich:1995bw,Bijnens:2006zp,Scherer:2012zzd}, determines the infrared behaviour of
$\Pi_{ud,V-A}^{L+T}(Q^2)$. Its low-energy expansion in powers of momenta is known to two loops, in terms of the pion decay constant and mass, and the $\chi$PT couplings $L_{10}$ [$\cO(p^4)$] and $C_{87}$ [$\cO(p^6)$] \cite{Amoros:1999dp}. The short- and long-distance regimes are related by analyticity through dispersion relations analogous to Eq.~\eqn{eq:weighted_integrals}. Using weight functions with inverse powers of $s$ (and accounting for the residue of the corresponding pole at $s=0$) one can then extract the values of $L_{10}$ and $C_{87}$ from the $\tau$ decay data, while positive powers of $s$
give access to the relevant vacuum condensates, {\it i.e.}, to the order parameters of chiral symmetry breaking \cite{Bordes:2005wv,Almasy:2008xu,GonzalezAlonso:2008rf,GonzalezAlonso:2010rn,GonzalezAlonso:2010xf,Boito:2012nt,Dominguez:2014fua,
Boito:2015fra,Rodriguez-Sanchez:2016jvw}. 

Denoting by $\cO_{D,\,\cJ}\equiv\sum_O
C^{L+T}_{ud,\,\cJ}(Q^2,\mu^2)\,\langle 0|O(\mu^2)|0\rangle$ the full coefficient of the $1/(Q^2)^{D/2}$ power correction in Eq.~\eqn{eq:OPE}, where the sum is over all possible operators with dimension $D$, and neglecting its small logarithmic dependence on $Q^2$,
the most recent fit to the ALEPH $\tau$ data gives \cite{Rodriguez-Sanchez:2016jvw}:
\bel{eq:V-Acondensates}
\cO_{6,V-A}\, =\, \left( -3.6\, {}^{+\,0.7}_{-\, 0.6}\right)\cdot 10^{-3}\;\mathrm{GeV}^6\, ,
\qquad\qquad\qquad
\cO_{8,V-A}\, =\, \left( -1.0\pm 0.4\right)\cdot 10^{-2}\;\mathrm{GeV}^8\, .
\ee
Additional estimates for the higher-order $\cO_{D,V-A}$ corrections with $10\le D\le 16$ can be found in Ref.~\cite{Rodriguez-Sanchez:2016jvw}, together with a compilation of results obtained in previous analyses. The $D=6$ contributions to the vector and axial-vector correlators are predicted to have opposite signs  \cite{Braaten:1991qm}, due to their different chiralities, which implies $|\cO_{6,V+A}| < |\cO_{6,V-A}|$. This expectation is fully compatible with Eq.~\eqn{eq:V-Acondensates} and the fitted  results from the $V$, $A$ and $V+A$ spectral functions \cite{Pich:2016bdg,Davier:2013sfa}. In spite of their larger uncertainties, the separate fits to the vector and axial-vector distributions exhibit a clear sign difference in their $D=6$ power correction and a sizeable numerical cancellation in $V+A$. Moreover, a similar cancellation seems to be operative in the $D=8$ terms.

The $\tau$ data also determine the two relevant $\chi$PT couplings (renormalized at the chiral scale $M_\rho$) \cite{Rodriguez-Sanchez:2016jvw}:\footnote{
From the $\tau$ data one directly extracts effective parameters $L_{10}^{\mathrm{eff}}$ and $C_{87}^{\mathrm{eff}}$ that include known logarithmic chiral corrections \cite{Amoros:1999dp,GonzalezAlonso:2008rf}. At the two-loop accuracy quoted in \eqn{eq:L10C87}, $L_{10}^{\mathrm{eff}}$ includes also small contributions from $\cO(p^6)$ $\chi$PT couplings that can be estimated in the large--$N_C$ limit  \cite{Rodriguez-Sanchez:2016jvw} or with lattice input \cite{Boito:2015fra}. More technical details and a compilation of previous determinations can be found in Ref.~\cite{Rodriguez-Sanchez:2016jvw}.
}
\bel{eq:L10C87}
L_{10}^r(M_\rho)\, =\, \left( -4.1\pm 0.4\right)\cdot 10^{-3}\, ,
\qquad\qquad\qquad
C_{87}^r(M_\rho)\, =\, \left( 5.10\pm 0.22\right)\cdot 10^{-3}\;\mathrm{GeV}^{-2}\, .
\ee
These values are in excellent agreement with the (less precise) theoretical predictions obtained at NLO in the $1/N_C$ expansion,
$L_{10}^r(M_\rho) =( -4.4\pm 0.9)\cdot 10^{-3}$ and
$C_{87}^r(M_\rho) = ( 3.6\pm 1.3)\cdot 10^{-3}\;\mathrm{GeV}^{-2}$
  \cite{Pich:2008jm}.

The hadronic $\tau$ decay data has also been used recently to extract upper bounds on new physics beyond the Standard Model, using low-energy effective Lagrangians to parametrize the unknown dynamics at high scales \cite{Cirigliano:2018dyk}. The inclusive $V-A$ constraints in Eq.~\eqn{eq:WSR} provide a quite powerful discriminating tool because they are valid in a very broad class of dynamical scenarios, which includes all asymptotically-free theories \cite{Bernard:1975cd}.

\subsection[Determination of $|V_{us}|$ from the Cabibbo-suppressed $\tau$ decay width]{\boldmath Determination of $|V_{us}|$ from the Cabibbo-suppressed $\tau$ decay width}

A very clean determination of the $V_{us}$ quark mixing can be obtained from the separate measurement of the Cabibbo-allowed and Cabibbo-suppressed inclusive $\tau$ decay widths 
\cite{Gamiz:2002nu,Gamiz:2004ar}. To a first approximation, the ratio $R_{\tau,S}/R_{\tau,V+A}$ directly measures $|V_{us}/V_{ud}|^2$.
The current experimental values \cite{Zyla:2020zbs,Pich:2013lsa,Amhis:2019ckw} of
\bel{eq:RtauSexp}
R_{\tau,S}\, =\, 0.1645\pm 0.0023\, ,
\ee
$R_{\tau,V+A}$ in Eq.~\eqn{eq:RtauV+Aexp} and $|V_{ud}| = 0.97370\pm 0.00014$ imply that 
$|V_{us}| = 0.212\pm 0.002$ in the SU(3) symmetry limit.
This result is slightly shifted by small SU(3)-breaking corrections induced by the strange quark mass \cite{Pich:1998yn,Pich:1999hc,Maltman:1998qz,Chetyrkin:1998ej,Kambor:2000dj,Korner:2000wd,Chen:2001qf,Maltman:2001sv} that can be theoretically estimated through a QCD analysis of \cite{Gamiz:2002nu,Gamiz:2004ar,Baikov:2004tk,Maltman:2006st,Gamiz:2006nj,Gamiz:2007qs,Gamiz:2013wn}
\bel{eq:deltaRtau}
\delta R_\tau\,\equiv\, \frac{R_{\tau,V+A}}{|V_{ud}|^2} - \frac{R_{\tau,S}}{|V_{us}|^2}
\;\approx\;
24\, S_{\mathrm{EW}}\left\{ \frac{m_s^2(m_\tau^2)}{m_\tau^2}\left( 1 - \epsilon_d^2\right) \Delta_{00}(a_\tau) - 2\pi^2\,\frac{\delta O_4}{m_\tau^4}\; Q_{00}(a_\tau)
\right\} ,
\ee
where $\epsilon_d\equiv m_d/m_s = 0.053\pm 0.002$ \cite{Leutwyler:1996qg} and
$a_\tau =\alpha_s(m_\tau^2)/\pi$. 
The difference $\delta R_\tau$ exactly vanishes in the SU(3) limit because the QCD interactions are flavour universal. The leading non-zero contributions are induced by the quark mass difference $m_s^2-m_d^2$ and the $D=4$ operator $\delta O_4 \equiv \langle 0| m_s\bar s s - m_d \bar d d |0\rangle = (-1.4\pm 0.4)\times 10^{-3}\;\mathrm{GeV}^4$
\cite{Pich:1998yn,Pich:1999hc,Gamiz:2002nu,Gamiz:2004ar,Gamiz:2007qs}. Since the dimensions of these two operators are compensated by the corresponding powers of the $\tau$ mass, $\delta R_\tau$ is a numerically small effect.

The perturbative QCD series $\Delta_{00}(a_\tau)$ is currently known to $\cO(\alpha_s^3)$, while the coefficient $Q_{00}(a_\tau)$ has been only computed to $\cO(\alpha_s^2)$
\cite{Pich:1998yn,Pich:1999hc,Baikov:2004tk}. It is convenient to separate the longitudinal (L) and transverse (L+T) components, following Eqs.~\eqn{eq:Adler_def_L_L+T} and \eqn{eq:circle}.
The longitudinal series $\Delta_{00}^{L}(a_\tau)$ exhibits a very bad perturbative behaviour. The already slow convergence of the scalar correlator at the $m_\tau$ scale is significantly deteriorated by the additional contributions generated by the contour integration, and the CIPT prescription does not seem to improve the situation. Fortunately, the total longitudinal contribution to $\delta R_\tau$ can be estimated phenomenologically with very good accuracy because it is dominated by the well-known $\tau\to\nu_\tau\pi$ and $\tau\to\nu_\tau K$ contributions:
$\delta R_\tau^L = 0.1544\pm 0.0037$  \cite{Gamiz:2002nu}. 
A very conservative estimate of the remaining transverse contribution, using 
as input the lattice world average of the strange quark mass \cite{Aoki:2019cca}
with an inflated uncertainty, gives $\delta R_{\tau,\mathrm{th}} = 0.240\pm 0.032$
\cite{Pich:2013lsa,Gamiz:2002nu,Gamiz:2004ar,Gamiz:2007qs,Gamiz:2013wn}. 

Inserting $\delta R_{\tau,\mathrm{th}}$ in Eq.~\eqn{eq:deltaRtau}, one gets the corrected determination of the Cabibbo quark mixing:
\bel{eq:Vus}
|V_{us}|\, =\, \left( \frac{R_{\tau,S}}{\frac{R_{\tau,V+A}}{|V_{ud}|^2}- \delta R_{\tau,\mathrm{th}}}\right)^{1/2}
\; =\; 0.2194\pm 0.0016_{\mathrm{exp}}\pm 0.0010_{\mathrm{th}}
\, =\, 0.2194\pm 0.0019\, .
\ee
This result is lower than the value extracted from $K\to\pi\ell\nu$ decays, $|V_{us}| = 0.2231\pm 0.0007$ \cite{Zyla:2020zbs}. Note however that the Cabibbo-suppressed $\tau$ data samples collected at LEP were statistically limited, while BaBar and Belle measure on average lower $\tau$ branching ratios \cite{Zyla:2020zbs}, a systematic effect that is still not well understood (although slowly improving). In fact, the experimental value of $R_{\tau,S}$ has recently increased \cite{Amhis:2019ckw}, shifting $|V_{us}|$ up by 0.0021  compared to its 2014 value \cite{Pich:2013lsa}. 
This is, however, still not enough.


Clearly, high-precision measurements of the Cabibbo-suppressed $\tau$ decays have the potential to provide a very accurate value of $|V_{us}|$, which does not involve any theoretical estimate of hadronic form factors or decay constants and, therefore, does not suffer the theoretical limitations of the kaon determinations. Precisely measured distributions of the final hadrons in these decays would also allow for many complementary tests \cite{Pich:1999hc,Gamiz:2002nu,Gamiz:2004ar,Chen:2001qf,Hudspith:2017vew,Boyle:2018ilm}.

\subsection{Finite-energy sum rules with electron--positron data}

High-precision measurements of the $\tau$ spectral functions, especially in the higher kinematically-allowed energy bins, are needed to improve the determinations of the strong constant and the Cabibbo quark mixing, and to perform more precise tests of non-perturbative aspects of QCD. Both higher statistics and a good control of experimental systematic uncertainties are needed, which could be possible at the Belle-II experiment \cite{Kou:2018nap} and, perhaps, at a super charm-tau factory \cite{Barniakov:2019zhx,Luo:2019xqt}. At long term, the TeraZ option of a future FCC-ee collider running at the $Z$ peak would produce an enormous data sample of $1.7\times 10^{11}$  $\tau^+\tau^-$ pairs in extremely clean kinematic (and background) conditions \cite{Abada:2019lih}, opening a broad range of interesting opportunities.

While $\tau$ decay data are kinematically limited to hadronic invariant masses below 
the $\tau$ mass, higher values of $s$ can be accessed in $e^+e^-$ annihilation. The onset of the QCD asymptotic behaviour is however also reached at larger invariant masses in the vector spectral function, as shown in figure~\ref{fig:SpectralFunction} for its isospin-one component. In spite of their currently larger experimental uncertainties, the $e^+e^-$ data provide useful complementary information that can be analysed through spectral moments in complete analogy to the $\tau$ decay studies \cite{Narison:1993sx,Eidelman:1978xy}. The integrated $e^+e^-$ distributions provide in fact a better sensitivity to the strong coupling than the ratio $R_{e^+e^-}(s)$. Notice also that the electromagnetic hadronic production gives access to the isospin-zero spectral function, which is absent in $\tau$ decays.

\begin{figure}[ht]
\centering
\includegraphics[width=0.5\textwidth,clip]{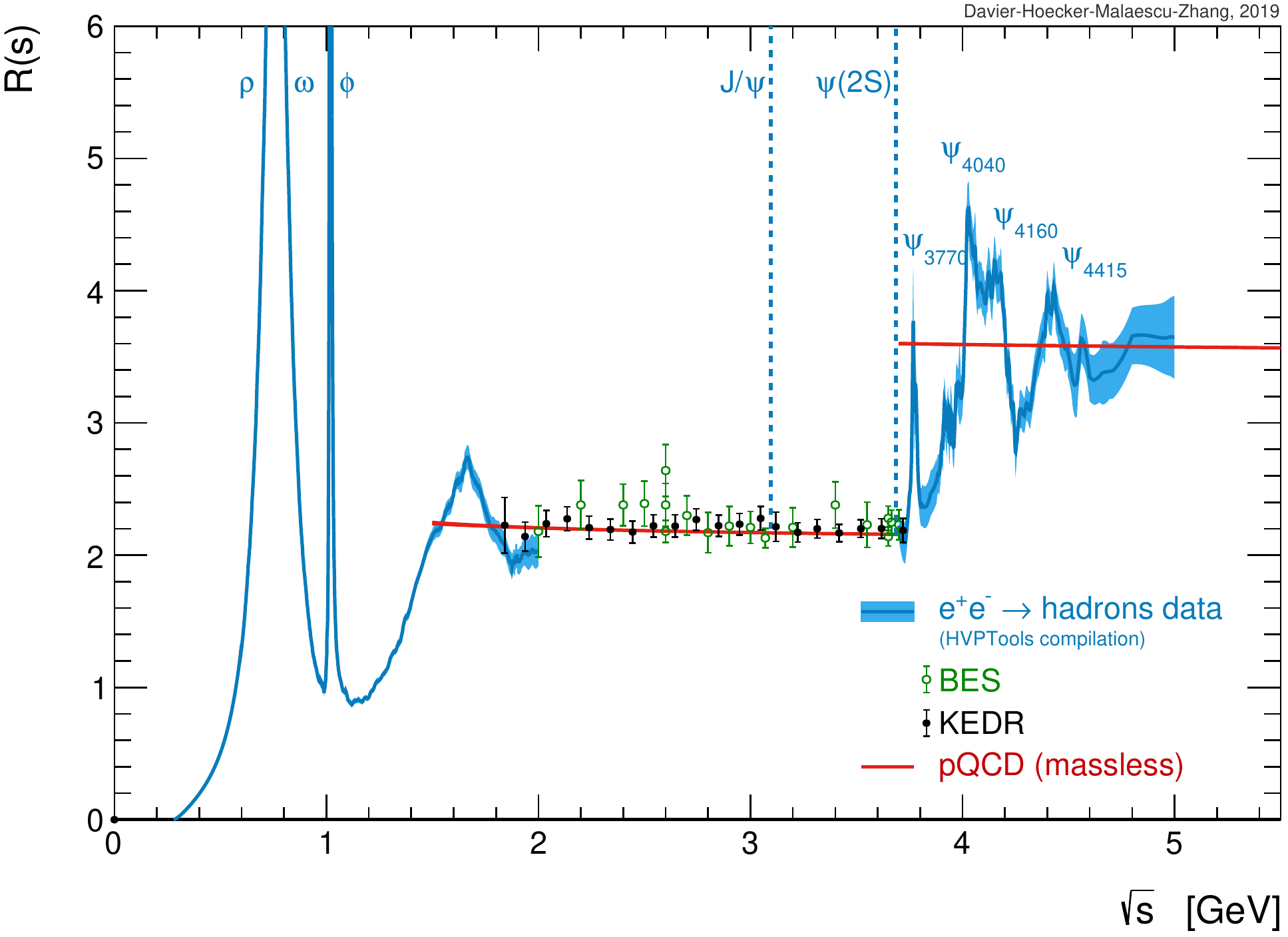}
\caption{Low-energy data on $R_{e^+e^-}(s)$. Inclusive measurements are shown as data points, while the sum of exclusive channels is given by the narrow blue bands. Figure taken from Ref.~\cite{Davier:2019can}.} 
\label{fig:Rlow}
\end{figure}

The determination of the $e^+e^-$ spectral function requires an experimental scanning over its full invariant-mass range, either directly taking data in a wide range of energies or through initial-state radiation \cite{Binner:1999bt,Rodrigo:2001kf}. This can only be completed combining data from experiments operating at different colliders, with different energies and with quite diverse experimental conditions (detector performances, acceptances, backgrounds), systematic errors and normalizations. Figure~\ref{fig:Ree} displays the currently available information \cite{Zyla:2020zbs}. A huge effort has been made in recent years to improve this data compilation \cite{Jegerlehner:2017lbd,Davier:2019can,Keshavarzi:2019abf} in order to refine the dispersive predictions of the hadronic vacuum polarization contribution to the anomalous muon magnetic moment and to the running of the electromagnetic coupling up to $M_Z$  \cite{Aoyama:2020ynm}. Nevertheless, significant experimental discrepancies remain still to be understood, specially in the $\rho$-$\omega$ interference region where the $2\pi$ data from the most precise BaBar and KLOE experiments differ, at the $\phi(1020)$ resonance peak where the differences among different $K^+K^-$ data sets largely exceed their quoted systematic uncertainties, and in the exclusive--inclusive transition region around 2~GeV with the inclusive results being slightly higher than the sum of the exclusive channels, due to unmeasured higher-multiplicity final states. 
The most recent compilation of low-energy $R_{e^+e^-}(s)$ data is shown in figure~\ref{fig:Rlow}~\cite{Davier:2019can}.

Worth mentioning at this point is the well-known discrepancy between the vector $\tau$ spectral function and its electromagnetic isospin-one counterpart (they are related by isospin), which lies systematically below the $\tau$ data. This discrepancy is easily visualized comparing the $e^+e^-$ predictions for the vector  $\tau$ branching ratios with their measured values. For instance, after properly accounting for isospin-breaking effects \cite{Cirigliano:2001er,Cirigliano:2002pv,Davier:2009ag}, the predicted $\tau$ branching fraction into two pions is $2.4\sigma$ lower \cite{Davier:2010nc} than the PDG average,
$\mathrm{Br}(\tau^-\to\nu_\tau\pi^-\pi^0) = (25.49\pm 0.09)\%$ \cite{Zyla:2020zbs}.
Since the main $\tau$ branching ratios were precisely measured at LEP in very clean experimental conditions and without any need for an external normalization, this discrepancy seems to signal unaccounted systematics in the $e^+e^-$ data. This conclusion is further reinforced by the most recent lattice determinations of the LO hadronic vacuum polarization contribution to the muon $g-2$ \cite{Borsanyi:2017zdw,Blum:2018mom,Giusti:2018mdh,Shintani:2019wai,Davies:2019efs,Gerardin:2019rua,Giusti:2019hkz,Borsanyi:2020mff,Lehner:2020crt}, which find values slightly higher than the dispersive $e^+e^-$ results and in better agreement with the $\tau$-based determination \cite{Davier:2010nc}.

The weighted integrals in Eq.~\eqn{eq:weighted_integrals} have been analysed with the exclusive $e^+e^-$ data compilation of Ref.~\cite{Keshavarzi:2018mgv} in a narrow window of $\sqrt{s_0}$ between 1.80 and 2~GeV,
just below the exclusive-inclusive data transition (see figure~\ref{fig:Rlow}). These values of the hadronic invariant mass are not much larger than $m_\tau$, but the current errors on the inclusive data above 2~GeV are unfortunately too large for a precision determination of $\alpha_s$. Taking as weight functions $\omega (x)= 1$ (very exposed to duality violations), $\omega_{00}(x)$ and $(1-x^2)^n$ with $n=1,2$, Ref.~\cite{Boito:2018yvl} finds $\alpha_s(m_\tau^2) = 0.301\pm 0.019$ (FOPT and CIPT combined) [$\alpha_s(M_Z^2) = 0.1162\pm 0.025$]. This should be compared with the value quoted before from the $\tau$ vector spectral function, $\alpha_s(m_\tau^2) = 0.328\pm 0.018$ [$\alpha_s(M_Z^2) = 0.1197\pm 0.021$] \cite{Pich:2016bdg}. The slightly lower value of the strong coupling obtained from the $e^+e^-$ data just reflects the current experimental discrepancy between the two sets of data. Given all caveats mentioned before, more detailed analyses are clearly needed. Nevertheless this extraction of $\alpha_s$ from  $e^+e^-$ data shows already that good sensitivity could be achieved once the current discrepancies get resolved and more precise inclusive measurements below the charm threshold become available.

\section{NNLO determinations of the strong coupling}
\label{sec:alphas}

The inclusive $Z$ and $\tau$ hadronic widths provide a very important test of the Standard Model at an impressive N${}^3$LO precision, where LO refers to the first nontrivial QCD contribution. The strong coupling is determined in two completely different energy regimes and with very different experimental systematics, but the theoretical description of the two observables is based on similar current correlators, being the four-loop calculation of the Adler function the basic ingredient in both cases. The lower sensitivity to $\alpha_s$ at higher energies is compensated by the higher experimental precision achieved at the $Z$ peak, so that comparable accuracies are finally reached. The $Z$ determination assumes local quark--hadron duality, {\it i.e.}, Eq.~\eqn{eq:LocalDuality} to be satisfied, and the absence of new physics contributions at the high scale $M_Z$, while the $\tau$ measurement is much more inclusive (only integrals of the spectral distribution are needed) but it has a higher sensitivity to inverse power corrections. The excellent agreement between the two determinations is then a highly non-trivial result that corroborates the predicted running of the QCD coupling with very high precision and puts a strong constraint on new-physics scenarios.

There are of course many other interesting QCD tests, based on less inclusive observables that have already reached a NNLO accuracy. For completeness, we present next a brief summary of the most precise determinations of the strong coupling, following the PDG organization of the results by subfields with their own intermediate averages \cite{HRZ:2020xxx}. A much more detailed discussion, including many analyses performed at a lower NLO precision, can be found in Refs.~\cite{Aoki:2019cca,HRZ:2020xxx,Bethke:2011tr,Pich:2013sqa,Moch:2014tta,dEnterria:2015kmd,BSG:2018xxx,Pich:2018lmu,dEnterria:2019its,Komijani:2020kst}. 

\subsection[Jets in $e^+e^-$ annihilation]{Jets in $\boldsymbol{e^+e^-}$ annihilation}

With a proper (infrared and collinear safe) jet definition, the study of jet production
in $e^+e^-$ annihilation provides a wealth of interesting dynamical information. The jet rates have a high sensitivity to the strong coupling, which increases with the jet multiplicity: the fraction of $n$-jet events grows as $R_n\sim \alpha_s^{n-2}$. In addition, there is a large variety of useful jet observables such as event shapes and energy correlations. Besides the global energy scale, the characterization of the observable jet unavoidably involves other physical scales ($E_{\mathrm{min}}$, $p_T$, $m_b$, $m_t$ \ldots), which makes necessary  to perform careful resummations of enhanced logarithmic corrections. Modern analyses include NNLO corrections \cite{Gehrmann-DeRidder:2007nzq,GehrmannDeRidder:2007jk,GehrmannDeRidder:2007hr,GehrmannDeRidder:2008ug,Weinzierl:2008iv,GehrmannDeRidder:2009dp,Weinzierl:2009ms,Weinzierl:2009yz,Weinzierl:2010cw,DelDuca:2016csb,DelDuca:2016ily},  matched to  next-to-leading logarithmic (NLL)  \cite{Catani:1992ua,Gehrmann:2008kh} or NNLL \cite{deFlorian:2004mp,Monni:2011gb,Becher:2012qc,Banfi:2014sua,Kardos:2018kqj} resummations. A higher N${}^3$LL accuracy has been achieved for thrust  \cite{Becher:2008cf}, C-parameter \cite{Hoang:2014wka,Hoang:2015hka}  and heavy-jet mass \cite{Chien:2010kc}, applying soft-collinear effective theory (SCET) \cite{Bauer:2000yr,Bauer:2001yt} techniques.
A good control of non-perturbative power corrections and hadronization effects is also needed. Monte Carlo models are usually employed for this purpose, but some recent works
perform analytic modelling with inverse power corrections, SCET factorization, dispersive methods or low-scale effective couplings \cite{Hoang:2014wka,Hoang:2015hka,Davison:2008vx,Abbate:2010xh,Gehrmann:2012sc}.

\begin{figure}[ht]
\centering
\includegraphics[width=0.7\textwidth,clip]{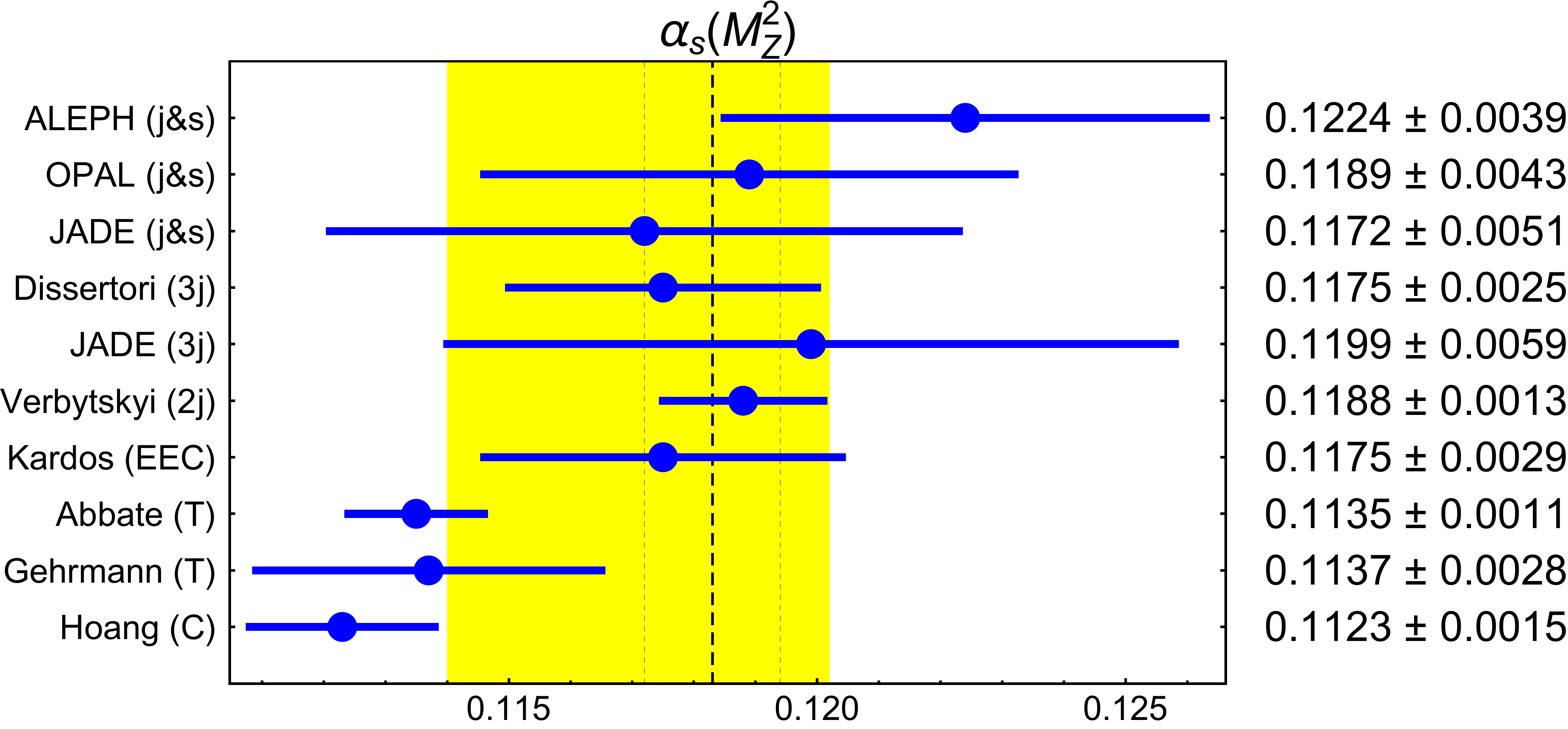}
\caption{NNLO determinations of $\alpha_s(M_Z^2)$ from $e^+e^-$ jets. The yellow band corresponds to the PDG recommended range in Eq.~\eqn{eq:jets}, while the vertical dashed lines show the world average value in Eq.~\eqn{eq:AlphaFinalWA}.} 
\label{fig:jets}
\end{figure}

The most precise determinations of the strong coupling with $e^+e^-$ jet data are shown in figure~\ref{fig:jets}. From top to bottom, the figure includes three NNLO+NLL re-analyses of event shapes (j{\&}s) at ALEPH \cite{Dissertori:2009ik}, 
OPAL  \cite{OPAL:2011aa}   
and JADE  \cite{Bethke:2008hf}, 
two NNLO analyses of 3-jet rates (3j) from 
ALEPH \cite{Dissertori:2009qa}  
and JADE data \cite{Schieck:2012mp},   
a precise NNLO+NNLL fit to the region of lower 3-jet rate (2j) \cite{Verbytskyi:2019zhh},
using LEP and PETRA data, and a fit to the available energy--energy correlation (EEC) data \cite{Kardos:2018kqj}.
All these analyses apply Monte Carlo models to estimate the transition from partons to hadrons.

The last three determinations employ analytic modelling of power corrections and hadronization, and are based on NNLO + NNLL \cite{Gehrmann:2012sc} or even N${}^3$LL \cite{Hoang:2015hka,Abbate:2010xh} analyses of the world data on thrust (T)
\cite{Abbate:2010xh,Gehrmann:2012sc} and C-parameter (C) \cite{Hoang:2015hka}
distributions. 
Ref.~\cite{Gehrmann:2012sc} takes into account hadronization effects through an effective coupling frozen in the infrared, while Refs.~\cite{Hoang:2015hka,Abbate:2010xh} incorporate explicitly the leading non-perturbative power corrections, which are also fitted to the data, together with a sophisticated infrared renormalon subtraction. The inclusion in the fit of an inverse power correction results in a large decrease of the central value ($\Delta\alpha_s\sim -0.01$), while the total uncertainty is reduced by a factor close to 3 after the renormalon subtraction. A study of the first moment of the thrust distribution by the same group (not included in the figure) gives a less precise but consistent result $\alpha_s(M_Z^2) = 0.1140\pm 0.0023$ \cite{Abbate:2012jh}. The rather low values obtained for the strong coupling are rather unexpected and the small quoted uncertainties should be better understood. In particular, the size of subleading power corrections remains to be investigated. Moreover, these determinations assume that the leading power correction is independent of the event-shape variable. A recent study of this correction for the C-parameter has pointed out a large variation of its coefficient, being over a factor of two smaller at $C=3/4$ than at $C=0$~\cite{Luisoni:2020efy}. Interpolating between these two singular configurations, the fitted value of $\alpha_s(M_Z^2)$ increases by about 3-4\%~\cite{Luisoni:2020efy}, becoming then compatible with the current world average in Eq.~\eqn{eq:AlphaFinalWA}, which is indicated by the vertical dashed lines in figure~\ref{fig:jets}.

In order to get a combined value of the strong coupling, avoiding that singular optimistic estimates of systematic uncertainties could bias the average, the PDG review on QCD adopts for each sub-field an unweighted average of all selected results and their quoted errors. From these $e^+e^-$ results, based on NNLO predictions, the PDG prescription gives the range \cite{HRZ:2020xxx}
\begin{equation}\label{eq:jets}
\alpha_s(M_Z^2)\; =\; 0.1171 \pm 0.0031\, ,
\end{equation}
which is displayed as a yellow band in figure~\ref{fig:jets}.

\subsection{Jets at hadron and electron--proton colliders}

\begin{figure}[ht]
\centering
\includegraphics[width=0.7\textwidth,clip]{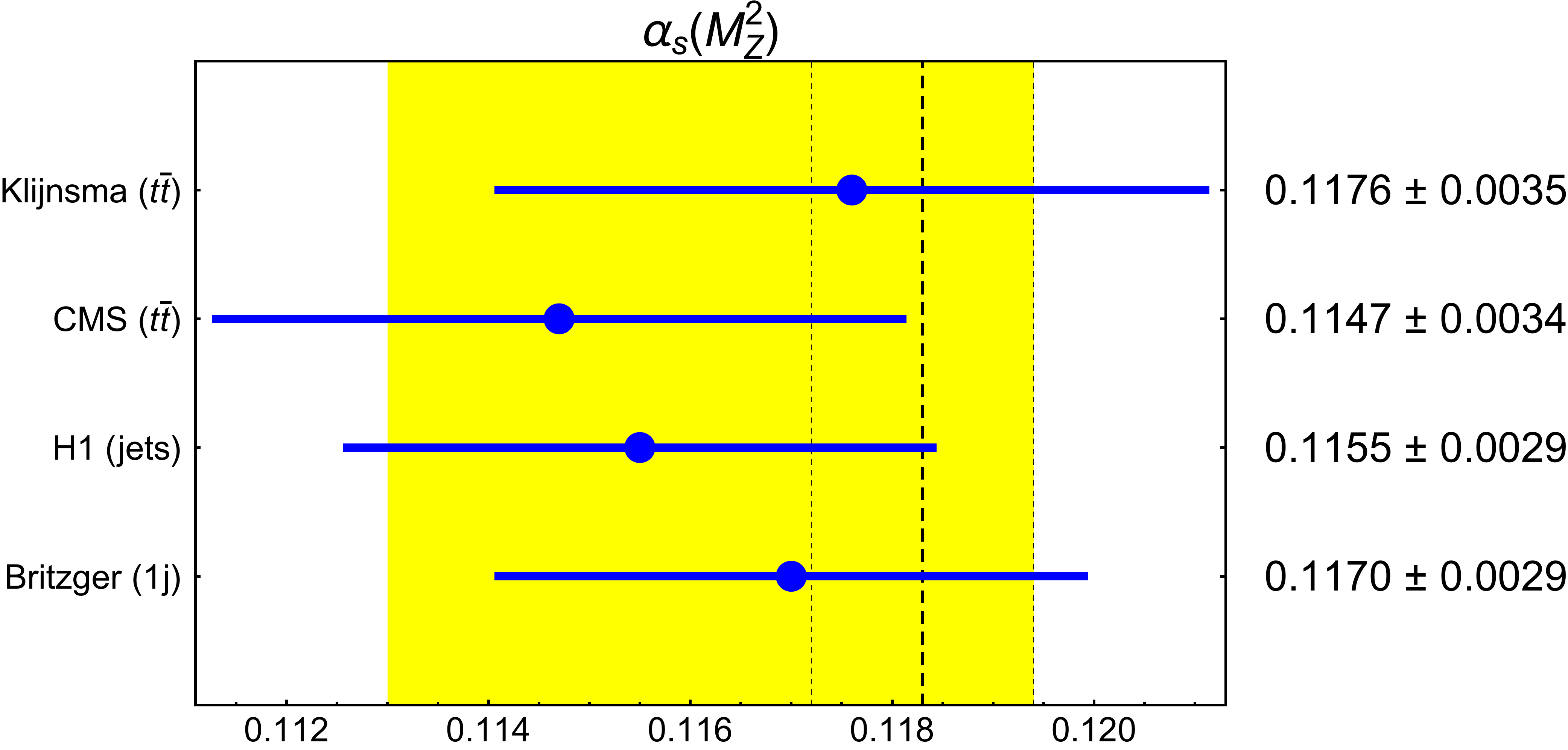}
\caption{NNLO determinations of $\alpha_s(M_Z^2)$ from jets at hadron and $ep$ colliders. The yellow band displays the unweighted average in Eq.~\eqn{eq:HadronJets}, while the vertical dashed lines show the world average value in Eq.~\eqn{eq:AlphaFinalWA}.} 
\label{fig:HadronJets}
\end{figure}

Several NNLO calculations have recently become available for some selected processes at hadron colliders: the production of $t\bar t$ pairs \cite{Czakon:2013goa,Czakon:2015owf,Catani:2019hip}, including  some NNLL resummations \cite{Czakon:2018nun},
inclusive jet \cite{Currie:2016bfm,Czakon:2019tmo} and dijet \cite{Currie:2017eqf} production, and $Z+$ 1-jet production \cite{Boughezal:2015ded,Ridder:2016nkl}.
The more important electroweak and mixed QCD-electroweak corrections to these processes have been also computed \cite{Dittmaier:2012kx,Frederix:2016ost,Czakon:2017wor}.
Figure~\ref{fig:HadronJets} displays the NNLO results for $\alpha_s(M_Z^2)$ obtained in high-energy hadronic collisions. A first determination by CMS \cite{Chatrchyan:2013haa}, from the $t\bar t$ production cross section at $\sqrt{7}$~TeV, has been already superseded by a more general analysis that includes additional $t\bar t$ data from the Tevatron and LHC (Klijnsma) \cite{Klijnsma:2017eqp}. More recently, an independent analysis of  new $t\bar t$ production data at $\sqrt{13}$~TeV has been presented by CMS \cite{Sirunyan:2018goh}. The figure shows the unweighted average of the CMS results obtained with four different sets of particle distribution functions (PDFs)~\cite{HRZ:2020xxx}.

The third entry in figure~\ref{fig:HadronJets} corresponds to a NNLO analysis of jet production at HERA, performed by the H1 collaboration \cite{Andreev:2017vxu}. The numerical value shown is an unweighted PDG average~\cite{HRZ:2020xxx} of the published H1 results, which combines the determination obtained from a joint PDF plus $\alpha_s$ fit with the most precise value of $\alpha_s$ obtained with pre-determined PDFs. The last entry in the figure 
is a NNLO global fit of single-jet production at HERA, which applies fast interpolation grid techniques \cite{Britzger:2019kkb}. Fitting only the H1 data gives  $\alpha_s(M_Z^2) = 0.1153\pm 0.0033$, while $0.1194\pm 0.0034$ is found with the ZEUS data \cite{Britzger:2019kkb}. The figure displays the combined value
including data from both experiments,  

The unweighted average of these four determinations,
\begin{equation}\label{eq:HadronJets}
\alpha_s(M_Z^2)\; =\; 0.1162 \pm 0.0032\, ,   
\end{equation}
is indicated by the yellow region in figure~\ref{fig:HadronJets}. 

\begin{figure}[t]
\centering
\begin{minipage}[c]{0.48\textwidth}
\includegraphics[width=\textwidth,clip]{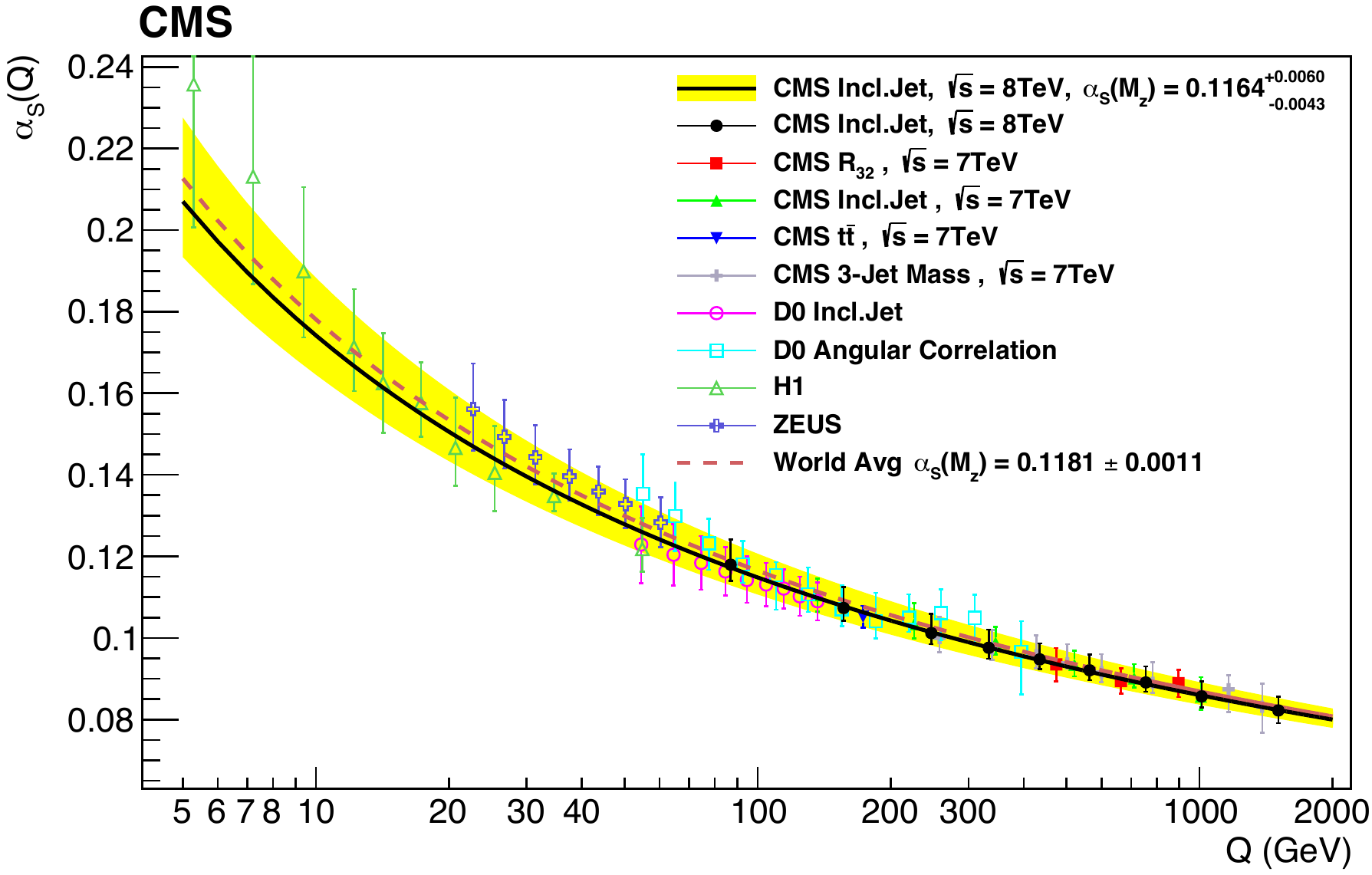}
\end{minipage}
\hskip .5cm
\begin{minipage}[c]{0.47\textwidth}
\mbox{}\vskip .23cm
\includegraphics[width=0.96\textwidth,clip]{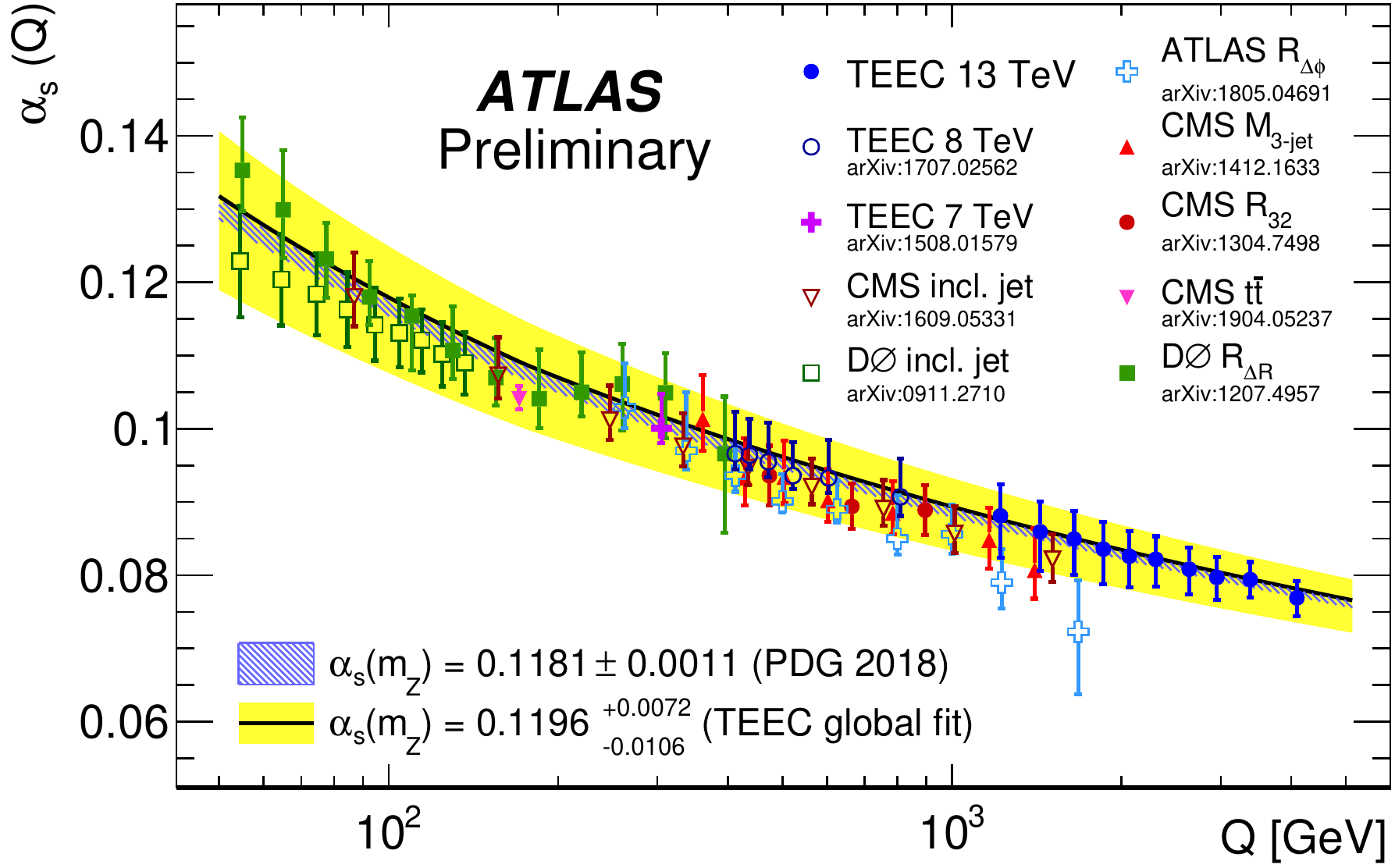}
\end{minipage}
\caption{LHC determinations of the strong coupling at different energy scales 
from CMS (left) and ATLAS (right), compared with previous results from the Tevatron and HERA. Figures taken from Refs.~\cite{Khachatryan:2016mlc,ATLAS:2020mee}.} 
\label{fig:HEjets}
\end{figure}

A very recent analysis of $t\bar t$ production data from ATLAS and CMS includes also
NNLO predictions of the differential distributions 
\cite{Cooper-Sarkar:2020twv}. Among the many extractions studied, using different data sets and PDFs, the most precise determination of the strong coupling is obtained from two ATLAS distributions and the CT14 PDF set, which give
$\alpha_s(M_Z^2)\; =\; 0.1159 \, {}^{+\, 0.0013}_{-\, 0.0014}$, showing the potentially high sensitivity of the differential distributions. This result is not displayed in the figure because theoretical uncertainties have not been yet included and the quoted errors do not account for the variations of the fitted results with different choices of input data and PDFs.

There are many other determinations of the strong coupling at hadron colliders, which do not comply with the requested NNLO theoretical accuracy. Nevertheless, they constitute an important test of QCD at the highest available energies. Figure~\ref{fig:HEjets}~\cite{Khachatryan:2016mlc,ATLAS:2020mee} compiles a large number of measurements of $\alpha_s$ performed at HERA, the Tevatron and the LHC, reaching energy scales up to 4~TeV. The agreement with the predicted running of the QCD coupling, also shown in the figure, is excellent over the whole range of energies explored.

\subsection{Particle distribution functions}

\begin{figure}[h]
\centering
\includegraphics[width=0.7\textwidth,clip]{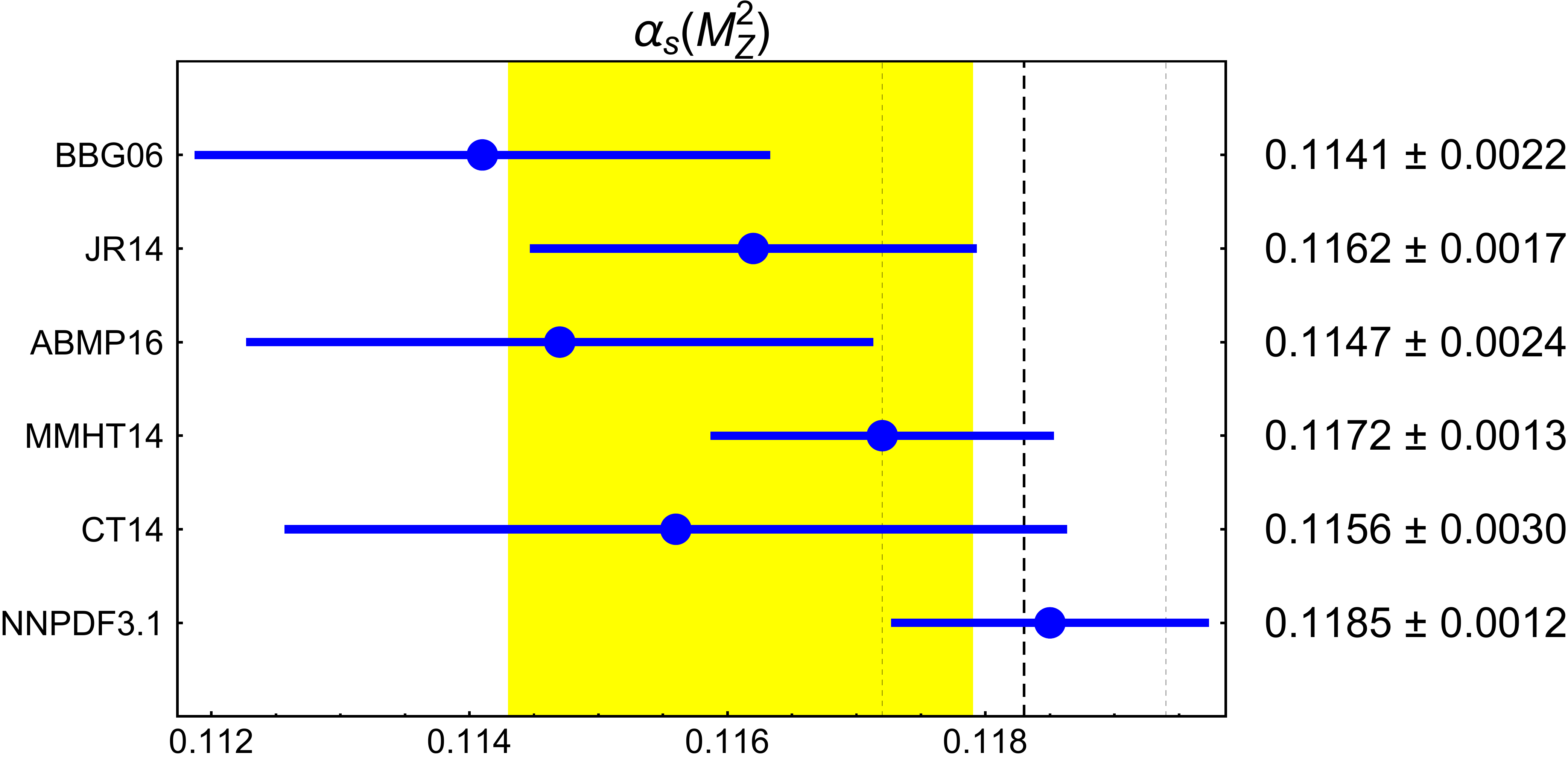}
\caption{NNLO determinations of $\alpha_s(M_Z^2)$ from particle distribution functions. The yellow band corresponds to the PDG recommended range in Eq.~\eqn{eq:pdf}, while the vertical dashed lines show the world average value in Eq.~\eqn{eq:AlphaFinalWA}.} 
\label{fig:PDFs}
\end{figure}

Precise determinations of $\alpha_s$ can also be obtained through the analysis of scaling violations in PDFs. Good deep-inelastic-scattering (DIS) data are available over a wide range of energies and, in particular, the HERA experiments provided a very accurate data set. The results from the most recent NNLO fits to the data are displayed in figure~\ref{fig:PDFs}, together with an older NNLO analysis of non-singlet structure functions in DIS (BBG06) \cite{Blumlein:2006be} that included some N${}^3$LO corrections, but neglected singlet contributions for $x> 0.3$ where the valence approximation was used (this has been claimed to have a negligible numerical effect \cite{Blumlein:2012se}). 
Both singlet and non-singlet structure functions are taken into account in the more recent studies of DIS  by the JR14\footnote{
A smaller central value $\alpha_s(M_Z^2) = 0.1136$ is obtained in
Ref.~\cite{Jimenez-Delgado:2014twa}, assuming 
valence-like PDFs at a low scale $Q_0^2=0.8\;\mathrm{GeV}^2$. The figure displays the result from a less-constrained standard fit with $Q_0^2=2\;\mathrm{GeV}^2$. The enlarged error accounts for the difference between both analyses.
} \cite{Jimenez-Delgado:2014twa} and ABMP16 \cite{Alekhin:2017kpj,Alekhin:2018pai} groups, together with Drell-Yan and di-muon data needed for a correct description of the sea-quark densities. The ABMP16 group also includes top-quark, $Z$ and $W$ production data from the Tevatron and LHC.

The global PDF analyses of the MMHT14 \cite{Harland-Lang:2015nxa}, CT14 \cite{Dulat:2015mca} and NNPDF3.1 \cite{Ball:2018iqk} groups incorporate into the fit a broader set of data from fixed-target experiments, HERA and the Tevatron and LHC colliders. Besides scaling violations, the dependence on $\alpha_s$ of the hard-scattering matrix elements of the different processes analysed is also exploited. The proton collider data allows for a better control of the gluon PDF, which turns out to be highly correlated with the fitted value of $\alpha_s(M_Z^2)$~\cite{Thorne:2011kq}. The NNPDF3.1 fit includes a much larger set of LHC data, being the first global analysis to simultaneously use differential top, inclusive jet, and $Z$ $p_T$ distribution data, all using exact NNLO theory.

It has been argued that the lower values of $\alpha_s(M_Z^2)$ emerging from the BBG06 and ABMP16 fits could be partly explained by the use of a fixed flavour-number scheme with $n_f = 3$ for the treatment of DIS data \cite{Ball:2013gsa,Thorne:2014toa}. 
The inclusion of LHC top, $W$ and $Z$ production data (described with $n_f = 5$) in the current AMBP16 analysis has in fact increased the fitted value of the strong coupling with respect to previous results from the same group~\cite{Alekhin:2013nda}. 
Lower values of $\alpha_s(M_Z^2)$ in DIS-only fits seem to be preferred by the oldest sets of data (BCDMS, E665, SLAC) \cite{Parente:1994bf,Shaikhatdenov:2009xd}, while higher values are favoured by the most recent experiments (NMC, HERA)~\cite{Martin:2009bu}. The improved constraints on the gluon PDF emerging from the collider data help to resolve this disagreement. Removing the precise BCDMS data increases significantly the result of the DIS-only fit, but has a marginal effect in the global fit~\cite{Thorne:2011kq}. Notice that NNLO fits result in slightly smaller values of $\alpha_s$ than NLO fits \cite{Alekhin:2012ig}.

On the other side, the higher value of $\alpha_s$ obtained in the NNPDF3.1 fit is driven by the high-precision LHC data, especially for gauge boson production
(including the $Z$ $p_T$ distribution) but also for top and jet production. This has been confirmed through separate statistical analyses of different sets of data, suggesting that the results from other groups would probably increase with the inclusion of the additional LHC data sets \cite{Ball:2018iqk}.

The unweighted average of all these determinations quoted by the PDG~\cite{HRZ:2020xxx},
\begin{equation}\label{eq:pdf}
\alpha_s(M_Z^2)\; =\; 0.1161 \pm 0.0018\, ,   
\end{equation}
is indicated with a yellow band in figure~\ref{fig:PDFs}. The world average value in Eq.~\eqn{eq:AlphaFinalWA} is also shown with dashed vertical lines.

\subsection{Quarkonium}

The bound states of a heavy quark and a heavy antiquark are rigorously described with
non-relativistic QCD (NRQCD) \cite{Caswell:1985ui,Bodwin:1994jh,Pineda:1997bj,Brambilla:1999xf,Luke:1999kz}, through a combined expansion in powers of $\alpha_s$ and the heavy-quark velocity $v$. Using these techniques, radiative $\Upsilon$ decays have been used to determine the strong coupling \cite{Brambilla:2007cz}, but only a NLO accuracy in $\alpha_s(m_b^2)$ and $v^2$ has been achieved so far.

\begin{figure}[h]
\centering
\includegraphics[width=0.7\textwidth,clip]{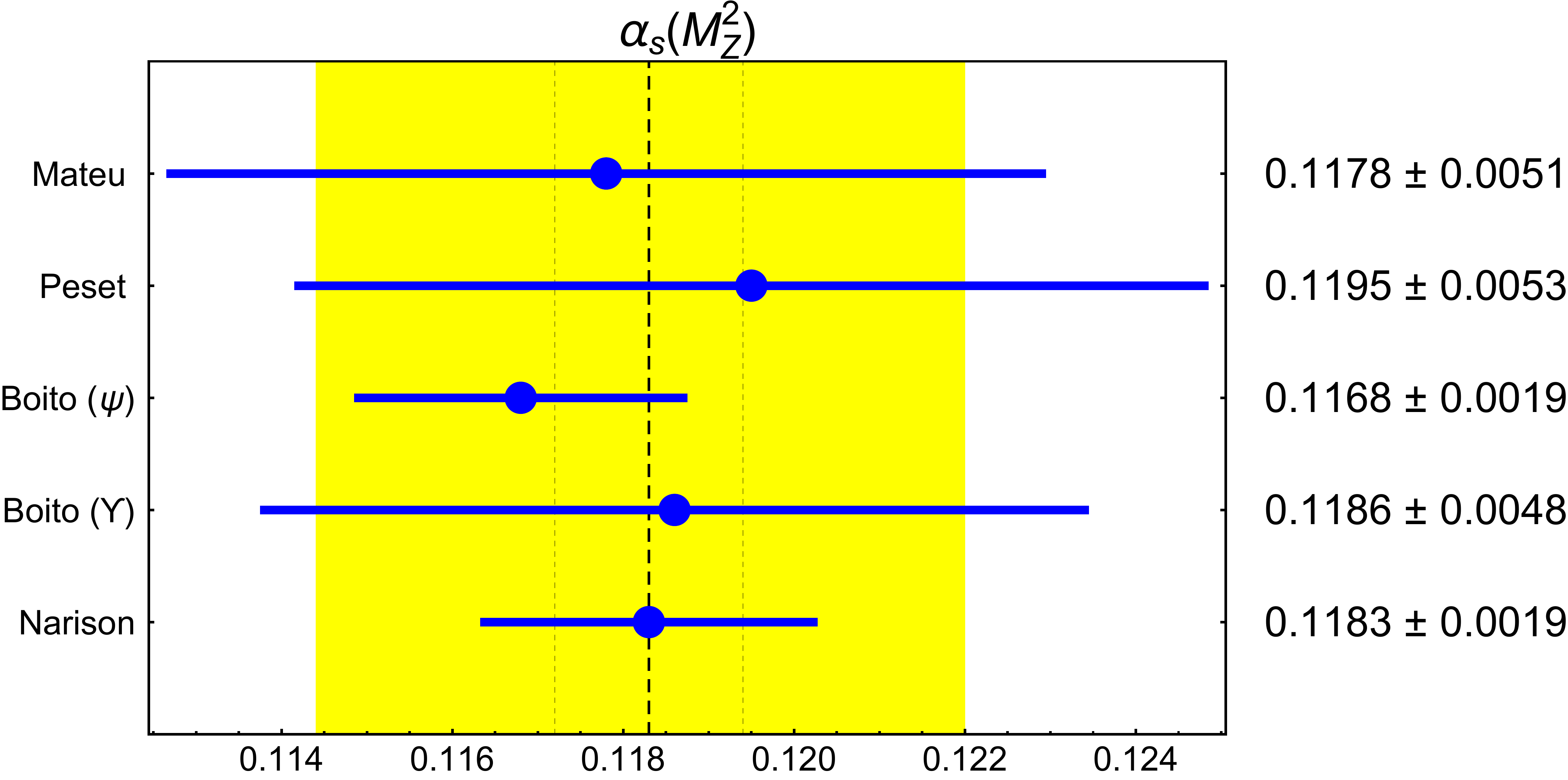}
\caption{NNLO determinations of $\alpha_s(M_Z^2)$ from quarkonium states. The yellow band indicates the unweighted average in Eq.~\eqn{eq:quarkonium}, while the vertical dashed lines show the world average value in Eq.~\eqn{eq:AlphaFinalWA}.} 
\label{fig:Quarkonium}
\end{figure}

Two determinations of the strong coupling and the heavy quark masses from quarkonium systems have been performed recently at N${}^3$LO in the NRQCD expansion, {\it i.e.}, including corrections to the quarkonium  spectrum  up to $\cO(m_Q\alpha_s^5)$ and 
$\cO(m_Q\alpha_s^5\log{\alpha_s})$ \cite{Kiyo:2013aea,Kiyo:2014uca}. They correspond to the top two entries in figure~\ref{fig:Quarkonium}. The first one (Mateu)  \cite{Mateu:2017hlz} performs a simultaneous fit of $\alpha_s$ and $m_b$ to the bottomonium states with principal quantum number $n\le 2$. 
The inclusion of both $n=1$ and 2 states is crucial to achieve a separate (but highly correlated) sensitivity to the bottom mass and the strong coupling.
The second determination (Peset)~\cite{Peset:2018ria} considers instead the renormalon-free combination of the $\eta_b$, $\eta_c$ and $B_c$ masses, $M_{B_c} - (M_{\eta_b} + M_{\eta_c})/2$, which is weakly dependent on the heavy-quark masses.

The figure includes two additional NNLO determinations from charmonium (Boito, $\psi$)~\cite{Boito:2019pqp} and bottomonium (Boito, $\Upsilon$)~\cite{Boito:2020lyp} sum rules, using ratios of moments of the electromagnetic charm and bottom spectral functions,
respectively. They combine the contribution from the narrow states below threshold and the available threshold data with a higher-energy continuum modelled with perturbative QCD, and require a background subtraction from non-charm or non-bottom states. 
The last entry in the figure (Narison) \cite{Narison:2018dcr} has been extracted from a study of the mass splitting $M_{\chi_{c0}}-M_{\eta_{c}}$, using Laplace Sum Rules with the two-point correlation functions of the charm scalar and pseudoscalar currents.

The unweighted combination of these five determinations,
\begin{equation}\label{eq:quarkonium}
\alpha_s(M_Z^2)\; =\; 0.1182 \pm 0.0038\, ,   
\end{equation}
agrees nicely with the results  from other physical systems, discussed previously. 

\subsection{Lattice determination}

The strong coupling can be determined non-perturbatively by measuring various Euclidean short-distance quantities, through a numerical evaluation of the QCD functional integral in a discretized space--time lattice, and comparing the results with the corresponding perturbative expansions in powers of $\alpha_s$. This involves using lattice QCD perturbation theory which introduces lattice-spacing artefacts. Most modern simulations include $2+1$ flavours of sea quarks (two taken with masses as small as possible for up and down and the other one tuned to the strange quark), 
and have a NNLO perturbative accuracy. At least one dimensionful physical quantity is needed to convert from lattice units to GeV, {\it i.e.}, to determine the scale at which $\alpha_s$ is measured. The hadron spectrum is normally used to fix the overall energy scale and the quark masses, but other intermediate scales related to dimensionful measured observables can also be employed.
The dominant sources of uncertainty in current lattice analyses of the strong coupling originate in the truncation of continuum/lattice perturbation theory and from discretization effects.

\begin{figure}[ht]\centering
\includegraphics[width=.6\textwidth]{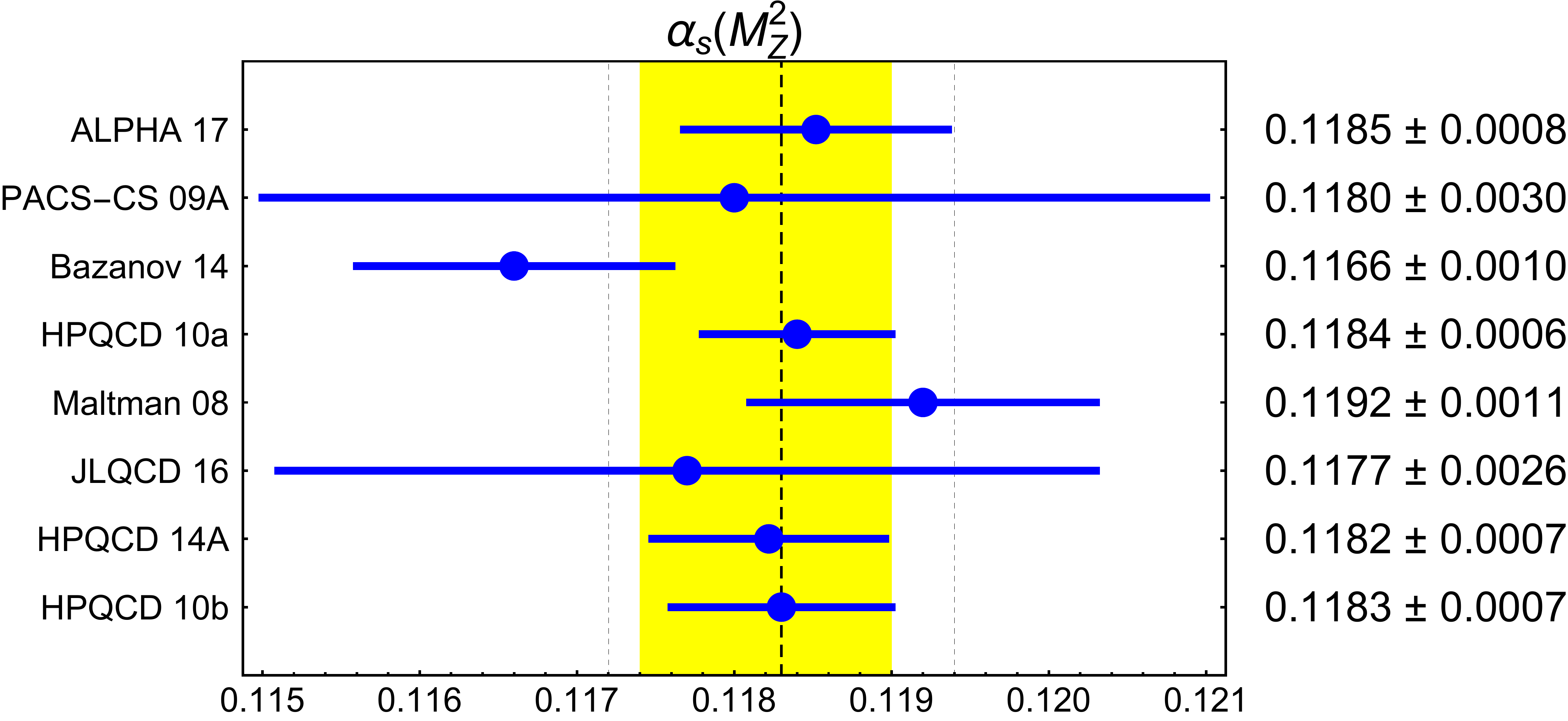}
\caption{Lattice determinations of $\alpha_s(M_Z^2)$ included in the FLAG average (yellow band)  \cite{Aoki:2019cca}. The vertical dashed lines show the world average value in Eq.~\eqn{eq:AlphaFinalWA}.}
\label{fig:LatticeAverage}
\end{figure}

The Flavour Lattice Averaging Group (FLAG) \cite{Aoki:2019cca} has reviewed the most reliable determinations of $\alpha_s$ and has combined them into a world lattice average. FLAG has established a set of quality requirements that a determination must satisfy in order to be included in the average, retaining only the eight lattice results displayed in figure~\ref{fig:LatticeAverage}.
Not yet considered in the FLAG compilation is a recent determination from the ghost-gluon vertex, obtained with $2+1$ flavours of domain-wall fermions and physical quark masses, which finds $\alpha_s(M_Z^2) = 0.1172\pm 0.0011$ \cite{Zafeiropoulos:2019flq}.

The computations of ALPHA 17 \cite{Bruno:2017gxd} and PACS-CS 09A \cite{Aoki:2009tf} are based on the so-called step-scaling method \cite{Luscher:1991wu}, running the coupling step-wise from the low energy region where the coupling scale is fixed to high energies around 50~GeV where the matching to the $\overline{\mathrm{MS}}$ scheme is performed.
Both collaborations adopt the Schr\"odinger functional scheme to carry out a non-perturbative running of the coupling, combined with the gradient flow scheme at higher energies in the ALPHA case.

The force between an infinitely massive quark and antiquark pair separated by a distance $r$ is analysed with lattice data in Bazanov~14 \cite{Bazavov:2014soa}, combining a tree-level improved gauge action with a highly-improved staggered quark action. Perturbative NRQCD calculations of the short-distance part of the static energy are used to determine the strong coupling (evaluated at $\mu\sim 1/r$). This determination has been slightly updated in \cite{Bazavov:2019qoo}. A recent re-analysis of these lattice data, using hyperasymptotic approximations (based on renormalon calculus) and a N${}^3$LL resummation of large logarithms, finds a larger value for the strong coupling: $\alpha_s(M_Z^2) = 0.1181\pm 0.0009$  \cite{Ayala:2020odx}.

The figure includes three different determinations of $\alpha_s$ by the HPQCD collaboration, which uses staggered fermions and fixes the lattice spacing with a wide variety of physical quantities. The HPQCD 10a result \cite{McNeile:2010ji} is extracted from 22 different simulations of small Wilson loops. An independent perturbative analysis of Wilson loops, using the results of a previous HPQCD-UKQCD simulation \cite{Mason:2005zx} already superseded by the new data, finds the slightly larger value tagged as Maltman 08~\cite{Maltman:2008bx}. The other two HPQCD determinations are obtained from moments of the correlation function of two heavy-quark currents. Eight different values of the heavy-quark mass between $m_c$ and $m_b$ and five different lattice spacings are analysed in
HPQCD 10b \cite{McNeile:2010ji}, while a $2+1+1$ lattice simulation is employed to derive the HPQCD 14A result \cite{Chakraborty:2014aca}. Finally, the JLQCD 16 \cite{Nakayama:2016atf} result is obtained from a simulation of the $c\bar c$ pseudoscalar two-point function, generated with  $2+1$ flavours of light sea quarks described with M\"obius domain-wall fermions.

The final FLAG average of these results \cite{Aoki:2019cca},
\begin{equation}\label{eq:lattice}
\alpha_s(M_Z^2)\; =\; 0.1182 \pm 0.0008\, ,
\end{equation}
corresponds to the yellow vertical region in figure~\ref{fig:LatticeAverage}.


\subsection{World average value of the strong coupling}

\begin{figure}[ht]\centering
\includegraphics[width=.6\textwidth]{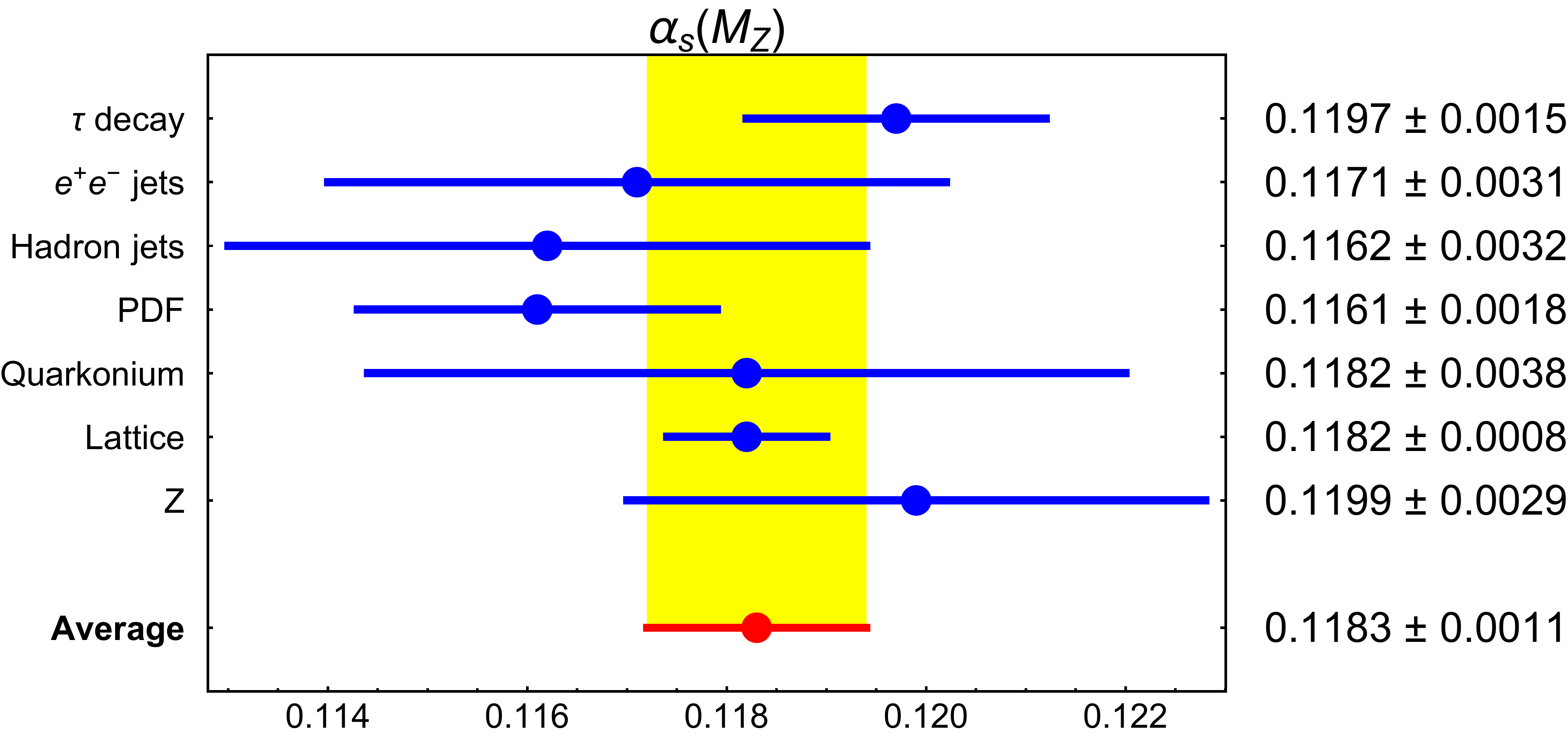}
\caption{Summary of $\alpha_s(M_Z^2)$ determinations from different physical systems.}
\label{fig:AlphaSummary}
\end{figure}

Determining a world average of $\alpha_s$ is a non-trivial and controversial task because systematic uncertainties dominate most measurements. The more or less conservative attitude adopted to estimate the errors of a given entry could easily bias the
result. Moreover, many theoretical and experimental inputs are highly correlated and the different observables analysed have different levels of theoretical precision.
Figure~\ref{fig:AlphaSummary} summarizes the pre-averages for each class of measurements, given in Eqs.~\eqn{eq:alpha_Z}, \eqn{eq:AlphaTauMZ}, \eqn{eq:jets}, \eqn{eq:HadronJets}, \eqn{eq:pdf}, \eqn{eq:quarkonium} and \eqn{eq:lattice}. 
Performing a weighted average of these seven pre-averages, which assumes them to be uncorrelated and of Gaussian nature, and adjusting the overall $\chi^2$ so that $\chi^2/\mathrm{d.o.f}$ equals unity \cite{Schmelling:1994pz} (this slightly increases the final error), one gets
\bel{eq:AlphaSummary}
\alpha_s(M_Z^2)\; =\; 0.1184 \pm 0.0007\, .
\end{equation}
%

Since the higher precision quoted by the lattice result dominates the final average, it is interesting to perform a separate non-lattice average of the other six entries in the figure. This gives
\bel{eq:AlphaNonLattice}
\alpha_s(M_Z^2)\; =\; 0.1184 \pm 0.0013\, ,
\qquad\qquad \text{(without lattice)},
\end{equation}
in astonishing agreement with the lattice result but with a much larger uncertainty.
The combined N${}^3$LO determinations from the $Z$ and $\tau$ hadronic widths prefer a slightly larger value, $\alpha_s(M_Z^2) = 0.1197 \pm 0.0013$, while a somewhat lower  range is obtained combining the four NNLO non-lattice determinations:
$\alpha_s(M_Z^2) = 0.1165 \pm 0.0013$. 

The PDG review on QCD~\cite{HRZ:2020xxx} performs an unweighted average of the lattice and non-lattice results. This prescription gives a world average strong coupling with a similar central value than \eqn{eq:AlphaSummary} and a larger uncertainty:
\bel{eq:AlphaFinalWA}
\alpha_s(M_Z^2)\; =\; 0.1183 \pm 0.0011\, . 
\ee
This more conservative world average, which is indicated by the yellow region in figure~\ref{fig:AlphaSummary}, is in good agreement with the PDG 2020 value,
$\alpha_s(M_Z^2) = 0.1179 \pm 0.0010$.
In figures \ref{fig:jets}, \ref{fig:HadronJets}, \ref{fig:PDFs}, \ref{fig:Quarkonium} and \ref{fig:LatticeAverage} the range \eqn{eq:AlphaFinalWA} has been indicated with vertical dashed lines, so that it can be easily compared with each individual determination. There is an excellent overall agreement, which provides a very significant consistency test among results extracted from different physical observables and with a large variety of techniques. Only a few results seem to deviate a bit too much towards lower values of $\alpha_s$, which could indicate unaccounted systematics that needs to be better understood.

\section{Summary}
\label{sec:summary}

A series of impressive four- and five-loop calculations has promoted the phenomenology of inclusive QCD processes into the realm of precision physics. The very accurate knowledge of the $\beta$ and $\gamma$ functions, which provides a powerful resummation of logarithmic corrections into the running coupling and quark masses, has been complemented with the $\cO(\alpha_s^4)$ computation of the two-point correlation functions of the vector, axial-vector, scalar and pseudoscalar currents, allowing us to obtain high-precision theoretical predictions for $\sigma(e^+e^-\to\mathrm{hadrons})$ and the hadronic decay widths of the electroweak $Z$, $W$ and Higgs bosons and the $\tau$ lepton. 

These fully-inclusive observables are free from hadronization uncertainties and infrared ambiguities. Non-pertur\-ba\-ti\-ve corrections are strongly suppressed by the heavy boson masses or the high centre-of-mass energy of the $e^+e^-$ collision. In spite of being a much lower energy process, non-perturbative contributions to the $\tau$ hadronic width are also small because they are suppressed by six powers of the $\tau$ mass. Thanks to the analyticity properties of the current correlators and the more inclusive nature of $R_\tau$, these corrections can be rigorously handled with the help of the OPE and their small effects directly extracted from the data themselves.

Combined with high-quality $Z$ and $\tau$ decay data, these calculations have made possible to perform N${}^3$LO determinations of the strong coupling at two broadly separated scales, $M_Z$ and $m_\tau$. The excellent agreement with the predicted QCD running at five loops, exhibited in figure~\ref{fig:runningTau}, constitutes a beautiful and highly non-trivial confirmation of asymptotic freedom. Together with recent LHC measurements reaching energy scales up to 4~TeV, this also puts very severe constraints on new-physics scenarios involving strongly-interacting particles.

In recent years we have also witnessed a spectacular progress in the perturbative calculation of other QCD observables that have reached a NNLO theoretical accuracy, including in some cases NNLL or even N${}^3$LL resummations. This has been complemented with corresponding improvements of the PDFs, and updated Monte Carlo generators with appropriate matching of matrix elements and parton showering and more efficient tools to address multi-particle interactions at higher orders. Many NNLO determinations of the strong coupling have been accomplished, with an excellent overall agreement, verifying with high precision that the QCD coupling is indeed unique. 
Moreover, a large number of cross sections for different processes, spanning a broad range from $10^{-3}$ to $10^6$ pb, have been measured to be in remarkable agreement with the Standard Model predictions.

The combined use of effective field theories, largely based on symmetry considerations, and lattice simulations has also made possible to achieve a considerably progress in the more difficult non-perturbative regime. Although many things remain still to be more deeply understood, all theoretical and experimental results confirm the correctness of the QCD predictions. The spectacular phenomenological success of QCD clearly establishes this elegant quantum field theory as the right description of the strong interactions over all energy scales investigated so-far.

\section*{Acknowledgements}

This work has been supported in part by the Spanish Government and ERDF funds from the EU Commission [Grant FPA2017-84445-P], by the Generalitat Valenciana [Grant Prometeo/2017/053], by the EU H2020 research and innovation programme [Grant Agreement 824093]
and by the EU COST Action CA16201 PARTICLEFACE.

\appendix

\section{Running and pole quark masses}
\label{app:masses}

Combining the renormalization-group equations \eqn{eq:beta-rge} and \eqn{eq:runningMass}, the running quark mass $m_q(\mu^2)$ can be related
to its reference value $m_q(\mu_0^2)$ at any other scale $\mu_0$ through the expression
\bel{eq:runningMassSol}
m_q(\mu^2)\; =\; m_q(\mu_0^2)\; \exp{\left\{-\int_{\alpha_s(\mu_0^2)}^{\alpha_s(\mu^2)}
\frac{d \alpha_s}{\alpha_s}\;\frac{\gamma(\alpha_s)}{\beta(\alpha_s)}\right\}}
\; =\; m_q(\mu_0^2)\;
\frac{\cG_m(\alpha_s(\mu^2)/\pi)}{\cG_m(\alpha_s(\mu_0^2)/\pi)}
\;\equiv\; \hat m_q\; \cG_m(\alpha_s(\mu^2)/\pi)
\, ,
\ee
which defines the renormalization-group invariant mass $\hat m_q$.
Expanding the integrand in powers of $a=\alpha_s/\pi$, one finds the following perturbative expression for the function $\cG_m(a)$:
\bea
\cG_m(a)&\!\! = &\!\! a^{-\gamma_1/\beta_1}\,\left\{ 1 \, +\,  A_1\, a \, +\, \left(A_1^2 + A_2\right)\, \frac{a^2}{2}\, +\, \left(\frac{1}{2}\, A_1^3 + \frac{3}{2}\, A_1 A_2 + A_3\right)\, \frac{a^3}{3} 
\right.\no\\
&&\hskip 1.05cm\left. +\,\left(  \frac{1}{6}\, A_1^4 + A_1^2 A_2 + \frac{4}{3}\, A_1 A_3 + \frac{1}{2}\, A_2^2 + A_4\right)\, \frac{a^4}{4}
\, +\, \cO(a^5)
\right\}\, ,
\eea
where
\bea
A_1 &\!\!\! = &\!\!\! \frac{\beta_2\gamma_1}{\beta_1^2} - \frac{\gamma_2}{\beta_1}\, ,
\no\\
A_2 &\!\!\! = &\!\!\! \frac{\gamma_1}{\beta_1} \left( \frac{\beta_3}{\beta_1} - \frac{\beta_2^2}{\beta_1^2}\right) +  \frac{\gamma_2\beta_2}{\beta_1^2} -  \frac{\gamma_3}{\beta_1}\, ,
\no\\
A_3 &\!\!\! = &\!\!\!  \frac{\gamma_1}{\beta_1} \left(  \frac{\beta_4}{\beta_1} - 2\, \frac{\beta_2\beta_3}{\beta_1^2} + \frac{\beta_2^3}{\beta_1^3}  \right)
+ \frac{\gamma_2}{\beta_1} \left( \frac{\beta_3}{\beta_1} - \frac{\beta_2^2}{\beta_1^2} \right) +  \frac{\gamma_3\beta_2}{\beta_1^2}
-  \frac{\gamma_4}{\beta_1}\, ,
\no\\
A_4 &\!\!\! = &\!\!\! 
\frac{\gamma_1}{\beta_1} \left(  \frac{\beta_5}{\beta_1} -2\,\frac{\beta_2\beta_4}{\beta_1^2} - \frac{\beta_3^2}{\beta_1^2} + 3\,\frac{\beta_2^2\beta_3}{\beta_1^3}
-\frac{\beta_2^4}{\beta_1^4} \right)
+ \frac{\gamma_2}{\beta_1} \left( \frac{\beta_4}{\beta_1} - 2\,\frac{\beta_2\beta_3}{\beta_1^2} +  \frac{\beta_2^3}{\beta_1^3}\right)
+ \frac{\gamma_3}{\beta_1} \left(  \frac{\beta_3}{\beta_1}-  \frac{\beta_2^2}{\beta_1^2
}\right)
+ \frac{\gamma_4\beta_2}{\beta_1^2} - \frac{\gamma_5}{\beta_1}\, .\qquad\;
\eea
Owing to the presence of an exponential function in \eqn{eq:runningMassSol}, the running has a very significant numerical effect. For instance,
$m_q(\mu^2)/m_q(M_Z^2) \approx 2.0$ at $\mu=2~\mathrm{GeV}$. Thus, the running quark masses become lighter when the scale increases. 

The light quark masses are usually given at the reference scale $\mu_m=2$~GeV \cite{Zyla:2020zbs}:
\bel{eq:lightMasses}
m_u(\mu_m^2)\, =\, \left(2.16\,{}^{+\,0.49}_{-\, 0.26}\right)\;\mathrm{MeV},
\qquad\qquad
m_d(\mu_m^2)\, =\, \left(4.67\,{}^{+\,0.48}_{-\, 0.17}\right)\;\mathrm{MeV},
\qquad\qquad
m_s(\mu_m^2)\, =\, \left(93\,{}^{+\,11}_{-\, 5}\right)\;\mathrm{MeV}.
\ee
For heavy quarks, the masses are better normalized at their own mass scale \cite{Zyla:2020zbs}: 
\bel{eq:heavyMasses}
m_c(m_c^2)\, =\, \left(1.27\pm 0.02\right)\;\mathrm{GeV},
\qquad\qquad
m_b(m_b^2)\, =\,  \left(4.18\,{}^{+\,0.03}_{-\, 0.02}\right)\;\mathrm{GeV},
\qquad\qquad
m_t(m_t^2)\, =\,  \left(162.5\,{}^{+\,2.1}_{-\, 1.5}\right)\;\mathrm{GeV}.
\ee

In some phenomenological applications the perturbative on-shell pole mass $M_q$ is also used for heavy quarks. The relation between these two definitions, 
\bel{eq:PoleMass}
M_q\, =\, m_q(m_q^2)\,\left\{ 1 + \sum_{n=1}\, c_m^{(n)}\; \left(\frac{\alpha_s(m_q^2)}{\pi}\right)^n\right\}\, ,
\ee
is currently known to $\cO(\alpha_s^4)$:
\bea\label{eq:PoleMassCoefficients}
c_m^{(1)} & = & \frac{4}{3}\qquad \mbox{\cite{Tarrach:1980up}},
\no\\
c_m^{(2)} & = & 
\; =\; 13.4434 - 1.04137\, n_\ell
\qquad \mbox{\cite{Gray:1990yh,Fleischer:1998dw}}, 
\no\\
c_m^{(3)} & = &  190.595 - 26.655\, n_\ell + 0.6527\, n_\ell^2
\qquad \mbox{\cite{Chetyrkin:1999ys,Chetyrkin:1999qi,Melnikov:2000qh}},
\no\\
c_m^{(4)} & = & (3567.60\pm 1.64) - (745.721\pm 0.040)\, n_\ell + 43.3963\, n_\ell^2 - 0.678141\, n_\ell^3
\qquad \mbox{\cite{Marquard:2015qpa,Marquard:2016dcn}} , 
\eea
with $n_\ell = n_f-1$ the number of light (massless) quarks. The perturbative coefficients of this expansion are rather large. The series has a reasonable convergence for $n_\ell = 5$ (top quark), but with $n_\ell=4$ (bottom) or 3 (charm) the $\cO(\alpha_s^2)$, $\cO(\alpha_s^3)$ and $\cO(\alpha_s^4)$ terms have the same order of magnitude. For charm, the four-loop term is even almost twice as large as the three-loop contribution. This bad perturbative behaviour is inherited by those observables written in terms of the pole mass, even if they are perfectly well behaved with other mass definitions such as the running quark mass.  Therefore, the use of pole masses in precision QCD tests should be discouraged.

The value of $m_t$ given in Eq.~\eqn{eq:heavyMasses} is the running top mass extracted from cross-section measurements, which is unambiguously defined within QCD. The Particle Data Group \cite{Zyla:2020zbs} quotes a more precise value for the top pole mass obtained from this type of measurements:
$M_t = (172.4\pm 0.7)$~GeV. A much more accurate value $M_t = (172.76\pm 0.30)$~GeV~ \cite{Zyla:2020zbs} can be determined from the kinematics of $t\bar t$ events, assuming that the top mass used in the Monte Carlo event generators corresponds to the pole mass; however, the theoretical uncertainty introduced by this interpretation is difficult to quantify \cite{Moch:2014tta}.

\section{Matching coefficients}
\label{app:matching}

The matching conditions for the running strong coupling and quark masses are currently known to four loops. 
In the $\overline{\mathrm{MS}}$ scheme, the two-loop perturbative coefficients for the matching of $\alpha_s$ are \cite{Larin:1994va,Bernreuther:1981sg,Wetzel:1981qg,Bernreuther:1983zp}
\be
d_{10} = 0\, ,
\qquad d_{11}= -\frac{1}{6}\, ,
\qquad\qquad
d_{20} = \frac{11}{72}\, ,
\qquad d_{21}= -\frac{11}{24}\, ,
\qquad d_{22}= \frac{1}{36}\, ,
\ee
the three-loop results are given by \cite{Chetyrkin:1997sg}
\be
d_{30} =\frac{564731}{124416} - \frac{82043}{27648}\,\zeta_3 - \frac{2633}{31104}\, n_\ell\, , \qquad
d_{31}= -\frac{955}{576} + \frac{67}{756}\, n_\ell\, ,
\qquad d_{32}= \frac{53}{576}-\frac{1}{36}\, n_\ell\, ,
\qquad d_{33}= -\frac{1}{216}\, ,
\ee
where $n_\ell = n_f-1$, while at four loops one finds \cite{Schroder:2005hy,Chetyrkin:2005ia}:
\be
d_{40} = d_{40}^a + d_{40}^b\, n_\ell
+ n_\ell^2\,\left[ -\frac{271883}{4478976} + \frac{167}{5184}\,\zeta_3\right]\, ,
\ee
with $d_{40}^a = 5.170346991\ldots$ and $d_{40}^b = -1.00993152\ldots$,
and
\be
d_{41} = \frac{7391699}{746496} - \frac{2529743}{165888}  \,\zeta_{3} +
\left[ -\frac{110341}{373248} +\frac{110779}{82944}  \,\zeta_{3} \right]\, n_\ell
+ \frac{6865}{186624} \,n^2_\ell\, ,
\ee
\be
d_{42} = \frac{2177}{3456} - \frac{1483}{10368} \, n_\ell - \frac{77}{20736} \, n^2_\ell\, ,
\qquad
d_{43} = -\frac{1883}{10368} -\frac{127}{5184} \, n_\ell + \frac{1}{324} \, n^2_\ell\, ,
\qquad
d_{44} = \frac{1}{1296} \, .
\ee

The matching corrections for the light quark masses start at the two-loop level \cite{Wetzel:1981qg,Bernreuther:1983zp}:
\be
h_{20} = \frac{89}{432}\, ,
\qquad
h_{21} = -\frac{5}{36}\, ,
\qquad
h_{22} = \frac{1}{12}\, .
\ee
The three-loop coefficients are given by \cite{Chetyrkin:1997un} 
\be
h_{30} = h_{30}^a + \left( \frac{1327}{11664} - \frac{2}{27}\,\zeta_3\right) n_\ell\, ,
\qquad
h_{31} = - \frac{311}{2592} + \frac{5}{6}\,\zeta_3 - \frac{53}{432}\, n_\ell\, ,
\qquad
h_{32} = \frac{175}{432}\, ,
\qquad
h_{33} = \frac{29}{216}-\frac{1}{108}\, n_\ell\, ,
\ee
with $h_{30}^a= 1.84762674\ldots$, 
while the four-loop ones read \cite{Liu:2015fxa}
\be
h_{40} = h_{40}^a + h_{40}^b\, n_\ell + \left(  \frac{17671}{124416}  
   - \frac{5}{864}\,\zeta_3  - \frac{7}{96}\,\zeta_4\right) n_\ell^2\, ,
   \qquad
h_{41} =  h_{41}^a + h_{41}^b\, n_\ell + \left( 
\frac{7}{108}\,\zeta_3 -\frac{3401}{46656} \right) n_\ell^2\, ,   
\ee
with $h_{40}^a = 6.8500649983\ldots$, $h_{40}^b = -1.465707690\ldots$, 
$h_{41}^a = -23.111711963\ldots$ and $h_{41}^b = -0.1547857294\ldots$, and
\be
h_{42} = \frac{51163}{10368} - \frac{155}{48}\,\zeta_3 - \frac{7825}{10368}\, n_\ell +\frac{31}{1296}\, n_\ell^2\, ,
\qquad
h_{43} = \frac{301}{324} -\frac{23}{288}\, n_\ell\, ,
\qquad
h_{44} = \frac{305}{1152} - \frac{5}{144}\, n_\ell + \frac{1}{864}\, n_\ell^2\, .
\ee

All these matching coefficients, together with the expansion parameters of the $\beta$ and $\gamma$ functions discussed in Section~\ref{sec:running} are implemented in the
RunDec package \cite{Chetyrkin:2000yt,Schmidt:2012az,Herren:2017osy}, which provides the five-loop evolution of the QCD coupling and quark masses.


\section{Inverse power corrections to the Adler correlators}

\subsection{Quark-mass contributions}
\label{app:QuarkMasses}

The $\cO(m_q^2)$ corrections to the non-singlet $J=L+T$ correlator in Eq.~\eqn{DLTij} are known to $\cO(\alpha_s^3)$.
In the $\overline{\mathrm{MS}}$ scheme, they take the values:\footnote{The $\cO(m_q^2\alpha_s^2)$ corrections to $\Pi^{L+T}_{ij,\,\cJ}(Q^2)$ with $m_i\not = m_j$ are given in Ref.~\cite{Chetyrkin:1993hi} for $n_f=3$. The explicit dependence on $n_f$ has been obtained combining the separate calculations of $\Pi^{T}_{ij,\,\cJ}(Q^2)$ \cite{Chetyrkin:1996hm} and $\Pi^{L}_{ij,\,\cJ}(Q^2)$ \cite{Chetyrkin:1996sr}. It agrees with the known result for the vector Adler function with equal quark masses \cite{Gorishnii:1986pz}, which provides the sum $2\, c_2^{L+T} + e^{L+T}_2$.} 
%
\bea
\lefteqn{
c^{L+T}_0 \, =\, 1 \, , \hspace*{0.75cm}
c^{L+T}_1 \, =\, \frac{13}{3} \quad\mbox{\cite{Generalis:1989hf}} , \hspace*{0.75cm}
c^{L+T}_2 \, =\, \frac{25291}{432} +\frac{215}{54} \, \zeta_3 \, - \frac{520}{27}\, \zeta_5
- n_f \left( \frac{41}{24} + \frac{2}{9}\,\zeta_3 \right)
\quad\mbox{\cite{Chetyrkin:1993hi}} , }&&
\nonumber\\ 
\lefteqn{
e^{L+T}_0 \, =\, 0 \, , \hspace*{0.75cm}
e^{L+T}_1 \, =\, \frac{2}{3} \quad\mbox{\cite{Generalis:1989hf}} , \hspace*{0.75cm}
e^{L+T}_2 \, =\, \frac{877}{54}  -\frac{91}{27} \, \zeta_3 \, - \frac{5}{27}\, \zeta_5
-  n_f\,\left( \frac{2}{3} - \frac{4}{9}\,\zeta_3 \right)
\quad\mbox{\cite{Chetyrkin:1993hi}} , }&&
\nonumber\\ 
\lefteqn{
f^{L+T}_0 \, =\, 0 \, , \hspace*{0.75cm}
f^{L+T}_1 \, =\, 0 \, , \hspace*{0.75cm}
f^{L+T}_2 \, =\, -\frac{32}{9} +\frac{8}{3} \, \zeta_3 \quad\mbox{\cite{Gorishnii:1986pz,Bernreuther:1981sp}} , }&&
\nonumber\\ 
\lefteqn{2\, c^{L+T}_3 + e^{L+T}_3 + f^{L+T}_3\, =\,
\frac{16828967}{7776} -\frac{12295}{81}\,\zeta_3 + \frac{7225}{108}\,\zeta_3^2 -\frac{93860}{81}\,\zeta_5 + \frac{1027019}{2592}\,\zeta_7}&&
\nonumber\\ 
\lefteqn{\hskip 3.45cm -\, n_f \left(\frac{33887}{216} +\frac{721}{486}\,\zeta_3 +\frac{106}{27}\,\zeta_3^2 +\frac{5}{3}\,\zeta_4 -\frac{10355}{243}\, \zeta_5 \right)
+ n_f^2 \left( \frac{9661}{5832} +\frac{2}{27}\,\zeta_3
\right)\quad\mbox{\cite{Baikov:2004ku},}}&&
\nonumber\\
\lefteqn{c^{L+T}_3[n_f=3]\, =\, 
\frac{3909929}{5184} -\frac{1541}{648}\,\zeta_3 +\frac{53}{2}\,\zeta_3^2 -\frac{5}{2}\,\zeta_4 -\frac{54265}{108}\,\zeta_5 + \frac{79835}{648}\,\zeta_7
\quad\mbox{\cite{Baikov:2004tk}} .}&& 
\eea

For the non-singlet $J=L$ correlator in Eq.~\eqn{DLij}, the $\cO(m_q^2)$ contributions have been computed to $\cO(\alpha_s^4)$:
%
\bea
d^{L}_0 &\!\!\! = &\!\!\! 1 \, , \hspace*{0.75cm}
d^{L}_1 \; =\;  \frac{17}{3} \quad\mbox{\cite{Becchi:1980vz,Broadhurst:1981jk}} , \hspace*{0.75cm}
d^{L}_2 \; =\;  \frac{10801}{144} -\frac{39}{2} \, \zeta_3 - n_f\, \left( \frac{65}{24} - \frac{2}{3}\,\zeta_3 \right)
\quad\mbox{\cite{Gorishnii:1990zu,Gorishnii:1991zr}} , 
\nonumber \\
d^L_3 &\!\!\! = &\!\!\! \frac{6163613}{5184} - \frac{109735}{216} \, \zeta_3 +\frac{815}{12} \, \zeta_5
- n_f\, \left( \frac{46147}{486} - \frac{262}{9}\,\zeta_3 + \frac{5}{6}\,\zeta_4 +\frac{25}{9} \, \zeta_5\right)
+ n_f^2\,\left( \frac{15511}{11664} - \frac{1}{3}\,\zeta_3 \right)
 \quad\mbox{\cite{Chetyrkin:1996sr}} , 
\nonumber \\ 
d^L_4 &\!\!\! = &\!\!\! 
\left[
\frac{10811054729}{497664} 
-\frac{3887351}{324}  \,\zeta_{3}
+\frac{458425}{432}  \,\zeta_3^2
+\frac{265}{18}  \,\zeta_{4}
+\frac{373975}{432}  \,\zeta_{5}
-\frac{1375}{32}  \,\zeta_{6}
-\frac{178045}{768}  \,\zeta_{7}
\right]
\nonumber \\
&\!\!\! + &\!\!\!  n_f 
\left[
-\frac{1045811915}{373248} 
+\frac{5747185}{5184}  \,\zeta_{3}
-\frac{955}{16}  \,\zeta_3^2
-\frac{9131}{576}  \,\zeta_{4}
+\frac{41215}{432}  \,\zeta_{5}
+\frac{2875}{288}  \,\zeta_{6}
+\frac{665}{72}  \,\zeta_{7}
\right]
\nonumber\\
&\!\!\! + &\!\!\!  n_f^2
\left[
\frac{220313525}{2239488} 
-\frac{11875}{432}  \,\zeta_{3}
+\frac{5}{6}  \,\zeta_3^2
+\frac{25}{96}  \,\zeta_{4}
-\frac{5015}{432}  \,\zeta_{5}
\right]
 + n_f^3
\left[
-\frac{520771}{559872} 
+\frac{65}{432}  \,\zeta_{3}
+\frac{1}{144}  \,\zeta_{4}
+\frac{5}{18}  \,\zeta_{5}
\right]
 \quad\mbox{\cite{Baikov:2005rw}} .
 \nonumber\\
\eea

In the general case with $m_i\not= m_j$, the $\cO(m_q^4)$ contributions to these correlators have only been computed to $\cO(\alpha_s)$ 
\cite{Broadhurst:1981jk,Generalis:1989hf,Chetyrkin:1985kn,Generalis:1990id,Jamin:1992se}: 
\bea
\lefteqn{h^{L+T}_0\, =\, g^{L+T}_0\, =\, h^{L}_0\, =\, k^{L}_0\, =\, 1 \, , \hspace*{1.5cm}
k^{L+T}_0\, =\, j^{L+T}_0\, =\, u^{L+T}_0\, =\, j^{L}_0 \, =\, 0\, ,} &&
\nonumber\\
\lefteqn{h^{L+T}_1\, =\, \frac{25}{4} -2\,\zeta_3  \, , \hspace*{1cm}
k^{L+T}_1\, =\, 1  \, , \hspace*{1cm}
g^{L+T}_1\, =\, \frac{94}{9}-\frac{4}{3} \, \zeta_3 \, , \hspace*{1cm}
j^{L+T}_1\, =\, u^{L+T}_1\, =\, 0 
\, ,} &&
\nonumber\\
\lefteqn{h^{L}_1\, =\, \frac{41}{6} -2\,\zeta_3 \, , \hspace*{1cm}
k^{L}_1\, =\, 8 - \frac{4}{3}\,\zeta_3 \, , \hspace*{1cm}
j^{L}_1\, =\, 0 
\, .}&&
\eea
These results are appropriate for light quarks, since additional corrections of the form $\alpha_s m_i^4 \log{(m_i^2/\mu^2)}$ have been reabsorbed into the quark condensate.
Mass corrections to the neutral vector-current correlator with $m_i=m_j$ are known up to $\cO(m_i^6\alpha_s^2)$ \cite{Chetyrkin:1997qi}, including also the  $\alpha_s^n m_i^{2m} \log^k{\! (m_i^2)}$ contributions.

Using the QCD renormalization-group equations, the calculation of the mass corrections to the correlators at a given order in the strong coupling allows one to reconstruct the $\log{(Q^2/\mu^2)}$ dependence of the next perturbative order. The relevant absorptive parts have been computed in this way at $\cO(m_i^2\alpha_s^3)$  \cite{Chetyrkin:1990kr}, $\cO(m_i^2\alpha_s^4)$ \cite{Baikov:2004ku}, $\cO(m_i^4\alpha_s^2)$ \cite{Chetyrkin:1994ex} and $\cO(m_i^4\alpha_s^3)$  \cite{Chetyrkin:2000zk}, for $m_i=m_j$. Detailed results for both the vector and axial-vector spectral functions, including the small non-singlet contributions, can be found in Ref.~\cite{Chetyrkin:2000zk}.

\subsection{Dimension-four OPE contributions}
\label{app:D=4}

For light quarks, the non-perturbative dimension-four contributions to the OPE of the correlators
$D^J_{ij,\,\cJ}(Q^2)$ can be written in the form \cite{Pich:1999hc}:
\bea
\label{DLT4ij}
D^{L+T}_{ij,\,\cJ}(Q^2) \bigg|_{D=4} &= & 
\frac{1}{(Q^2)^2}\,\sum_{n=0}\, \Omega_n^{L+T}\, a^n(Q^2)
\, , \\ \label{DL4ij}
D^{L}_{ij,\,\cJ}(Q^2) \bigg|_{D=4} &= & 
-\frac{1}{(Q^2)^2}\, \langle (m_i\mp m_j) \, (\bar q_i q_i \mp \bar q_j q_j)\rangle\, ,
\eea
where $a(Q^2)\equiv \alpha_s(Q^2)/\pi$ and
\bel{eq:OmegaLT} 
\Omega_n^{L+T}\, = \,
\frac{1}{6} <G^2> \, p_n^{L+T}
\, +\, \biggl( \sum_k\, \langle m_k\, \bar q_k q_k\rangle\biggr) \,\,  r_n^{L+T}
\, +\, 2 \,\langle m_i\,\bar q_i q_i + m_j\, \bar q_j q_j\rangle\,\, q_n^{L+T}
\,\pm\, \frac{8}{3}\,
\langle m_j\,\bar q_i q_i + m_i\, \bar q_j q_j\rangle\,\, t_n^{L+T}\, .
\ee
The upper signs correspond to $\cJ=V$ and the lower ones to $\cJ=A$.
The vacuum condensates appearing in these expressions are minimally subtracted operators, defined in  the $\overline{\rm MS}$ scheme at the scale $\mu^2=Q^2$: 
\be\label{MScond}
\langle G^2\rangle\,\equiv\, \langle 0|G^2|0\rangle\,(Q^2)\, ,
\qquad\qquad\qquad
\langle m_i\,\bar q_j q_j\rangle \,\equiv\, 
\langle 0|m_i\,\bar q_j q_j|0\rangle\,(Q^2)\, .
\ee
Together with the genuine non-perturbative contributions, these condensates reabsorb light-quark mass singularities of the form $m_i^4(\mu^2)\, \alpha_s^n(\mu^2)\, \log^k{(m_i^2/\mu^2)}$ \cite{BroadhurstGen1984}, so that a clear separation between short- and long-distance contributions can be enforced, and depend non-trivially on the chosen renormalization scale.\footnote{The explicit renormalization-scale dependence of the different coefficients can be found in Ref.~\cite{Pich:1999hc}. Other condensate choices, such as the scale-invariant condensates \cite{Generalis:1989hf,BroadhurstGen1984,Spiridonov:1988md} adopted in Ref.~\cite{Braaten:1991qm}, lead to $\cO(m_q^4)$ corrections slightly different than the ones given in Eqs.~\eqn{DLTij} and \eqn{DLij}. }

The quark condensate contribution to the longitudinal correlator (\ref{DL4ij})
is fixed to all orders in perturbation theory by the Ward identity \eqn{eq:WardIdentity} \cite{Becchi:1980vz,Broadhurst:1981jk}. 
The perturbative expansion coefficients in Eq.~(\ref{eq:OmegaLT}) have been computed to $\cO(\alpha_s^2)$ \cite{Shifman:1978bx,Shifman:1978by,Becchi:1980vz,Generalis:1989hf,Chetyrkin:1985kn,Pascual:1981jr,Generalis:1983hb,Loladze:1985qk,Bagan:1985zp,Generalis:1990iy,Surguladze:1990sp}:
\bea
\lefteqn{p^{L+T}_0 \, =\, 0 \, , \hspace*{0.75cm} 
p^{L+T}_1 \, =\, 1 \, , \hspace*{0.75cm}
p^{L+T}_2 \, =\, \frac{7}{6} 
\,  , }&&
\nonumber\\
\lefteqn{r^{L+T}_0 \, =\, 0 \, , \hspace*{0.75cm} 
r^{L+T}_1 \, =\, 0 \, , \hspace*{0.75cm}
r^{L+T}_2  \, =\, - \frac{5}{3} +
\frac{8}{3}  \, \zeta_3 
\,  , }&&
\nonumber\\
\lefteqn{ q^{L+T}_0 \, =\, 1 \, , \hspace*{0.75cm} 
 q^{L+T}_1 \, =\, -1 \, , \hspace*{0.75cm}
q^{L+T}_2  \, =\, -\frac{149}{24} + \frac{1}{4}\, n_F
\,  , }&&
\nonumber\\
\lefteqn{t^{L+T}_0 \, =\, 0 \, , \hspace*{0.75cm} 
t^{L+T}_1 \, =\, 1 \, , \hspace*{0.75cm}
t^{L+T}_2  \, =\, \frac{28}{3} - \frac{5}{18}\, n_F
\, . }&&
\eea
A compilation of the most important dimension-six contributions to the OPE of the light-quark current correlators can be found in Ref.~\cite{Braaten:1991qm}.

The vacuum condensates parametrize large-distance fluctuations of the fields. For heavy quarks these fluctuations correspond to momentum scales that are much smaller than the heavy quark mass $M_Q$ \cite{Shifman:1978bx}. Therefore, they are usually expanded in inverse powers of $M_Q$ and get reduced to vacuum condensates of gluon operators. For instance, the lowest-dimension quark condensate can be expressed in the form \cite{Broadhurst:1994qj}
\bel{eq:HeavyQuarkCondensate}
\langle 0|\bar Q Q|0\rangle\, =\, -\frac{1}{12\pi\, M_Q}\; \left\{ 1 + \frac{11}{4}\,\frac{\alpha_s}{\pi} +\cO(\alpha_s^2)\right\}
\;\langle 0|\frac{\alpha_s}{\pi}\, G_{\mu\nu}^a G^{a\mu\nu}| 0\rangle
\; +\; \cO(1/M_Q^2)\, .
\ee

\section{Chiral Ward identity}
\label{app:WardIdentity}

Contracting with $q_\mu q_\nu$ the current correlation function in Eq.~\eqn{eq:correlators} and using invariance under space--time translations, one gets 
\bea\label{eq:WardIdentityDerivation}
\lefteqn{
(q^2)^2\,\Pi_{ij,\,\cJ}^L(q^2) \; = \;
 i q_\mu q_\nu \int d^4x \; \mathrm{e}^{iq(x-y)}\,
\langle 0|T(\cJ^{\mu}_{ij}(x)\, \cJ^{\nu}_{ij}(y)^\dagger)|0\rangle
\; =\;
q_\nu \int d^4x \; \partial_\mu^x\left(\mathrm{e}^{iq(x-y)}\right)\,
\langle 0|T(\cJ^{\mu}_{ij}(x)\, \cJ^{\nu}_{ij}(y)^\dagger)|0\rangle
}&&
\no\\ && =\;
- q_\nu \int d^4x \; \mathrm{e}^{iq(x-y)}\,
\left\{
\langle 0|T(\partial_\mu^x\cJ^{\mu}_{ij}(x)\, \cJ^{\nu}_{ij}(y)^\dagger)|0\rangle
+ \delta(x^0-y^0)\, \langle 0|\left[\cJ^{0}_{ij}(x)\, , \cJ^{\nu}_{ij}(y)^\dagger\right]|0\rangle
\right\}
\no\\ && =\;
 i \int d^4x \; \mathrm{e}^{iq(x-y)}\,
\langle 0|T(\partial_\mu^x\cJ^{\mu}_{ij}(x)\, \partial_\nu^y\cJ^{\nu}_{ij}(y)^\dagger)|0\rangle
-i \int d^4x \; \mathrm{e}^{iq(x-y)}\,\delta(x^0-y^0)\, \langle 0|\left[\partial_\mu^x\cJ^{\mu}_{ij}(x)\, , \cJ^{0}_{ij}(y)^\dagger\right]|0\rangle
\hskip 1.cm\mbox{}
\no\\ &&\hskip .3cm
-\, q_\nu \int d^4x \; \mathrm{e}^{iq(x-y)}\,\delta(x^0-y^0)\, \langle 0|\left[\cJ^{0}_{ij}(x)\, , \cJ^{\nu}_{ij}(y)^\dagger\right]|0\rangle\, .
\eea
Inserting in the first and second terms the current divergences in Eq.~\eqn{eq:CurrentDivergences}, applying the equal-time commutation relations
\bea
\delta(x^0-y^0)\,\left[\cJ^{S}_{ij}(x)\, , V^{0}_{ij}(y)^\dagger\right]
&\!\!\! = &\!\!\!
\delta^{(4)}(x-y)\; \left( \cJ^{S}_{jj}(x) - \cJ^{S}_{ii}(x)\right) ,
\no\\
\delta(x^0-y^0)\,\left[\cJ^{P}_{ij}(x)\, , A^{0}_{ij}(y)^\dagger\right]
&\!\!\! = &\!\!\!
\delta^{(4)}(x-y)\; \left( \cJ^{S}_{jj}(x) + \cJ^{S}_{ii}(x)\right) ,
\no\\
\delta(x^0-y^0)\,\left[V^0_{ij}(x)\, , V^{\nu}_{ij}(y)^\dagger\right]
&\!\!\! = &\!\!\!
\delta^{(4)}(x-y)\; \left( V^\nu_{jj}(x) - V^\nu_{ii}(x)\right) ,
\no\\
\delta(x^0-y^0)\,\left[A^0_{ij}(x)\, , A^{\nu}_{ij}(y)^\dagger\right]
&\!\!\! = &\!\!\!
\delta^{(4)}(x-y)\; \left( V^\nu_{jj}(x) - V^\nu_{ii}(x)\right) ,
\eea
and using again space--time translation invariance to bring back $y=0$, the Ward identity \eqn{eq:WardIdentity} follows. The last term in \eqn{eq:WardIdentityDerivation} does not contribute because, owing to Lorentz invariance, the vacuum expectation value of the vector current is identically zero.


\bibliographystyle{elsarticle-num}

\bibliography{myQCDrefs}

\begin{thebibliography}{100}
\expandafter\ifx\csname url\endcsname\relax
  \def\url#1{\texttt{#1}}\fi
\expandafter\ifx\csname urlprefix\endcsname\relax\def\urlprefix{URL }\fi
\expandafter\ifx\csname href\endcsname\relax
  \def\href#1#2{#2} \def\path#1{#1}\fi

\bibitem{Fritzsch:1972jv}
H.~Fritzsch, M.~Gell-Mann, {Current algebra: Quarks and what else?, in: XVI
  Intern. Conf. on High Energy Physics (Fermilab, Chicago, 6--13 September
  1972)}, eConf C720906V2 (1972) 135--165.
\newblock \href {http://arxiv.org/abs/hep-ph/0208010}
  {\path{arXiv:hep-ph/0208010}}.

\bibitem{Fritzsch:1973pi}
H.~Fritzsch, M.~Gell-Mann, H.~Leutwyler, {Advantages of the Color Octet Gluon
  Picture}, Phys. Lett. B 47 (1973) 365--368.
\newblock \href {http://dx.doi.org/10.1016/0370-2693(73)90625-4}
  {\path{doi:10.1016/0370-2693(73)90625-4}}.

\bibitem{Wilson:1969zs}
K.~G. Wilson, {Nonlagrangian models of current algebra}, Phys. Rev. 179 (1969)
  1499--1512.
\newblock \href {http://dx.doi.org/10.1103/PhysRev.179.1499}
  {\path{doi:10.1103/PhysRev.179.1499}}.

\bibitem{Bardeen:1978yd}
W.~A. Bardeen, A.~Buras, D.~Duke, T.~Muta, {Deep Inelastic Scattering Beyond
  the Leading Order in Asymptotically Free Gauge Theories}, Phys. Rev. D 18
  (1978) 3998.
\newblock \href {http://dx.doi.org/10.1103/PhysRevD.18.3998}
  {\path{doi:10.1103/PhysRevD.18.3998}}.

\bibitem{Espriu:1981eh}
D.~Espriu, R.~Tarrach, {On Prescription Dependence of Renormalization Group
  Functions}, Phys. Rev. D 25 (1982) 1073.
\newblock \href {http://dx.doi.org/10.1103/PhysRevD.25.1073}
  {\path{doi:10.1103/PhysRevD.25.1073}}.

\bibitem{Gross:1973id}
D.~J. Gross, F.~Wilczek, {Ultraviolet Behavior of Nonabelian Gauge Theories},
  Phys. Rev. Lett. 30 (1973) 1343--1346.
\newblock \href {http://dx.doi.org/10.1103/PhysRevLett.30.1343}
  {\path{doi:10.1103/PhysRevLett.30.1343}}.

\bibitem{Politzer:1973fx}
H.~Politzer, {Reliable Perturbative Results for Strong Interactions?}, Phys.
  Rev. Lett. 30 (1973) 1346--1349.
\newblock \href {http://dx.doi.org/10.1103/PhysRevLett.30.1346}
  {\path{doi:10.1103/PhysRevLett.30.1346}}.

\bibitem{Caswell:1974gg}
W.~E. Caswell, {Asymptotic Behavior of Nonabelian Gauge Theories to Two Loop
  Order}, Phys. Rev. Lett. 33 (1974) 244.
\newblock \href {http://dx.doi.org/10.1103/PhysRevLett.33.244}
  {\path{doi:10.1103/PhysRevLett.33.244}}.

\bibitem{Jones:1974mm}
D.~Jones, {Two Loop Diagrams in Yang-Mills Theory}, Nucl. Phys. B 75 (1974)
  531.
\newblock \href {http://dx.doi.org/10.1016/0550-3213(74)90093-5}
  {\path{doi:10.1016/0550-3213(74)90093-5}}.

\bibitem{Tarasov:1980au}
O.~Tarasov, A.~Vladimirov, A.~Zharkov, {The Gell-Mann-Low Function of QCD in
  the Three Loop Approximation}, Phys. Lett. B 93 (1980) 429--432.
\newblock \href {http://dx.doi.org/10.1016/0370-2693(80)90358-5}
  {\path{doi:10.1016/0370-2693(80)90358-5}}.

\bibitem{vanRitbergen:1997va}
T.~van Ritbergen, J.~Vermaseren, S.~Larin, {The Four loop beta function in
  quantum chromodynamics}, Phys. Lett. B 400 (1997) 379--384.
\newblock \href {http://arxiv.org/abs/hep-ph/9701390}
  {\path{arXiv:hep-ph/9701390}}, \href
  {http://dx.doi.org/10.1016/S0370-2693(97)00370-5}
  {\path{doi:10.1016/S0370-2693(97)00370-5}}.

\bibitem{Czakon:2004bu}
M.~Czakon, {The Four-loop QCD beta-function and anomalous dimensions}, Nucl.
  Phys. B 710 (2005) 485--498.
\newblock \href {http://arxiv.org/abs/hep-ph/0411261}
  {\path{arXiv:hep-ph/0411261}}, \href
  {http://dx.doi.org/10.1016/j.nuclphysb.2005.01.012}
  {\path{doi:10.1016/j.nuclphysb.2005.01.012}}.

\bibitem{Baikov:2016tgj}
P.~Baikov, K.~Chetyrkin, J.~Kühn, {Five-Loop Running of the QCD coupling
  constant}, Phys. Rev. Lett. 118~(8) (2017) 082002.
\newblock \href {http://arxiv.org/abs/1606.08659} {\path{arXiv:1606.08659}},
  \href {http://dx.doi.org/10.1103/PhysRevLett.118.082002}
  {\path{doi:10.1103/PhysRevLett.118.082002}}.

\bibitem{Luthe:2016ima}
T.~Luthe, A.~Maier, P.~Marquard, Y.~Schröder, {Towards the five-loop Beta
  function for a general gauge group}, JHEP 07 (2016) 127.
\newblock \href {http://arxiv.org/abs/1606.08662} {\path{arXiv:1606.08662}},
  \href {http://dx.doi.org/10.1007/JHEP07(2016)127}
  {\path{doi:10.1007/JHEP07(2016)127}}.

\bibitem{Herzog:2017ohr}
F.~Herzog, B.~Ruijl, T.~Ueda, J.~Vermaseren, A.~Vogt, {The five-loop beta
  function of Yang-Mills theory with fermions}, JHEP 02 (2017) 090.
\newblock \href {http://arxiv.org/abs/1701.01404} {\path{arXiv:1701.01404}},
  \href {http://dx.doi.org/10.1007/JHEP02(2017)090}
  {\path{doi:10.1007/JHEP02(2017)090}}.

\bibitem{Luthe:2017ttc}
T.~Luthe, A.~Maier, P.~Marquard, Y.~Schröder, {Complete renormalization of QCD
  at five loops}, JHEP 03 (2017) 020.
\newblock \href {http://arxiv.org/abs/1701.07068} {\path{arXiv:1701.07068}},
  \href {http://dx.doi.org/10.1007/JHEP03(2017)020}
  {\path{doi:10.1007/JHEP03(2017)020}}.

\bibitem{Luthe:2017ttg}
T.~Luthe, A.~Maier, P.~Marquard, Y.~Schröder, {The five-loop Beta function for
  a general gauge group and anomalous dimensions beyond Feynman gauge}, JHEP 10
  (2017) 166.
\newblock \href {http://arxiv.org/abs/1709.07718} {\path{arXiv:1709.07718}},
  \href {http://dx.doi.org/10.1007/JHEP10(2017)166}
  {\path{doi:10.1007/JHEP10(2017)166}}.

\bibitem{Chetyrkin:2017bjc}
K.~Chetyrkin, G.~Falcioni, F.~Herzog, J.~Vermaseren, {Five-loop renormalisation
  of QCD in covariant gauges}, JHEP 10 (2017) 179, [Addendum: JHEP 12, 006
  (2017)].
\newblock \href {http://arxiv.org/abs/1709.08541} {\path{arXiv:1709.08541}},
  \href {http://dx.doi.org/10.1007/JHEP10(2017)179}
  {\path{doi:10.1007/JHEP10(2017)179}}.

\bibitem{Rodrigo:1997zd}
G.~Rodrigo, A.~Pich, A.~Santamaria, {$\alpha_s(m_Z)$ from $\tau$ decays with
  matching conditions at three loops}, Phys. Lett. B 424 (1998) 367--374.
\newblock \href {http://arxiv.org/abs/hep-ph/9707474}
  {\path{arXiv:hep-ph/9707474}}, \href
  {http://dx.doi.org/10.1016/S0370-2693(98)00219-6}
  {\path{doi:10.1016/S0370-2693(98)00219-6}}.

\bibitem{Tarrach:1980up}
R.~Tarrach, {The Pole Mass in Perturbative QCD}, Nucl. Phys. B 183 (1981)
  384--396.
\newblock \href {http://dx.doi.org/10.1016/0550-3213(81)90140-1}
  {\path{doi:10.1016/0550-3213(81)90140-1}}.

\bibitem{Tarasov:1982gk}
O.~Tarasov, {Anomalous dimensions of quark masses in three loop approximation.
  JINR-P2-82-900 (1982)}.

\bibitem{Larin:1993tq}
S.~Larin, {The Renormalization of the axial anomaly in dimensional
  regularization}, Phys. Lett. B 303 (1993) 113--118.
\newblock \href {http://arxiv.org/abs/hep-ph/9302240}
  {\path{arXiv:hep-ph/9302240}}, \href
  {http://dx.doi.org/10.1016/0370-2693(93)90053-K}
  {\path{doi:10.1016/0370-2693(93)90053-K}}.

\bibitem{Tarasov:2019rwk}
O.~Tarasov, {Anomalous dimensions of quark masses in the three-loop
  approximation}, Phys. Part. Nucl. Lett. 17~(2) (2020) 109--115.
\newblock \href {http://arxiv.org/abs/1910.12231} {\path{arXiv:1910.12231}},
  \href {http://dx.doi.org/10.1134/S1547477120020223}
  {\path{doi:10.1134/S1547477120020223}}.

\bibitem{Chetyrkin:1997dh}
K.~Chetyrkin, {Quark mass anomalous dimension to $O (\alpha_s^4)$}, Phys. Lett.
  B 404 (1997) 161--165.
\newblock \href {http://arxiv.org/abs/hep-ph/9703278}
  {\path{arXiv:hep-ph/9703278}}, \href
  {http://dx.doi.org/10.1016/S0370-2693(97)00535-2}
  {\path{doi:10.1016/S0370-2693(97)00535-2}}.

\bibitem{Vermaseren:1997fq}
J.~Vermaseren, S.~Larin, T.~van Ritbergen, {The four loop quark mass anomalous
  dimension and the invariant quark mass}, Phys. Lett. B 405 (1997) 327--333.
\newblock \href {http://arxiv.org/abs/hep-ph/9703284}
  {\path{arXiv:hep-ph/9703284}}, \href
  {http://dx.doi.org/10.1016/S0370-2693(97)00660-6}
  {\path{doi:10.1016/S0370-2693(97)00660-6}}.

\bibitem{Baikov:2014qja}
P.~Baikov, K.~Chetyrkin, J.~Kühn, {Quark Mass and Field Anomalous Dimensions
  to ${\cal O}(\alpha_s^5)$}, JHEP 10 (2014) 076.
\newblock \href {http://arxiv.org/abs/1402.6611} {\path{arXiv:1402.6611}},
  \href {http://dx.doi.org/10.1007/JHEP10(2014)076}
  {\path{doi:10.1007/JHEP10(2014)076}}.

\bibitem{Luthe:2016xec}
T.~Luthe, A.~Maier, P.~Marquard, Y.~Schröder, {Five-loop quark mass and field
  anomalous dimensions for a general gauge group}, JHEP 01 (2017) 081.
\newblock \href {http://arxiv.org/abs/1612.05512} {\path{arXiv:1612.05512}},
  \href {http://dx.doi.org/10.1007/JHEP01(2017)081}
  {\path{doi:10.1007/JHEP01(2017)081}}.

\bibitem{Baikov:2017ujl}
P.~Baikov, K.~Chetyrkin, J.~Kühn, {Five-loop fermion anomalous dimension for a
  general gauge group from four-loop massless propagators}, JHEP 04 (2017) 119.
\newblock \href {http://arxiv.org/abs/1702.01458} {\path{arXiv:1702.01458}},
  \href {http://dx.doi.org/10.1007/JHEP04(2017)119}
  {\path{doi:10.1007/JHEP04(2017)119}}.

\bibitem{Appelquist:1974tg}
T.~Appelquist, J.~Carazzone, {Infrared Singularities and Massive Fields}, Phys.
  Rev. D 11 (1975) 2856.
\newblock \href {http://dx.doi.org/10.1103/PhysRevD.11.2856}
  {\path{doi:10.1103/PhysRevD.11.2856}}.

\bibitem{Weinberg:1980wa}
S.~Weinberg, {Effective Gauge Theories}, Phys. Lett. B 91 (1980) 51--55.
\newblock \href {http://dx.doi.org/10.1016/0370-2693(80)90660-7}
  {\path{doi:10.1016/0370-2693(80)90660-7}}.

\bibitem{Hall:1980kf}
L.~J. Hall, {Grand Unification of Effective Gauge Theories}, Nucl. Phys. B 178
  (1981) 75--124.
\newblock \href {http://dx.doi.org/10.1016/0550-3213(81)90498-3}
  {\path{doi:10.1016/0550-3213(81)90498-3}}.

\bibitem{Ovrut:1980uv}
B.~A. Ovrut, H.~J. Schnitzer, {Gauge Theories With Minimal Subtraction and the
  Decoupling Theorem}, Nucl. Phys. B 179 (1981) 381--416.
\newblock \href {http://dx.doi.org/10.1016/0550-3213(81)90011-0}
  {\path{doi:10.1016/0550-3213(81)90011-0}}.

\bibitem{Ovrut:1981ue}
B.~A. Ovrut, H.~J. Schnitzer, {Gauge Theory and Effective Lagrangian}, Nucl.
  Phys. B 189 (1981) 509--534.
\newblock \href {http://dx.doi.org/10.1016/0550-3213(81)90578-2}
  {\path{doi:10.1016/0550-3213(81)90578-2}}.

\bibitem{Schroder:2005hy}
Y.~Schröder, M.~Steinhauser, {Four-loop decoupling relations for the strong
  coupling}, JHEP 01 (2006) 051.
\newblock \href {http://arxiv.org/abs/hep-ph/0512058}
  {\path{arXiv:hep-ph/0512058}}, \href
  {http://dx.doi.org/10.1088/1126-6708/2006/01/051}
  {\path{doi:10.1088/1126-6708/2006/01/051}}.

\bibitem{Chetyrkin:2005ia}
K.~Chetyrkin, J.~H. Kühn, C.~Sturm, {QCD decoupling at four loops}, Nucl.
  Phys. B 744 (2006) 121--135.
\newblock \href {http://arxiv.org/abs/hep-ph/0512060}
  {\path{arXiv:hep-ph/0512060}}, \href
  {http://dx.doi.org/10.1016/j.nuclphysb.2006.03.020}
  {\path{doi:10.1016/j.nuclphysb.2006.03.020}}.

\bibitem{Kniehl:2006bg}
B.~Kniehl, A.~Kotikov, A.~Onishchenko, O.~Veretin, {Strong-coupling constant
  with flavor thresholds at five loops in the $\overline{\mathrm{MS}}$ scheme},
  Phys. Rev. Lett. 97 (2006) 042001.
\newblock \href {http://arxiv.org/abs/hep-ph/0607202}
  {\path{arXiv:hep-ph/0607202}}, \href
  {http://dx.doi.org/10.1103/PhysRevLett.97.042001}
  {\path{doi:10.1103/PhysRevLett.97.042001}}.

\bibitem{Liu:2015fxa}
T.~Liu, M.~Steinhauser, {Decoupling of heavy quarks at four loops and effective
  Higgs-fermion coupling}, Phys. Lett. B 746 (2015) 330--334.
\newblock \href {http://arxiv.org/abs/1502.04719} {\path{arXiv:1502.04719}},
  \href {http://dx.doi.org/10.1016/j.physletb.2015.05.023}
  {\path{doi:10.1016/j.physletb.2015.05.023}}.

\bibitem{Chetyrkin:1994js}
K.~Chetyrkin, J.~H. Kühn, A.~Kwiatkowski, {QCD corrections to the $e^{+}
  e^{-}$ cross-section and the $Z$ boson decay rate}, Phys. Rept. 277 (1996)
  189--281.
\newblock \href {http://arxiv.org/abs/hep-ph/9503396}
  {\path{arXiv:hep-ph/9503396}}, \href
  {http://dx.doi.org/10.1016/S0370-1573(96)00012-9}
  {\path{doi:10.1016/S0370-1573(96)00012-9}}.

\bibitem{Adler:1974gd}
S.~L. Adler, {Some Simple Vacuum Polarization Phenomenology: $e^+ e^- \to$
  Hadrons: The $\mu$ - Mesic Atom x-Ray Discrepancy and $g_{\mu}-2$}, Phys.
  Rev. D 10 (1974) 3714.
\newblock \href {http://dx.doi.org/10.1103/PhysRevD.10.3714}
  {\path{doi:10.1103/PhysRevD.10.3714}}.

\bibitem{Appelquist:1973uz}
T.~Appelquist, H.~Georgi, {$e^+ e^-$ annihilation in gauge theories of strong
  interactions}, Phys. Rev. D 8 (1973) 4000--4002.
\newblock \href {http://dx.doi.org/10.1103/PhysRevD.8.4000}
  {\path{doi:10.1103/PhysRevD.8.4000}}.

\bibitem{Zee:1973sr}
A.~Zee, {Electron positron annihilation in stagnant field theories}, Phys. Rev.
  D 8 (1973) 4038--4041.
\newblock \href {http://dx.doi.org/10.1103/PhysRevD.8.4038}
  {\path{doi:10.1103/PhysRevD.8.4038}}.

\bibitem{Chetyrkin:1979bj}
K.~Chetyrkin, A.~Kataev, F.~Tkachov, {Higher Order Corrections to
  $\sigma_{\mathrm{tot}} (e^+ e^-\to \mathrm{Hadrons}$) in Quantum
  Chromodynamics}, Phys. Lett. B 85 (1979) 277--279.
\newblock \href {http://dx.doi.org/10.1016/0370-2693(79)90596-3}
  {\path{doi:10.1016/0370-2693(79)90596-3}}.

\bibitem{Dine:1979qh}
M.~Dine, J.~Sapirstein, {Higher Order QCD Corrections in $e^+ e^-$
  Annihilation}, Phys. Rev. Lett. 43 (1979) 668.
\newblock \href {http://dx.doi.org/10.1103/PhysRevLett.43.668}
  {\path{doi:10.1103/PhysRevLett.43.668}}.

\bibitem{Gorishnii:1990vf}
S.~Gorishnii, A.~Kataev, S.~Larin, {The $O(\alpha^{3}_{s})$-corrections to
  $\sigma_{tot}(e^{+}e^{-}\rightarrow hadrons)$ and $\Gamma(\tau^{-}
  \rightarrow \nu_{\tau} + hadrons)$ in QCD}, Phys. Lett. B 259 (1991)
  144--150.
\newblock \href {http://dx.doi.org/10.1016/0370-2693(91)90149-K}
  {\path{doi:10.1016/0370-2693(91)90149-K}}.

\bibitem{Surguladze:1990tg}
L.~R. Surguladze, M.~A. Samuel, {Total hadronic cross-section in $e^+ e^-$
  annihilation at the four loop level of perturbative QCD}, Phys. Rev. Lett. 66
  (1991) 560--563, [Erratum: Phys.Rev.Lett. 66, 2416 (1991)].
\newblock \href {http://dx.doi.org/10.1103/PhysRevLett.66.560}
  {\path{doi:10.1103/PhysRevLett.66.560}}.

\bibitem{Chetyrkin:1996ez}
K.~Chetyrkin, {Corrections of order $\alpha_s^3$ to $R_{\mathrm{had}}$ in pQCD
  with light gluinos}, Phys. Lett. B 391 (1997) 402--412.
\newblock \href {http://arxiv.org/abs/hep-ph/9608480}
  {\path{arXiv:hep-ph/9608480}}, \href
  {http://dx.doi.org/10.1016/S0370-2693(96)01478-5}
  {\path{doi:10.1016/S0370-2693(96)01478-5}}.

\bibitem{Baikov:2008jh}
P.~Baikov, K.~Chetyrkin, J.~H. Kühn, {Order $\alpha_s^4$ QCD Corrections to Z
  and $\tau$ Decays}, Phys. Rev. Lett. 101 (2008) 012002.
\newblock \href {http://arxiv.org/abs/0801.1821} {\path{arXiv:0801.1821}},
  \href {http://dx.doi.org/10.1103/PhysRevLett.101.012002}
  {\path{doi:10.1103/PhysRevLett.101.012002}}.

\bibitem{Baikov:2010je}
P.~Baikov, K.~Chetyrkin, J.~Kühn, {Adler Function, Bjorken Sum Rule, and the
  Crewther Relation to Order $\alpha_s^4$ in a General Gauge Theory}, Phys.
  Rev. Lett. 104 (2010) 132004.
\newblock \href {http://arxiv.org/abs/1001.3606} {\path{arXiv:1001.3606}},
  \href {http://dx.doi.org/10.1103/PhysRevLett.104.132004}
  {\path{doi:10.1103/PhysRevLett.104.132004}}.

\bibitem{Herzog:2017dtz}
F.~Herzog, B.~Ruijl, T.~Ueda, J.~Vermaseren, A.~Vogt, {On Higgs decays to
  hadrons and the R-ratio at N$^{4}$LO}, JHEP 08 (2017) 113.
\newblock \href {http://arxiv.org/abs/1707.01044} {\path{arXiv:1707.01044}},
  \href {http://dx.doi.org/10.1007/JHEP08(2017)113}
  {\path{doi:10.1007/JHEP08(2017)113}}.

\bibitem{Pennington:1981cw}
M.~Pennington, G.~G. Ross, {Perturbative {QCD} for Timelike Processes: What Is
  the Best Expansion Parameter?}, Phys. Lett. B 102 (1981) 167--171.
\newblock \href {http://dx.doi.org/10.1016/0370-2693(81)91055-8}
  {\path{doi:10.1016/0370-2693(81)91055-8}}.

\bibitem{Baikov:2012zn}
P.~Baikov, K.~Chetyrkin, J.~Kühn, J.~Rittinger, {Adler Function, Sum Rules and
  Crewther Relation of Order $O(\alpha_s^4)$: the Singlet Case}, Phys. Lett. B
  714 (2012) 62--65.
\newblock \href {http://arxiv.org/abs/1206.1288} {\path{arXiv:1206.1288}},
  \href {http://dx.doi.org/10.1016/j.physletb.2012.06.052}
  {\path{doi:10.1016/j.physletb.2012.06.052}}.

\bibitem{Pich:1998yn}
A.~Pich, J.~Prades, {Perturbative quark mass corrections to the tau hadronic
  width}, JHEP 06 (1998) 013.
\newblock \href {http://arxiv.org/abs/hep-ph/9804462}
  {\path{arXiv:hep-ph/9804462}}, \href
  {http://dx.doi.org/10.1088/1126-6708/1998/06/013}
  {\path{doi:10.1088/1126-6708/1998/06/013}}.

\bibitem{Pich:1999hc}
A.~Pich, J.~Prades, {Strange quark mass determination from Cabibbo suppressed
  tau decays}, JHEP 10 (1999) 004.
\newblock \href {http://arxiv.org/abs/hep-ph/9909244}
  {\path{arXiv:hep-ph/9909244}}, \href
  {http://dx.doi.org/10.1088/1126-6708/1999/10/004}
  {\path{doi:10.1088/1126-6708/1999/10/004}}.

\bibitem{Shifman:1978bx}
M.~A. Shifman, A.~Vainshtein, V.~I. Zakharov, {QCD and Resonance Physics.
  Theoretical Foundations}, Nucl. Phys. B 147 (1979) 385--447.
\newblock \href {http://dx.doi.org/10.1016/0550-3213(79)90022-1}
  {\path{doi:10.1016/0550-3213(79)90022-1}}.

\bibitem{Shifman:1978by}
M.~A. Shifman, A.~Vainshtein, V.~I. Zakharov, {QCD and Resonance Physics:
  Applications}, Nucl. Phys. B 147 (1979) 448--518.
\newblock \href {http://dx.doi.org/10.1016/0550-3213(79)90023-3}
  {\path{doi:10.1016/0550-3213(79)90023-3}}.

\bibitem{Shifman:1978bw}
M.~A. Shifman, A.~Vainshtein, V.~I. Zakharov, {QCD and Resonance Physics. The
  rho-omega Mixing}, Nucl. Phys. B 147 (1979) 519--534.
\newblock \href {http://dx.doi.org/10.1016/0550-3213(79)90024-5}
  {\path{doi:10.1016/0550-3213(79)90024-5}}.

\bibitem{Novikov:1980uj}
V.~Novikov, M.~A. Shifman, A.~Vainshtein, V.~I. Zakharov, {Operator expansion
  in Quantum Chromodynamics beyond perturbation theory}, Nucl. Phys. B 174
  (1980) 378--396.
\newblock \href {http://dx.doi.org/10.1016/0550-3213(80)90290-4}
  {\path{doi:10.1016/0550-3213(80)90290-4}}.

\bibitem{Aoki:2019cca}
S.~Aoki, et~al., {FLAG Review 2019: Flavour Lattice Averaging Group (FLAG)},
  Eur. Phys. J. C 80~(2) (2020) 113.
\newblock \href {http://arxiv.org/abs/1902.08191} {\path{arXiv:1902.08191}},
  \href {http://dx.doi.org/10.1140/epjc/s10052-019-7354-7}
  {\path{doi:10.1140/epjc/s10052-019-7354-7}}.

\bibitem{Sakurai:1973rh}
J.~Sakurai, {Duality in $e^+ e^-\to \mathrm{hadrons}$}, Phys. Lett. B 46 (1973)
  207--210.
\newblock \href {http://dx.doi.org/10.1016/0370-2693(73)90685-0}
  {\path{doi:10.1016/0370-2693(73)90685-0}}.

\bibitem{Poggio:1975af}
E.~Poggio, H.~R. Quinn, S.~Weinberg, {Smearing the Quark Model}, Phys. Rev. D
  13 (1976) 1958.
\newblock \href {http://dx.doi.org/10.1103/PhysRevD.13.1958}
  {\path{doi:10.1103/PhysRevD.13.1958}}.

\bibitem{Becchi:1980vz}
C.~Becchi, S.~Narison, E.~de~Rafael, F.~Yndurain, {Light Quark Masses in
  Quantum Chromodynamics and Chiral Symmetry Breaking}, Z. Phys. C 8 (1981)
  335.
\newblock \href {http://dx.doi.org/10.1007/BF01546328}
  {\path{doi:10.1007/BF01546328}}.

\bibitem{Broadhurst:1981jk}
D.~J. Broadhurst, {Chiral Symmetry Breaking and Perturbative QCD}, Phys. Lett.
  B 101 (1981) 423--426.
\newblock \href {http://dx.doi.org/10.1016/0370-2693(81)90167-2}
  {\path{doi:10.1016/0370-2693(81)90167-2}}.

\bibitem{GellMann:1968rz}
M.~Gell-Mann, R.~Oakes, B.~Renner, {Behavior of current divergences under SU(3)
  x SU(3)}, Phys. Rev. 175 (1968) 2195--2199.
\newblock \href {http://dx.doi.org/10.1103/PhysRev.175.2195}
  {\path{doi:10.1103/PhysRev.175.2195}}.

\bibitem{Pich:2018ltt}
A.~Pich, {Effective Field Theory with Nambu-Goldstone Modes}, Les Houches Lect.
  Notes 108, 137--219 (Oxford University Press, 2020).
\newblock \href {http://arxiv.org/abs/1804.05664} {\path{arXiv:1804.05664}},
  \href {http://dx.doi.org/10.1093/oso/9780198855743.003.0003}
  {\path{doi:10.1093/oso/9780198855743.003.0003}}.

\bibitem{Chetyrkin:1996sr}
K.~Chetyrkin, {Correlator of the quark scalar currents and
  $\Gamma_{\mathrm{tot}} (H \to \mathrm{hadrons})$ at $O (\alpha_s^3)$ in
  pQCD}, Phys. Lett. B 390 (1997) 309--317.
\newblock \href {http://arxiv.org/abs/hep-ph/9608318}
  {\path{arXiv:hep-ph/9608318}}, \href
  {http://dx.doi.org/10.1016/S0370-2693(96)01368-8}
  {\path{doi:10.1016/S0370-2693(96)01368-8}}.

\bibitem{Baikov:2005rw}
P.~Baikov, K.~Chetyrkin, J.~H. Kühn, {Scalar correlator at $O(\alpha_s^4)$,
  Higgs decay into b-quarks and bounds on the light quark masses}, Phys. Rev.
  Lett. 96 (2006) 012003.
\newblock \href {http://arxiv.org/abs/hep-ph/0511063}
  {\path{arXiv:hep-ph/0511063}}, \href
  {http://dx.doi.org/10.1103/PhysRevLett.96.012003}
  {\path{doi:10.1103/PhysRevLett.96.012003}}.

\bibitem{Baikov:2012er}
P.~Baikov, K.~Chetyrkin, J.~Kühn, J.~Rittinger, {Complete ${\cal
  O}(\alpha_s^4)$ QCD Corrections to Hadronic $Z$-Decays}, Phys. Rev. Lett. 108
  (2012) 222003.
\newblock \href {http://arxiv.org/abs/1201.5804} {\path{arXiv:1201.5804}},
  \href {http://dx.doi.org/10.1103/PhysRevLett.108.222003}
  {\path{doi:10.1103/PhysRevLett.108.222003}}.

\bibitem{Zyla:2020zbs}
P.~Zyla, et~al., {Review of Particle Physics}, PTEP 2020~(8) (2020) 083C01.
\newblock \href {http://dx.doi.org/10.1093/ptep/ptaa104}
  {\path{doi:10.1093/ptep/ptaa104}}.

\bibitem{Chetyrkin:1993jm}
K.~Chetyrkin, J.~H. Kühn, {Complete QCD corrections of order $\alpha_s^2$ to
  the Z decay rate}, Phys. Lett. B 308 (1993) 127--136.
\newblock \href {http://dx.doi.org/10.1016/0370-2693(93)90613-M}
  {\path{doi:10.1016/0370-2693(93)90613-M}}.

\bibitem{Larin:1993ju}
S.~Larin, T.~van Ritbergen, J.~Vermaseren, {The $\alpha_s^3$ correction to
  $\Gamma (Z^0 \to\mathrm{hadrons})$}, Phys. Lett. B 320 (1994) 159--164.
\newblock \href {http://arxiv.org/abs/hep-ph/9310378}
  {\path{arXiv:hep-ph/9310378}}, \href
  {http://dx.doi.org/10.1016/0370-2693(94)90840-0}
  {\path{doi:10.1016/0370-2693(94)90840-0}}.

\bibitem{Chetyrkin:1993ug}
K.~Chetyrkin, O.~Tarasov, {The $\alpha_s^3$ corrections to the effective
  neutral current and to the Z decay rate in the heavy top quark limit}, Phys.
  Lett. B 327 (1994) 114--122.
\newblock \href {http://arxiv.org/abs/hep-ph/9312323}
  {\path{arXiv:hep-ph/9312323}}, \href
  {http://dx.doi.org/10.1016/0370-2693(94)91538-5}
  {\path{doi:10.1016/0370-2693(94)91538-5}}.

\bibitem{Veltman:1977kh}
M.~Veltman, {Limit on Mass Differences in the Weinberg Model}, Nucl. Phys. B
  123 (1977) 89--99.
\newblock \href {http://dx.doi.org/10.1016/0550-3213(77)90342-X}
  {\path{doi:10.1016/0550-3213(77)90342-X}}.

\bibitem{Bernabeu:1987me}
J.~Bernabeu, A.~Pich, A.~Santamaria, {$\Gamma (Z \to b \bar b)$: A Signature of
  Hard Mass Terms for a Heavy Top}, Phys. Lett. B 200 (1988) 569--574.
\newblock \href {http://dx.doi.org/10.1016/0370-2693(88)90173-6}
  {\path{doi:10.1016/0370-2693(88)90173-6}}.

\bibitem{Bernabeu:1990ws}
J.~Bernabeu, A.~Pich, A.~Santamaria, {Top quark mass from radiative corrections
  to the $Z \to b \bar b$ decay}, Nucl. Phys. B 363 (1991) 326--344.
\newblock \href {http://dx.doi.org/10.1016/0550-3213(91)80023-F}
  {\path{doi:10.1016/0550-3213(91)80023-F}}.

\bibitem{Kniehl:1989bb}
B.~A. Kniehl, J.~H. Kühn, {QCD Corrections to the Axial Part of the Z Decay
  Rate}, Phys. Lett. B 224 (1989) 229--232.
\newblock \href {http://dx.doi.org/10.1016/0370-2693(89)91079-4}
  {\path{doi:10.1016/0370-2693(89)91079-4}}.

\bibitem{Kniehl:1989qu}
B.~A. Kniehl, J.~H. Kühn, {QCD Corrections to the Z Decay Rate}, Nucl. Phys. B
  329 (1990) 547--573.
\newblock \href {http://dx.doi.org/10.1016/0550-3213(90)90070-T}
  {\path{doi:10.1016/0550-3213(90)90070-T}}.

\bibitem{Baikov:2004ku}
P.~Baikov, K.~Chetyrkin, J.~H. Kühn, {Vacuum polarization in pQCD: First
  complete $O(\alpha_s^4)$ result}, Nucl. Phys. B Proc. Suppl. 135 (2004)
  243--246.
\newblock \href {http://dx.doi.org/10.1016/j.nuclphysbps.2004.09.013}
  {\path{doi:10.1016/j.nuclphysbps.2004.09.013}}.

\bibitem{Chetyrkin:2000zk}
K.~Chetyrkin, R.~Harlander, J.~H. Kühn, {Quartic mass corrections to $R_{had}$
  at $\mathcal O(\alpha^3_s)$}, Nucl. Phys. B 586 (2000) 56--72, [Erratum:
  Nucl.Phys.B 634, 413--414 (2002)].
\newblock \href {http://arxiv.org/abs/hep-ph/0005139}
  {\path{arXiv:hep-ph/0005139}}, \href
  {http://dx.doi.org/10.1016/S0550-3213(00)00393-X}
  {\path{doi:10.1016/S0550-3213(00)00393-X}}.

\bibitem{Chetyrkin:1993tt}
K.~Chetyrkin, {Power suppressed heavy quark mass corrections to the tau lepton
  and Z boson decay rates}, Phys. Lett. B 307 (1993) 169--176.
\newblock \href {http://dx.doi.org/10.1016/0370-2693(93)90207-X}
  {\path{doi:10.1016/0370-2693(93)90207-X}}.

\bibitem{Larin:1994va}
S.~Larin, T.~van Ritbergen, J.~Vermaseren, {The large quark mass expansion of
  $\Gamma (Z^0\to\mathrm{hadrons})$ and $\Gamma (\tau^-\to\nu_\tau
  +\mathrm{hadrons})$ in the order $\alpha_s^3$}, Nucl. Phys. B 438 (1995)
  278--306.
\newblock \href {http://arxiv.org/abs/hep-ph/9411260}
  {\path{arXiv:hep-ph/9411260}}, \href
  {http://dx.doi.org/10.1016/0550-3213(94)00574-X}
  {\path{doi:10.1016/0550-3213(94)00574-X}}.

\bibitem{Akhundov:1985fc}
A.~Akhundov, D.~Bardin, T.~Riemann, {Electroweak One Loop Corrections to the
  Decay of the Neutral Vector Boson}, Nucl. Phys. B 276 (1986) 1--13.
\newblock \href {http://dx.doi.org/10.1016/0550-3213(86)90014-3}
  {\path{doi:10.1016/0550-3213(86)90014-3}}.

\bibitem{Beenakker:1988pv}
W.~Beenakker, W.~Hollik, {The Width of the Z Boson}, Z. Phys. C 40 (1988) 141.
\newblock \href {http://dx.doi.org/10.1007/BF01559728}
  {\path{doi:10.1007/BF01559728}}.

\bibitem{Dubovyk:2018rlg}
I.~Dubovyk, A.~Freitas, J.~Gluza, T.~Riemann, J.~Usovitsch, {Complete
  electroweak two-loop corrections to Z boson production and decay}, Phys.
  Lett. B 783 (2018) 86--94.
\newblock \href {http://arxiv.org/abs/1804.10236} {\path{arXiv:1804.10236}},
  \href {http://dx.doi.org/10.1016/j.physletb.2018.06.037}
  {\path{doi:10.1016/j.physletb.2018.06.037}}.

\bibitem{Dubovyk:2019szj}
I.~Dubovyk, A.~Freitas, J.~Gluza, T.~Riemann, J.~Usovitsch, {Electroweak
  pseudo-observables and Z-boson form factors at two-loop accuracy}, JHEP 08
  (2019) 113.
\newblock \href {http://arxiv.org/abs/1906.08815} {\path{arXiv:1906.08815}},
  \href {http://dx.doi.org/10.1007/JHEP08(2019)113}
  {\path{doi:10.1007/JHEP08(2019)113}}.

\bibitem{Chen:2020xzx}
L.~Chen, A.~Freitas, {Leading fermionic three-loop corrections to electroweak
  precision observables}, JHEP 07 (2020) 210.
\newblock \href {http://arxiv.org/abs/2002.05845} {\path{arXiv:2002.05845}},
  \href {http://dx.doi.org/10.1007/JHEP07(2020)210}
  {\path{doi:10.1007/JHEP07(2020)210}}.

\bibitem{Czarnecki:1996ei}
A.~Czarnecki, J.~H. Kühn, {Nonfactorizable QCD and electroweak corrections to
  the hadronic Z boson decay rate}, Phys. Rev. Lett. 77 (1996) 3955--3958.
\newblock \href {http://arxiv.org/abs/hep-ph/9608366}
  {\path{arXiv:hep-ph/9608366}}, \href
  {http://dx.doi.org/10.1103/PhysRevLett.77.3955}
  {\path{doi:10.1103/PhysRevLett.77.3955}}.

\bibitem{Fleischer:1999iq}
J.~Fleischer, F.~Jegerlehner, M.~Tentyukov, O.~Veretin, {Nonfactorizable
  $\cO(\alpha \alpha_s)$ corrections to the process $Z \to b \bar b$}, Phys.
  Lett. B 459 (1999) 625--630.
\newblock \href {http://arxiv.org/abs/hep-ph/9904256}
  {\path{arXiv:hep-ph/9904256}}, \href
  {http://dx.doi.org/10.1016/S0370-2693(99)00716-9}
  {\path{doi:10.1016/S0370-2693(99)00716-9}}.

\bibitem{Freitas:2014hra}
A.~Freitas, {Higher-order electroweak corrections to the partial widths and
  branching ratios of the Z boson}, JHEP 04 (2014) 070.
\newblock \href {http://arxiv.org/abs/1401.2447} {\path{arXiv:1401.2447}},
  \href {http://dx.doi.org/10.1007/JHEP04(2014)070}
  {\path{doi:10.1007/JHEP04(2014)070}}.

\bibitem{dEnterria:2020cpv}
D.~d'Enterria, V.~Jacobsen, {Improved strong coupling determinations from
  hadronic decays of electroweak bosons at N$^3$LO accuracy, }\href
  {http://arxiv.org/abs/2005.04545} {\path{arXiv:2005.04545}}.

\bibitem{ALEPH:2005ab}
S.~Schael, et~al., {Precision electroweak measurements on the $Z$ resonance},
  Phys. Rept. 427 (2006) 257--454.
\newblock \href {http://arxiv.org/abs/hep-ex/0509008}
  {\path{arXiv:hep-ex/0509008}}, \href
  {http://dx.doi.org/10.1016/j.physrep.2005.12.006}
  {\path{doi:10.1016/j.physrep.2005.12.006}}.

\bibitem{Baak:2014ora}
M.~Baak, J.~Cúth, J.~Haller, A.~Hoecker, R.~Kogler, K.~Mönig, M.~Schott,
  J.~Stelzer, {The global electroweak fit at NNLO and prospects for the LHC and
  ILC}, Eur. Phys. J. C 74 (2014) 3046.
\newblock \href {http://arxiv.org/abs/1407.3792} {\path{arXiv:1407.3792}},
  \href {http://dx.doi.org/10.1140/epjc/s10052-014-3046-5}
  {\path{doi:10.1140/epjc/s10052-014-3046-5}}.

\bibitem{Haller:2018nnx}
J.~Haller, A.~Hoecker, R.~Kogler, K.~Mönig, T.~Peiffer, J.~Stelzer, {Update of
  the global electroweak fit and constraints on two-Higgs-doublet models}, Eur.
  Phys. J. C 78~(8) (2018) 675.
\newblock \href {http://arxiv.org/abs/1803.01853} {\path{arXiv:1803.01853}},
  \href {http://dx.doi.org/10.1140/epjc/s10052-018-6131-3}
  {\path{doi:10.1140/epjc/s10052-018-6131-3}}.

\bibitem{Cabibbo:1963yz}
N.~Cabibbo, {Unitary Symmetry and Leptonic Decays}, Phys. Rev. Lett. 10 (1963)
  531--533.
\newblock \href {http://dx.doi.org/10.1103/PhysRevLett.10.531}
  {\path{doi:10.1103/PhysRevLett.10.531}}.

\bibitem{Kobayashi:1973fv}
M.~Kobayashi, T.~Maskawa, {CP Violation in the Renormalizable Theory of Weak
  Interaction}, Prog. Theor. Phys. 49 (1973) 652--657.
\newblock \href {http://dx.doi.org/10.1143/PTP.49.652}
  {\path{doi:10.1143/PTP.49.652}}.

\bibitem{Chang:1981qq}
T.~Chang, K.~Gaemers, W.~van Neerven, {QCD Corrections to the Mass and Width of
  the Intermediate Vector Bosons}, Nucl. Phys. B 202 (1982) 407--436.
\newblock \href {http://dx.doi.org/10.1016/0550-3213(82)90407-2}
  {\path{doi:10.1016/0550-3213(82)90407-2}}.

\bibitem{Bardin:1986fi}
D.~Bardin, S.~Riemann, T.~Riemann, {Electroweak One Loop Corrections to the
  Decay of the Charged Vector Boson}, Z. Phys. C 32 (1986) 121--125.
\newblock \href {http://dx.doi.org/10.1007/BF01441360}
  {\path{doi:10.1007/BF01441360}}.

\bibitem{Denner:1990tx}
A.~Denner, T.~Sack, {The W boson width}, Z. Phys. C 46 (1990) 653--663.
\newblock \href {http://dx.doi.org/10.1007/BF01560267}
  {\path{doi:10.1007/BF01560267}}.

\bibitem{Denner:1991kt}
A.~Denner, {Techniques for calculation of electroweak radiative corrections at
  the one loop level and results for W physics at LEP-200}, Fortsch. Phys. 41
  (1993) 307--420.
\newblock \href {http://arxiv.org/abs/0709.1075} {\path{arXiv:0709.1075}},
  \href {http://dx.doi.org/10.1002/prop.2190410402}
  {\path{doi:10.1002/prop.2190410402}}.

\bibitem{Kniehl:2000rb}
B.~A. Kniehl, F.~Madricardo, M.~Steinhauser, {Gauge independent W boson partial
  decay widths}, Phys. Rev. D 62 (2000) 073010.
\newblock \href {http://arxiv.org/abs/hep-ph/0005060}
  {\path{arXiv:hep-ph/0005060}}, \href
  {http://dx.doi.org/10.1103/PhysRevD.62.073010}
  {\path{doi:10.1103/PhysRevD.62.073010}}.

\bibitem{Kara:2013dua}
D.~Kara, {Corrections of Order $\alpha \alpha_s$ to W Boson Decays}, Nucl.
  Phys. B 877 (2013) 683--718.
\newblock \href {http://arxiv.org/abs/1307.7190} {\path{arXiv:1307.7190}},
  \href {http://dx.doi.org/10.1016/j.nuclphysb.2013.10.024}
  {\path{doi:10.1016/j.nuclphysb.2013.10.024}}.

\bibitem{dEnterria:2016rbf}
D.~d'Enterria, M.~Srebre, {$\alpha_s$ and $\rm V_{cs}$ determination, and CKM
  unitarity test, from W decays at NNLO}, Phys. Lett. B 763 (2016) 465--471.
\newblock \href {http://arxiv.org/abs/1603.06501} {\path{arXiv:1603.06501}},
  \href {http://dx.doi.org/10.1016/j.physletb.2016.10.012}
  {\path{doi:10.1016/j.physletb.2016.10.012}}.

\bibitem{Chetyrkin:1997un}
K.~Chetyrkin, B.~A. Kniehl, M.~Steinhauser, {Decoupling relations to
  $\mathcal{O} (\alpha_s^3)$ and their connection to low-energy theorems},
  Nucl. Phys. B 510 (1998) 61--87.
\newblock \href {http://arxiv.org/abs/hep-ph/9708255}
  {\path{arXiv:hep-ph/9708255}}, \href
  {http://dx.doi.org/10.1016/S0550-3213(97)00649-4}
  {\path{doi:10.1016/S0550-3213(97)00649-4}}.

\bibitem{Harlander:1997xa}
R.~Harlander, M.~Steinhauser, {Higgs decay to top quarks at $O (\alpha_s^2)$},
  Phys. Rev. D 56 (1997) 3980--3990.
\newblock \href {http://arxiv.org/abs/hep-ph/9704436}
  {\path{arXiv:hep-ph/9704436}}, \href
  {http://dx.doi.org/10.1103/PhysRevD.56.3980}
  {\path{doi:10.1103/PhysRevD.56.3980}}.

\bibitem{Chetyrkin:1997mb}
K.~Chetyrkin, J.~H. Kühn, M.~Steinhauser, {Heavy quark current correlators to
  $\cO (\alpha_s^2)$}, Nucl. Phys. B 505 (1997) 40--64.
\newblock \href {http://arxiv.org/abs/hep-ph/9705254}
  {\path{arXiv:hep-ph/9705254}}, \href
  {http://dx.doi.org/10.1016/S0550-3213(97)00481-1}
  {\path{doi:10.1016/S0550-3213(97)00481-1}}.

\bibitem{Chetyrkin:1998ix}
K.~Chetyrkin, R.~Harlander, M.~Steinhauser, {Singlet polarization functions at
  $O (\alpha_s^2)$}, Phys. Rev. D 58 (1998) 014012.
\newblock \href {http://arxiv.org/abs/hep-ph/9801432}
  {\path{arXiv:hep-ph/9801432}}, \href
  {http://dx.doi.org/10.1103/PhysRevD.58.014012}
  {\path{doi:10.1103/PhysRevD.58.014012}}.

\bibitem{Inami:1982xt}
T.~Inami, T.~Kubota, Y.~Okada, {Effective Gauge Theory and the Effect of Heavy
  Quarks in Higgs Boson Decays}, Z. Phys. C 18 (1983) 69--80.
\newblock \href {http://dx.doi.org/10.1007/BF01571710}
  {\path{doi:10.1007/BF01571710}}.

\bibitem{Djouadi:1991tka}
A.~Djouadi, M.~Spira, P.~Zerwas, {Production of Higgs bosons in proton
  colliders: QCD corrections}, Phys. Lett. B 264 (1991) 440--446.
\newblock \href {http://dx.doi.org/10.1016/0370-2693(91)90375-Z}
  {\path{doi:10.1016/0370-2693(91)90375-Z}}.

\bibitem{Chetyrkin:1997iv}
K.~Chetyrkin, B.~A. Kniehl, M.~Steinhauser, {Hadronic Higgs decay to order
  $\alpha_s^4$}, Phys. Rev. Lett. 79 (1997) 353--356.
\newblock \href {http://arxiv.org/abs/hep-ph/9705240}
  {\path{arXiv:hep-ph/9705240}}, \href
  {http://dx.doi.org/10.1103/PhysRevLett.79.353}
  {\path{doi:10.1103/PhysRevLett.79.353}}.

\bibitem{Chetyrkin:1997vj}
K.~Chetyrkin, M.~Steinhauser, {Complete QCD corrections of order $O
  (\alpha_s^3)$ to the hadronic Higgs decay}, Phys. Lett. B 408 (1997)
  320--324.
\newblock \href {http://arxiv.org/abs/hep-ph/9706462}
  {\path{arXiv:hep-ph/9706462}}, \href
  {http://dx.doi.org/10.1016/S0370-2693(97)00779-X}
  {\path{doi:10.1016/S0370-2693(97)00779-X}}.

\bibitem{Baikov:2006ch}
P.~Baikov, K.~Chetyrkin, {Top Quark Mediated Higgs Boson Decay into Hadrons to
  Order $\alpha_s^5$}, Phys. Rev. Lett. 97 (2006) 061803.
\newblock \href {http://arxiv.org/abs/hep-ph/0604194}
  {\path{arXiv:hep-ph/0604194}}, \href
  {http://dx.doi.org/10.1103/PhysRevLett.97.061803}
  {\path{doi:10.1103/PhysRevLett.97.061803}}.

\bibitem{Chetyrkin:1995pd}
K.~Chetyrkin, A.~Kwiatkowski, {Second order QCD corrections to scalar and
  pseudoscalar Higgs decays into massive bottom quarks}, Nucl. Phys. B 461
  (1996) 3--18.
\newblock \href {http://arxiv.org/abs/hep-ph/9505358}
  {\path{arXiv:hep-ph/9505358}}, \href
  {http://dx.doi.org/10.1016/0550-3213(95)00616-8}
  {\path{doi:10.1016/0550-3213(95)00616-8}}.

\bibitem{Larin:1995sq}
S.~Larin, T.~van Ritbergen, J.~Vermaseren, {The Large top quark mass expansion
  for Higgs boson decays into bottom quarks and into gluons}, Phys. Lett. B 362
  (1995) 134--140.
\newblock \href {http://arxiv.org/abs/hep-ph/9506465}
  {\path{arXiv:hep-ph/9506465}}, \href
  {http://dx.doi.org/10.1016/0370-2693(95)01192-S}
  {\path{doi:10.1016/0370-2693(95)01192-S}}.

\bibitem{Schreck:2007um}
M.~Schreck, M.~Steinhauser, {Higgs Decay to Gluons at NNLO}, Phys. Lett. B 655
  (2007) 148--155.
\newblock \href {http://arxiv.org/abs/0708.0916} {\path{arXiv:0708.0916}},
  \href {http://dx.doi.org/10.1016/j.physletb.2007.08.080}
  {\path{doi:10.1016/j.physletb.2007.08.080}}.

\bibitem{Davies:2017xsp}
J.~Davies, M.~Steinhauser, D.~Wellmann, {Completing the hadronic Higgs boson
  decay at order $\alpha_s^4$}, Nucl. Phys. B 920 (2017) 20--31.
\newblock \href {http://arxiv.org/abs/1703.02988} {\path{arXiv:1703.02988}},
  \href {http://dx.doi.org/10.1016/j.nuclphysb.2017.04.012}
  {\path{doi:10.1016/j.nuclphysb.2017.04.012}}.

\bibitem{Braaten:1980yq}
E.~Braaten, J.~Leveille, {Higgs Boson Decay and the Running Mass}, Phys. Rev. D
  22 (1980) 715.
\newblock \href {http://dx.doi.org/10.1103/PhysRevD.22.715}
  {\path{doi:10.1103/PhysRevD.22.715}}.

\bibitem{deFlorian:2016spz}
D.~de~Florian, et~al., {Handbook of LHC Higgs Cross Sections: 4. Deciphering
  the Nature of the Higgs Sector} 2/2017.
\newblock \href {http://arxiv.org/abs/1610.07922} {\path{arXiv:1610.07922}},
  \href {http://dx.doi.org/10.23731/CYRM-2017-002}
  {\path{doi:10.23731/CYRM-2017-002}}.

\bibitem{Spira:2016ztx}
M.~Spira, {Higgs Boson Production and Decay at Hadron Colliders}, Prog. Part.
  Nucl. Phys. 95 (2017) 98--159.
\newblock \href {http://arxiv.org/abs/1612.07651} {\path{arXiv:1612.07651}},
  \href {http://dx.doi.org/10.1016/j.ppnp.2017.04.001}
  {\path{doi:10.1016/j.ppnp.2017.04.001}}.

\bibitem{Fleischer:1980ub}
J.~Fleischer, F.~Jegerlehner, {Radiative Corrections to Higgs Decays in the
  Extended Weinberg-Salam Model}, Phys. Rev. D 23 (1981) 2001--2026.
\newblock \href {http://dx.doi.org/10.1103/PhysRevD.23.2001}
  {\path{doi:10.1103/PhysRevD.23.2001}}.

\bibitem{Bardin:1990zj}
D.~Bardin, B.~Vilensky, P.~Khristova, {Calculation of the Higgs boson decay
  width into fermion pairs}, Sov. J. Nucl. Phys. 53 (1991) 152--158.

\bibitem{Dabelstein:1991ky}
A.~Dabelstein, W.~Hollik, {Electroweak corrections to the fermionic decay width
  of the standard Higgs boson}, Z. Phys. C 53 (1992) 507--516.
\newblock \href {http://dx.doi.org/10.1007/BF01625912}
  {\path{doi:10.1007/BF01625912}}.

\bibitem{Kniehl:1991ze}
B.~A. Kniehl, {Radiative corrections for $H \to f \bar f (\gamma)$ in the
  standard model}, Nucl. Phys. B 376 (1992) 3--28.
\newblock \href {http://dx.doi.org/10.1016/0550-3213(92)90065-J}
  {\path{doi:10.1016/0550-3213(92)90065-J}}.

\bibitem{Kwiatkowski:1994cu}
A.~Kwiatkowski, M.~Steinhauser, {Corrections of order ${\cal O}(G_F \alpha_s
  m_t^2)$ to the Higgs decay rate $\Gamma (H \to b \bar b)$}, Phys. Lett. B 338
  (1994) 66--70, [Erratum: Phys.Lett.B 342, 455--455 (1995)].
\newblock \href {http://arxiv.org/abs/hep-ph/9405308}
  {\path{arXiv:hep-ph/9405308}}, \href
  {http://dx.doi.org/10.1016/0370-2693(94)91345-5}
  {\path{doi:10.1016/0370-2693(94)91345-5}}.

\bibitem{Kniehl:1994ju}
B.~A. Kniehl, M.~Spira, {Two loop $O (\alpha_s G_F m_t^2)$ correction to the $H
  \to b \bar{b}$ decay rate}, Nucl. Phys. B 432 (1994) 39--48.
\newblock \href {http://arxiv.org/abs/hep-ph/9410319}
  {\path{arXiv:hep-ph/9410319}}, \href
  {http://dx.doi.org/10.1016/0550-3213(94)90592-4}
  {\path{doi:10.1016/0550-3213(94)90592-4}}.

\bibitem{Chetyrkin:1996wr}
K.~Chetyrkin, B.~A. Kniehl, M.~Steinhauser, {Virtual top quark effects on the
  $H \to b \bar b$ decay at next-to-leading order in QCD}, Phys. Rev. Lett. 78
  (1997) 594--597.
\newblock \href {http://arxiv.org/abs/hep-ph/9610456}
  {\path{arXiv:hep-ph/9610456}}, \href
  {http://dx.doi.org/10.1103/PhysRevLett.78.594}
  {\path{doi:10.1103/PhysRevLett.78.594}}.

\bibitem{Mihaila:2015lwa}
L.~Mihaila, B.~Schmidt, M.~Steinhauser, {$\Gamma(H\to b\bar{b})$ to order
  $\alpha\alpha_s$}, Phys. Lett. B 751 (2015) 442--447.
\newblock \href {http://arxiv.org/abs/1509.02294} {\path{arXiv:1509.02294}},
  \href {http://dx.doi.org/10.1016/j.physletb.2015.10.078}
  {\path{doi:10.1016/j.physletb.2015.10.078}}.

\bibitem{Pich:2013lsa}
A.~Pich, {Precision Tau Physics}, Prog. Part. Nucl. Phys. 75 (2014) 41--85.
\newblock \href {http://arxiv.org/abs/1310.7922} {\path{arXiv:1310.7922}},
  \href {http://dx.doi.org/10.1016/j.ppnp.2013.11.002}
  {\path{doi:10.1016/j.ppnp.2013.11.002}}.

\bibitem{Amhis:2019ckw}
Y.~S. Amhis, et~al., {Averages of $b$-hadron, $c$-hadron, and $\tau$-lepton
  properties as of 2018, }\href {http://arxiv.org/abs/1909.12524}
  {\path{arXiv:1909.12524}}.

\bibitem{Braaten:1991qm}
E.~Braaten, S.~Narison, A.~Pich, {QCD analysis of the tau hadronic width},
  Nucl. Phys. B 373 (1992) 581--612.
\newblock \href {http://dx.doi.org/10.1016/0550-3213(92)90267-F}
  {\path{doi:10.1016/0550-3213(92)90267-F}}.

\bibitem{Marciano:1988vm}
W.~Marciano, A.~Sirlin, {Electroweak Radiative Corrections to tau Decay}, Phys.
  Rev. Lett. 61 (1988) 1815--1818.
\newblock \href {http://dx.doi.org/10.1103/PhysRevLett.61.1815}
  {\path{doi:10.1103/PhysRevLett.61.1815}}.

\bibitem{Braaten:1990ef}
E.~Braaten, C.-S. Li, {Electroweak radiative corrections to the semihadronic
  decay rate of the tau lepton}, Phys. Rev. D 42 (1990) 3888--3891.
\newblock \href {http://dx.doi.org/10.1103/PhysRevD.42.3888}
  {\path{doi:10.1103/PhysRevD.42.3888}}.

\bibitem{Erler:2002mv}
J.~Erler, {Electroweak radiative corrections to semileptonic tau decays}, Rev.
  Mex. Fis. 50 (2004) 200--202.
\newblock \href {http://arxiv.org/abs/hep-ph/0211345}
  {\path{arXiv:hep-ph/0211345}}.

\bibitem{Braaten:1988hc}
E.~Braaten, {QCD Predictions for the Decay of the tau Lepton}, Phys. Rev. Lett.
  60 (1988) 1606--1609.
\newblock \href {http://dx.doi.org/10.1103/PhysRevLett.60.1606}
  {\path{doi:10.1103/PhysRevLett.60.1606}}.

\bibitem{Narison:1988ni}
S.~Narison, A.~Pich, {QCD Formulation of the tau Decay and Determination of
  $\Lambda_{\overline{MS}}$}, Phys. Lett. B 211 (1988) 183--188.
\newblock \href {http://dx.doi.org/10.1016/0370-2693(88)90830-1}
  {\path{doi:10.1016/0370-2693(88)90830-1}}.

\bibitem{Braaten:1988ea}
E.~Braaten, {The Perturbative QCD Corrections to the Ratio R for tau Decay},
  Phys. Rev. D 39 (1989) 1458.
\newblock \href {http://dx.doi.org/10.1103/PhysRevD.39.1458}
  {\path{doi:10.1103/PhysRevD.39.1458}}.

\bibitem{LeDiberder:1992jjr}
F.~Le~Diberder, A.~Pich, {The perturbative QCD prediction to $R_\tau$
  revisited}, Phys. Lett. B 286 (1992) 147--152.
\newblock \href {http://dx.doi.org/10.1016/0370-2693(92)90172-Z}
  {\path{doi:10.1016/0370-2693(92)90172-Z}}.

\bibitem{Pivovarov:1991rh}
A.~Pivovarov, {Renormalization group analysis of the tau lepton decay within
  QCD}, Sov. J. Nucl. Phys. 54 (1991) 676--678.
\newblock \href {http://arxiv.org/abs/hep-ph/0302003}
  {\path{arXiv:hep-ph/0302003}}, \href {http://dx.doi.org/10.1007/BF01625906}
  {\path{doi:10.1007/BF01625906}}.

\bibitem{LovettTurner:1994hx}
C.~Lovett-Turner, C.~Maxwell, {Renormalon singularities of the QCD vacuum
  polarization function to leading order in $1/N_f$}, Nucl. Phys. B 432 (1994)
  147--162.
\newblock \href {http://arxiv.org/abs/hep-ph/9407268}
  {\path{arXiv:hep-ph/9407268}}, \href
  {http://dx.doi.org/10.1016/0550-3213(94)90597-5}
  {\path{doi:10.1016/0550-3213(94)90597-5}}.

\bibitem{Ball:1995ni}
P.~Ball, M.~Beneke, V.~M. Braun, {Resummation of $(\beta_ 0 \alpha_s)^n$
  corrections in QCD: Techniques and applications to the tau hadronic width and
  the heavy quark pole mass}, Nucl. Phys. B 452 (1995) 563--625.
\newblock \href {http://arxiv.org/abs/hep-ph/9502300}
  {\path{arXiv:hep-ph/9502300}}, \href
  {http://dx.doi.org/10.1016/0550-3213(95)00392-6}
  {\path{doi:10.1016/0550-3213(95)00392-6}}.

\bibitem{Kataev:1995vh}
A.~L. Kataev, V.~V. Starshenko, {Estimates of the higher order QCD corrections
  to $R(s)$, $R_\tau$ and deep inelastic scattering sum rules}, Mod. Phys.
  Lett. A 10 (1995) 235--250.
\newblock \href {http://arxiv.org/abs/hep-ph/9502348}
  {\path{arXiv:hep-ph/9502348}}, \href
  {http://dx.doi.org/10.1142/S0217732395000272}
  {\path{doi:10.1142/S0217732395000272}}.

\bibitem{LovettTurner:1995ti}
C.~Lovett-Turner, C.~Maxwell, {All orders renormalon resummations for some QCD
  observables}, Nucl. Phys. B 452 (1995) 188--212.
\newblock \href {http://arxiv.org/abs/hep-ph/9505224}
  {\path{arXiv:hep-ph/9505224}}, \href
  {http://dx.doi.org/10.1016/0550-3213(95)00383-4}
  {\path{doi:10.1016/0550-3213(95)00383-4}}.

\bibitem{Neubert:1995gd}
M.~Neubert, {QCD analysis of hadronic tau decays revisited}, Nucl. Phys. B 463
  (1996) 511--546.
\newblock \href {http://arxiv.org/abs/hep-ph/9509432}
  {\path{arXiv:hep-ph/9509432}}, \href
  {http://dx.doi.org/10.1016/0550-3213(96)00002-8}
  {\path{doi:10.1016/0550-3213(96)00002-8}}.

\bibitem{Maxwell:1996he}
C.~Maxwell, D.~Tonge, {RS invariant all orders renormalon resummations for some
  QCD observables}, Nucl. Phys. B 481 (1996) 681--703.
\newblock \href {http://arxiv.org/abs/hep-ph/9606392}
  {\path{arXiv:hep-ph/9606392}}, \href
  {http://dx.doi.org/10.1016/S0550-3213(96)00532-9}
  {\path{doi:10.1016/S0550-3213(96)00532-9}}.

\bibitem{Maxwell:1997yw}
C.~Maxwell, D.~Tonge, {The Uncertainty in $\alpha_s (M_Z^2)$ determined from
  hadronic tau decay measurements}, Nucl. Phys. B 535 (1998) 19--40.
\newblock \href {http://arxiv.org/abs/hep-ph/9705314}
  {\path{arXiv:hep-ph/9705314}}, \href
  {http://dx.doi.org/10.1016/S0550-3213(98)00562-8}
  {\path{doi:10.1016/S0550-3213(98)00562-8}}.

\bibitem{Maxwell:2001he}
C.~Maxwell, A.~Mirjalili, {Renormalon inspired resummations for vector and
  scalar correlators: Estimating the uncertainty in $\alpha_s(M^2_\tau)$ and
  $\alpha(M^2_Z)$}, Nucl. Phys. B 611 (2001) 423--446.
\newblock \href {http://arxiv.org/abs/hep-ph/0103164}
  {\path{arXiv:hep-ph/0103164}}, \href
  {http://dx.doi.org/10.1016/S0550-3213(01)00327-3}
  {\path{doi:10.1016/S0550-3213(01)00327-3}}.

\bibitem{Davier:2005xq}
M.~Davier, A.~Höcker, Z.~Zhang, {The Physics of Hadronic Tau Decays}, Rev.
  Mod. Phys. 78 (2006) 1043--1109.
\newblock \href {http://arxiv.org/abs/hep-ph/0507078}
  {\path{arXiv:hep-ph/0507078}}, \href
  {http://dx.doi.org/10.1103/RevModPhys.78.1043}
  {\path{doi:10.1103/RevModPhys.78.1043}}.

\bibitem{Jamin:2005ip}
M.~Jamin, {Contour-improved versus fixed-order perturbation theory in hadronic
  tau decays}, JHEP 09 (2005) 058.
\newblock \href {http://arxiv.org/abs/hep-ph/0509001}
  {\path{arXiv:hep-ph/0509001}}, \href
  {http://dx.doi.org/10.1088/1126-6708/2005/09/058}
  {\path{doi:10.1088/1126-6708/2005/09/058}}.

\bibitem{Davier:2008sk}
M.~Davier, S.~Descotes-Genon, A.~Höcker, B.~Malaescu, Z.~Zhang, {The
  Determination of $\alpha_s$ from Tau Decays Revisited}, Eur. Phys. J. C 56
  (2008) 305--322.
\newblock \href {http://arxiv.org/abs/0803.0979} {\path{arXiv:0803.0979}},
  \href {http://dx.doi.org/10.1140/epjc/s10052-008-0666-7}
  {\path{doi:10.1140/epjc/s10052-008-0666-7}}.

\bibitem{Pich:2010xb}
A.~Pich, {$\alpha_s$ Determination from $\tau$ Decays: Theoretical Status},
  Acta Phys. Polon. Supp. 3 (2010) 165--170.
\newblock \href {http://arxiv.org/abs/1001.0389} {\path{arXiv:1001.0389}}.

\bibitem{Cvetic:2010ut}
G.~Cvetic, M.~Loewe, C.~Martinez, C.~Valenzuela, {Modified Contour-Improved
  Perturbation Theory}, Phys. Rev. D 82 (2010) 093007.
\newblock \href {http://arxiv.org/abs/1005.4444} {\path{arXiv:1005.4444}},
  \href {http://dx.doi.org/10.1103/PhysRevD.82.093007}
  {\path{doi:10.1103/PhysRevD.82.093007}}.

\bibitem{Pich:2011cj}
A.~Pich, {QCD Description of Hadronic Tau Decays}, Nucl. Phys. B Proc. Suppl.
  218 (2011) 89--97.
\newblock \href {http://arxiv.org/abs/1101.2107} {\path{arXiv:1101.2107}},
  \href {http://dx.doi.org/10.1016/j.nuclphysbps.2011.06.016}
  {\path{doi:10.1016/j.nuclphysbps.2011.06.016}}.

\bibitem{Groote:2012jq}
S.~Groote, J.~Korner, A.~Pivovarov, {Understanding PT results for decays of
  $\tau$ leptons into hadrons}, Phys. Part. Nucl. 44 (2013) 285--298.
\newblock \href {http://arxiv.org/abs/1212.5346} {\path{arXiv:1212.5346}},
  \href {http://dx.doi.org/10.1134/S1063779613020147}
  {\path{doi:10.1134/S1063779613020147}}.

\bibitem{Boito:2016pwf}
D.~Boito, M.~Jamin, R.~Miravitllas, {Scheme Variations of the QCD Coupling and
  Hadronic \ensuremath{\tau} Decays}, Phys. Rev. Lett. 117~(15) (2016) 152001.
\newblock \href {http://arxiv.org/abs/1606.06175} {\path{arXiv:1606.06175}},
  \href {http://dx.doi.org/10.1103/PhysRevLett.117.152001}
  {\path{doi:10.1103/PhysRevLett.117.152001}}.

\bibitem{Boito:2018rwt}
D.~Boito, P.~Masjuan, F.~Oliani, {Higher-order QCD corrections to hadronic
  $\tau$ decays from Pad\'e approximants}, JHEP 08 (2018) 075.
\newblock \href {http://arxiv.org/abs/1807.01567} {\path{arXiv:1807.01567}},
  \href {http://dx.doi.org/10.1007/JHEP08(2018)075}
  {\path{doi:10.1007/JHEP08(2018)075}}.

\bibitem{Wu:2019mky}
X.-G. Wu, J.-M. Shen, B.-L. Du, X.-D. Huang, S.-Q. Wang, S.~J. Brodsky, {The
  QCD renormalization group equation and the elimination of fixed-order
  scheme-and-scale ambiguities using the principle of maximum conformality},
  Prog. Part. Nucl. Phys. 108 (2019) 103706.
\newblock \href {http://arxiv.org/abs/1903.12177} {\path{arXiv:1903.12177}},
  \href {http://dx.doi.org/10.1016/j.ppnp.2019.05.003}
  {\path{doi:10.1016/j.ppnp.2019.05.003}}.

\bibitem{Beneke:1998ui}
M.~Beneke, {Renormalons}, Phys. Rept. 317 (1999) 1--142.
\newblock \href {http://arxiv.org/abs/hep-ph/9807443}
  {\path{arXiv:hep-ph/9807443}}, \href
  {http://dx.doi.org/10.1016/S0370-1573(98)00130-6}
  {\path{doi:10.1016/S0370-1573(98)00130-6}}.

\bibitem{Altarelli:1994vz}
G.~Altarelli, P.~Nason, G.~Ridolfi, {A Study of ultraviolet renormalon
  ambiguities in the determination of $\alpha_s$ from tau decay}, Z. Phys. C 68
  (1995) 257--268.
\newblock \href {http://arxiv.org/abs/hep-ph/9501240}
  {\path{arXiv:hep-ph/9501240}}, \href {http://dx.doi.org/10.1007/BF01566673}
  {\path{doi:10.1007/BF01566673}}.

\bibitem{Cvetic:2001ws}
G.~Cvetic, C.~Dib, T.~Lee, I.~Schmidt, {Resummation of the hadronic tau decay
  width with modified Borel transform method}, Phys. Rev. D 64 (2001) 093016.
\newblock \href {http://arxiv.org/abs/hep-ph/0106024}
  {\path{arXiv:hep-ph/0106024}}, \href
  {http://dx.doi.org/10.1103/PhysRevD.64.093016}
  {\path{doi:10.1103/PhysRevD.64.093016}}.

\bibitem{Beneke:2008ad}
M.~Beneke, M.~Jamin, {$\alpha_s$ and the $\tau$ hadronic width: fixed-order,
  contour-improved and higher-order perturbation theory}, JHEP 09 (2008) 044.
\newblock \href {http://arxiv.org/abs/0806.3156} {\path{arXiv:0806.3156}},
  \href {http://dx.doi.org/10.1088/1126-6708/2008/09/044}
  {\path{doi:10.1088/1126-6708/2008/09/044}}.

\bibitem{Menke:2009vg}
S.~Menke, {On the determination of $\alpha_s$ from hadronic tau decays with
  contour-improved, fixed order and renormalon-chain perturbation theory,
  }\href {http://arxiv.org/abs/0904.1796} {\path{arXiv:0904.1796}}.

\bibitem{Caprini:2009vf}
I.~Caprini, J.~Fischer, {$\alpha_s$ from tau decays: Contour-improved versus
  fixed-order summation in a new QCD perturbation expansion}, Eur. Phys. J. C
  64 (2009) 35--45.
\newblock \href {http://arxiv.org/abs/0906.5211} {\path{arXiv:0906.5211}},
  \href {http://dx.doi.org/10.1140/epjc/s10052-009-1142-8}
  {\path{doi:10.1140/epjc/s10052-009-1142-8}}.

\bibitem{DescotesGenon:2010cr}
S.~Descotes-Genon, B.~Malaescu, {A Note on Renormalon Models for the
  Determination of $\alpha_s(M_\tau)$, }\href {http://arxiv.org/abs/1002.2968}
  {\path{arXiv:1002.2968}}.

\bibitem{Caprini:2011ya}
I.~Caprini, J.~Fischer, {Expansion functions in perturbative QCD and the
  determination of $\alpha_s(M_\tau^2)$}, Phys. Rev. D 84 (2011) 054019.
\newblock \href {http://arxiv.org/abs/1106.5336} {\path{arXiv:1106.5336}},
  \href {http://dx.doi.org/10.1103/PhysRevD.84.054019}
  {\path{doi:10.1103/PhysRevD.84.054019}}.

\bibitem{Abbas:2012py}
G.~Abbas, B.~Ananthanarayan, I.~Caprini, {Determination of
  $\alpha_s(M_{\tau}^2)$ from Improved Fixed Order Perturbation Theory}, Phys.
  Rev. D 85 (2012) 094018.
\newblock \href {http://arxiv.org/abs/1202.2672} {\path{arXiv:1202.2672}},
  \href {http://dx.doi.org/10.1103/PhysRevD.85.094018}
  {\path{doi:10.1103/PhysRevD.85.094018}}.

\bibitem{Beneke:2012vb}
M.~Beneke, D.~Boito, M.~Jamin, {Perturbative expansion of tau hadronic spectral
  function moments and $\alpha_s$ extractions}, JHEP 01 (2013) 125.
\newblock \href {http://arxiv.org/abs/1210.8038} {\path{arXiv:1210.8038}},
  \href {http://dx.doi.org/10.1007/JHEP01(2013)125}
  {\path{doi:10.1007/JHEP01(2013)125}}.

\bibitem{Abbas:2012fi}
G.~Abbas, B.~Ananthanarayan, I.~Caprini, J.~Fischer, {Perturbative expansion of
  the QCD Adler function improved by renormalization-group summation and
  analytic continuation in the Borel plane}, Phys. Rev. D 87~(1) (2013) 014008.
\newblock \href {http://arxiv.org/abs/1211.4316} {\path{arXiv:1211.4316}},
  \href {http://dx.doi.org/10.1103/PhysRevD.87.014008}
  {\path{doi:10.1103/PhysRevD.87.014008}}.

\bibitem{Abbas:2013usa}
G.~Abbas, B.~Ananthanarayan, I.~Caprini, J.~Fischer, {Expansions of $\tau$
  hadronic spectral function moments in a nonpower QCD perturbation theory with
  tamed large order behavior}, Phys. Rev. D 88~(3) (2013) 034026.
\newblock \href {http://arxiv.org/abs/1307.6323} {\path{arXiv:1307.6323}},
  \href {http://dx.doi.org/10.1103/PhysRevD.88.034026}
  {\path{doi:10.1103/PhysRevD.88.034026}}.

\bibitem{Caprini:2019kwp}
I.~Caprini, {Higher-order perturbative coefficients in QCD from series
  acceleration by conformal mappings}, Phys. Rev. D 100~(5) (2019) 056019.
\newblock \href {http://arxiv.org/abs/1908.06632} {\path{arXiv:1908.06632}},
  \href {http://dx.doi.org/10.1103/PhysRevD.100.056019}
  {\path{doi:10.1103/PhysRevD.100.056019}}.

\bibitem{Caprini:2020lff}
I.~Caprini, {Conformal mapping of the Borel plane: going beyond perturbative
  QCD}, Phys. Rev. D 102~(5) (2020) 054017.
\newblock \href {http://arxiv.org/abs/2006.16605} {\path{arXiv:2006.16605}},
  \href {http://dx.doi.org/10.1103/PhysRevD.102.054017}
  {\path{doi:10.1103/PhysRevD.102.054017}}.

\bibitem{Hoang:2020mkw}
A.~H. Hoang, C.~Regner, {Borel Representation of $\tau$ Hadronic Spectral
  Function Moments in Contour-Improved Perturbation Theory, }\href
  {http://arxiv.org/abs/2008.00578} {\path{arXiv:2008.00578}}.

\bibitem{Pich:2016bdg}
A.~Pich, A.~Rodríguez-Sánchez, {Determination of the QCD coupling from ALEPH
  $\tau$ decay data}, Phys. Rev. D 94~(3) (2016) 034027.
\newblock \href {http://arxiv.org/abs/1605.06830} {\path{arXiv:1605.06830}},
  \href {http://dx.doi.org/10.1103/PhysRevD.94.034027}
  {\path{doi:10.1103/PhysRevD.94.034027}}.

\bibitem{Davier:2013sfa}
M.~Davier, A.~Höcker, B.~Malaescu, C.-Z. Yuan, Z.~Zhang, {Update of the ALEPH
  non-strange spectral functions from hadronic $\tau$ decays}, Eur. Phys. J. C
  74~(3) (2014) 2803.
\newblock \href {http://arxiv.org/abs/1312.1501} {\path{arXiv:1312.1501}},
  \href {http://dx.doi.org/10.1140/epjc/s10052-014-2803-9}
  {\path{doi:10.1140/epjc/s10052-014-2803-9}}.

\bibitem{LeDiberder:1992zhd}
F.~Le~Diberder, A.~Pich, {Testing QCD with tau decays}, Phys. Lett. B 289
  (1992) 165--175.
\newblock \href {http://dx.doi.org/10.1016/0370-2693(92)91380-R}
  {\path{doi:10.1016/0370-2693(92)91380-R}}.

\bibitem{Pich:1989pq}
A.~Pich, {QCD Tests from Tau Decay Data, in: Tau-Charm Factory Workshop (SLAC,
  23--27 May 1989)}, Conf. Proc. C 890523 (1989) 416--435.

\bibitem{Buskulic:1993sv}
D.~Buskulic, et~al., {Measurement of the strong coupling constant using tau
  decays}, Phys. Lett. B 307 (1993) 209--220.
\newblock \href {http://dx.doi.org/10.1016/0370-2693(93)90212-Z}
  {\path{doi:10.1016/0370-2693(93)90212-Z}}.

\bibitem{Barate:1998uf}
R.~Barate, et~al., {Measurement of the axial-vector $\tau$ spectral functions
  and determination of $\alpha_s(M^2_\tau)$ from hadronic $\tau$ decays}, Eur.
  Phys. J. C 4 (1998) 409--431.
\newblock \href {http://dx.doi.org/10.1007/s100520050217}
  {\path{doi:10.1007/s100520050217}}.

\bibitem{Schael:2005am}
S.~Schael, et~al., {Branching ratios and spectral functions of tau decays:
  Final ALEPH measurements and physics implications}, Phys. Rept. 421 (2005)
  191--284.
\newblock \href {http://arxiv.org/abs/hep-ex/0506072}
  {\path{arXiv:hep-ex/0506072}}, \href
  {http://dx.doi.org/10.1016/j.physrep.2005.06.007}
  {\path{doi:10.1016/j.physrep.2005.06.007}}.

\bibitem{Coan:1995nk}
T.~Coan, et~al., {Measurement of $\alpha_s$ from tau decays}, Phys. Lett. B 356
  (1995) 580--588.
\newblock \href {http://dx.doi.org/10.1016/0370-2693(95)00824-5}
  {\path{doi:10.1016/0370-2693(95)00824-5}}.

\bibitem{Ackerstaff:1998yj}
K.~Ackerstaff, et~al., {Measurement of the strong coupling constant $\alpha_s$
  and the vector and axial vector spectral functions in hadronic tau decays},
  Eur. Phys. J. C 7 (1999) 571--593.
\newblock \href {http://arxiv.org/abs/hep-ex/9808019}
  {\path{arXiv:hep-ex/9808019}}, \href
  {http://dx.doi.org/10.1007/s100529901061} {\path{doi:10.1007/s100529901061}}.

\bibitem{Pich:2016mgv}
A.~Pich, A.~Rodríguez-Sánchez, {Updated determination of $\alpha_s(m_\tau^2)$
  from tau decays}, Mod. Phys. Lett. A 31~(30) (2016) 1630032.
\newblock \href {http://arxiv.org/abs/1606.07764} {\path{arXiv:1606.07764}},
  \href {http://dx.doi.org/10.1142/S0217732316300329}
  {\path{doi:10.1142/S0217732316300329}}.

\bibitem{Narison:1993sx}
S.~Narison, A.~Pich, {Semi-inclusive tau decays involving the vector or
  axial-vector hadronic currents}, Phys. Lett. B 304 (1993) 359--365.
\newblock \href {http://dx.doi.org/10.1016/0370-2693(93)90309-6}
  {\path{doi:10.1016/0370-2693(93)90309-6}}.

\bibitem{Girone:1995xb}
M.~Girone, M.~Neubert, {Test of the running of $\alpha_s$ in tau decays}, Phys.
  Rev. Lett. 76 (1996) 3061--3064.
\newblock \href {http://arxiv.org/abs/hep-ph/9511392}
  {\path{arXiv:hep-ph/9511392}}, \href
  {http://dx.doi.org/10.1103/PhysRevLett.76.3061}
  {\path{doi:10.1103/PhysRevLett.76.3061}}.

\bibitem{Cata:2008ru}
O.~Cata, M.~Golterman, S.~Peris, {Possible duality violations in tau decay and
  their impact on the determination of $\alpha_s$}, Phys. Rev. D 79 (2009)
  053002.
\newblock \href {http://arxiv.org/abs/0812.2285} {\path{arXiv:0812.2285}},
  \href {http://dx.doi.org/10.1103/PhysRevD.79.053002}
  {\path{doi:10.1103/PhysRevD.79.053002}}.

\bibitem{Boito:2014sta}
D.~Boito, M.~Golterman, K.~Maltman, J.~Osborne, S.~Peris, {Strong coupling from
  the revised ALEPH data for hadronic $\tau$ decays}, Phys. Rev. D 91~(3)
  (2015) 034003.
\newblock \href {http://arxiv.org/abs/1410.3528} {\path{arXiv:1410.3528}},
  \href {http://dx.doi.org/10.1103/PhysRevD.91.034003}
  {\path{doi:10.1103/PhysRevD.91.034003}}.

\bibitem{Boito:2012cr}
D.~Boito, M.~Golterman, M.~Jamin, A.~Mahdavi, K.~Maltman, J.~Osborne, S.~Peris,
  {An Updated determination of $\alpha_s$ from $\tau$ decays}, Phys. Rev. D 85
  (2012) 093015.
\newblock \href {http://arxiv.org/abs/1203.3146} {\path{arXiv:1203.3146}},
  \href {http://dx.doi.org/10.1103/PhysRevD.85.093015}
  {\path{doi:10.1103/PhysRevD.85.093015}}.

\bibitem{Pich:2018jiy}
A.~Pich, {Tau-decay determination of the strong coupling}, SciPost Phys. Proc.
  1 (2019) 036.
\newblock \href {http://arxiv.org/abs/1811.10067} {\path{arXiv:1811.10067}},
  \href {http://dx.doi.org/10.21468/SciPostPhysProc.1.036}
  {\path{doi:10.21468/SciPostPhysProc.1.036}}.

\bibitem{Nesterenko:2013vja}
A.~Nesterenko, {Dispersive approach to QCD and inclusive $\tau$ lepton hadronic
  decay}, Phys. Rev. D 88~(5) (2013) 056009.
\newblock \href {http://arxiv.org/abs/1306.4970} {\path{arXiv:1306.4970}},
  \href {http://dx.doi.org/10.1103/PhysRevD.88.056009}
  {\path{doi:10.1103/PhysRevD.88.056009}}.

\bibitem{Deur:2016tte}
A.~Deur, S.~J. Brodsky, G.~F. de~Teramond, {The QCD Running Coupling}, Prog.
  Part. Nucl. Phys. 90 (2016) 1--74.
\newblock \href {http://arxiv.org/abs/1604.08082} {\path{arXiv:1604.08082}},
  \href {http://dx.doi.org/10.1016/j.ppnp.2016.04.003}
  {\path{doi:10.1016/j.ppnp.2016.04.003}}.

\bibitem{Ayala:2017tco}
C.~Ayala, G.~Cvetic, R.~Kogerler, I.~Kondrashuk, {Nearly perturbative
  lattice-motivated QCD coupling with zero IR limit}, J. Phys. G 45~(3) (2018)
  035001.
\newblock \href {http://arxiv.org/abs/1703.01321} {\path{arXiv:1703.01321}},
  \href {http://dx.doi.org/10.1088/1361-6471/aa9ecc}
  {\path{doi:10.1088/1361-6471/aa9ecc}}.

\bibitem{Cvetic:2020naz}
G.~Cvetic, R.~Kogerler, {Lattice-motivated QCD coupling and hadronic
  contribution to muon $g-2$, }\href {http://arxiv.org/abs/2009.13742}
  {\path{arXiv:2009.13742}}.

\bibitem{Weinberg:1967kj}
S.~Weinberg, {Precise relations between the spectra of vector and axial vector
  mesons}, Phys. Rev. Lett. 18 (1967) 507--509.
\newblock \href {http://dx.doi.org/10.1103/PhysRevLett.18.507}
  {\path{doi:10.1103/PhysRevLett.18.507}}.

\bibitem{Floratos:1978jb}
E.~G. Floratos, S.~Narison, E.~de~Rafael, {Spectral Function Sum Rules in
  Quantum Chromodynamics. 1. Charged Currents Sector}, Nucl. Phys. B 155 (1979)
  115--149.
\newblock \href {http://dx.doi.org/10.1016/0550-3213(79)90359-6}
  {\path{doi:10.1016/0550-3213(79)90359-6}}.

\bibitem{Weinberg:1978kz}
S.~Weinberg, {Phenomenological Lagrangians}, Physica A 96~(1-2) (1979)
  327--340.
\newblock \href {http://dx.doi.org/10.1016/0378-4371(79)90223-1}
  {\path{doi:10.1016/0378-4371(79)90223-1}}.

\bibitem{Gasser:1983yg}
J.~Gasser, H.~Leutwyler, {Chiral Perturbation Theory to One Loop}, Annals Phys.
  158 (1984) 142.
\newblock \href {http://dx.doi.org/10.1016/0003-4916(84)90242-2}
  {\path{doi:10.1016/0003-4916(84)90242-2}}.

\bibitem{Gasser:1984gg}
J.~Gasser, H.~Leutwyler, {Chiral Perturbation Theory: Expansions in the Mass of
  the Strange Quark}, Nucl. Phys. B 250 (1985) 465--516.
\newblock \href {http://dx.doi.org/10.1016/0550-3213(85)90492-4}
  {\path{doi:10.1016/0550-3213(85)90492-4}}.

\bibitem{Ecker:1994gg}
G.~Ecker, {Chiral perturbation theory}, Prog. Part. Nucl. Phys. 35 (1995)
  1--80.
\newblock \href {http://arxiv.org/abs/hep-ph/9501357}
  {\path{arXiv:hep-ph/9501357}}, \href
  {http://dx.doi.org/10.1016/0146-6410(95)00041-G}
  {\path{doi:10.1016/0146-6410(95)00041-G}}.

\bibitem{Pich:1995bw}
A.~Pich, {Chiral perturbation theory}, Rept. Prog. Phys. 58 (1995) 563--610.
\newblock \href {http://arxiv.org/abs/hep-ph/9502366}
  {\path{arXiv:hep-ph/9502366}}, \href
  {http://dx.doi.org/10.1088/0034-4885/58/6/001}
  {\path{doi:10.1088/0034-4885/58/6/001}}.

\bibitem{Bijnens:2006zp}
J.~Bijnens, {Chiral perturbation theory beyond one loop}, Prog. Part. Nucl.
  Phys. 58 (2007) 521--586.
\newblock \href {http://arxiv.org/abs/hep-ph/0604043}
  {\path{arXiv:hep-ph/0604043}}, \href
  {http://dx.doi.org/10.1016/j.ppnp.2006.08.002}
  {\path{doi:10.1016/j.ppnp.2006.08.002}}.

\bibitem{Scherer:2012zzd}
S.~Scherer, M.~R. Schindler, {Quantum chromodynamics and chiral symmetry},
  Lect. Notes Phys. 830 (2012) 1--48.
\newblock \href {http://dx.doi.org/10.1007/978-3-642-19254-8_1}
  {\path{doi:10.1007/978-3-642-19254-8_1}}.

\bibitem{Amoros:1999dp}
G.~Amorós, J.~Bijnens, P.~Talavera, {Two point functions at two loops in three
  flavor chiral perturbation theory}, Nucl. Phys. B 568 (2000) 319--363.
\newblock \href {http://arxiv.org/abs/hep-ph/9907264}
  {\path{arXiv:hep-ph/9907264}}, \href
  {http://dx.doi.org/10.1016/S0550-3213(99)00674-4}
  {\path{doi:10.1016/S0550-3213(99)00674-4}}.

\bibitem{Bordes:2005wv}
J.~Bordes, C.~A. Dominguez, J.~Penarrocha, K.~Schilcher, {Chiral condensates
  from tau decay: A Critical reappraisal}, JHEP 02 (2006) 037.
\newblock \href {http://arxiv.org/abs/hep-ph/0511293}
  {\path{arXiv:hep-ph/0511293}}, \href
  {http://dx.doi.org/10.1088/1126-6708/2006/02/037}
  {\path{doi:10.1088/1126-6708/2006/02/037}}.

\bibitem{Almasy:2008xu}
A.~Almasy, K.~Schilcher, H.~Spiesberger, {Determination of QCD condensates from
  tau-decay data}, Eur. Phys. J. C 55 (2008) 237--248.
\newblock \href {http://arxiv.org/abs/0802.0980} {\path{arXiv:0802.0980}},
  \href {http://dx.doi.org/10.1140/epjc/s10052-008-0579-5}
  {\path{doi:10.1140/epjc/s10052-008-0579-5}}.

\bibitem{GonzalezAlonso:2008rf}
M.~González-Alonso, A.~Pich, J.~Prades, {Determination of the Chiral Couplings
  $L_{10}$ and $C_{87}$ from Semileptonic $\tau$ Decays}, Phys. Rev. D 78
  (2008) 116012.
\newblock \href {http://arxiv.org/abs/0810.0760} {\path{arXiv:0810.0760}},
  \href {http://dx.doi.org/10.1103/PhysRevD.78.116012}
  {\path{doi:10.1103/PhysRevD.78.116012}}.

\bibitem{GonzalezAlonso:2010rn}
M.~González-Alonso, A.~Pich, J.~Prades, {Violation of Quark-Hadron Duality and
  Spectral Chiral Moments in QCD}, Phys. Rev. D 81 (2010) 074007.
\newblock \href {http://arxiv.org/abs/1001.2269} {\path{arXiv:1001.2269}},
  \href {http://dx.doi.org/10.1103/PhysRevD.81.074007}
  {\path{doi:10.1103/PhysRevD.81.074007}}.

\bibitem{GonzalezAlonso:2010xf}
M.~González-Alonso, A.~Pich, J.~Prades, {Pinched weights and Duality Violation
  in QCD Sum Rules: a critical analysis}, Phys. Rev. D 82 (2010) 014019.
\newblock \href {http://arxiv.org/abs/1004.4987} {\path{arXiv:1004.4987}},
  \href {http://dx.doi.org/10.1103/PhysRevD.82.014019}
  {\path{doi:10.1103/PhysRevD.82.014019}}.

\bibitem{Boito:2012nt}
D.~Boito, M.~Golterman, M.~Jamin, K.~Maltman, S.~Peris, {Low-energy constants
  and condensates from the $\tau$ hadronic spectral functions}, Phys. Rev. D
  87~(9) (2013) 094008.
\newblock \href {http://arxiv.org/abs/1212.4471} {\path{arXiv:1212.4471}},
  \href {http://dx.doi.org/10.1103/PhysRevD.87.094008}
  {\path{doi:10.1103/PhysRevD.87.094008}}.

\bibitem{Dominguez:2014fua}
C.~Dominguez, L.~Hernandez, K.~Schilcher, H.~Spiesberger, {Chiral sum rules and
  vacuum condensates from tau-lepton decay data}, JHEP 03 (2015) 053.
\newblock \href {http://arxiv.org/abs/1410.3779} {\path{arXiv:1410.3779}},
  \href {http://dx.doi.org/10.1007/JHEP03(2015)053}
  {\path{doi:10.1007/JHEP03(2015)053}}.

\bibitem{Boito:2015fra}
D.~Boito, A.~Francis, M.~Golterman, R.~Hudspith, R.~Lewis, K.~Maltman,
  S.~Peris, {Low-energy constants and condensates from ALEPH hadronic
  \ensuremath{\tau} decay data}, Phys. Rev. D 92~(11) (2015) 114501.
\newblock \href {http://arxiv.org/abs/1503.03450} {\path{arXiv:1503.03450}},
  \href {http://dx.doi.org/10.1103/PhysRevD.92.114501}
  {\path{doi:10.1103/PhysRevD.92.114501}}.

\bibitem{Rodriguez-Sanchez:2016jvw}
M.~González-Alonso, A.~Pich, A.~Rodríguez-Sánchez, {Updated determination of
  chiral couplings and vacuum condensates from hadronic $\tau$ decay data},
  Phys. Rev. D 94~(1) (2016) 014017.
\newblock \href {http://arxiv.org/abs/1602.06112} {\path{arXiv:1602.06112}},
  \href {http://dx.doi.org/10.1103/PhysRevD.94.014017}
  {\path{doi:10.1103/PhysRevD.94.014017}}.

\bibitem{Pich:2008jm}
A.~Pich, I.~Rosell, J.~Sanz-Cillero, {Form-factors and current correlators:
  Chiral couplings $L_{10}^r(\mu)$ and $C_{87}^r(\mu)$ at NLO in $1/N_C$}, JHEP
  07 (2008) 014.
\newblock \href {http://arxiv.org/abs/0803.1567} {\path{arXiv:0803.1567}},
  \href {http://dx.doi.org/10.1088/1126-6708/2008/07/014}
  {\path{doi:10.1088/1126-6708/2008/07/014}}.

\bibitem{Cirigliano:2018dyk}
V.~Cirigliano, A.~Falkowski, M.~Gonz\'alez-Alonso, A.~Rodr\'\i{}guez-S\'anchez,
  {Hadronic \ensuremath{\tau} Decays as New Physics Probes in the LHC Era},
  Phys. Rev. Lett. 122~(22) (2019) 221801.
\newblock \href {http://arxiv.org/abs/1809.01161} {\path{arXiv:1809.01161}},
  \href {http://dx.doi.org/10.1103/PhysRevLett.122.221801}
  {\path{doi:10.1103/PhysRevLett.122.221801}}.

\bibitem{Bernard:1975cd}
C.~W. Bernard, A.~Duncan, J.~LoSecco, S.~Weinberg, {Exact Spectral Function Sum
  Rules}, Phys. Rev. D 12 (1975) 792.
\newblock \href {http://dx.doi.org/10.1103/PhysRevD.12.792}
  {\path{doi:10.1103/PhysRevD.12.792}}.

\bibitem{Gamiz:2002nu}
E.~Gámiz, M.~Jamin, A.~Pich, J.~Prades, F.~Schwab, {Determination of $m_s$ and
  $|V_{us}|$ from hadronic $\tau$ decays}, JHEP 01 (2003) 060.
\newblock \href {http://arxiv.org/abs/hep-ph/0212230}
  {\path{arXiv:hep-ph/0212230}}, \href
  {http://dx.doi.org/10.1088/1126-6708/2003/01/060}
  {\path{doi:10.1088/1126-6708/2003/01/060}}.

\bibitem{Gamiz:2004ar}
E.~Gámiz, M.~Jamin, A.~Pich, J.~Prades, F.~Schwab, {$V_{us}$ and $m_s$ from
  hadronic $\tau$ decays}, Phys. Rev. Lett. 94 (2005) 011803.
\newblock \href {http://arxiv.org/abs/hep-ph/0408044}
  {\path{arXiv:hep-ph/0408044}}, \href
  {http://dx.doi.org/10.1103/PhysRevLett.94.011803}
  {\path{doi:10.1103/PhysRevLett.94.011803}}.

\bibitem{Maltman:1998qz}
K.~Maltman, {Problems with extracting $m_s$ from flavor breaking in hadronic
  tau decays}, Phys. Rev. D 58 (1998) 093015.
\newblock \href {http://arxiv.org/abs/hep-ph/9804298}
  {\path{arXiv:hep-ph/9804298}}, \href
  {http://dx.doi.org/10.1103/PhysRevD.58.093015}
  {\path{doi:10.1103/PhysRevD.58.093015}}.

\bibitem{Chetyrkin:1998ej}
K.~Chetyrkin, J.~H. Kühn, A.~Pivovarov, {Determining the strange quark mass in
  Cabibbo suppressed tau lepton decays}, Nucl. Phys. B 533 (1998) 473--493.
\newblock \href {http://arxiv.org/abs/hep-ph/9805335}
  {\path{arXiv:hep-ph/9805335}}, \href
  {http://dx.doi.org/10.1016/S0550-3213(98)00511-2}
  {\path{doi:10.1016/S0550-3213(98)00511-2}}.

\bibitem{Kambor:2000dj}
J.~Kambor, K.~Maltman, {The Strange quark mass from flavor breaking in hadronic
  tau decays}, Phys. Rev. D 62 (2000) 093023.
\newblock \href {http://arxiv.org/abs/hep-ph/0005156}
  {\path{arXiv:hep-ph/0005156}}, \href
  {http://dx.doi.org/10.1103/PhysRevD.62.093023}
  {\path{doi:10.1103/PhysRevD.62.093023}}.

\bibitem{Korner:2000wd}
J.~Korner, F.~Krajewski, A.~Pivovarov, {Determination of the strange quark mass
  from Cabibbo suppressed tau decays with resummed perturbation theory in an
  effective scheme}, Eur. Phys. J. C 20 (2001) 259--269.
\newblock \href {http://arxiv.org/abs/hep-ph/0003165}
  {\path{arXiv:hep-ph/0003165}}, \href
  {http://dx.doi.org/10.1007/s100520100662} {\path{doi:10.1007/s100520100662}}.

\bibitem{Chen:2001qf}
S.~Chen, M.~Davier, E.~Gámiz, A.~Höcker, A.~Pich, J.~Prades, {Strange quark
  mass from the invariant mass distribution of Cabibbo suppressed tau decays},
  Eur. Phys. J. C 22 (2001) 31--38.
\newblock \href {http://arxiv.org/abs/hep-ph/0105253}
  {\path{arXiv:hep-ph/0105253}}, \href
  {http://dx.doi.org/10.1007/s100520100791} {\path{doi:10.1007/s100520100791}}.

\bibitem{Maltman:2001sv}
K.~Maltman, J.~Kambor, {On the longitudinal contributions to hadronic tau
  decay}, Phys. Rev. D 64 (2001) 093014.
\newblock \href {http://arxiv.org/abs/hep-ph/0107187}
  {\path{arXiv:hep-ph/0107187}}, \href
  {http://dx.doi.org/10.1103/PhysRevD.64.093014}
  {\path{doi:10.1103/PhysRevD.64.093014}}.

\bibitem{Baikov:2004tk}
P.~Baikov, K.~Chetyrkin, J.~H. Kühn, {Strange quark mass from tau lepton
  decays with $O(\alpha_s^3)$ accuracy}, Phys. Rev. Lett. 95 (2005) 012003.
\newblock \href {http://arxiv.org/abs/hep-ph/0412350}
  {\path{arXiv:hep-ph/0412350}}, \href
  {http://dx.doi.org/10.1103/PhysRevLett.95.012003}
  {\path{doi:10.1103/PhysRevLett.95.012003}}.

\bibitem{Maltman:2006st}
K.~Maltman, C.~E. Wolfe, {$V_{us}$ from hadronic tau decays}, Phys. Lett. B 639
  (2006) 283--289.
\newblock \href {http://arxiv.org/abs/hep-ph/0603215}
  {\path{arXiv:hep-ph/0603215}}, \href
  {http://dx.doi.org/10.1016/j.physletb.2006.05.062}
  {\path{doi:10.1016/j.physletb.2006.05.062}}.

\bibitem{Gamiz:2006nj}
E.~Gámiz, M.~Jamin, A.~Pich, J.~Prades, F.~Schwab, {$|V_{us}|$ from strange
  hadronic tau data}, Conf. Proc. C 060726 (2006) 786--789.
\newblock \href {http://arxiv.org/abs/hep-ph/0610246}
  {\path{arXiv:hep-ph/0610246}}, \href
  {http://dx.doi.org/10.1142/9789812790873\_0161}
  {\path{doi:10.1142/9789812790873\_0161}}.

\bibitem{Gamiz:2007qs}
E.~Gámiz, M.~Jamin, A.~Pich, J.~Prades, F.~Schwab, {Theoretical progress on
  the $V_{us}$ determination from $\tau$ decays}, PoS KAON (2008) 008.
\newblock \href {http://arxiv.org/abs/0709.0282} {\path{arXiv:0709.0282}},
  \href {http://dx.doi.org/10.22323/1.046.0008}
  {\path{doi:10.22323/1.046.0008}}.

\bibitem{Gamiz:2013wn}
E.~Gámiz, {$|V_{us}|$ from hadronic $\tau$ decays}, in: {7th International
  Workshop on the CKM Unitarity Triangle}, 2013.
\newblock \href {http://arxiv.org/abs/1301.2206} {\path{arXiv:1301.2206}}.

\bibitem{Leutwyler:1996qg}
H.~Leutwyler, {The Ratios of the light quark masses}, Phys. Lett. B 378 (1996)
  313--318.
\newblock \href {http://arxiv.org/abs/hep-ph/9602366}
  {\path{arXiv:hep-ph/9602366}}, \href
  {http://dx.doi.org/10.1016/0370-2693(96)00386-3}
  {\path{doi:10.1016/0370-2693(96)00386-3}}.

\bibitem{Hudspith:2017vew}
R.~J. Hudspith, R.~Lewis, K.~Maltman, J.~Zanotti, {A resolution of the
  inclusive flavor-breaking $\tau$ $|V_{us}|$ puzzle}, Phys. Lett. B 781 (2018)
  206--212.
\newblock \href {http://arxiv.org/abs/1702.01767} {\path{arXiv:1702.01767}},
  \href {http://dx.doi.org/10.1016/j.physletb.2018.03.074}
  {\path{doi:10.1016/j.physletb.2018.03.074}}.

\bibitem{Boyle:2018ilm}
P.~Boyle, R.~J. Hudspith, T.~Izubuchi, A.~J\"uttner, C.~Lehner, R.~Lewis,
  K.~Maltman, H.~Ohki, A.~Portelli, M.~Spraggs, {Novel $|V_{us}|$ Determination
  Using Inclusive Strange $\tau$ Decay and Lattice Hadronic Vacuum Polarization
  Functions}, Phys. Rev. Lett. 121~(20) (2018) 202003.
\newblock \href {http://arxiv.org/abs/1803.07228} {\path{arXiv:1803.07228}},
  \href {http://dx.doi.org/10.1103/PhysRevLett.121.202003}
  {\path{doi:10.1103/PhysRevLett.121.202003}}.

\bibitem{Kou:2018nap}
W.~Altmannshofer, et~al., {The Belle II Physics Book}, PTEP 2019~(12) (2019)
  123C01, [Erratum: PTEP 2020, 029201 (2020)].
\newblock \href {http://arxiv.org/abs/1808.10567} {\path{arXiv:1808.10567}},
  \href {http://dx.doi.org/10.1093/ptep/ptz106}
  {\path{doi:10.1093/ptep/ptz106}}.

\bibitem{Barniakov:2019zhx}
A.~Y. Barniakov, {The Super Charm-Tau Factory in Novosibirsk}, PoS
  LeptonPhoton2019 (2019) 062.
\newblock \href {http://dx.doi.org/10.22323/1.367.0062}
  {\path{doi:10.22323/1.367.0062}}.

\bibitem{Luo:2019xqt}
Q.~Luo, W.~Gao, J.~Lan, W.~Li, D.~Xu, {Progress of Conceptual Study for the
  Accelerators of a 2-7~GeV Super Tau Charm Facility at China}, in: {10th
  International Particle Accelerator Conference}, 2019, p. MOPRB031.
\newblock \href {http://dx.doi.org/10.18429/JACoW-IPAC2019-MOPRB031}
  {\path{doi:10.18429/JACoW-IPAC2019-MOPRB031}}.

\bibitem{Abada:2019lih}
A.~Abada, et~al., {FCC Physics Opportunities}: {Future Circular Collider
  Conceptual Design Report Volume 1}, Eur. Phys. J. C 79~(6) (2019) 474.
\newblock \href {http://dx.doi.org/10.1140/epjc/s10052-019-6904-3}
  {\path{doi:10.1140/epjc/s10052-019-6904-3}}.

\bibitem{Eidelman:1978xy}
S.~Eidelman, L.~Kurdadze, A.~Vainshtein, {$e^+ e^-$ Annihilation Into Hadrons
  Below 2~GeV. Test of QCD Predictions}, Phys. Lett. B 82 (1979) 278--280.
\newblock \href {http://dx.doi.org/10.1016/0370-2693(79)90755-X}
  {\path{doi:10.1016/0370-2693(79)90755-X}}.

\bibitem{Binner:1999bt}
S.~Binner, J.~H. Kühn, K.~Melnikov, {Measuring $\sigma (e^+
  e^-\to\mathrm{hadrons})$ using tagged photon}, Phys. Lett. B 459 (1999)
  279--287.
\newblock \href {http://arxiv.org/abs/hep-ph/9902399}
  {\path{arXiv:hep-ph/9902399}}, \href
  {http://dx.doi.org/10.1016/S0370-2693(99)00658-9}
  {\path{doi:10.1016/S0370-2693(99)00658-9}}.

\bibitem{Rodrigo:2001kf}
G.~Rodrigo, H.~Czyz, J.~H. Kühn, M.~Szopa, {Radiative return at NLO and the
  measurement of the hadronic cross-section in electron positron annihilation},
  Eur. Phys. J. C 24 (2002) 71--82.
\newblock \href {http://arxiv.org/abs/hep-ph/0112184}
  {\path{arXiv:hep-ph/0112184}}, \href
  {http://dx.doi.org/10.1007/s100520200912} {\path{doi:10.1007/s100520200912}}.

\bibitem{Jegerlehner:2017lbd}
F.~Jegerlehner, {Muon $g - 2$ theory: The hadronic part}, EPJ Web Conf. 166
  (2018) 00022.
\newblock \href {http://arxiv.org/abs/1705.00263} {\path{arXiv:1705.00263}},
  \href {http://dx.doi.org/10.1051/epjconf/201816600022}
  {\path{doi:10.1051/epjconf/201816600022}}.

\bibitem{Davier:2019can}
M.~Davier, A.~Hoecker, B.~Malaescu, Z.~Zhang, {A new evaluation of the hadronic
  vacuum polarisation contributions to the muon anomalous magnetic moment and
  to $\alpha(m_Z^2)$}, Eur. Phys. J. C 80~(3) (2020) 241, [Erratum:
  Eur.Phys.J.C 80, 410 (2020)].
\newblock \href {http://arxiv.org/abs/1908.00921} {\path{arXiv:1908.00921}},
  \href {http://dx.doi.org/10.1140/epjc/s10052-020-7792-2}
  {\path{doi:10.1140/epjc/s10052-020-7792-2}}.

\bibitem{Keshavarzi:2019abf}
A.~Keshavarzi, D.~Nomura, T.~Teubner, {$g-2$ of charged leptons, $\alpha
  (M^2_Z)$ , and the hyperfine splitting of muonium}, Phys. Rev. D 101~(1)
  (2020) 014029.
\newblock \href {http://arxiv.org/abs/1911.00367} {\path{arXiv:1911.00367}},
  \href {http://dx.doi.org/10.1103/PhysRevD.101.014029}
  {\path{doi:10.1103/PhysRevD.101.014029}}.

\bibitem{Aoyama:2020ynm}
T.~Aoyama, et~al., {The anomalous magnetic moment of the muon in the Standard
  Model}, Phys. Rept. 887 (2020) 1--166.
\newblock \href {http://arxiv.org/abs/2006.04822} {\path{arXiv:2006.04822}},
  \href {http://dx.doi.org/10.1016/j.physrep.2020.07.006}
  {\path{doi:10.1016/j.physrep.2020.07.006}}.

\bibitem{Cirigliano:2001er}
V.~Cirigliano, G.~Ecker, H.~Neufeld, {Isospin violation and the magnetic moment
  of the muon}, Phys. Lett. B 513 (2001) 361--370.
\newblock \href {http://arxiv.org/abs/hep-ph/0104267}
  {\path{arXiv:hep-ph/0104267}}, \href
  {http://dx.doi.org/10.1016/S0370-2693(01)00764-X}
  {\path{doi:10.1016/S0370-2693(01)00764-X}}.

\bibitem{Cirigliano:2002pv}
V.~Cirigliano, G.~Ecker, H.~Neufeld, {Radiative $\tau$ decay and the magnetic
  moment of the muon}, JHEP 08 (2002) 002.
\newblock \href {http://arxiv.org/abs/hep-ph/0207310}
  {\path{arXiv:hep-ph/0207310}}, \href
  {http://dx.doi.org/10.1088/1126-6708/2002/08/002}
  {\path{doi:10.1088/1126-6708/2002/08/002}}.

\bibitem{Davier:2009ag}
M.~Davier, A.~Hoecker, G.~Lopez~Castro, B.~Malaescu, X.~Mo, G.~Toledo~Sanchez,
  P.~Wang, C.~Yuan, Z.~Zhang, {The Discrepancy Between $\tau$ and $e^+e^-$
  Spectral Functions Revisited and the Consequences for the Muon Magnetic
  Anomaly}, Eur. Phys. J. C 66 (2010) 127--136.
\newblock \href {http://arxiv.org/abs/0906.5443} {\path{arXiv:0906.5443}},
  \href {http://dx.doi.org/10.1140/epjc/s10052-009-1219-4}
  {\path{doi:10.1140/epjc/s10052-009-1219-4}}.

\bibitem{Davier:2010nc}
M.~Davier, A.~Hoecker, B.~Malaescu, Z.~Zhang, {Reevaluation of the Hadronic
  Contributions to the Muon $g-2$ and to $\alpha(M_Z)$}, Eur. Phys. J. C 71
  (2011) 1515, [Erratum: Eur.Phys.J.C 72, 1874 (2012)].
\newblock \href {http://arxiv.org/abs/1010.4180} {\path{arXiv:1010.4180}},
  \href {http://dx.doi.org/10.1140/epjc/s10052-012-1874-8}
  {\path{doi:10.1140/epjc/s10052-012-1874-8}}.

\bibitem{Borsanyi:2017zdw}
S.~Borsanyi, et~al., {Hadronic vacuum polarization contribution to the
  anomalous magnetic moments of leptons from first principles}, Phys. Rev.
  Lett. 121~(2) (2018) 022002.
\newblock \href {http://arxiv.org/abs/1711.04980} {\path{arXiv:1711.04980}},
  \href {http://dx.doi.org/10.1103/PhysRevLett.121.022002}
  {\path{doi:10.1103/PhysRevLett.121.022002}}.

\bibitem{Blum:2018mom}
T.~Blum, P.~Boyle, V.~G\"ulpers, T.~Izubuchi, L.~Jin, C.~Jung, A.~J\"uttner,
  C.~Lehner, A.~Portelli, J.~Tsang, {Calculation of the hadronic vacuum
  polarization contribution to the muon anomalous magnetic moment}, Phys. Rev.
  Lett. 121~(2) (2018) 022003.
\newblock \href {http://arxiv.org/abs/1801.07224} {\path{arXiv:1801.07224}},
  \href {http://dx.doi.org/10.1103/PhysRevLett.121.022003}
  {\path{doi:10.1103/PhysRevLett.121.022003}}.

\bibitem{Giusti:2018mdh}
D.~Giusti, F.~Sanfilippo, S.~Simula, {Light-quark contribution to the leading
  hadronic vacuum polarization term of the muon $g-2$ from twisted-mass
  fermions}, Phys. Rev. D 98~(11) (2018) 114504.
\newblock \href {http://arxiv.org/abs/1808.00887} {\path{arXiv:1808.00887}},
  \href {http://dx.doi.org/10.1103/PhysRevD.98.114504}
  {\path{doi:10.1103/PhysRevD.98.114504}}.

\bibitem{Shintani:2019wai}
E.~Shintani, Y.~Kuramashi, {Hadronic vacuum polarization contribution to the
  muon $g-2$ with $2+1$ flavor lattice QCD on a larger than (10 fm$)^4$ lattice
  at the physical point}, Phys. Rev. D 100~(3) (2019) 034517.
\newblock \href {http://arxiv.org/abs/1902.00885} {\path{arXiv:1902.00885}},
  \href {http://dx.doi.org/10.1103/PhysRevD.100.034517}
  {\path{doi:10.1103/PhysRevD.100.034517}}.

\bibitem{Davies:2019efs}
C.~Davies, et~al., {Hadronic-vacuum-polarization contribution to the muon's
  anomalous magnetic moment from four-flavor lattice QCD}, Phys. Rev. D 101~(3)
  (2020) 034512.
\newblock \href {http://arxiv.org/abs/1902.04223} {\path{arXiv:1902.04223}},
  \href {http://dx.doi.org/10.1103/PhysRevD.101.034512}
  {\path{doi:10.1103/PhysRevD.101.034512}}.

\bibitem{Gerardin:2019rua}
A.~G\'erardin, M.~C\`e, G.~von Hippel, B.~H\"orz, H.~B. Meyer, D.~Mohler,
  K.~Ottnad, J.~Wilhelm, H.~Wittig, {The leading hadronic contribution to
  $(g-2)_\mu$ from lattice QCD with $N_{\rm f}=2+1$ flavours of O($a$) improved
  Wilson quarks}, Phys. Rev. D 100~(1) (2019) 014510.
\newblock \href {http://arxiv.org/abs/1904.03120} {\path{arXiv:1904.03120}},
  \href {http://dx.doi.org/10.1103/PhysRevD.100.014510}
  {\path{doi:10.1103/PhysRevD.100.014510}}.

\bibitem{Giusti:2019hkz}
D.~Giusti, S.~Simula, {Lepton anomalous magnetic moments in Lattice QCD+QED},
  PoS LATTICE2019 (2019) 104.
\newblock \href {http://arxiv.org/abs/1910.03874} {\path{arXiv:1910.03874}},
  \href {http://dx.doi.org/10.22323/1.363.0104}
  {\path{doi:10.22323/1.363.0104}}.

\bibitem{Borsanyi:2020mff}
S.~Borsanyi, et~al., {Leading-order hadronic vacuum polarization contribution
  to the muon magnetic momentfrom lattice QCD, }\href
  {http://arxiv.org/abs/2002.12347} {\path{arXiv:2002.12347}}.

\bibitem{Lehner:2020crt}
C.~Lehner, A.~S. Meyer, {Consistency of hadronic vacuum polarization between
  lattice QCD and the R-ratio}, Phys. Rev. D 101 (2020) 074515.
\newblock \href {http://arxiv.org/abs/2003.04177} {\path{arXiv:2003.04177}},
  \href {http://dx.doi.org/10.1103/PhysRevD.101.074515}
  {\path{doi:10.1103/PhysRevD.101.074515}}.

\bibitem{Keshavarzi:2018mgv}
A.~Keshavarzi, D.~Nomura, T.~Teubner, {Muon $g-2$ and $\alpha(M_Z^2)$: a new
  data-based analysis}, Phys. Rev. D 97~(11) (2018) 114025.
\newblock \href {http://arxiv.org/abs/1802.02995} {\path{arXiv:1802.02995}},
  \href {http://dx.doi.org/10.1103/PhysRevD.97.114025}
  {\path{doi:10.1103/PhysRevD.97.114025}}.

\bibitem{Boito:2018yvl}
D.~Boito, M.~Golterman, A.~Keshavarzi, K.~Maltman, D.~Nomura, S.~Peris,
  T.~Teubner, {Strong coupling from $e^+e^-\to$ hadrons below charm}, Phys.
  Rev. D 98~(7) (2018) 074030.
\newblock \href {http://arxiv.org/abs/1805.08176} {\path{arXiv:1805.08176}},
  \href {http://dx.doi.org/10.1103/PhysRevD.98.074030}
  {\path{doi:10.1103/PhysRevD.98.074030}}.

\bibitem{HRZ:2020xxx}
J.~Huston, K.~Rabbertz, G.~Zanderighi, {Quantum Chromodynamics, in: PDG 2020},
  PTEP 2020~(8) (2020) 083C01, 153--179.

\bibitem{Bethke:2011tr}
S.~Bethke, et~al., {Workshop on Precision Measurements of $\alpha_s$ (Munich,
  9--11 February, 2011), }\href {http://arxiv.org/abs/1110.0016}
  {\path{arXiv:1110.0016}}.

\bibitem{Pich:2013sqa}
A.~Pich, {Review of $\alpha_s$ determinations}, PoS ConfinementX (2012) 022.
\newblock \href {http://arxiv.org/abs/1303.2262} {\path{arXiv:1303.2262}},
  \href {http://dx.doi.org/10.22323/1.171.0022}
  {\path{doi:10.22323/1.171.0022}}.

\bibitem{Moch:2014tta}
S.~Moch, et~al., {High precision fundamental constants at the TeV scale, }\href
  {http://arxiv.org/abs/1405.4781} {\path{arXiv:1405.4781}}.

\bibitem{dEnterria:2015kmd}
S.~Alekhin, et~al., {High-Precision $\alpha_s$ Measurements from LHC to FCC-ee
  (CERN, Geneva, October 2-13, 2015), }\href {http://arxiv.org/abs/1512.05194}
  {\path{arXiv:1512.05194}}.

\bibitem{BSG:2018xxx}
S.~Bethke, G.~Dissertori, G.~P. Salam, {Quantum Chromodynamics, in: PDG2018},
  Phys. Rev. D 98~(3) (2018) 141--160.

\bibitem{Pich:2018lmu}
A.~Pich, J.~Rojo, R.~Sommer, A.~Vairo, {Determining the strong coupling: status
  and challenges}, PoS Confinement2018 (2018) 035.
\newblock \href {http://arxiv.org/abs/1811.11801} {\path{arXiv:1811.11801}},
  \href {http://dx.doi.org/10.22323/1.336.0035}
  {\path{doi:10.22323/1.336.0035}}.

\bibitem{dEnterria:2019its}
D.~d'Enterria, et~al., {$\alpha_s$(2019): Precision measurements of the QCD
  coupling,} ALPHAS2019 (Trento, 11--15 February, 2019).
\newblock \href {http://arxiv.org/abs/1907.01435} {\path{arXiv:1907.01435}}.

\bibitem{Komijani:2020kst}
J.~Komijani, P.~Petreczky, J.~H. Weber, {Strong coupling constant and quark
  masses from lattice QCD}, Prog. Part. Nucl. Phys. 113 (2020) 103788.
\newblock \href {http://arxiv.org/abs/2003.11703} {\path{arXiv:2003.11703}},
  \href {http://dx.doi.org/10.1016/j.ppnp.2020.103788}
  {\path{doi:10.1016/j.ppnp.2020.103788}}.

\bibitem{Gehrmann-DeRidder:2007nzq}
A.~Gehrmann-De~Ridder, T.~Gehrmann, E.~Glover, G.~Heinrich, {Second-order QCD
  corrections to the thrust distribution}, Phys. Rev. Lett. 99 (2007) 132002.
\newblock \href {http://arxiv.org/abs/0707.1285} {\path{arXiv:0707.1285}},
  \href {http://dx.doi.org/10.1103/PhysRevLett.99.132002}
  {\path{doi:10.1103/PhysRevLett.99.132002}}.

\bibitem{GehrmannDeRidder:2007jk}
A.~Gehrmann-De~Ridder, T.~Gehrmann, E.~Glover, G.~Heinrich, {Infrared structure
  of $e^+ e^-\to 3$ jets at NNLO}, JHEP 11 (2007) 058.
\newblock \href {http://arxiv.org/abs/0710.0346} {\path{arXiv:0710.0346}},
  \href {http://dx.doi.org/10.1088/1126-6708/2007/11/058}
  {\path{doi:10.1088/1126-6708/2007/11/058}}.

\bibitem{GehrmannDeRidder:2007hr}
A.~Gehrmann-De~Ridder, T.~Gehrmann, E.~Glover, G.~Heinrich, {NNLO corrections
  to event shapes in $e^+ e^-$ annihilation}, JHEP 12 (2007) 094.
\newblock \href {http://arxiv.org/abs/0711.4711} {\path{arXiv:0711.4711}},
  \href {http://dx.doi.org/10.1088/1126-6708/2007/12/094}
  {\path{doi:10.1088/1126-6708/2007/12/094}}.

\bibitem{GehrmannDeRidder:2008ug}
A.~Gehrmann-De~Ridder, T.~Gehrmann, E.~Glover, G.~Heinrich, {Jet rates in
  electron-positron annihilation at $O(\alpha_s^3)$ in QCD}, Phys. Rev. Lett.
  100 (2008) 172001.
\newblock \href {http://arxiv.org/abs/0802.0813} {\path{arXiv:0802.0813}},
  \href {http://dx.doi.org/10.1103/PhysRevLett.100.172001}
  {\path{doi:10.1103/PhysRevLett.100.172001}}.

\bibitem{Weinzierl:2008iv}
S.~Weinzierl, {NNLO corrections to 3-jet observables in electron-positron
  annihilation}, Phys. Rev. Lett. 101 (2008) 162001.
\newblock \href {http://arxiv.org/abs/0807.3241} {\path{arXiv:0807.3241}},
  \href {http://dx.doi.org/10.1103/PhysRevLett.101.162001}
  {\path{doi:10.1103/PhysRevLett.101.162001}}.

\bibitem{GehrmannDeRidder:2009dp}
A.~Gehrmann-De~Ridder, T.~Gehrmann, E.~Glover, G.~Heinrich, {NNLO moments of
  event shapes in $e^+e^-$ annihilation}, JHEP 05 (2009) 106.
\newblock \href {http://arxiv.org/abs/0903.4658} {\path{arXiv:0903.4658}},
  \href {http://dx.doi.org/10.1088/1126-6708/2009/05/106}
  {\path{doi:10.1088/1126-6708/2009/05/106}}.

\bibitem{Weinzierl:2009ms}
S.~Weinzierl, {Event shapes and jet rates in electron-positron annihilation at
  NNLO}, JHEP 06 (2009) 041.
\newblock \href {http://arxiv.org/abs/0904.1077} {\path{arXiv:0904.1077}},
  \href {http://dx.doi.org/10.1088/1126-6708/2009/06/041}
  {\path{doi:10.1088/1126-6708/2009/06/041}}.

\bibitem{Weinzierl:2009yz}
S.~Weinzierl, {Moments of event shapes in electron-positron annihilation at
  NNLO}, Phys. Rev. D 80 (2009) 094018.
\newblock \href {http://arxiv.org/abs/0909.5056} {\path{arXiv:0909.5056}},
  \href {http://dx.doi.org/10.1103/PhysRevD.80.094018}
  {\path{doi:10.1103/PhysRevD.80.094018}}.

\bibitem{Weinzierl:2010cw}
S.~Weinzierl, {Jet algorithms in electron-positron annihilation: Perturbative
  higher order predictions}, Eur. Phys. J. C 71 (2011) 1565, [Erratum:
  Eur.Phys.J.C 71, 1717 (2011)].
\newblock \href {http://arxiv.org/abs/1011.6247} {\path{arXiv:1011.6247}},
  \href {http://dx.doi.org/10.1140/epjc/s10052-011-1717-z}
  {\path{doi:10.1140/epjc/s10052-011-1717-z}}.

\bibitem{DelDuca:2016csb}
V.~Del~Duca, C.~Duhr, A.~Kardos, G.~Somogyi, Z.~Tr\'ocs\'anyi, {Three-Jet
  Production in Electron-Positron Collisions at Next-to-Next-to-Leading Order
  Accuracy}, Phys. Rev. Lett. 117~(15) (2016) 152004.
\newblock \href {http://arxiv.org/abs/1603.08927} {\path{arXiv:1603.08927}},
  \href {http://dx.doi.org/10.1103/PhysRevLett.117.152004}
  {\path{doi:10.1103/PhysRevLett.117.152004}}.

\bibitem{DelDuca:2016ily}
V.~Del~Duca, C.~Duhr, A.~Kardos, G.~Somogyi, Z.~Sz\H{o}r, Z.~Tr\'ocs\'anyi,
  Z.~Tulip\'ant, {Jet production in the CoLoRFulNNLO method: event shapes in
  electron-positron collisions}, Phys. Rev. D 94~(7) (2016) 074019.
\newblock \href {http://arxiv.org/abs/1606.03453} {\path{arXiv:1606.03453}},
  \href {http://dx.doi.org/10.1103/PhysRevD.94.074019}
  {\path{doi:10.1103/PhysRevD.94.074019}}.

\bibitem{Catani:1992ua}
S.~Catani, L.~Trentadue, G.~Turnock, B.~Webber, {Resummation of large
  logarithms in $e^+ e^-$ event shape distributions}, Nucl. Phys. B 407 (1993)
  3--42.
\newblock \href {http://dx.doi.org/10.1016/0550-3213(93)90271-P}
  {\path{doi:10.1016/0550-3213(93)90271-P}}.

\bibitem{Gehrmann:2008kh}
T.~Gehrmann, G.~Luisoni, H.~Stenzel, {Matching NLLA + NNLO for event shape
  distributions}, Phys. Lett. B 664 (2008) 265--273.
\newblock \href {http://arxiv.org/abs/0803.0695} {\path{arXiv:0803.0695}},
  \href {http://dx.doi.org/10.1016/j.physletb.2008.05.023}
  {\path{doi:10.1016/j.physletb.2008.05.023}}.

\bibitem{deFlorian:2004mp}
D.~de~Florian, M.~Grazzini, {The Back-to-back region in $e^+ e^-$ energy-energy
  correlation}, Nucl. Phys. B 704 (2005) 387--403.
\newblock \href {http://arxiv.org/abs/hep-ph/0407241}
  {\path{arXiv:hep-ph/0407241}}, \href
  {http://dx.doi.org/10.1016/j.nuclphysb.2004.10.051}
  {\path{doi:10.1016/j.nuclphysb.2004.10.051}}.

\bibitem{Monni:2011gb}
P.~F. Monni, T.~Gehrmann, G.~Luisoni, {Two-Loop Soft Corrections and
  Resummation of the Thrust Distribution in the Dijet Region}, JHEP 08 (2011)
  010.
\newblock \href {http://arxiv.org/abs/1105.4560} {\path{arXiv:1105.4560}},
  \href {http://dx.doi.org/10.1007/JHEP08(2011)010}
  {\path{doi:10.1007/JHEP08(2011)010}}.

\bibitem{Becher:2012qc}
T.~Becher, G.~Bell, {NNLL Resummation for Jet Broadening}, JHEP 11 (2012) 126.
\newblock \href {http://arxiv.org/abs/1210.0580} {\path{arXiv:1210.0580}},
  \href {http://dx.doi.org/10.1007/JHEP11(2012)126}
  {\path{doi:10.1007/JHEP11(2012)126}}.

\bibitem{Banfi:2014sua}
A.~Banfi, H.~McAslan, P.~F. Monni, G.~Zanderighi, {A general method for the
  resummation of event-shape distributions in $e^+ e^-$ annihilation}, JHEP 05
  (2015) 102.
\newblock \href {http://arxiv.org/abs/1412.2126} {\path{arXiv:1412.2126}},
  \href {http://dx.doi.org/10.1007/JHEP05(2015)102}
  {\path{doi:10.1007/JHEP05(2015)102}}.

\bibitem{Kardos:2018kqj}
A.~Kardos, S.~Kluth, G.~Somogyi, Z.~Tulip\'ant, A.~Verbytskyi, {Precise
  determination of $\alpha _{s}(M_Z)$ from a global fit of energy-energy
  correlation to NNLO+NNLL predictions}, Eur. Phys. J. C 78~(6) (2018) 498.
\newblock \href {http://arxiv.org/abs/1804.09146} {\path{arXiv:1804.09146}},
  \href {http://dx.doi.org/10.1140/epjc/s10052-018-5963-1}
  {\path{doi:10.1140/epjc/s10052-018-5963-1}}.

\bibitem{Becher:2008cf}
T.~Becher, M.~D. Schwartz, {A precise determination of $\alpha_s$ from LEP
  thrust data using effective field theory}, JHEP 07 (2008) 034.
\newblock \href {http://arxiv.org/abs/0803.0342} {\path{arXiv:0803.0342}},
  \href {http://dx.doi.org/10.1088/1126-6708/2008/07/034}
  {\path{doi:10.1088/1126-6708/2008/07/034}}.

\bibitem{Hoang:2014wka}
A.~H. Hoang, D.~W. Kolodrubetz, V.~Mateu, I.~W. Stewart, {$C$-parameter
  distribution at N$^3$LL' including power corrections}, Phys. Rev. D 91~(9)
  (2015) 094017.
\newblock \href {http://arxiv.org/abs/1411.6633} {\path{arXiv:1411.6633}},
  \href {http://dx.doi.org/10.1103/PhysRevD.91.094017}
  {\path{doi:10.1103/PhysRevD.91.094017}}.

\bibitem{Hoang:2015hka}
A.~H. Hoang, D.~W. Kolodrubetz, V.~Mateu, I.~W. Stewart, {Precise determination
  of $\alpha_s$ from the $C$-parameter distribution}, Phys. Rev. D 91~(9)
  (2015) 094018.
\newblock \href {http://arxiv.org/abs/1501.04111} {\path{arXiv:1501.04111}},
  \href {http://dx.doi.org/10.1103/PhysRevD.91.094018}
  {\path{doi:10.1103/PhysRevD.91.094018}}.

\bibitem{Chien:2010kc}
Y.-T. Chien, M.~D. Schwartz, {Resummation of heavy jet mass and comparison to
  LEP data}, JHEP 08 (2010) 058.
\newblock \href {http://arxiv.org/abs/1005.1644} {\path{arXiv:1005.1644}},
  \href {http://dx.doi.org/10.1007/JHEP08(2010)058}
  {\path{doi:10.1007/JHEP08(2010)058}}.

\bibitem{Bauer:2000yr}
C.~W. Bauer, S.~Fleming, D.~Pirjol, I.~W. Stewart, {An Effective field theory
  for collinear and soft gluons: Heavy to light decays}, Phys. Rev. D 63 (2001)
  114020.
\newblock \href {http://arxiv.org/abs/hep-ph/0011336}
  {\path{arXiv:hep-ph/0011336}}, \href
  {http://dx.doi.org/10.1103/PhysRevD.63.114020}
  {\path{doi:10.1103/PhysRevD.63.114020}}.

\bibitem{Bauer:2001yt}
C.~W. Bauer, D.~Pirjol, I.~W. Stewart, {Soft collinear factorization in
  effective field theory}, Phys. Rev. D 65 (2002) 054022.
\newblock \href {http://arxiv.org/abs/hep-ph/0109045}
  {\path{arXiv:hep-ph/0109045}}, \href
  {http://dx.doi.org/10.1103/PhysRevD.65.054022}
  {\path{doi:10.1103/PhysRevD.65.054022}}.

\bibitem{Davison:2008vx}
R.~Davison, B.~Webber, {Non-Perturbative Contribution to the Thrust
  Distribution in $e^+ e^-$ Annihilation}, Eur. Phys. J. C 59 (2009) 13--25.
\newblock \href {http://arxiv.org/abs/0809.3326} {\path{arXiv:0809.3326}},
  \href {http://dx.doi.org/10.1140/epjc/s10052-008-0836-7}
  {\path{doi:10.1140/epjc/s10052-008-0836-7}}.

\bibitem{Abbate:2010xh}
R.~Abbate, M.~Fickinger, A.~H. Hoang, V.~Mateu, I.~W. Stewart, {Thrust at
  N${}^3$LL with Power Corrections and a Precision Global Fit for $\alpha_s
  (m_Z)$}, Phys. Rev. D 83 (2011) 074021.
\newblock \href {http://arxiv.org/abs/1006.3080} {\path{arXiv:1006.3080}},
  \href {http://dx.doi.org/10.1103/PhysRevD.83.074021}
  {\path{doi:10.1103/PhysRevD.83.074021}}.

\bibitem{Gehrmann:2012sc}
T.~Gehrmann, G.~Luisoni, P.~F. Monni, {Power corrections in the dispersive
  model for a determination of the strong coupling constant from the thrust
  distribution}, Eur. Phys. J. C 73~(1) (2013) 2265.
\newblock \href {http://arxiv.org/abs/1210.6945} {\path{arXiv:1210.6945}},
  \href {http://dx.doi.org/10.1140/epjc/s10052-012-2265-x}
  {\path{doi:10.1140/epjc/s10052-012-2265-x}}.

\bibitem{Dissertori:2009ik}
G.~Dissertori, A.~Gehrmann-De~Ridder, T.~Gehrmann, E.~Glover, G.~Heinrich,
  G.~Luisoni, H.~Stenzel, {Determination of the strong coupling constant using
  matched NNLO+NLLA predictions for hadronic event shapes in $e^+ e^-$
  annihilations}, JHEP 08 (2009) 036.
\newblock \href {http://arxiv.org/abs/0906.3436} {\path{arXiv:0906.3436}},
  \href {http://dx.doi.org/10.1088/1126-6708/2009/08/036}
  {\path{doi:10.1088/1126-6708/2009/08/036}}.

\bibitem{OPAL:2011aa}
G.~Abbiendi, et~al., {Determination of $\alpha_s$ using OPAL hadronic event
  shapes at $\sqrt{s}=91$ - 209 GeV and resummed NNLO calculations}, Eur. Phys.
  J. C 71 (2011) 1733.
\newblock \href {http://arxiv.org/abs/1101.1470} {\path{arXiv:1101.1470}},
  \href {http://dx.doi.org/10.1140/epjc/s10052-011-1733-z}
  {\path{doi:10.1140/epjc/s10052-011-1733-z}}.

\bibitem{Bethke:2008hf}
S.~Bethke, S.~Kluth, C.~Pahl, J.~Schieck, {Determination of the Strong Coupling
  $\alpha_s$ from hadronic Event Shapes with $O(\alpha_s^3)$ and resummed QCD
  predictions using JADE Data}, Eur. Phys. J. C 64 (2009) 351--360.
\newblock \href {http://arxiv.org/abs/0810.1389} {\path{arXiv:0810.1389}},
  \href {http://dx.doi.org/10.1140/epjc/s10052-009-1149-1}
  {\path{doi:10.1140/epjc/s10052-009-1149-1}}.

\bibitem{Dissertori:2009qa}
G.~Dissertori, A.~Gehrmann-De~Ridder, T.~Gehrmann, E.~Glover, G.~Heinrich,
  H.~Stenzel, {Precise determination of the strong coupling constant at NNLO in
  QCD from the three-jet rate in electron--positron annihilation at LEP}, Phys.
  Rev. Lett. 104 (2010) 072002.
\newblock \href {http://arxiv.org/abs/0910.4283} {\path{arXiv:0910.4283}},
  \href {http://dx.doi.org/10.1103/PhysRevLett.104.072002}
  {\path{doi:10.1103/PhysRevLett.104.072002}}.

\bibitem{Schieck:2012mp}
J.~Schieck, S.~Bethke, S.~Kluth, C.~Pahl, Z.~Trocsanyi, {Measurement of the
  strong coupling $\alpha_s$ from the three-jet rate in $e^+e^-$~ -
  annihilation using JADE data}, Eur. Phys. J. C 73~(3) (2013) 2332.
\newblock \href {http://arxiv.org/abs/1205.3714} {\path{arXiv:1205.3714}},
  \href {http://dx.doi.org/10.1140/epjc/s10052-013-2332-y}
  {\path{doi:10.1140/epjc/s10052-013-2332-y}}.

\bibitem{Verbytskyi:2019zhh}
A.~Verbytskyi, A.~Banfi, A.~Kardos, P.~F. Monni, S.~Kluth, G.~Somogyi,
  Z.~Sz\H{o}r, Z.~Tr\'ocs\'anyi, Z.~Tulip\'ant, G.~Zanderighi, {High precision
  determination of $\alpha_s$ from a global fit of jet rates}, JHEP 08 (2019)
  129.
\newblock \href {http://arxiv.org/abs/1902.08158} {\path{arXiv:1902.08158}},
  \href {http://dx.doi.org/10.1007/JHEP08(2019)129}
  {\path{doi:10.1007/JHEP08(2019)129}}.

\bibitem{Abbate:2012jh}
R.~Abbate, M.~Fickinger, A.~H. Hoang, V.~Mateu, I.~W. Stewart, {Precision
  Thrust Cumulant Moments at $N^3$LL}, Phys. Rev. D 86 (2012) 094002.
\newblock \href {http://arxiv.org/abs/1204.5746} {\path{arXiv:1204.5746}},
  \href {http://dx.doi.org/10.1103/PhysRevD.86.094002}
  {\path{doi:10.1103/PhysRevD.86.094002}}.

\bibitem{Luisoni:2020efy}
G.~Luisoni, P.~F. Monni, G.~P. Salam, {$C$-parameter hadronisation in the
  symmetric 3-jet limit and impact on $\alpha_s$ fit, }\href
  {http://arxiv.org/abs/2012.00622} {\path{arXiv:2012.00622}}.

\bibitem{Czakon:2013goa}
M.~Czakon, P.~Fiedler, A.~Mitov, {Total Top-Quark Pair-Production Cross Section
  at Hadron Colliders Through $O(\alpha^4_S)$}, Phys. Rev. Lett. 110 (2013)
  252004.
\newblock \href {http://arxiv.org/abs/1303.6254} {\path{arXiv:1303.6254}},
  \href {http://dx.doi.org/10.1103/PhysRevLett.110.252004}
  {\path{doi:10.1103/PhysRevLett.110.252004}}.

\bibitem{Czakon:2015owf}
M.~Czakon, D.~Heymes, A.~Mitov, {High-precision differential predictions for
  top-quark pairs at the LHC}, Phys. Rev. Lett. 116~(8) (2016) 082003.
\newblock \href {http://arxiv.org/abs/1511.00549} {\path{arXiv:1511.00549}},
  \href {http://dx.doi.org/10.1103/PhysRevLett.116.082003}
  {\path{doi:10.1103/PhysRevLett.116.082003}}.

\bibitem{Catani:2019hip}
S.~Catani, S.~Devoto, M.~Grazzini, S.~Kallweit, J.~Mazzitelli, {Top-quark pair
  production at the LHC: Fully differential QCD predictions at NNLO}, JHEP 07
  (2019) 100.
\newblock \href {http://arxiv.org/abs/1906.06535} {\path{arXiv:1906.06535}},
  \href {http://dx.doi.org/10.1007/JHEP07(2019)100}
  {\path{doi:10.1007/JHEP07(2019)100}}.

\bibitem{Czakon:2018nun}
M.~Czakon, A.~Ferroglia, D.~Heymes, A.~Mitov, B.~D. Pecjak, D.~J. Scott,
  X.~Wang, L.~L. Yang, {Resummation for (boosted) top-quark pair production at
  NNLO+NNLL' in QCD}, JHEP 05 (2018) 149.
\newblock \href {http://arxiv.org/abs/1803.07623} {\path{arXiv:1803.07623}},
  \href {http://dx.doi.org/10.1007/JHEP05(2018)149}
  {\path{doi:10.1007/JHEP05(2018)149}}.

\bibitem{Currie:2016bfm}
J.~Currie, E.~Glover, J.~Pires, {Next-to-Next-to Leading Order QCD Predictions
  for Single Jet Inclusive Production at the LHC}, Phys. Rev. Lett. 118~(7)
  (2017) 072002.
\newblock \href {http://arxiv.org/abs/1611.01460} {\path{arXiv:1611.01460}},
  \href {http://dx.doi.org/10.1103/PhysRevLett.118.072002}
  {\path{doi:10.1103/PhysRevLett.118.072002}}.

\bibitem{Czakon:2019tmo}
M.~Czakon, A.~van Hameren, A.~Mitov, R.~Poncelet, {Single-jet inclusive rates
  with exact color at $\mathcal{O}(\alpha_s^4)$}, JHEP 10 (2019) 262.
\newblock \href {http://arxiv.org/abs/1907.12911} {\path{arXiv:1907.12911}},
  \href {http://dx.doi.org/10.1007/JHEP10(2019)262}
  {\path{doi:10.1007/JHEP10(2019)262}}.

\bibitem{Currie:2017eqf}
J.~Currie, A.~Gehrmann-De~Ridder, T.~Gehrmann, E.~Glover, A.~Huss, J.~Pires,
  {Precise predictions for dijet production at the LHC}, Phys. Rev. Lett.
  119~(15) (2017) 152001.
\newblock \href {http://arxiv.org/abs/1705.10271} {\path{arXiv:1705.10271}},
  \href {http://dx.doi.org/10.1103/PhysRevLett.119.152001}
  {\path{doi:10.1103/PhysRevLett.119.152001}}.

\bibitem{Boughezal:2015ded}
R.~Boughezal, J.~M. Campbell, R.~Ellis, C.~Focke, W.~T. Giele, X.~Liu,
  F.~Petriello, {Z-boson production in association with a jet at
  next-to-next-to-leading order in perturbative QCD}, Phys. Rev. Lett. 116~(15)
  (2016) 152001.
\newblock \href {http://arxiv.org/abs/1512.01291} {\path{arXiv:1512.01291}},
  \href {http://dx.doi.org/10.1103/PhysRevLett.116.152001}
  {\path{doi:10.1103/PhysRevLett.116.152001}}.

\bibitem{Ridder:2016nkl}
A.~Gehrmann-De~Ridder, T.~Gehrmann, E.~Glover, A.~Huss, T.~Morgan, {The NNLO
  QCD corrections to Z boson production at large transverse momentum}, JHEP 07
  (2016) 133.
\newblock \href {http://arxiv.org/abs/1605.04295} {\path{arXiv:1605.04295}},
  \href {http://dx.doi.org/10.1007/JHEP07(2016)133}
  {\path{doi:10.1007/JHEP07(2016)133}}.

\bibitem{Dittmaier:2012kx}
S.~Dittmaier, A.~Huss, C.~Speckner, {Weak radiative corrections to dijet
  production at hadron colliders}, JHEP 11 (2012) 095.
\newblock \href {http://arxiv.org/abs/1210.0438} {\path{arXiv:1210.0438}},
  \href {http://dx.doi.org/10.1007/JHEP11(2012)095}
  {\path{doi:10.1007/JHEP11(2012)095}}.

\bibitem{Frederix:2016ost}
R.~Frederix, S.~Frixione, V.~Hirschi, D.~Pagani, H.-S. Shao, M.~Zaro, {The
  complete NLO corrections to dijet hadroproduction}, JHEP 04 (2017) 076.
\newblock \href {http://arxiv.org/abs/1612.06548} {\path{arXiv:1612.06548}},
  \href {http://dx.doi.org/10.1007/JHEP04(2017)076}
  {\path{doi:10.1007/JHEP04(2017)076}}.

\bibitem{Czakon:2017wor}
M.~Czakon, D.~Heymes, A.~Mitov, D.~Pagani, I.~Tsinikos, M.~Zaro, {Top-pair
  production at the LHC through NNLO QCD and NLO EW}, JHEP 10 (2017) 186.
\newblock \href {http://arxiv.org/abs/1705.04105} {\path{arXiv:1705.04105}},
  \href {http://dx.doi.org/10.1007/JHEP10(2017)186}
  {\path{doi:10.1007/JHEP10(2017)186}}.

\bibitem{Chatrchyan:2013haa}
S.~Chatrchyan, et~al., {Determination of the Top-Quark Pole Mass and Strong
  Coupling Constant from the $t \bar{t}$ Production Cross Section in $pp$
  Collisions at $\sqrt{s}$ = 7 TeV}, Phys. Lett. B 728 (2014) 496--517,
  [Erratum: Phys.Lett.B 738, 526--528 (2014)].
\newblock \href {http://arxiv.org/abs/1307.1907} {\path{arXiv:1307.1907}},
  \href {http://dx.doi.org/10.1016/j.physletb.2013.12.009}
  {\path{doi:10.1016/j.physletb.2013.12.009}}.

\bibitem{Klijnsma:2017eqp}
T.~Klijnsma, S.~Bethke, G.~Dissertori, G.~P. Salam, {Determination of the
  strong coupling constant $\alpha_s(m_Z)$ from measurements of the total cross
  section for top-antitop quark production}, Eur. Phys. J. C 77~(11) (2017)
  778.
\newblock \href {http://arxiv.org/abs/1708.07495} {\path{arXiv:1708.07495}},
  \href {http://dx.doi.org/10.1140/epjc/s10052-017-5340-5}
  {\path{doi:10.1140/epjc/s10052-017-5340-5}}.

\bibitem{Sirunyan:2018goh}
A.~M. Sirunyan, et~al., {Measurement of the $\mathrm{t}\overline{\mathrm{t}}$
  production cross section, the top quark mass, and the strong coupling
  constant using dilepton events in pp collisions at $\sqrt{s} =$ 13 TeV}, Eur.
  Phys. J. C 79~(5) (2019) 368.
\newblock \href {http://arxiv.org/abs/1812.10505} {\path{arXiv:1812.10505}},
  \href {http://dx.doi.org/10.1140/epjc/s10052-019-6863-8}
  {\path{doi:10.1140/epjc/s10052-019-6863-8}}.

\bibitem{Andreev:2017vxu}
V.~Andreev, et~al., {Determination of the strong coupling constant
  $\alpha_s(m_Z)$ in next-to-next-to-leading order QCD using H1 jet cross
  section measurements}, Eur. Phys. J. C 77~(11) (2017) 791.
\newblock \href {http://arxiv.org/abs/1709.07251} {\path{arXiv:1709.07251}},
  \href {http://dx.doi.org/10.1140/epjc/s10052-017-5314-7}
  {\path{doi:10.1140/epjc/s10052-017-5314-7}}.

\bibitem{Britzger:2019kkb}
D.~Britzger, et~al., {Calculations for deep inelastic scattering using fast
  interpolation grid techniques at NNLO in QCD and the extraction of $\alpha_s$
  from HERA data}, Eur. Phys. J. C 79~(10) (2019) 845.
\newblock \href {http://arxiv.org/abs/1906.05303} {\path{arXiv:1906.05303}},
  \href {http://dx.doi.org/10.1140/epjc/s10052-019-7351-x}
  {\path{doi:10.1140/epjc/s10052-019-7351-x}}.

\bibitem{Khachatryan:2016mlc}
V.~Khachatryan, et~al., {Measurement and QCD analysis of double-differential
  inclusive jet cross sections in pp collisions at $ \sqrt{s}=8 $ TeV and cross
  section ratios to 2.76 and 7 TeV}, JHEP 03 (2017) 156.
\newblock \href {http://arxiv.org/abs/1609.05331} {\path{arXiv:1609.05331}},
  \href {http://dx.doi.org/10.1007/JHEP03(2017)156}
  {\path{doi:10.1007/JHEP03(2017)156}}.

\bibitem{ATLAS:2020mee}
{The ATLAS Collaboration}, {Determination of the strong coupling constant and
  test of asymptotic freedom from Transverse Energy-Energy Correlations in
  multijet events at $\sqrt{s} = 13$ TeV with the ATLAS detector.
  ATLAS-CONF-2020-025}.

\bibitem{Cooper-Sarkar:2020twv}
A.~M. Cooper-Sarkar, M.~Czakon, M.~A. Lim, A.~Mitov, A.~S. Papanastasiou,
  {Simultaneous extraction of $\alpha_s$ and $m_t$ from LHC $t\bar{t}$
  differential distributions, }\href {http://arxiv.org/abs/2010.04171}
  {\path{arXiv:2010.04171}}.

\bibitem{Blumlein:2006be}
{Blümlein, Johannes and B\"{o}ttcher, Helmut and Guffanti, Alberto},
  {Non-singlet QCD analysis of deep inelastic world data at $O(\alpha_s^3)$},
  Nucl. Phys. B 774 (2007) 182--207.
\newblock \href {http://arxiv.org/abs/hep-ph/0607200}
  {\path{arXiv:hep-ph/0607200}}, \href
  {http://dx.doi.org/10.1016/j.nuclphysb.2007.03.035}
  {\path{doi:10.1016/j.nuclphysb.2007.03.035}}.

\bibitem{Blumlein:2012se}
{Bl\"umlein, Johannes and B\"ottcher, Helmut}, {Higher Twist contributions to
  the Structure Functions $F_2(x,Q^2)$ and $g_2(x,Q^2)$}, in: {20th
  International Workshop on Deep-Inelastic Scattering and Related Subjects},
  2012, pp. 237--241.
\newblock \href {http://arxiv.org/abs/1207.3170} {\path{arXiv:1207.3170}},
  \href {http://dx.doi.org/10.3204/DESY-PROC-2012-02/252}
  {\path{doi:10.3204/DESY-PROC-2012-02/252}}.

\bibitem{Jimenez-Delgado:2014twa}
P.~Jimenez-Delgado, E.~Reya, {Delineating parton distributions and the strong
  coupling}, Phys. Rev. D 89~(7) (2014) 074049.
\newblock \href {http://arxiv.org/abs/1403.1852} {\path{arXiv:1403.1852}},
  \href {http://dx.doi.org/10.1103/PhysRevD.89.074049}
  {\path{doi:10.1103/PhysRevD.89.074049}}.

\bibitem{Alekhin:2017kpj}
S.~Alekhin, J.~Bl\"umlein, S.~Moch, R.~Placakyte, {Parton distribution
  functions, $\alpha_s$, and heavy-quark masses for LHC Run II}, Phys. Rev. D
  96~(1) (2017) 014011.
\newblock \href {http://arxiv.org/abs/1701.05838} {\path{arXiv:1701.05838}},
  \href {http://dx.doi.org/10.1103/PhysRevD.96.014011}
  {\path{doi:10.1103/PhysRevD.96.014011}}.

\bibitem{Alekhin:2018pai}
S.~Alekhin, J.~Bl\"umlein, S.~Moch, {NLO PDFs from the ABMP16 fit}, Eur. Phys.
  J. C 78~(6) (2018) 477.
\newblock \href {http://arxiv.org/abs/1803.07537} {\path{arXiv:1803.07537}},
  \href {http://dx.doi.org/10.1140/epjc/s10052-018-5947-1}
  {\path{doi:10.1140/epjc/s10052-018-5947-1}}.

\bibitem{Harland-Lang:2015nxa}
L.~Harland-Lang, A.~Martin, P.~Motylinski, R.~Thorne, {Uncertainties on $\alpha
  _s$ in the MMHT2014 global PDF analysis and implications for SM predictions},
  Eur. Phys. J. C 75~(9) (2015) 435.
\newblock \href {http://arxiv.org/abs/1506.05682} {\path{arXiv:1506.05682}},
  \href {http://dx.doi.org/10.1140/epjc/s10052-015-3630-3}
  {\path{doi:10.1140/epjc/s10052-015-3630-3}}.

\bibitem{Dulat:2015mca}
S.~Dulat, T.-J. Hou, J.~Gao, M.~Guzzi, J.~Huston, P.~Nadolsky, J.~Pumplin,
  C.~Schmidt, D.~Stump, C.~Yuan, {New parton distribution functions from a
  global analysis of quantum chromodynamics}, Phys. Rev. D 93~(3) (2016)
  033006.
\newblock \href {http://arxiv.org/abs/1506.07443} {\path{arXiv:1506.07443}},
  \href {http://dx.doi.org/10.1103/PhysRevD.93.033006}
  {\path{doi:10.1103/PhysRevD.93.033006}}.

\bibitem{Ball:2018iqk}
R.~D. Ball, S.~Carrazza, L.~Del~Debbio, S.~Forte, Z.~Kassabov, J.~Rojo,
  E.~Slade, M.~Ubiali, {Precision determination of the strong coupling constant
  within a global PDF analysis}, Eur. Phys. J. C 78~(5) (2018) 408.
\newblock \href {http://arxiv.org/abs/1802.03398} {\path{arXiv:1802.03398}},
  \href {http://dx.doi.org/10.1140/epjc/s10052-018-5897-7}
  {\path{doi:10.1140/epjc/s10052-018-5897-7}}.

\bibitem{Thorne:2011kq}
R.~Thorne, G.~Watt, {PDF dependence of Higgs cross sections at the Tevatron and
  LHC: Response to recent criticism}, JHEP 08 (2011) 100.
\newblock \href {http://arxiv.org/abs/1106.5789} {\path{arXiv:1106.5789}},
  \href {http://dx.doi.org/10.1007/JHEP08(2011)100}
  {\path{doi:10.1007/JHEP08(2011)100}}.

\bibitem{Ball:2013gsa}
R.~D. Ball, V.~Bertone, L.~Del~Debbio, S.~Forte, A.~Guffanti, J.~Rojo,
  M.~Ubiali, {Theoretical issues in PDF determination and associated
  uncertainties}, Phys. Lett. B 723 (2013) 330--339.
\newblock \href {http://arxiv.org/abs/1303.1189} {\path{arXiv:1303.1189}},
  \href {http://dx.doi.org/10.1016/j.physletb.2013.05.019}
  {\path{doi:10.1016/j.physletb.2013.05.019}}.

\bibitem{Thorne:2014toa}
R.~Thorne, {The effect on PDFs and $\alpha _S(M_Z^2)$ due to changes in flavour
  scheme and higher twist contributions}, Eur. Phys. J. C 74~(7) (2014) 2958.
\newblock \href {http://arxiv.org/abs/1402.3536} {\path{arXiv:1402.3536}},
  \href {http://dx.doi.org/10.1140/epjc/s10052-014-2958-4}
  {\path{doi:10.1140/epjc/s10052-014-2958-4}}.

\bibitem{Alekhin:2013nda}
S.~Alekhin, J.~Blümlein, S.~Moch, {The ABM parton distributions tuned to LHC
  data}, Phys. Rev. D 89~(5) (2014) 054028.
\newblock \href {http://arxiv.org/abs/1310.3059} {\path{arXiv:1310.3059}},
  \href {http://dx.doi.org/10.1103/PhysRevD.89.054028}
  {\path{doi:10.1103/PhysRevD.89.054028}}.

\bibitem{Parente:1994bf}
G.~Parente, A.~Kotikov, V.~Krivokhizhin, {Next to next-to-leading order QCD
  analysis of DIS structure functions}, Phys. Lett. B 333 (1994) 190--195.
\newblock \href {http://arxiv.org/abs/hep-ph/9405290}
  {\path{arXiv:hep-ph/9405290}}, \href
  {http://dx.doi.org/10.1016/0370-2693(94)91028-6}
  {\path{doi:10.1016/0370-2693(94)91028-6}}.

\bibitem{Shaikhatdenov:2009xd}
B.~Shaikhatdenov, A.~Kotikov, V.~Krivokhizhin, G.~Parente, {QCD coupling
  constant at NNLO from DIS data}, Phys. Rev. D 81 (2010) 034008, [Erratum:
  Phys.Rev.D 81, 079904 (2010)].
\newblock \href {http://arxiv.org/abs/0912.4672} {\path{arXiv:0912.4672}},
  \href {http://dx.doi.org/10.1103/PhysRevD.81.079904}
  {\path{doi:10.1103/PhysRevD.81.079904}}.

\bibitem{Martin:2009bu}
A.~Martin, W.~Stirling, R.~Thorne, G.~Watt, {Uncertainties on $\alpha_s$ in
  global PDF analyses and implications for predicted hadronic cross sections},
  Eur. Phys. J. C 64 (2009) 653--680.
\newblock \href {http://arxiv.org/abs/0905.3531} {\path{arXiv:0905.3531}},
  \href {http://dx.doi.org/10.1140/epjc/s10052-009-1164-2}
  {\path{doi:10.1140/epjc/s10052-009-1164-2}}.

\bibitem{Alekhin:2012ig}
S.~Alekhin, J.~Blümlein, S.~Moch, {Parton Distribution Functions and Benchmark
  Cross Sections at NNLO}, Phys. Rev. D 86 (2012) 054009.
\newblock \href {http://arxiv.org/abs/1202.2281} {\path{arXiv:1202.2281}},
  \href {http://dx.doi.org/10.1103/PhysRevD.86.054009}
  {\path{doi:10.1103/PhysRevD.86.054009}}.

\bibitem{Caswell:1985ui}
W.~Caswell, G.~Lepage, {Effective Lagrangians for Bound State Problems in QED,
  QCD, and Other Field Theories}, Phys. Lett. B 167 (1986) 437--442.
\newblock \href {http://dx.doi.org/10.1016/0370-2693(86)91297-9}
  {\path{doi:10.1016/0370-2693(86)91297-9}}.

\bibitem{Bodwin:1994jh}
G.~T. Bodwin, E.~Braaten, G.~Lepage, {Rigorous QCD analysis of inclusive
  annihilation and production of heavy quarkonium}, Phys. Rev. D 51 (1995)
  1125--1171, [Erratum: Phys.Rev.D 55, 5853 (1997)].
\newblock \href {http://arxiv.org/abs/hep-ph/9407339}
  {\path{arXiv:hep-ph/9407339}}, \href
  {http://dx.doi.org/10.1103/PhysRevD.55.5853}
  {\path{doi:10.1103/PhysRevD.55.5853}}.

\bibitem{Pineda:1997bj}
A.~Pineda, J.~Soto, {Effective field theory for ultrasoft momenta in NRQCD and
  NRQED}, Nucl. Phys. B Proc. Suppl. 64 (1998) 428--432.
\newblock \href {http://arxiv.org/abs/hep-ph/9707481}
  {\path{arXiv:hep-ph/9707481}}, \href
  {http://dx.doi.org/10.1016/S0920-5632(97)01102-X}
  {\path{doi:10.1016/S0920-5632(97)01102-X}}.

\bibitem{Brambilla:1999xf}
N.~Brambilla, A.~Pineda, J.~Soto, A.~Vairo, {Potential NRQCD: An Effective
  theory for heavy quarkonium}, Nucl. Phys. B 566 (2000) 275.
\newblock \href {http://arxiv.org/abs/hep-ph/9907240}
  {\path{arXiv:hep-ph/9907240}}, \href
  {http://dx.doi.org/10.1016/S0550-3213(99)00693-8}
  {\path{doi:10.1016/S0550-3213(99)00693-8}}.

\bibitem{Luke:1999kz}
M.~E. Luke, A.~V. Manohar, I.~Z. Rothstein, {Renormalization group scaling in
  nonrelativistic QCD}, Phys. Rev. D 61 (2000) 074025.
\newblock \href {http://arxiv.org/abs/hep-ph/9910209}
  {\path{arXiv:hep-ph/9910209}}, \href
  {http://dx.doi.org/10.1103/PhysRevD.61.074025}
  {\path{doi:10.1103/PhysRevD.61.074025}}.

\bibitem{Brambilla:2007cz}
N.~Brambilla, X.~Garcia~i Tormo, J.~Soto, A.~Vairo, {Extraction of $\alpha_s$
  from radiative $\Upsilon (1S)$ decays}, Phys. Rev. D 75 (2007) 074014.
\newblock \href {http://arxiv.org/abs/hep-ph/0702079}
  {\path{arXiv:hep-ph/0702079}}, \href
  {http://dx.doi.org/10.1103/PhysRevD.75.074014}
  {\path{doi:10.1103/PhysRevD.75.074014}}.

\bibitem{Kiyo:2013aea}
Y.~Kiyo, Y.~Sumino, {Perturbative heavy quarkonium spectrum at
  next-to-next-to-next-to-leading order}, Phys. Lett. B 730 (2014) 76--80.
\newblock \href {http://arxiv.org/abs/1309.6571} {\path{arXiv:1309.6571}},
  \href {http://dx.doi.org/10.1016/j.physletb.2014.01.030}
  {\path{doi:10.1016/j.physletb.2014.01.030}}.

\bibitem{Kiyo:2014uca}
Y.~Kiyo, Y.~Sumino, {Full Formula for Heavy Quarkonium Energy Levels at
  Next-to-next-to-next-to-leading Order}, Nucl. Phys. B 889 (2014) 156--191.
\newblock \href {http://arxiv.org/abs/1408.5590} {\path{arXiv:1408.5590}},
  \href {http://dx.doi.org/10.1016/j.nuclphysb.2014.10.010}
  {\path{doi:10.1016/j.nuclphysb.2014.10.010}}.

\bibitem{Mateu:2017hlz}
V.~Mateu, P.~G. Ortega, {Bottom and Charm Mass determinations from global fits
  to $Q\bar{Q}$ bound states at N$^3$LO}, JHEP 01 (2018) 122.
\newblock \href {http://arxiv.org/abs/1711.05755} {\path{arXiv:1711.05755}},
  \href {http://dx.doi.org/10.1007/JHEP01(2018)122}
  {\path{doi:10.1007/JHEP01(2018)122}}.

\bibitem{Peset:2018ria}
C.~Peset, A.~Pineda, J.~Segovia, {The charm/bottom quark mass from heavy
  quarkonium at N$^{3}$LO}, JHEP 09 (2018) 167.
\newblock \href {http://arxiv.org/abs/1806.05197} {\path{arXiv:1806.05197}},
  \href {http://dx.doi.org/10.1007/JHEP09(2018)167}
  {\path{doi:10.1007/JHEP09(2018)167}}.

\bibitem{Boito:2019pqp}
D.~Boito, V.~Mateu, {Precise $\alpha_s$ determination from charmonium sum
  rules}, Phys. Lett. B 806 (2020) 135482.
\newblock \href {http://arxiv.org/abs/1912.06237} {\path{arXiv:1912.06237}},
  \href {http://dx.doi.org/10.1016/j.physletb.2020.135482}
  {\path{doi:10.1016/j.physletb.2020.135482}}.

\bibitem{Boito:2020lyp}
D.~Boito, V.~Mateu, {Precise determination of $\alpha_s$ from relativistic
  quarkonium sum rules}, JHEP 03 (2020) 094.
\newblock \href {http://arxiv.org/abs/2001.11041} {\path{arXiv:2001.11041}},
  \href {http://dx.doi.org/10.1007/JHEP03(2020)094}
  {\path{doi:10.1007/JHEP03(2020)094}}.

\bibitem{Narison:2018dcr}
S.~Narison, {QCD parameter correlations from heavy quarkonia}, Int. J. Mod.
  Phys. A 33~(10) (2018) 1850045, [Addendum: Int.J.Mod.Phys.A 33, 1850045
  (2018)].
\newblock \href {http://arxiv.org/abs/1801.00592} {\path{arXiv:1801.00592}},
  \href {http://dx.doi.org/10.1142/S0217751X18500458}
  {\path{doi:10.1142/S0217751X18500458}}.

\bibitem{Zafeiropoulos:2019flq}
S.~Zafeiropoulos, P.~Boucaud, F.~De~Soto, J.~Rodr\'\i{}guez-Quintero,
  J.~Segovia, {Strong Running Coupling from the Gauge Sector of Domain Wall
  Lattice QCD with Physical Quark Masses}, Phys. Rev. Lett. 122~(16) (2019)
  162002.
\newblock \href {http://arxiv.org/abs/1902.08148} {\path{arXiv:1902.08148}},
  \href {http://dx.doi.org/10.1103/PhysRevLett.122.162002}
  {\path{doi:10.1103/PhysRevLett.122.162002}}.

\bibitem{Bruno:2017gxd}
M.~Bruno, M.~Dalla~Brida, P.~Fritzsch, T.~Korzec, A.~Ramos, S.~Schaefer,
  H.~Simma, S.~Sint, R.~Sommer, {QCD Coupling from a Nonperturbative
  Determination of the Three-Flavor $\Lambda$ Parameter}, Phys. Rev. Lett.
  119~(10) (2017) 102001.
\newblock \href {http://arxiv.org/abs/1706.03821} {\path{arXiv:1706.03821}},
  \href {http://dx.doi.org/10.1103/PhysRevLett.119.102001}
  {\path{doi:10.1103/PhysRevLett.119.102001}}.

\bibitem{Aoki:2009tf}
S.~Aoki, et~al., {Precise determination of the strong coupling constant in $N_f
  = 2+1$ lattice QCD with the Schrodinger functional scheme}, JHEP 10 (2009)
  053.
\newblock \href {http://arxiv.org/abs/0906.3906} {\path{arXiv:0906.3906}},
  \href {http://dx.doi.org/10.1088/1126-6708/2009/10/053}
  {\path{doi:10.1088/1126-6708/2009/10/053}}.

\bibitem{Luscher:1991wu}
M.~Luscher, P.~Weisz, U.~Wolff, {A Numerical method to compute the running
  coupling in asymptotically free theories}, Nucl. Phys. B 359 (1991) 221--243.
\newblock \href {http://dx.doi.org/10.1016/0550-3213(91)90298-C}
  {\path{doi:10.1016/0550-3213(91)90298-C}}.

\bibitem{Bazavov:2014soa}
A.~Bazavov, N.~Brambilla, I.~Tormo, Xavier~Garcia, P.~Petreczky, J.~Soto,
  A.~Vairo, {Determination of $\alpha_s$ from the QCD static energy: An
  update}, Phys. Rev. D 90~(7) (2014) 074038, [Erratum: Phys.Rev.D 101, 119902
  (2020)].
\newblock \href {http://arxiv.org/abs/1407.8437} {\path{arXiv:1407.8437}},
  \href {http://dx.doi.org/10.1103/PhysRevD.90.074038}
  {\path{doi:10.1103/PhysRevD.90.074038}}.

\bibitem{Bazavov:2019qoo}
A.~Bazavov, N.~Brambilla, X.~Garcia~i Tormo, P.~Petreczky, J.~Soto, A.~Vairo,
  J.~H. Weber, {Determination of the QCD coupling from the static energy and
  the free energy}, Phys. Rev. D 100~(11) (2019) 114511.
\newblock \href {http://arxiv.org/abs/1907.11747} {\path{arXiv:1907.11747}},
  \href {http://dx.doi.org/10.1103/PhysRevD.100.114511}
  {\path{doi:10.1103/PhysRevD.100.114511}}.

\bibitem{Ayala:2020odx}
C.~Ayala, X.~Lobregat, A.~Pineda, {Determination of $\alpha(M_z)$ from an
  hyperasymptotic approximation to the energy of a static quark-antiquark
  pair}, JHEP 09 (2020) 016.
\newblock \href {http://arxiv.org/abs/2005.12301} {\path{arXiv:2005.12301}},
  \href {http://dx.doi.org/10.1007/JHEP09(2020)016}
  {\path{doi:10.1007/JHEP09(2020)016}}.

\bibitem{McNeile:2010ji}
C.~McNeile, C.~Davies, E.~Follana, K.~Hornbostel, G.~Lepage, {High-Precision c
  and b Masses, and QCD Coupling from Current-Current Correlators in Lattice
  and Continuum QCD}, Phys. Rev. D 82 (2010) 034512.
\newblock \href {http://arxiv.org/abs/1004.4285} {\path{arXiv:1004.4285}},
  \href {http://dx.doi.org/10.1103/PhysRevD.82.034512}
  {\path{doi:10.1103/PhysRevD.82.034512}}.

\bibitem{Mason:2005zx}
Q.~Mason, H.~Trottier, C.~Davies, K.~Foley, A.~Gray, G.~Lepage, M.~Nobes,
  J.~Shigemitsu, {Accurate determinations of $\alpha_s$ from realistic lattice
  QCD}, Phys. Rev. Lett. 95 (2005) 052002.
\newblock \href {http://arxiv.org/abs/hep-lat/0503005}
  {\path{arXiv:hep-lat/0503005}}, \href
  {http://dx.doi.org/10.1103/PhysRevLett.95.052002}
  {\path{doi:10.1103/PhysRevLett.95.052002}}.

\bibitem{Maltman:2008bx}
K.~Maltman, D.~Leinweber, P.~Moran, A.~Sternbeck, {The Realistic Lattice
  Determination of $\alpha_s(M_Z)$ Revisited}, Phys. Rev. D 78 (2008) 114504.
\newblock \href {http://arxiv.org/abs/0807.2020} {\path{arXiv:0807.2020}},
  \href {http://dx.doi.org/10.1103/PhysRevD.78.114504}
  {\path{doi:10.1103/PhysRevD.78.114504}}.

\bibitem{Chakraborty:2014aca}
B.~Chakraborty, C.~Davies, B.~Galloway, P.~Knecht, J.~Koponen, G.~C. Donald,
  R.~J. Dowdall, G.~P. Lepage, C.~McNeile, {High-precision quark masses and QCD
  coupling from $n_f=4$ lattice QCD}, Phys. Rev. D 91~(5) (2015) 054508.
\newblock \href {http://arxiv.org/abs/1408.4169} {\path{arXiv:1408.4169}},
  \href {http://dx.doi.org/10.1103/PhysRevD.91.054508}
  {\path{doi:10.1103/PhysRevD.91.054508}}.

\bibitem{Nakayama:2016atf}
K.~Nakayama, B.~Fahy, S.~Hashimoto, {Short-distance charmonium correlator on
  the lattice with M\"obius domain-wall fermion and a determination of charm
  quark mass}, Phys. Rev. D 94~(5) (2016) 054507.
\newblock \href {http://arxiv.org/abs/1606.01002} {\path{arXiv:1606.01002}},
  \href {http://dx.doi.org/10.1103/PhysRevD.94.054507}
  {\path{doi:10.1103/PhysRevD.94.054507}}.

\bibitem{Schmelling:1994pz}
M.~Schmelling, {Averaging correlated data}, Phys. Scripta 51 (1995) 676--679.
\newblock \href {http://dx.doi.org/10.1088/0031-8949/51/6/002}
  {\path{doi:10.1088/0031-8949/51/6/002}}.

\bibitem{Gray:1990yh}
N.~Gray, D.~J. Broadhurst, W.~Grafe, K.~Schilcher, {Three Loop Relation of
  Quark $\overline{\mathrm{MS}}$ and Pole Masses}, Z. Phys. C 48 (1990)
  673--680.
\newblock \href {http://dx.doi.org/10.1007/BF01614703}
  {\path{doi:10.1007/BF01614703}}.

\bibitem{Fleischer:1998dw}
J.~Fleischer, F.~Jegerlehner, O.~Tarasov, O.~Veretin, {Two loop QCD corrections
  of the massive fermion propagator}, Nucl. Phys. B 539 (1999) 671--690,
  [Erratum: Nucl.Phys.B 571, 511--512 (2000)].
\newblock \href {http://arxiv.org/abs/hep-ph/9803493}
  {\path{arXiv:hep-ph/9803493}}, \href
  {http://dx.doi.org/10.1016/S0550-3213(98)00705-6}
  {\path{doi:10.1016/S0550-3213(98)00705-6}}.

\bibitem{Chetyrkin:1999ys}
K.~Chetyrkin, M.~Steinhauser, {Short distance mass of a heavy quark at order
  $\alpha_s^3$}, Phys. Rev. Lett. 83 (1999) 4001--4004.
\newblock \href {http://arxiv.org/abs/hep-ph/9907509}
  {\path{arXiv:hep-ph/9907509}}, \href
  {http://dx.doi.org/10.1103/PhysRevLett.83.4001}
  {\path{doi:10.1103/PhysRevLett.83.4001}}.

\bibitem{Chetyrkin:1999qi}
K.~Chetyrkin, M.~Steinhauser, {The relation between the
  $\overline{\mathrm{MS}}$ and the on-shell quark mass at order $\alpha_s^3$},
  Nucl. Phys. B 573 (2000) 617--651.
\newblock \href {http://arxiv.org/abs/hep-ph/9911434}
  {\path{arXiv:hep-ph/9911434}}, \href
  {http://dx.doi.org/10.1016/S0550-3213(99)00784-1}
  {\path{doi:10.1016/S0550-3213(99)00784-1}}.

\bibitem{Melnikov:2000qh}
K.~Melnikov, T.~v. Ritbergen, {The Three loop relation between the
  $\overline{\mathrm{MS}}$ and the pole quark masses}, Phys. Lett. B 482 (2000)
  99--108.
\newblock \href {http://arxiv.org/abs/hep-ph/9912391}
  {\path{arXiv:hep-ph/9912391}}, \href
  {http://dx.doi.org/10.1016/S0370-2693(00)00507-4}
  {\path{doi:10.1016/S0370-2693(00)00507-4}}.

\bibitem{Marquard:2015qpa}
P.~Marquard, A.~V. Smirnov, V.~A. Smirnov, M.~Steinhauser, {Quark Mass
  Relations to Four-Loop Order in Perturbative QCD}, Phys. Rev. Lett. 114~(14)
  (2015) 142002.
\newblock \href {http://arxiv.org/abs/1502.01030} {\path{arXiv:1502.01030}},
  \href {http://dx.doi.org/10.1103/PhysRevLett.114.142002}
  {\path{doi:10.1103/PhysRevLett.114.142002}}.

\bibitem{Marquard:2016dcn}
P.~Marquard, A.~V. Smirnov, V.~A. Smirnov, M.~Steinhauser, D.~Wellmann,
  {$\overline{\rm MS}$-on-shell quark mass relation up to four loops in QCD and
  a general SU$(N)$ gauge group}, Phys. Rev. D 94~(7) (2016) 074025.
\newblock \href {http://arxiv.org/abs/1606.06754} {\path{arXiv:1606.06754}},
  \href {http://dx.doi.org/10.1103/PhysRevD.94.074025}
  {\path{doi:10.1103/PhysRevD.94.074025}}.

\bibitem{Bernreuther:1981sg}
W.~Bernreuther, W.~Wetzel, {Decoupling of Heavy Quarks in the Minimal
  Subtraction Scheme}, Nucl. Phys. B 197 (1982) 228--236, [Erratum: Nucl.Phys.B
  513, 758--758 (1998)].
\newblock \href {http://dx.doi.org/10.1016/0550-3213(82)90288-7}
  {\path{doi:10.1016/0550-3213(82)90288-7}}.

\bibitem{Wetzel:1981qg}
W.~Wetzel, {Minimal Subtraction and the Decoupling of Heavy Quarks for
  Arbitrary Values of the Gauge Parameter}, Nucl. Phys. B 196 (1982) 259--272.
\newblock \href {http://dx.doi.org/10.1016/0550-3213(82)90038-4}
  {\path{doi:10.1016/0550-3213(82)90038-4}}.

\bibitem{Bernreuther:1983zp}
W.~Bernreuther, {Decoupling of Heavy Quarks in Quantum Chromodynamics}, Annals
  Phys. 151 (1983) 127.
\newblock \href {http://dx.doi.org/10.1016/0003-4916(83)90317-2}
  {\path{doi:10.1016/0003-4916(83)90317-2}}.

\bibitem{Chetyrkin:1997sg}
K.~Chetyrkin, B.~A. Kniehl, M.~Steinhauser, {Strong coupling constant with
  flavor thresholds at four loops in the modified minimal-subtraction scheme},
  Phys. Rev. Lett. 79 (1997) 2184--2187.
\newblock \href {http://arxiv.org/abs/hep-ph/9706430}
  {\path{arXiv:hep-ph/9706430}}, \href
  {http://dx.doi.org/10.1103/PhysRevLett.79.2184}
  {\path{doi:10.1103/PhysRevLett.79.2184}}.

\bibitem{Chetyrkin:2000yt}
K.~Chetyrkin, J.~H. Kühn, M.~Steinhauser, {RunDec: A Mathematica package for
  running and decoupling of the strong coupling and quark masses}, Comput.
  Phys. Commun. 133 (2000) 43--65.
\newblock \href {http://arxiv.org/abs/hep-ph/0004189}
  {\path{arXiv:hep-ph/0004189}}, \href
  {http://dx.doi.org/10.1016/S0010-4655(00)00155-7}
  {\path{doi:10.1016/S0010-4655(00)00155-7}}.

\bibitem{Schmidt:2012az}
B.~Schmidt, M.~Steinhauser, {CRunDec: a C++ package for running and decoupling
  of the strong coupling and quark masses}, Comput. Phys. Commun. 183 (2012)
  1845--1848.
\newblock \href {http://arxiv.org/abs/1201.6149} {\path{arXiv:1201.6149}},
  \href {http://dx.doi.org/10.1016/j.cpc.2012.03.023}
  {\path{doi:10.1016/j.cpc.2012.03.023}}.

\bibitem{Herren:2017osy}
F.~Herren, M.~Steinhauser, {Version 3 of RunDec and CRunDec}, Comput. Phys.
  Commun. 224 (2018) 333--345.
\newblock \href {http://arxiv.org/abs/1703.03751} {\path{arXiv:1703.03751}},
  \href {http://dx.doi.org/10.1016/j.cpc.2017.11.014}
  {\path{doi:10.1016/j.cpc.2017.11.014}}.

\bibitem{Chetyrkin:1993hi}
K.~Chetyrkin, A.~Kwiatkowski, {Mass corrections to the tau decay rate}, Z.
  Phys. C 59 (1993) 525--532.
\newblock \href {http://arxiv.org/abs/hep-ph/9805232}
  {\path{arXiv:hep-ph/9805232}}, \href {http://dx.doi.org/10.1007/BF01498634}
  {\path{doi:10.1007/BF01498634}}.

\bibitem{Chetyrkin:1996hm}
K.~Chetyrkin, J.~H. Kühn, {Quadratic mass corrections of order $O (\alpha_s^3
  m_q^2/s)$ to the decay rate of Z and W bosons}, Phys. Lett. B 406 (1997)
  102--109.
\newblock \href {http://arxiv.org/abs/hep-ph/9609202}
  {\path{arXiv:hep-ph/9609202}}, \href
  {http://dx.doi.org/10.1016/S0370-2693(97)00631-X}
  {\path{doi:10.1016/S0370-2693(97)00631-X}}.

\bibitem{Gorishnii:1986pz}
S.~Gorishnii, A.~Kataev, S.~Larin, {Three Loop Corrections of Order $O (m^2)$
  to the Correlator of Electromagnetic Quark Currents}, Nuovo Cim. A 92 (1986)
  119--131.
\newblock \href {http://dx.doi.org/10.1007/BF02727185}
  {\path{doi:10.1007/BF02727185}}.

\bibitem{Generalis:1989hf}
S.~Generalis, {Improved two loop quark mass corrections}, J. Phys. G 15 (1989)
  L225--L229.
\newblock \href {http://dx.doi.org/10.1088/0954-3899/15/11/001}
  {\path{doi:10.1088/0954-3899/15/11/001}}.

\bibitem{Bernreuther:1981sp}
W.~Bernreuther, W.~Wetzel, {Order $\alpha_s^2$ Massive Quark Contribution to
  the Vacuum Polarization of Massless Quarks}, Z. Phys. C 11 (1981) 113.
\newblock \href {http://dx.doi.org/10.1007/BF01573992}
  {\path{doi:10.1007/BF01573992}}.

\bibitem{Gorishnii:1990zu}
S.~Gorishnii, A.~Kataev, S.~Larin, L.~Surguladze, {Corrected Three Loop {QCD}
  Correction to the Correlator of the Quark Scalar Currents and
  $\Gamma_{\mathrm{Tot}}(H^0 \to\mathrm{Hadrons})$}, Mod. Phys. Lett. A 5
  (1990) 2703--2712.
\newblock \href {http://dx.doi.org/10.1142/S0217732390003152}
  {\path{doi:10.1142/S0217732390003152}}.

\bibitem{Gorishnii:1991zr}
S.~Gorishnii, A.~Kataev, S.~Larin, L.~Surguladze, {Scheme dependence of the
  next to next-to-leading QCD corrections to $\Gamma_{\mathrm{tot}}(H^0
  \to\mathrm{hadrons})$ and the spurious QCD infrared fixed point}, Phys. Rev.
  D 43 (1991) 1633--1640.
\newblock \href {http://dx.doi.org/10.1103/PhysRevD.43.1633}
  {\path{doi:10.1103/PhysRevD.43.1633}}.

\bibitem{Chetyrkin:1985kn}
K.~Chetyrkin, V.~Spiridonov, S.~Gorishnii, {Wilson expansion for correlators of
  vector currents at the two loop level: dimension four operators}, Phys. Lett.
  B 160 (1985) 149--153.
\newblock \href {http://dx.doi.org/10.1016/0370-2693(85)91482-0}
  {\path{doi:10.1016/0370-2693(85)91482-0}}.

\bibitem{Generalis:1990id}
S.~Generalis, {QCD sum rules. 1: Perturbative results for current correlators},
  J. Phys. G 16 (1990) 785--793.
\newblock \href {http://dx.doi.org/10.1088/0954-3899/16/6/002}
  {\path{doi:10.1088/0954-3899/16/6/002}}.

\bibitem{Jamin:1992se}
M.~Jamin, M.~Munz, {Current correlators to all orders in the quark masses}, Z.
  Phys. C 60 (1993) 569--578.
\newblock \href {http://arxiv.org/abs/hep-ph/9208201}
  {\path{arXiv:hep-ph/9208201}}, \href {http://dx.doi.org/10.1007/BF01560056}
  {\path{doi:10.1007/BF01560056}}.

\bibitem{Chetyrkin:1997qi}
K.~Chetyrkin, R.~Harlander, J.~H. Kühn, M.~Steinhauser, {Mass corrections to
  the vector current correlator}, Nucl. Phys. B 503 (1997) 339--353.
\newblock \href {http://arxiv.org/abs/hep-ph/9704222}
  {\path{arXiv:hep-ph/9704222}}, \href
  {http://dx.doi.org/10.1016/S0550-3213(97)00383-0}
  {\path{doi:10.1016/S0550-3213(97)00383-0}}.

\bibitem{Chetyrkin:1990kr}
K.~Chetyrkin, J.~H. Kühn, {Mass corrections to the Z decay rate}, Phys. Lett.
  B 248 (1990) 359--364.
\newblock \href {http://dx.doi.org/10.1016/0370-2693(90)90306-Q}
  {\path{doi:10.1016/0370-2693(90)90306-Q}}.

\bibitem{Chetyrkin:1994ex}
K.~Chetyrkin, J.~H. Kühn, {Quartic mass corrections to $R_{\mathrm{had}}$},
  Nucl. Phys. B 432 (1994) 337--350.
\newblock \href {http://arxiv.org/abs/hep-ph/9406299}
  {\path{arXiv:hep-ph/9406299}}, \href
  {http://dx.doi.org/10.1016/0550-3213(94)90605-X}
  {\path{doi:10.1016/0550-3213(94)90605-X}}.

\bibitem{BroadhurstGen1984}
D.~Broadhurst, S.~Generalis, open University preprint OUT-4102-12 (unpublished)
  (1984).

\bibitem{Spiridonov:1988md}
V.~Spiridonov, K.~Chetyrkin, {Nonleading mass corrections and renormalization
  of the operators $m \psi \bar\psi$ and $G^2_{\mu\nu}$}, Sov. J. Nucl. Phys.
  47 (1988) 522--527.

\bibitem{Pascual:1981jr}
P.~Pascual, E.~de~Rafael, {Gluonic Corrections to Quark Vacuum Condensate
  Contributions to Two Point Functions in {QCD}}, Z. Phys. C 12 (1982) 127.
\newblock \href {http://dx.doi.org/10.1007/BF01548609}
  {\path{doi:10.1007/BF01548609}}.

\bibitem{Generalis:1983hb}
S.~Generalis, D.~J. Broadhurst, {The Heavy Quark Expansion and {QCD} Sum Rules
  for Light Quarks}, Phys. Lett. B 139 (1984) 85--89.
\newblock \href {http://dx.doi.org/10.1016/0370-2693(84)90040-6}
  {\path{doi:10.1016/0370-2693(84)90040-6}}.

\bibitem{Loladze:1985qk}
G.~Loladze, L.~Surguladze, F.~Tkachov, {Two loop corrections to coefficient
  functions of condensates $\langle G^2\rangle_0$ and $\langle m \bar q
  q\rangle_0$ in the QCD sum rules for the $\varphi$ meson}, Phys. Lett. B 162
  (1985) 363--366.
\newblock \href {http://dx.doi.org/10.1016/0370-2693(85)90940-2}
  {\path{doi:10.1016/0370-2693(85)90940-2}}.

\bibitem{Bagan:1985zp}
E.~Bagan, J.~Latorre, P.~Pascual, {Heavy and Heavy to Light Quark Expansions},
  Z. Phys. C 32 (1986) 43.
\newblock \href {http://dx.doi.org/10.1007/BF01441349}
  {\path{doi:10.1007/BF01441349}}.

\bibitem{Generalis:1990iy}
S.~Generalis, {Light quark current correlators}, J. Phys. G 16 (1990) 367--373.
\newblock \href {http://dx.doi.org/10.1088/0954-3899/16/3/008}
  {\path{doi:10.1088/0954-3899/16/3/008}}.

\bibitem{Surguladze:1990sp}
L.~Surguladze, F.~Tkachov, {Two Loop Effects in {QCD} Sum Rules for Light
  Mesons}, Nucl. Phys. B 331 (1990) 35.
\newblock \href {http://dx.doi.org/10.1016/0550-3213(90)90017-8}
  {\path{doi:10.1016/0550-3213(90)90017-8}}.

\bibitem{Broadhurst:1994qj}
D.~J. Broadhurst, P.~Baikov, V.~Ilyin, J.~Fleischer, O.~Tarasov, V.~A. Smirnov,
  {Two loop gluon condensate contributions to heavy quark current correlators:
  Exact results and approximations}, Phys. Lett. B 329 (1994) 103--110.
\newblock \href {http://arxiv.org/abs/hep-ph/9403274}
  {\path{arXiv:hep-ph/9403274}}, \href
  {http://dx.doi.org/10.1016/0370-2693(94)90524-X}
  {\path{doi:10.1016/0370-2693(94)90524-X}}.

\end{thebibliography}

\end{document}